%% file: main.tex
\documentclass[]{article}
\usepackage[utf8]{inputenc}
\usepackage[english]{babel}
\usepackage[driver=pdftex]{geometry}
\usepackage[table]{xcolor}	% table row shading
\usepackage{lineno}   %% <- no mathlines option
\usepackage{amsmath}  %% <- after lineno
\usepackage{etoolbox} %% <- for \cspreto, \csappto
\usepackage{authblk}
\usepackage{natbib} % this is used by elsarticle
\usepackage{ulem}
\usepackage{tikz}
\usepackage{orcidlink}
\usepackage[table]{xcolor}	% table row shading
\usepackage{epsfig}
\usepackage{graphics}  %\usepackage{graphicx}
\usepackage{color}
\usepackage{amsmath}
\usepackage{amsthm}
\usepackage{longtable}
\usepackage{rotating}
\usepackage{epstopdf}
\usepackage{amssymb}
\usepackage{color}	    % colored text
\usepackage{rotating}		% landscape table
\usepackage{caption}
\usepackage{subcaption}		% subtables
\usepackage{multirow}
\usepackage{booktabs}
\usepackage{adjustbox}
\usepackage[shortlabels]{enumitem}

\newcommand*\linenomathpatch[1]{%
  \cspreto{#1}{\linenomath}%
  \cspreto{#1*}{\linenomath}%
  \csappto{end#1}{\endlinenomath}%
  \csappto{end#1*}{\endlinenomath}%
}

\linenomathpatch{equation}
\linenomathpatch{gather}
\linenomathpatch{multline}
\linenomathpatch{align}
\linenomathpatch{alignat}
\linenomathpatch{flalign}

\newcommand{\tcr}{\textcolor{red}}
\newcommand{\tcb}{\textcolor{blue}}

\graphicspath{
  {converted_graphics/}% inserted by PCTeX
}

\title{A Comparison for Non-Specialists of Workflow Steps and Similarity of Factor Rankings for Several Global Sensitivity Analysis Methods}
\title{Global sensitivity analysis workflows and rankings: a practical comparison for researchers}
\author{...}
\author[1,2]{Ken B. Newman\, \orcidlink{0000-0003-1734-5833}}
\author[4]{Shaini Naha\, \orcidlink{0000-0002-2737-1739}}
\author[3]{Leah A. Jackson-Blake\, \orcidlink{0000-0002-4915-8779}}
\author[5]{Cairistiona Topp\, \orcidlink{0000-0002-7064-638X}}
\author[4]{Miriam Glendell\,\orcidlink{0000-0003-0110-9879}}
\author[1]{Adam Butler\, \orcidlink{0000-0002-0860-6475}}
\affil[1]{Biomathematics and Statistics Scotland, Scotland}
\affil[2]{School of Mathematics, University of Edinburgh, Scotland}
\affil[3]{Norwegian Institute for Water Research, Oslo, Norway}
\affil[4]{The James Hutton Institute, Scotland}
\affil[5]{Scotland's Rural College, Scotland}
\date{\today} 

\begin{document}
% \title{Global sensitivity analysis workflows and rankings: a practical comparison for researchers}
% \author[1,2]{Ken B. Newman}
% \ead{ken.newman@bioss.ac.uk}
% \author[4]{Shaini Naha\, \orcidlink{0000-0002-2737-1739}}
% \affiliation[1]{Biomathematics and Statistics Scotland, Scotland}
% \affiliation[2]{School of Mathematics, University of Edinburgh, Scotland}
% \affiliation[3]{Norwegian Institute for Water Research, Oslo, Norway}
% \affiliation[4]{The James Hutton Institute, Scotland}
% \affiliation[5]{Scotland's Rural College, Scotland}

 \maketitle

%**************************************************************
 
%\begin{document}

% 20 August w/ 149 words
\begin{abstract}
 Global sensitivity analysis (GSA) is a recommended step in the use of computer simulation models. GSA quantifies the relative importance of model inputs on outputs (Factor Ranking), identifies inputs that could be fixed, thus simplifying model calibration (Factor Fixing), and pinpointing areas for future data collection (Factor Prioritization). Given the wide variety of GSA methods, choosing between methods can be challenging for non-GSA experts. Issues include workflow steps and complexity, interpretation of GSA outputs, and the degree of similarity between methods in Factor Ranking. We conducted a study of both widely and less commonly used GSA methods applied to three simulators of differing complexity. All methods share common issues around implementation with specification of parameter ranges particularly critical. Similarities in Factor Rankings were generally high based on Kendall's W. Sobol' first order and total sensitivity indices were easy to interpret and informative with regression trees providing additional insight into interactions. 
\end{abstract}

 
\maketitle
%*************************************************
\newpage 
\input{1_Introduction/1_Introduction}

\input{2_Methods/2_Methods_SA}

\clearpage 
\newpage  
\input{3_Simulators/3_Simulators}

\input{4_Results/4_Results}

\clearpage 
\newpage 
\input{5_Discussion/5_Discussion}

%*************************************************************
\section*{Acknowledgements} 
KN, SN, KT, MG, and AB were supported by the Scottish Government’s Rural and Environment Science and Analytical Services Division (RESAS).   Magnus Norling, Norwegian Institute for Water Research, provided much assistance in the running of SimplyP including providing an \texttt{R} wrapper for running the simulator.

\section*{Data Availability Statement}
The exogenous forcing factors and the default parameter values used for each of the three simulators are available by contacting the lead author.

\bibliographystyle{biom}
\bibliography{SA_Comparison}

%\section*{Supporting Information}
\newpage 
\appendix
% \setcounter{table}{0}
% \setcounter{figure}{0}
% \renewcommand\thetable{\Alph{section}.\arabic{table}}
% \renewcommand\thefigure{\Alph{section}.\arabic{figure}}
% \input{6_Appendices/6A_Methods_Details}
% \clearpage 
% \newpage 

\setcounter{table}{0}
\setcounter{figure}{0}
\renewcommand\thetable{\Alph{section}.\arabic{table}}
\renewcommand\thefigure{\Alph{section}.\arabic{figure}}
\input{6_Appendices/1_GSA_Details/6A_GSA_Details}

\clearpage 
\newpage 

\setcounter{table}{0}
\setcounter{figure}{0}
\renewcommand\thetable{\Alph{section}.\arabic{table}}
\renewcommand\thefigure{\Alph{section}.\arabic{figure}}
\input{6_Appendices/2_Simulator_Details/6B_Simulator_Details}

\clearpage 
\newpage 

\setcounter{table}{0}
\setcounter{figure}{0}
\renewcommand\thetable{\Alph{section}.\arabic{table}}
\renewcommand\thefigure{\Alph{section}.\arabic{figure}}
\input{6_Appendices/3_Results_Details/6C_Results_Details}

\end{document}

%% file: 1_Introduction/1_Introduction.tex
\section{\label{sec:Intro}Introduction}

Scientists in many disciplines, including environmental science, geosciences, life sciences, and systems science, increasingly use computer models, simulators, to characterize complex systems, to examine how the systems change over time, and to assess and to predict the effects of environmental changes and management actions.  Such computer experiments are surrogates for, and potentially complements to, physical experiments \citep{2018_Santner_etal} that might be practically impossible or prohibitively expensive to implement. Additionally,  simulators are used to address questions at a larger spatial scale (e.g., regional, national) than is possible via experiment.

The quality of the simulators used to carry out such experiments is critical to the quality of the decisions and policy making that are based on these simulators. \citet{2020_Saltelli_etal} issued a manifesto aimed at ensuring that models serve society, and part of that manifesto is a demand that uncertainty analysis (UA) and sensitivity analysis (SA) be conducted for models that are used to guide policy decision making.  There is also a strong advisement from the European Commission (EC) regarding the need for such analyses.  The EC includes in its guidelines for improving the creation of regulations an entry in its ``tool box'', tool \#65, for uncertainty and sensitivity analysis \citep{2023_EC_Toolbox} (see also \citet{2020_Azzini_etal}).  In particular they state ``A transparent and high-quality impact assessment should acknowledge and,
to the extent relevant or possible, attempt to quantify the uncertainty in results
as it could change the ranking and conclusions about the policy options.''  Thus the incentive and impetus for carrying out such analyses is strong.

 Sensitivity analysis is one part of the discipline of the design and analysis of computer experiments (DACE) that has developed over the last thirty years, which now plays a role similar to the role of statistical design of physical experiments.  \citet{2002_Saltelli} defines sensitivity analysis as ``the study of how the uncertainty in the output of a model (numerical or otherwise) can be apportioned to different sources of uncertainty in the model input''.  Sensitivity analysis can also be viewed as a means of quantifying the relative importance of simulator inputs.  While SA can be carried out on multiple features of a simulator such as initial conditions, boundary constraints, and exogenous forcing factors,  we consider only fixed, scalar input parameters. 

\citet{2007_Cariboni_etal} discuss different applications or uses of SA along with associated different questions to be answered regarding model factors. One output from SA is \textit{Factor Ranking}, where the relative influence of each parameter on the variability in a given output is quantified \citep{2016_Pianosi_etal}.  Another use of SA is for \textit{Factor Prioritization} where parameters are identified, which if set equal to their true values, would most reduce the variation in the model output and this then guides future data collection and prioritizes future research \citep{2013_Cosenza_etal}. Note that quantitative output from SA is being used in a relative sense with Factor Ranking while it is being used in an absolute sense with Factor Prioritization. SA can also be used for \textit{Factor Screening}, or dimensionality reduction \citep{2021_Razavi_etal}, where SA identifies non-influential parameters that can be removed from the estimation process because they have, within their specified range, so little effect on the output. This is related to \textit{Factor Fixing} \citep{2016_Pianosi_etal},  where the values for such non-influential parameters are set at constant values deemed reasonable and are then not estimated  \citep{2007_Cariboni_etal, 2013_Cosenza_etal, 2015_Song_etal, 2016_Pianosi_etal}.  Thus SA potentially reduces calibration difficulty by focusing on a reduced parameter set \citep{2019_Saltelli_etal}.  Sensitivity analysis is used for other purposes as well such as ``diagnostic evaluation, dominant control analysis and robust decision-making'' \citep{2016_Pianosi_etal}.

\citet{2021_Razavi_etal} is a recent perspective paper on the status of sensitivity analysis that includes classification of SA methods. They dichotomize SA procedures into local SA and global SA. Local SA examines sensitivity of output to variations around a given parameter value, i.e., the ``sensitivity of the problem is assessed only around a ‘nominal point’ in the problem space'' while global SA (GSA) provides measures that  characterize sensitivity across a range of parameter values,  they attempt[] to provide a ‘global’ representation of how the different factors work and interact across the full problem space to influence some function of the system output'' \citep{2021_Razavi_etal}, where full problem space means all parameters. GSA is generally preferred over local SA \citep{2008_Saltelli_etal} and one-at-a-time assessment, with the latter failing to detect interactions, and GSA methods are our focus.  \citet{2021_Razavi_etal} categorize GSA methods as (i) derivative-based, (ii) distribution-based, (iii) variogram-based, and (iv) regression-based and they discuss particular methods in each category. 

While \citet{2021_Razavi_etal} advocate for more widespread and routine use of GSA and discuss  challenges for doing so, they do not provide guidance for selecting methods, nor describe the typical workflow involved in carrying out a GSA, nor evaluate the degree of similarity in results between different GSA methods.

 Our work was motivated by our experience of statisticians collaborating with scientists in hydrology, water quality, crop and soil systems, who use simulators of widely varying complexity. The simulators range from those with a handful of input parameters and one or two scalar outputs of primary interest to 100s of input parameters and 100s of output variables. In the course of this collaboration, the following questions arose.
 \begin{itemize}
     \item ``What are advantages and disadvantages of different GSA procedures?''
     \item ``How do implementation difficulties compare between the procedures?''
     \item ``How do computational costs differ?''
     \item ``How similar are the conclusions for different GSA methods, specifically, how similar are Factor Rankings?''
     \item ``Do simulators with multiple outputs, particularly time series outputs, pose unique challenges?''
 \end{itemize}   

 To answer these questions, we conducted a comparative study of the following different GSA methods that included examples from  \citet{2021_Razavi_etal}'s categories. 
 \begin{itemize}
     \item Derivative-based: Morris elementary effects
     \item Distribution-based: Sobol' sensitivities
     \item Variogram-based: VARS-TO
     \item Regression-Based
     \begin{itemize}
         \item Multiple regression  
          \item Regression trees
         \item Random forests
         \item Gaussian Process regression
     \end{itemize}
 \end{itemize}
 Each GSA method was applied to three deterministic simulators that differed considerably in terms of model complexity, number of input parameters, and number of outputs, but all had relatively low computational costs. However, the number of input parameters included in the final GSA's was not that large, from six to 13, after the subject matter experts saw initial GSA results and re-evaluated what parameters should be focused on. Likewise, the number of outputs considered was only three or four, though all were points in a time series.
 
 The intended audience for this paper is users of deterministic simulators who are not specialists in GSA, and our aim is to provide such users with a practitioner's perspective on using GSA methods. This perspective includes the choice of methods, common workflow, implementation issues, computational expense, differences between GSA methods in terms of their outputs and interpretation, and some personal and context-specific preferences for one SA method over another.
 We also examine similarity in Factor Rankings using pairwise comparisons (graphical and analytical) as well as a single measure of overall inter-rater reliability or degree of agreement, Kendall's W.  
  
 Before continuing, we acknowledge that several comparative studies of GSA procedures have been carried out previously that partially address our questions. We draw attention to three papers with a general focus on GSA and view this paper as complementary to them in that we examine GSA methods not discussed in those papers and provide a somewhat different workflow.  \citet{2007_Cariboni_etal}  apply local and global SA methods, including multiple regression, Morris and Sobol' methods, in the analysis of two simulators of population dynamics.  \citet{2016_Pianosi_etal} provide a systematic review of five classes of GSA methods (including ones we discuss: Morris elementary effects, multiple regression, Sobol'), contrast their strengths and weaknesses, and discuss some practical implementation issues.  \citet{2019_Wagener_Pianosi} focus on multiple GSA methods being applied to a single but general class of simulators, earth system models, and, similar to \citet{2016_Pianosi_etal}, provide thoughts about practical implementation, workflow, and strengths and weaknesses of different methods.  
 % Other recent papers comparing popular global SA procedures include  \citet{2013_Cosenza_etal} (three SA methods applied to a single wastewater simulator),  \cite{2015_Song_etal} (five SA methods applied to over 40 hydrology simulators),  \citet{2021_Ujjwal_etal} (four SA methods applied to two fire spread simulators), and \citet{2022_Dela_etal} (six SA methods applied to two very different simulators, one for HIV and another for tumor growth). 
 What we add to this literature is that our selection of GSA methods includes less commonly used methods of regression trees, random forests and Gaussian Process regression, the application of three simulators of varying complexity from a range of environmental science disciplines, our selection of various practical implementation and interpretation details in real applications, and a proposed workflow.
 
%-------------------------------------------------
\subsection{Simulators and Computer Experiments}

To facilitate discussion, we introduce the following notation for simulators. $F$ represents the operator, the ``simulator'', which links inputs to outputs. Generic inputs to $F$ are denoted $\mathbf{x}$ = $(x_1,x_2,\ldots,x_p)$. A subset of inputs, which are viewed as fixed but unknown parameters, are denoted by $\Theta$ = $(\theta_1, \theta_2, \ldots, \theta_K)$.  For the simulators we will discuss here, model inputs also include one or more time series of environmental variables, $\mathbf{w}_{1:T}$ = $(w_1,w_2,\ldots,w_T)$, exogenous forcing factors such as daily precipitation and daily mean air temperature.  Simulator output ranges from simple scalars, $y$, at one extreme to multivariate time series at another extreme, and can include output at intermediate points in the internal processing of the simulator.  Output is generically denoted $\bf{y}$=($y_1$,$y_2$,$\ldots$,$y_n)$, including single and multiple time series and scalar and vector non-time series. The connection between simulator input and output is then
\begin{align}
\label{eq:simulator}
   \mathbf{y} &= F(\Theta,\mathbf{w})
\end{align}
Equation \ref{eq:simulator} denotes a functional relationship that is deterministic, i.e., inputting the same values for $\Theta$ and $\mathbf{w}$ multiple times yields the same output $\mathbf{y}$, thus $F$ is a \textit{deterministic} simulator. Such simulators are distinguished from stochastic simulators, where inputs include random components, $\mathbf{y} = F(\Theta,\mathbf{w}, \mathbf{\epsilon})$, where $\epsilon$ is a random variable, and repeated runs of the simulator with identical fixed inputs yield different, random outputs.  Stochastic simulation is a crucial component of most agent-based models \citep{2019_Railsback_Volker}.   While a variety of sensitivity analysis approaches have been developed for stochastic simulators \citep{2013_Damiani_etal, 2016_Broeke_etal, 2022_Borgonovo_etal}, here we only consider deterministic simulators.

Simulators vary considerably in terms of the number and types of inputs and outputs, as well as the complexity of the intermediate computations, which can involve sequencing of multiple sub-processes with outputs from one being inputs to another. For example, one simulator  we  discuss is a hydrology and water chemistry model \textit{SimplyP}  \citep{2017_Jackson_etal}, which has 13 (non-fixed) input parameters, $\Theta=(\theta_1,\ldots,\theta_{13})$, requires three time series of environmental data, daily precipitation, potential evapotranspiration, and  temperature, $\mathbf{w}_{i,1:T}$, $i$=1,2,3, and outputs approximately 60 time series of water flows, sediments, and phosphorous levels, $\mathbf{y}_{j,1:T}$, $j$=1,2,$\ldots,\sim 60.$  

For simulators with many parameters and vector-valued outputs, the field data needed to provide information about parameter values, namely to fit or calibrate, such models may be inadequate. In some cases calibration to all parameters can be problematic due to non-identifiability; e.g., one set of values for a parameter pair ($\theta_1,\theta_2$) produces output identical to that for a different set of values for those parameters \citep{2006_Beven}. However, even without non-identifiability problems and with adequate field data, model calibration can still be extremely difficult and time-consuming. Knowing the relative importance of different inputs or parameters to simulator output can help focus calibration and sensitivity analysis is one means of quantifying relative importance. 

%--------------------------------------------------
\subsection{Aleatory and epistemic uncertainty and uncertainty analysis}
 Before proceeding with the discussion of sensitivity analysis, we discuss the general topic of uncertainty and briefly mention uncertainty analysis. Uncertainty is a feature of nearly all simulators and includes uncertainty about the model to use to characterize a natural system, uncertainty about the values of factors in the model (e.g., parameters, boundary conditions, initial conditions), and inherent stochasticity \citep{2007_Cariboni_etal}. One categorization of uncertainty relevant here is into either aleatory uncertainty or epistemic uncertainty \citep{2021_Hullermeier_Waegeman}; see also \citet{2017_Beven_Lamb}. 

 \textit{Aleatory} uncertainty arises from so-called true randomness or natural or environmental variation. Which side of a six-sided die that will lie face up after rolling the die is an example of aleatory uncertainty.  Stochastic simulators such as agent-based models explicitly mimic aleatory uncertainty. Even with deterministic simulators, aleatory uncertainty can arise from variation in exogenous forcing factors,  e.g., simulator output based on daily rainfall for June 1990 will differ from that based on daily rainfall for June 1991. Note that, conditional on the exogenous inputs, such simulators can still be viewed as deterministic. A key point is that collecting more data will not reduce aleatory uncertainty. However, the inclusion of new predictors in a modified simulator can potentially reduce that aleatory uncertainty. 
 
\textit{Epistemic} uncertainty, on the other hand, arises from a lack of knowledge or ignorance; e.g., the ``true'' values for input parameters are unknown. Referring to the hydrology-water quality simulator, \textit{SimplyP}, the input parameter $f_{quick}$ is a fixed but unknown number related to surface water runoff, and uncertainty about its true value is epistemic uncertainty. The determination of such values using real-world data is model calibration, or parameter estimation, which typically yields an estimated value or point estimate, e.g., $\hat{f}_{quick}$, which could be a maximum likelihood estimate in frequentist inference or the mode of a posterior distribution in Bayesian inference. Quantification of epistemic uncertainty about the parameter is manifested by a standard error or confidence interval for that estimate or by a posterior distribution and credible interval.  In contrast to aleatory uncertainty, the collection of more data, or different types of data, can reduce epistemic uncertainty \citep{2007_Cariboni_etal}. 

 Uncertainty analysis (UA) quantifies the magnitude of uncertainty of model output as a function of uncertainty in model input parameters and in other factors such as model assumptions. SA differs from UA in that SA quantifies the \textit{relative importance or effect} of different model inputs while UA quantifies total variation in output without necessarily attributing the sources of that variation.  UA can incorporate both aleatory and epistemic uncertainty.  For example, simulating from prior distributions for parameters in the Bayesian framework is one means of quantifying epistemic uncertainty; in fact, this part of UA can be viewed as simply simulating from the priors. UA can also incorporate aleatory uncertainty in stochastic simulators with random generation of environmental uncertainty. Both UA and SA are essential parts of the design and analysis of computer experiments, and widespread usage of both has long been advocated as an essential, and what should be routine, aspect of simulator development, calibration and usage \citep{2000_Saltelli_Tarantola_Campolongo, 2002_Saltelli}. With the simulators we discuss herein our focus is solely on SA.  

%--------------------------------------------------
\subsection{Organization of remainder of the paper}

  In Section \ref{sec:Methods} we present a workflow for sensitivity analysis that we think will be helpful for non-GSA specialists. Then we describe the different GSA methods used, which we group as \citet{2021_Razavi_etal} does: derivative-based Morris mean and variance of elementary effects, distribution-based Sobol' 1st order and total sensitivities, variogram-based total sensitivity, and four regression-based approaches: multiple regression with Standardized Regression Coefficients, regression trees with variable importance, random forests with variable importance and Gaussian process regression models with slope parameters and inverse ranges. Section \ref{sec:Simulators} outlines the three simulators that are, in order of complexity:  \textit{GR6J},   a six parameter hydrology model; \textit{SimplyP},  a hydrology and water quality model with up to 30 input parameters; and  \textit{STICS},  an individual crop plant growth and soil biogeochemistry model with around 200 input parameters. Section \ref{sec:Results} presents the results of applying the GSA procedures to each of the models, and we conclude with a discussion in Section \ref{sec:Discussion}.  Additional technical details of the GSA procedures, the simulators and the results are available in supplemental materials.

%% file: 2_Methods/2_Methods_SA.tex
% \newpage 

\section{\label{sec:Methods}Global Sensitivity Analysis Methods}
Section \ref{sec:Methods.Implementation} discusses general workflow issues, the implementation features of all of the methods, and decisions that need to be made. Such a framework is useful for those relatively new to GSA as it provides both a sequence of steps to take and identifies places where iterations are likely.  Brief descriptions of each of the GSA methods are provided in Section \ref{GSA.Methods.Subsection}. 

\input{2_Methods/2A_Methods_SAProcedures}

\input{2_Methods/2B_Methods_Derivative}

\input{2_Methods/2C_Methods_Distribution}

\input{2_Methods/2D_Methods_Variogram}

\input{2_Methods/2E_Methods_Reg_MultReg}

\input{2_Methods/2F_Methods_Reg_RegTree}

\input{2_Methods/2G_Methods_Reg_RandomForest}

\input{2_Methods/2H_Methods_Reg_GPR}

%% file: 2_Methods/2A_Methods_SAProcedures.tex
%--------------------------------------------------
\subsection{\label{sec:Methods.Implementation}GSA Workflow: Implementation components and decisions}
 Some aspects of implementation and necessary decisions are common to all GSA methods. Our experiences led to a workflow structure that is summarized in Table \ref{T:Methods.Workflow}. This workflow overlaps with that presented by \citet{2016_Pianosi_etal}, but the emphases differ somewhat; e.g., we found the determination of parameter range limits an extremely crucial step.  The workflow steps assume that the GSA method or methods to be used have been determined, but if the method is not pre-determined, that determination would be a step and then affects the ordering of steps.  We found the performance of this workflow to be very much an iterative process and a team effort between the subject matter experts who use the simulators and those carrying out the GSA, in this case, statisticians.

 \begin{table}[h]
     \centering
     \begin{tabular}{ll}
      Step & \multicolumn{1}{c}{Action} \\ \hline 
         1. & Identify Simulator Outputs of Interest.  \\
         2. & Specify Simulator Input Parameters to be Evaluated. \\
         3. & Specify Input Parameter Space. \\
         4. & Determine $N$. \\
         5. & Generate $N$ Space-Filling Input Parameter Combinations. \\
         6. & Run Simulator $N$ Times. \\ 
         7. & Examine Sensibility of Outputs. \\
         8. & Calculate Sensitivity Measures. \\
         9. & Summarize GSA Measures. \\
         10. & Potentially Iterate.  
     \end{tabular}
     \caption{Workflow for the implementation of a given GSA method.}
     \label{T:Methods.Workflow}
 \end{table}

% ***************************************************************
%--------------------------------------------------------------------
\subsubsection{Step 1: Identify Simulator Outputs of Interest}
 The simplest situation is when interest is solely on a single scalar-valued output, for example, the nitrogen content in a grain of barley at the time of harvest. A more common situation, however, is that the simulator produces multiple and categorically different outputs of interest, e.g., for the grain at the time of harvest, N level, C level, and mass. For categorically different outputs, separate parameter rankings for each output will often be desired.  Collective assessment of parameters' importance for multiple outputs is similar to the problem of model calibration for multiple outputs, multi-criterion optimization, and Pareto Frontiers are one means of displaying the combined importance of the inputs across multiple outputs \citep{2024_Giannelos_etal}.
   
 Many simulators, including the three we consider,  produce time series of outputs, e.g., daily N levels in the grain for some period of time up to and including the day of harvest. One analysis approach is to calculate $T$ GSA measures for the individual outputs at a sequence of time points, $t$=1,2,$\ldots$,$T$, what \citet{2016_Pianosi_Wagener} call Time Varying Sensitivity Analysis (TVSA); \citet{2018_Gupta_Razavi} call this time-varying parameter importance.  An alternative approach is to calculate a summary measure over a particular interval of time, e.g., average over a single time interval, multiple averages over a set of non-overlapping time intervals, or a sequence of averages from a moving time series window \citep{2016_Pianosi_Wagener}. \citet{2018_Gupta_Razavi} discuss the application of such time period time-aggregation for assessing parameter importance. Note that the choice between these will presumably depend a lot on the key questions of interest, and hence on context.

 Another distinction is between outputs that are based solely on the model results, which we label Internal GSA, and those based on both model results and real-world data, External GSA. External GSA often uses performance measures that are variations of model goodness-of-fit, such as squared error between model output and a corresponding field observation \citep{1999_Legates_McCabe}.  For simulators with time series outputs, popular performance measures are Nash-Sutcliffe efficiency (NSE) and Kling-Gupta efficiency (KGE) \citep{2019_Knoben_etal, 2006_McCuen_etal}. \citet{2018_Gupta_Razavi} discuss the differences between Internal GSA and External GSA (with performance-based measures), and argue that performance-based GSA is actually a parameter identifiability analysis and is conceptually flawed for interpreting the sensitivity of model outputs to parameter perturbations. 
 
 Another issue with regard to the distinction between Internal and External GSA is generalizability, as the real-world data used for External GSA (often model calibration) are location-specific and the results are potentially relevant only to that location. Internal GSA has the potential to be more generalizable, although in both cases the selection of the exogenous forcing factors, $\mathbf{w}$, is location-specific.  Another disadvantage of External GSA is that the theoretical methodology underpinning standard SA measures, e.g. Sobol' sensitivities, may not hold. A pragmatic response is to conduct both Internal and External GSA and look for similarities and differences. Our focus here is primarily on Internal GSA, but we briefly mention examples of both.  
 
 A final point on Internal versus External GSA is that there are often situations where simulators produce outputs, perhaps intermediate outputs, for which no real-world data can be collected, and thus External GSA is impossible. For such outputs, it can still be informative to carry out Internal GSA to gain insight into the simulator's inner workings.  

It was our experience that the choice of outputs was an iterative procedure, as the subject matter specialist sometimes narrowed their focus after seeing initial results, e.g., two outputs having a high degree of correlation.  

%--------------------------------------------------------
\subsubsection{Step 2: Specify Simulator Input Parameters to be Evaluated}
The choice of input parameters depends on the degree of uncertainty about parameter values, the \textit{a priori} perceived relevance of the parameters to the outputs of interest, the dimension of the input parameter space ($K$), and simulation  computation time (which relates to $N$). For example, if the simulator has $K$=10 parameters, then including all 10 parameters may be an easy choice. With 100s or 1000s of parameters, however, including all may be too computationally expensive as $N$ generally increases with $K$.  As the relevance of a parameter is generally output-specific, with different outputs depending on different inputs, the choice of outputs in Step 1 can play a role in the selection of input parameters in Step 2. The larger the set of categorically different outputs, the larger the set of input parameters to consider. Uncertainty analysis plays a part in determining the set of input parameters as well, and UA work can serve as a pilot study for GSA. Selection of input parameters can be an iterative process with some initial simulator runs and analysis guiding parameter set selection. For example, with   SimplyP, which has up to 30 parameters, initial exploratory runs focused on four time series outputs led to a reduced set of 13 parameters. Similarly, for STICS, which has 100s of parameters, an initial selection of over 50 parameters for GSA was reduced to 12 parameters for four time series outputs. In all cases, the selection of the input parameters was strongly dependent on the subject-matter expert's judgment.

%--------------------------------------------------------------------
\subsubsection{Step 3: Specify Input Parameter Space}
Parameter ranges, $[\theta_{k,L},\theta_{k,U}]$, $k$=1,2,$\ldots$,$K$, can have a considerable effect on GSA results.  Lower bounds or upper bounds may be too extreme in the sense that the simulator outputs are deemed unreasonable for the phenomena of interest, and selecting parameter values from such a range may yield parameter importance rankings considerably different from rankings resulting from narrower ranges. We found, for example, with the STICS model that some parameter combinations yielded  Leaf Area Index (LAI) values deemed unacceptable and even though LAI was not an output of primary interest for GSA, parameter bounds were adjusted to yield acceptable LAI values. 

To address the problem of parameter importance rankings being affected by the chosen parameter ranges \citet{2016_Paleari_Confalonieri} recommend a sensitivity analysis of the sensitivity analysis. This includes a literature review of applications of a given model to construct a set of parameter ranges, the formulation of probability distributions for those ranges (which could be viewed as similar to prior distributions), followed by sampling from those distributions, and carrying out GSA repeatedly for each sample.  For our three simulators and specific applications, the available literature was considered too limited or not applicable and instead range specification, like input parameter specification, was an iterative process guided by the subject-matter specialists.    

This discussion of ranges ignores the fact that parameter combination generation (Step 5) is often done as if the parameters are drawn from independent Uniform  probability distributions for each parameter.  The theoretical results of Sobol' sensitivities   assume that, for example.  A deviation  from this assumption is to use non-Uniform prior distributions, such as Normal distributions. The assumption of independence between input parameters is also debatable, and \citet{2006_Jacques_etal} discuss that point and present alternatives for dependent input parameters. A related point is that while the implied parameter space for many GSA methods is a K-dimensional unit cube (with parameters transformed to the interval [0,1]), more realistic parameter spaces may be very different, with portions of such hypercubes empty of points.

%--------------------------------------------------------------------
\subsubsection{Step 4: Determine $N$}
Most GSA methods use Monte Carlo procedures to yield sensitivity measures which are then random variables. Factor rankings based on one set of $N$ parameter combinations will potentially differ from a second set due to Monte Carlo (MC) variation. The size of $N$ affects MC variation and thus one would like to choose $N$ sufficiently large to ensure that the GSA results are relatively stable or robust. The degree of MC variation is not only a function of $N$ but also the specific GSA method and the dimension $K$ of the input parameter space.

 \citet{2016_Pianosi_etal} discuss  sample size determination  for different GSA methods, e,g., Morris elementary effects and Sobol' sensitivities (see also \citet{2013_Cosenza_etal}). They also provide some approximate rules of thumb; e.g., for Morris,  10$K$ $\le$ $N$ $\le$ 100$K$, and for Sobol' 5500$(K+2)$ $\le$ $N$ $\le$ 1000$(K+2)$.  Such rules of thumb are helpful starting points, and perhaps essential for computationally expensive simulators. A pragmatic alternative is to repeat the analysis with multiple sets of $N$ and examine the degree of variation in the GSA measures.  Fewer trial-and-error procedures that yield estimates of the Monte Carlo based on bootstrapping are discussed by \citet{2016_Pianosi_etal}, and some of the software used herein included such measures. 

If $N$ is considered inadequate, i.e., unacceptably large MC error, for some GSA methods one can select an additional set of parameter combinations, make additional simulator runs, and simply append the outputs to the original set of results.   For example, an initial $N_1$=100 parameter combinations are generated and 100 simulator runs are made, and GSA values are calculated. Then a new set of $N_2$=100 parameter combinations are generated with 100 additional simulator runs, and GSA values based on those runs are calculated. Differences in the sensitivities between the two sets provide an indication of MC variation. For some GSA methods, particularly the regression-based ones, and parameter generation methods, the combined outputs of the $N_1$+$N+2$ runs can be used to calculate sensitivity measures. For one particular  parameter generation approach, Latin Hypercube Sampling, one could generate a large initial oversample, e.g., $N$=1000, and then pull off the first 100 for $N_1$, and then pull off the next 100 for $N_2$, and so on.  However, for some GSA methods, Morris, Sobol, and variogram-based, such an appending of runs cannot be readily done because the algorithms for the generation of combinations are dependent on $N$; if one wants an ``eventual'' $N$=200 run outputs, then an initial and sole generation of 200 combinations must be done.  

%--------------------------------------------------------------------
\subsubsection{Step 5: Generate $N$ Space-Filling Input Parameter Combinations.}
The selection of the $N$ combinations is done using one of several possible space-filling designs (SFD, \citet{2020_Gramacy}). In most cases, the SFD samples are taken from a $K$-dimensional unit hypercube, $[0,1]^K$, and then the resulting sampled values per parameter are mapped from $U_k$ $\in$ (0,1) to the $(\theta_{k,L},\theta_{k,U})$, $k$=$1,\ldots,K$. The mapping of the $N$ generated vectors is $\theta_{i,k}$  = $\theta_{k,L} + (\theta_{k,U}-\theta_{k,L})*U_{i,k}$, $i=1,\ldots,N$ where $U_{i,k}$ is the row $i$ and column $k$ component in the $N$ by $K$ matrix of generated values. For example, if $\theta_{k,L}$=-3 and $\theta_{k,U}$=2, and $U_{i,k}$=0.37, $\theta_{i,k}$= -3 + (2- -3)*0.37 =  -1.15. The end result is an $N$ by $K$ matrix of parameter combinations.
\begin{align*}
    (\Theta^1, \Theta^2, \ldots, \Theta^N)^T &=
\left [ 
 \begin{array}{cccc} 
 \theta_{1,1} & \theta_{1,2} & \ldots & \theta_{1,K} \\
 \theta_{2,1} & \theta_{2,2} & \ldots & \theta_{2,K} \\
 \vdots \\
 \theta_{N,1} & \theta_{N,2} & \ldots & \theta_{N,K} \\ 
 \end{array}
\right ]
\end{align*}

As indicated previously, some of the GSA methods, e.g., Morris elementary effects, Sobol' sensitivities, and variogram-based total sensitivity, have customized procedures for sampling the space. For Sobol' measures in particular, there is a lengthy literature on the development of computationally efficient procedures \citep{1993_Sobol, 1994_Jansen, 2002a_Saltelli, 2021_Azzini_etal}. There are several general space-filling designs with Latin Hypercube Sampling (LHS; \citet{1987_Stein}) and variations of LHS being quite popular, one such LHS variation was used for all the regression-based procedures.  

%--------------------------------------------------------------------
\subsubsection{Step 6: Run Simulator $N$ Times}
The simulator is run $N$ times with each input parameter combination 
thus yielding a matrix of outputs:
\begin{align*}
\left [ 
\begin{array}{c}
 F(\theta_{1,1}, \theta_{1,2}, \ldots, \theta_{1,K}) \\
 F(\theta_{2,1}, \theta_{2,2}, \ldots, \theta_{2,K}) \\
 \vdots \\
  F(\theta_{N,1}, \theta_{N,2}, \ldots, \theta_{N,K}) 
 \end{array}
 \right ]
 & = 
 \left [ 
 \begin{array}{cccc}
 y_{1,1} & y_{1,2} & \ldots & y_{1,p} \\
 y_{2,1} & y_{2,2} & \ldots & y_{2,p} \\
 \vdots \\
  y_{N,1} & y_{N,2} & \ldots & y_{N,p} 
 \end{array}
 \right ]
\end{align*}

%--------------------------------------------------------------------
\subsubsection{Step 7: Examine Sensibility of Outputs}
 Certain combinations of parameter inputs may yield unacceptable outputs, and this can include outputs that are not of primary interest. For example, with STICS, Leaf Area Index (LAI) values above 6 are unreasonable. Part of the iterative GSA process was to adjust some parameter lower bounds and/or upper bounds to avoid such values (return to Step 3); in this case, the parameters dlaimax $\le$ 1.5 and adens $\le$ -1  were found acceptable.
 
 An alternative to re-running the simulators using a different set of input parameter ranges is to remove outputs that reflect ``unacceptable model behaviour'', a process \citet{2016_Pianosi_etal} call \textit{filtering}, and then calculate the sensitivity measures. Studying the effects of such after-the-fact filtering on GSA results is still an area of research. 
 
%--------------------------------------------------------------------
\subsubsection{Step 8: Calculate Sensitivity Measures} 
For each GSA method, carry out multiple sensitivity analyses. For example, for each of the $K$ output components, calculate $K$ sensitivity measures. Some GSA methods yield multiple measures, e.g., the Sobol' approach can produce multiple sensitivity measures, including total and first order. With time series outputs and time-varying GSA multiple sensitivities over different time points or intervals are calculated. Ideally, measures of the uncertainty, the MC variation, in the sensitivity measures are calculated as well, which provide a measure of the robustness of the results. 
 
%--------------------------------------------------------------------
\subsubsection{Step 9: Summarize GSA Measures}
Summaries of GSA results can be numerical and graphical, and an assessment of the robustness of the results should be included.  Our focus is primarily on Factor Ranking, and for each of the GSA methods, we calculated scaled measures of importance per parameter that fell between 0 and 1 and summed to 1.0 over all the parameters.  Graphical displays and summaries differed between the GSA methods and will be discussed in detail later. 

 While Factor Rankings  provide an ordering of importance, the  ordering alone does not necessarily guide Factor Prioritization, Screening, nor Fixing as the magnitude of the measures matter.  For example, suppose $K$=3 and the scaled measures are 0.33, 0.37, and 0.30 for $\theta_1$, $\theta_2$, and $\theta_3$. The magnitudes are so similar that one could reasonably argue that all three are nearly equally important (particularly if the MC variation is relatively small).  At the other extreme, if the scaled measures are 0.29, 0.69, and 0.02, then $\theta_2$ seems most important and should be prioritized for additional data collection, followed by $\theta_1$ and $\theta_3$ could be fixed at a reasonable value, Factor Fixing, or perhaps removed in a revision of the simulator, Factor Screening \citep{2007_Cariboni_etal}. A caveat to this is the degree of a priori uncertainty in the parameter values, as those with low uncertainty do not need to be prioritized.

 Factor Rankings and scaled sensitivity measures alone  often fail to tell the entire story, particularly when the parameters interact in their effects on the outputs. For example, in the hypothetical example given above with nearly identical scaled values, $\theta_1$ and $\theta_2$ may be interacting or be an indication of a parameter identifiability problem (an example of equi-finality \citet{2006_Beven}). Some of the GSA methods provide two or more sensitivity measures that can provide insight about both the main effects of the parameters, e.g., first order effects, as well as interactions and higher order or nonlinear effects. Thus comparison of the relative value of different GSA methods includes their ability to provide insight into the nature of the input parameters' effects on the outputs.

%---------------------------------------------------------
\subsubsection{Step 10: Potentially Iterate}
Potentially return to Step 1 to revise outputs and inputs, adjust input ranges, and/or change $N$, and repeat subsequent steps. If MC variation is unacceptably high, then the GSA procedure may need to be repeated with a larger $N$, which, as mentioned above, might require only generating a sufficient number of additional combinations and runs and combining or appending results.  Scrutiny of model outputs (Step 7) may indicate the need to adjust the parameter ranges or, more generally, to generate parameter combinations in a more restricted parameter space.

% ***************************************************************
\subsubsection{Remarks on simulator runs and software}
 Running the simulator $N$ times is conceptually simple in that one loops over the $N$ combinations, passing the input parameters to the simulator. On the other hand, the passing of new parameter values can be somewhat complex, depending on the simulator. For example, with STICS $N$ different so-called plant files must be created prior to each run, where only a subset of 100s of input parameters are selectively altered.  

 It is important to note that for all three simulators, the input data time series of exogenous forcing factors, $\mathbf{w}$, e.g., daily temperatures and daily precipitation, were imported as well, but these did not change between iterations.  A broader application of sensitivity analysis is to allow for and account for variation in the forcing factors. 
 
  The computing of indices as well as running of the three simulators was all done within the \texttt{R} computing environment \citep{Rpackage}. Calculations of the GSA measures given simulator output were relatively simple thanks to existing \texttt{R} packages that are described in the next section.

%% file: 2_Methods/2B_Methods_Derivative.tex
\subsection{\label{GSA.Methods.Subsection}GSA Methods}
Each of the seven different GSA methods is briefly described below, with further technical details provided in Appendix \ref{app.methods}. Within each method's description, the key \texttt{R} packages used are shown, and Table \ref{T:SA.methods.R.code} is a summary listing for all methods. To display output results from each of the methods, the same example simulator, SimplyP, and output, (log) Outflow on a given date, were used.

\subsubsection{\label{subsec:Morris}Derivative-based: Morris elementary effects}
 
\begin{figure}[h]
    \centering
    \begin{subfigure}[b]{0.45\textwidth}
    \includegraphics[width=\columnwidth,height=0.28\textheight]{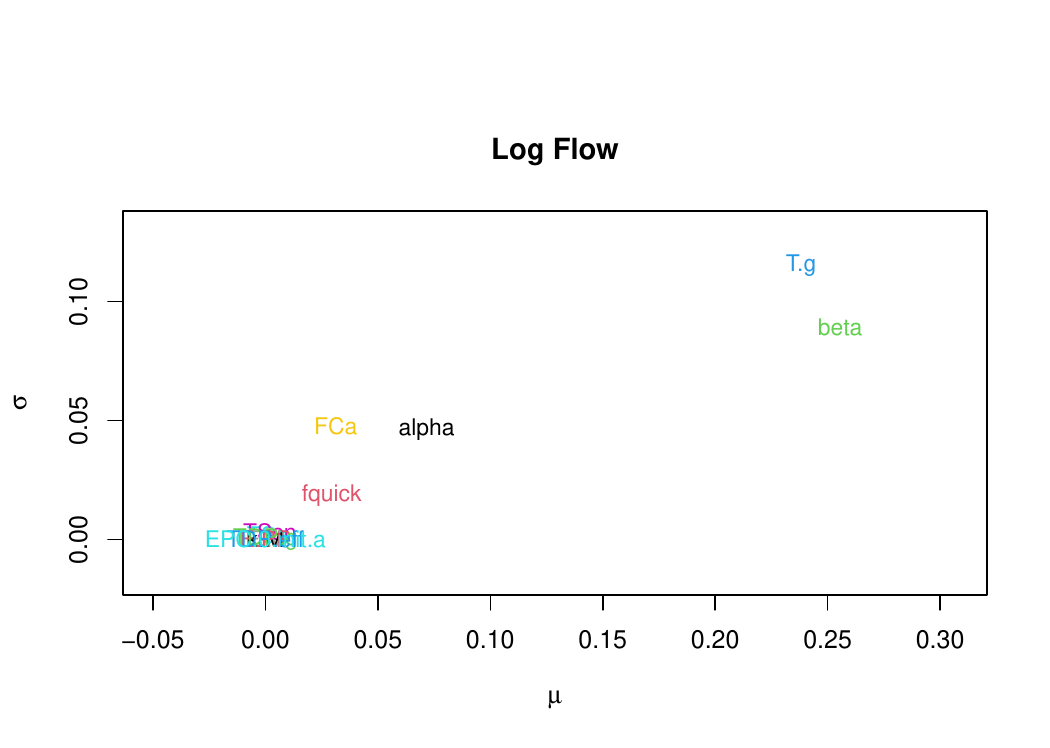}
    \caption{Derivative-based: scatterplot of Morris measures $\sigma$ versus $\mu^*$. Larger values of $\mu^*$ can generally reflect larger 1st order effects with larger values of $\sigma$ indicating interactions and/or higher order effects.}
    \label{F:SimplyP.Morris.Demo}
    \end{subfigure}
    \hfill
    \begin{subfigure}[b]{0.45\textwidth}
    \centering 
        \includegraphics[width=\columnwidth,height=0.28\textheight]{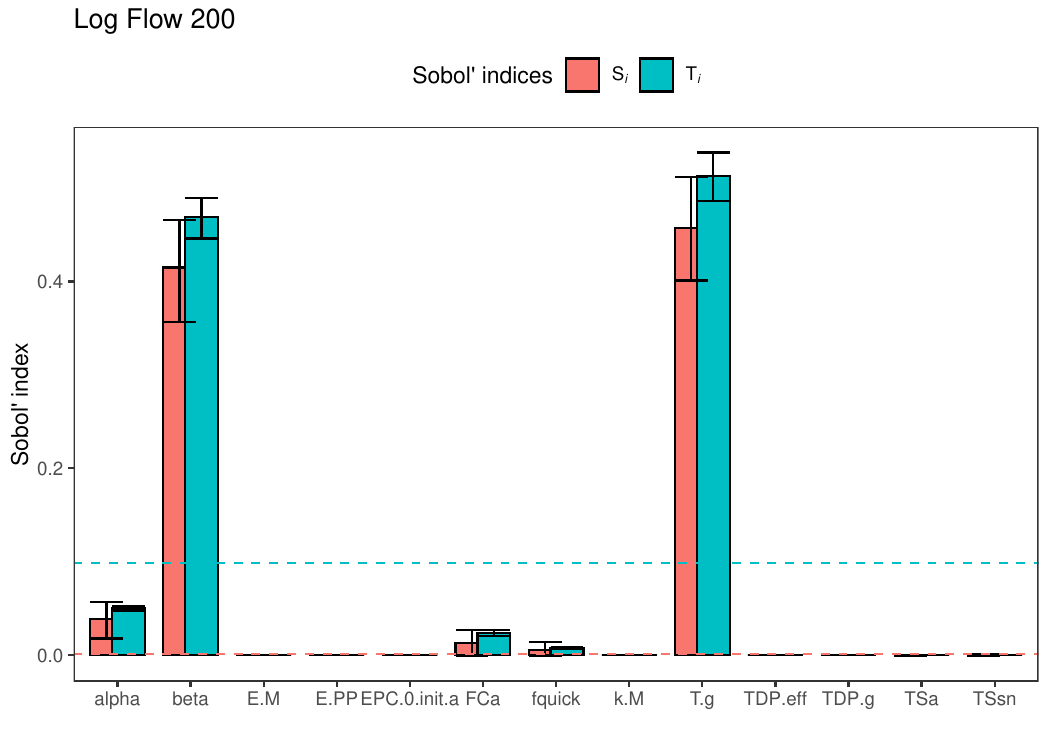}
     \caption{Distribution-based: Sobol $S_i$ (red bars) and $T_i$ (blue bars) with 95\% confidence intervals drawn over each bar.}
     \label{F:SimplyP.Sobol.Demo}
    \end{subfigure}
% --------------------------------------------------------------

   \begin{subfigure}[b]{0.45\textwidth}
    \centering
    \includegraphics[width=\columnwidth,height=0.28\textheight]{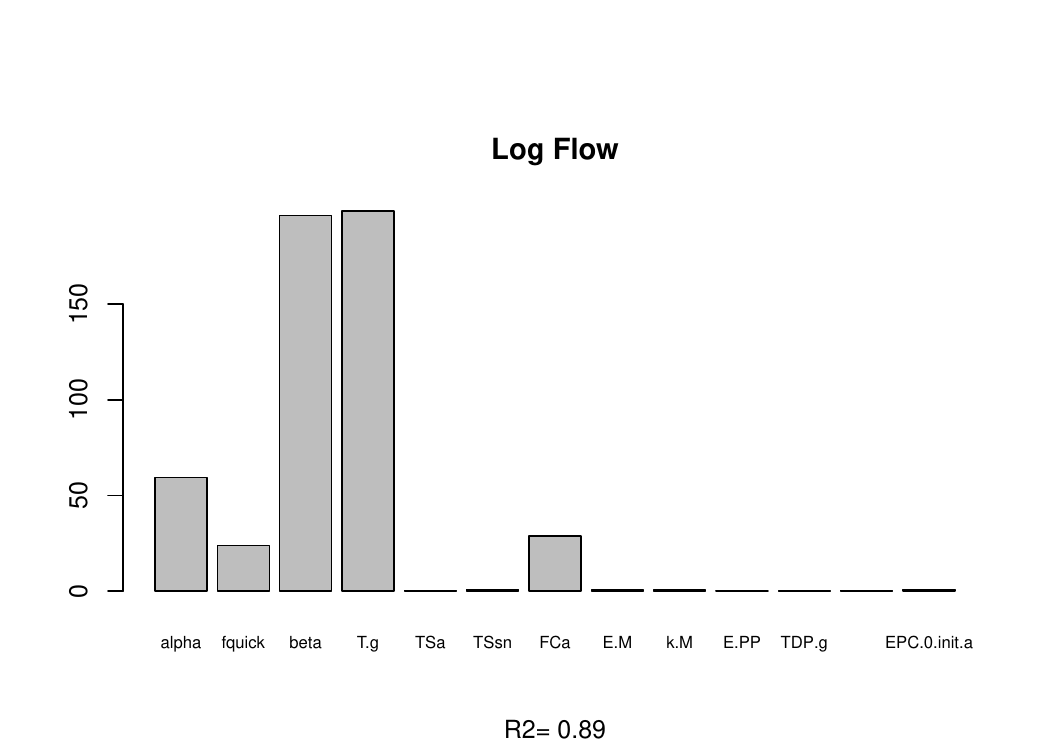}
    \caption{Multiple Regression standardized regression coefficients.}
    \label{F:SimplyP.Regression.Demo}
 \end{subfigure}   
\hfill
    \begin{subfigure}[b]{0.45\textwidth}
    \centering
\includegraphics[width=\columnwidth,height=0.28\textheight]{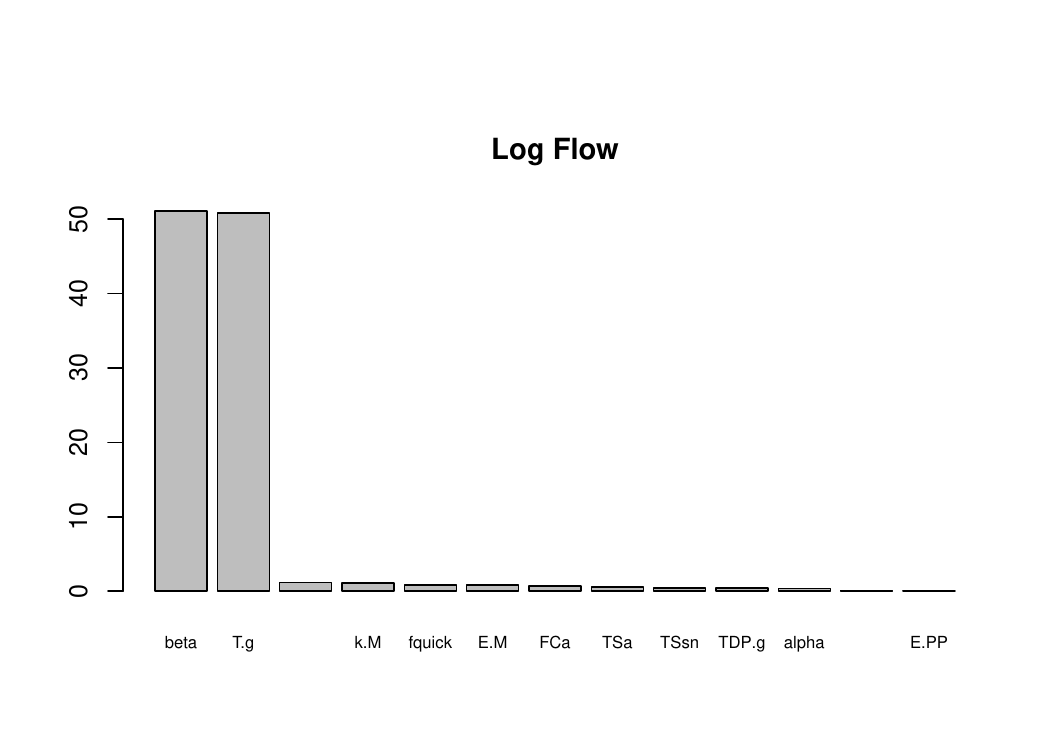} 
     \caption{Regression Tree parameter importance.}
     \label{F:SimplyP.RegTree.Demo}
    \end{subfigure}   
% --------------------------------------------------------------

   \begin{subfigure}[b]{0.45\textwidth}
    \centering
    \includegraphics[width=\columnwidth,height=0.28\textheight]{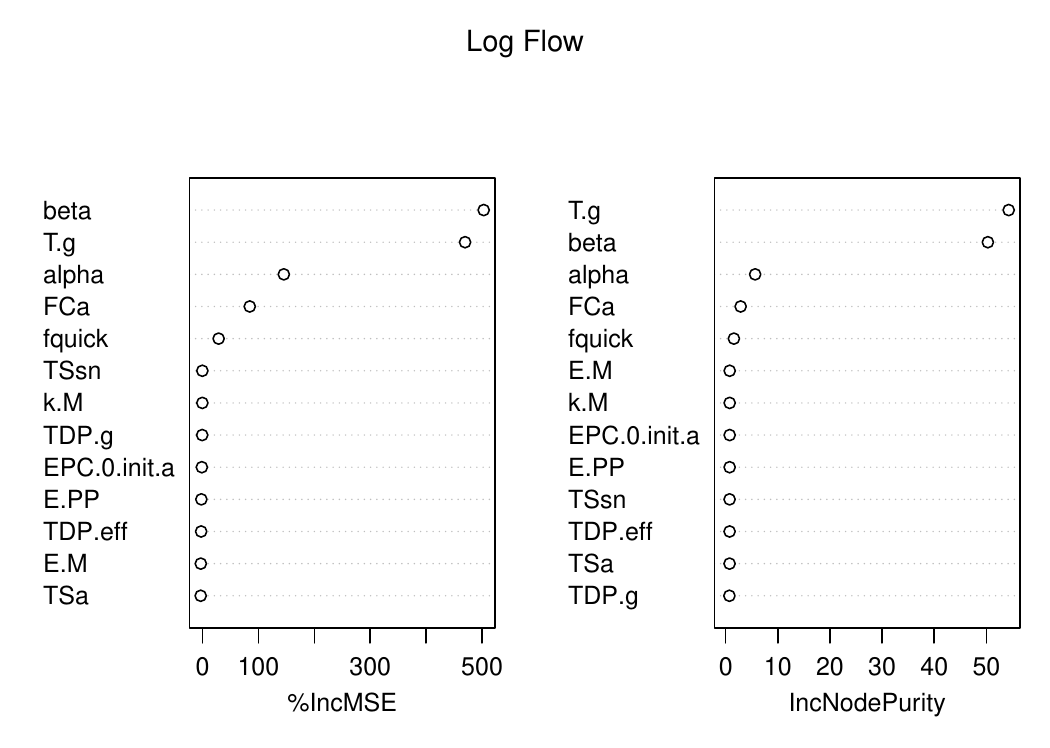}
    \caption{Random Forests results.}
    \label{F:SimplyP.RF.Demo}
 \end{subfigure}   
 \hfill
   \begin{subfigure}[b]{0.45\textwidth}
    \centering
    \includegraphics[width=\columnwidth,height=0.28\textheight]{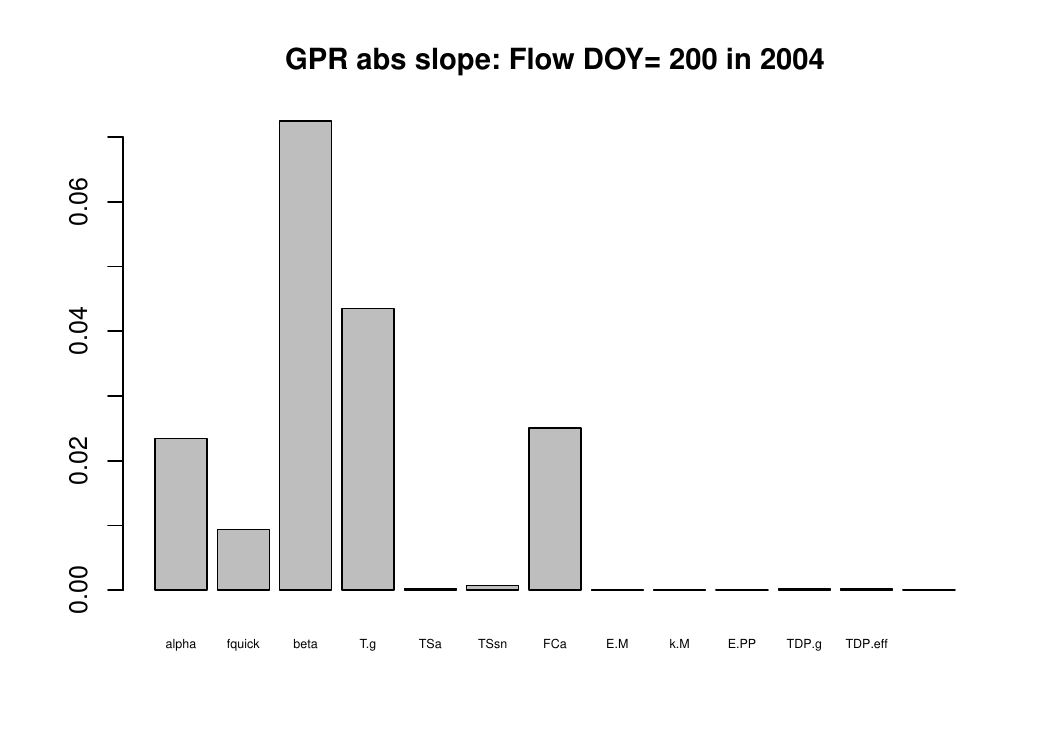}
    \caption{GPR standardized regression coefficients.}
    \label{F:SimplyP.GPR.Slope.Demo}
 \end{subfigure}   
 \caption{SimplyP: Six methods of Sensitivity Analyses applied to Log Flow on Day 200 in 2004.}
 \label{F:SimplyP.Q.Demonstration}
\end{figure}

The derivative-based procedure used is the Morris elementary effects method due to \citet{1991_Morris} (also called Morris screening \citep{2013_Cosenza_etal}). The essence of the method is to construct $K$ distributions of numerically calculated partial derivatives for each of the $K$ parameters, where values for the other parameters are at fixed values.  Summaries of these distributions, namely the means, $\mu_k$, and the variances, $\sigma^2_k$, of the derivatives, are measures of each parameter's relative influence on the output. We used a slight modification of the mean, $\mu^*$, due to \citet{2007_Campolongo_etal}, which is always positive valued, and calculated a measure that combines the mean and variance, called the derivative global sensitivity measure (DGSM, \citet{2009_Kucherenko_etal, 2022_Dela_etal}): 
\begin{align*}
    G_k & = \sqrt{\mu^{* 2}_k + \sigma^2_k}
\end{align*}
 For the case examples, we examined both the scatterplots of $\sigma_k$ vs $\mu^*_k$ as well as calculated DGSM. An example of the scatterplot display for outflow for a single day from  SimplyP is shown in Figure \ref{F:SimplyP.Morris.Demo} where two parameters (beta, baseflow index, and Tg, groundwater time constant; see Table \ref{T:SimplyP.parameters} for descriptions of the parameters) clearly dominate with larger values of $\mu^*$ and $\sigma^2$. The larger values of $\sigma$ reflect likely interactions between the input parameters. Further details of the Morris method are given in Section \ref{app.Derivative.details}.

 \paragraph{Software.}   From the \texttt{R} \texttt{sensitivity} package,  the function \texttt{morris} was used to generate the parameter combinations, and the function  \texttt{tell} was used to calculate $\mu_k^*$ and $\sigma^2_k$. 

% *********************************************************************

%% file: 2_Methods/2C_Methods_Distribution.tex
\subsubsection{\label{subsec:Sobol}Distribution-based: Sobol' sensitivities}

The distribution-based GSA measures, also known as variance-based, used were the Sobol' 1st order, $S_{1,k}$, and total, $T_{k}$, $k$=1,2,$\ldots$,$K$, sensitivities \citep{1993_Sobol, 2021_Razavi_etal}. Both $S_{1,k}$ and $T_k$ can be viewed as byproducts of the law of total variance \citep{2015_Bickel_Doksum} when applied to functions of random variables. Section \ref{app.Distribution.details} provides a sketch of the main ideas, but for brevity, here we simply present the general definitions of $S_{1,k}$ and $T_k$. 

The output from the simulator with input parameters $F(\theta_1,\ldots,\theta_K)$ is denoted $Y$. The notation for the conditioning subscript $\not{\theta}_k$ means evaluation (integration) over all input variables \textit{but} $\theta_k$.  First order sensitivity is  
\begin{align}
\label{eq:Sobol.S1}
S_{1,k} &= \frac{V_{\theta_k} [E_{\not{\theta}_k}(Y|\theta_k)]}{V(Y)}
\end{align}
and total sensitivity is
\begin{align}
\label{eq:Sobol.T}
T_k &=  \frac{E_{\not{\theta}_k} \left [ V_{\theta_k}(Y|\not{\theta}_i) \right ]}{V(Y)}  
 = 1- \frac{V_{\not{\theta}_k} \left [ E_{\theta_k}(Y|\not{\theta}_k) \right ]}{V(Y)}
\end{align}
Exact or analytical calculation of $S_{1,k}$ and $T_k$, assuming independent and identical probability distributions for the $\theta_k$s, is practically impossible for most simulators and a variety of Monte Carlo integration procedures have been developed \citep{2000_Saltelli_etal, 2008_Saltelli_etal}.

An example of resulting estimates of Sobol' 1st order and total sensitivities is shown in Figure \ref{F:SimplyP.Sobol.Demo}. As with the Morris results,  beta and Tg clearly dominate; the dashed blue and dotted red horizontal lines are estimates of significance cut-offs for $T_k$ and $S_{1,k}$, respectively. \citet{2022_Puy_etal} provides more detail on interpreting these plots. $T_k$ is necessarily at least as large as $S_{1,k}$ with the magnitude of $T_k-S_{1,k}$ a measure of the influence due to higher order effects and interactions.  We note that when $S_{1,k}$ is close to $T_k$, then interpretation becomes relatively simple and the relative contribution to the total variance of the simulator output of each input parameter can be read almost directly from the scaled 1st order sensitivities: $S_{1,k}/\sum_{j=1}^K S_{1,j}$. 

\paragraph{Software.}   From the \texttt{R} \texttt{sensobol} package,   the function \texttt{sobol\_matrices} was used to generate the parameter combinations and the function  \texttt{sobol\_indices} was used to calculate $S_{1,k}$ and $T_k$. 

% \begin{figure}[h]
%     \centering
%     \includegraphics[width=0.6\linewidth]{0_Figures/SimplyP/SimplyP_Sobol_Q.pdf}
%     \caption{Example of variance-based procedure with estimates of Sobol' $S_i$ and $T_i$ values.}
%     \label{F:Method.Sobol.Example}
% \end{figure}

%% file: 2_Methods/2D_Methods_Variogram.tex
\subsubsection{\label{subsec:Variogram}Variogram-based: VARS-TO}
The variogram-based approach, developed by \citet{2016_Razavi_Gupta}, has its origins in spatial statistics where model output, $Y$=$F(\theta_1,\ldots,\theta_K)$, is viewed as $K$-dimensional response surface. 

\citet{2016_Razavi_Gupta} define a continuous interval of sensitivity measures labelled IVARS$_{k,p}$, where $p$ is a proportion of parameter $k$ interval length. IVARS$_{k,p}$ is the integral of the directional semi-variogram in the $k$th dimension of the parameter space, $(\theta_1,\ldots,\theta_K)$, and the larger this value, the more sensitive the output $Y$ is to that parameter.  We worked with another variogram-based measure \citet{2016_Razavi_Gupta} developed called VARS-TO, which stands for variance-based total order effects.  It uses the directional variogram, $\gamma(h_k)$, without integration. The user specifies the value of $h$ (we used $h$=0.1).  VARS-TO is proportional to Sobol' total sensitivity $T_k$:
\begin{align*}
   \mbox{VARS-TO}_k =  \gamma(h_k)+f(Cov_{\not k},h_k) & \propto T_k,
\end{align*}
where $f(Cov_{\not k},h_k)$ is a function of the covariance of $Y$ excluding input parameter $k$ and the distance $h_k$.  Further details are given in Section \ref{app.Variogram.details}.

\begin{figure}[h]
    \centering
    \includegraphics[width=0.8\linewidth]{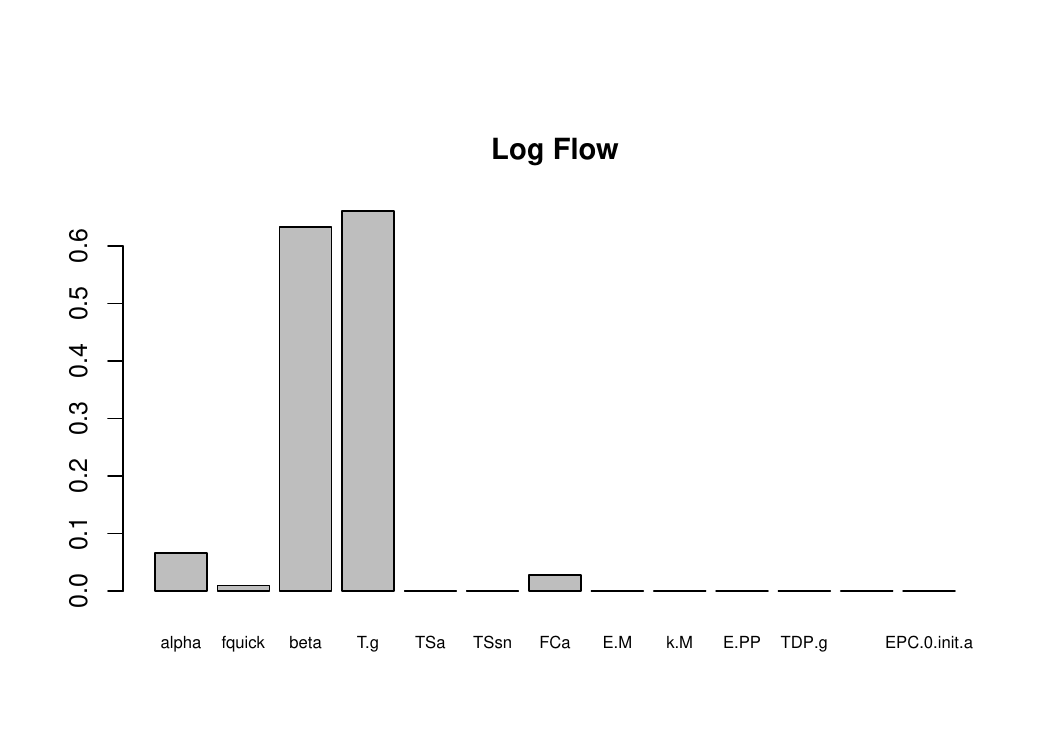}
 \caption{Example of VARS-TO calculations for SimplyP log outflow.}
    \label{F:Method.VARS-TO.Demo}
\end{figure}

Figure \ref{F:Method.VARS-TO.Demo} shows the VARS-TO values for SimplyP outflow. Given the proportional relationship with $T_k$, one expects the results for VARS-TO and $T_k$ to be quite similar (see Figure \ref{F:SimplyP.Sobol.Demo}). And as for Sobol' $T_k$, the two parameters  beta and Tg  dominate. 

\paragraph{Software.} From the \texttt{R} \texttt{sensobol} package,  the function \texttt{vars\_matrices} was used to generate the parameter combinations and the function  \texttt{vars\_to} was used to calculate VARS-TO.

%% file: 2_Methods/2E_Methods_Reg_MultReg.tex
 \subsubsection{\label{subsubsec:Reg.MR}Regression-based: Multiple linear regression}
  For all four regression-based methods, the space-filling parameter combinations were generated using Latin Hypercube sampling. Given model output $y$ and parameters $\theta_1$, $\ldots$, $\theta_K$, the following linear regression is fit.
 \begin{align*}
y &= \beta_0 + \sum_{k=1}^K \beta_k \theta_k
 \end{align*}
 The relative importance of each input parameter was calculated as the absolute value of the $t$-statistic for the corresponding coefficient, a Standardised Regression Coefficient (SRC).
 \begin{align}
    SRC_k & = |t_k| = \left | \frac{\hat{\beta}_k}{se(\hat{\beta}_k)} \right |
 \end{align}
 The $SRC_k$ with the largest values are deemed most influential.  This is a slight variation on the usual SRC measure \citep{2013_Cosenza_etal}, with the distinction here being the use of the absolute value.  With highly nonlinear simulators, the linear model is expected to fit the data poorly, and a rule of thumb for using SRC for ordering parameter influence is that the $R^2$ be at least 0.7 \citep{2007_Cariboni_etal,2015_Song_etal}.  In cases where the $R^2$ was less than that, quadratic models with all possible pairwise interactions were tried to improve the $R^2$, but gains were often slight. However, even if improvements did result, the interpretation of the relative importance of a given parameter becomes more complicated as coefficients for squared terms and interaction terms need to be somehow combined.   

Figure \ref{F:SimplyP.Regression.Demo} shows the SRC values for SimplyP Outflow with beta and Tg again dominating.  In this particular case, the $R^2$ was relatively high at 0.89 but this was somewhat exceptional compared to other outputs for this and the other simulators and examination of the residuals indicate departures from linearity and normality.

\paragraph{Software.} Parameter combinations were generated using the function \texttt{maximinLHS} in the \texttt{R} package \texttt{lhs}, and the \texttt{lm} function was used for fitting the regression models and calculating the SRCs. 

%% file: 2_Methods/2F_Methods_Reg_RegTree.tex
\subsubsection{Regression-based: Regression tree} Regression trees \citep{1983_Breiman_etal, 2021_James_etal} are a very flexible and somewhat intuitive approach to fitting a nonlinear model to a response variable. The essence of the method is to exhaustively partition a $K$-dimensional input parameter space into multiple hypercubes, e.g., for $K$=3 hypercube $i$ defined as 
\begin{align*}
 [\theta_{A,Low_i} \le \theta_A < \theta_{A,Upper_i}, ~\theta_{B,Low_i} \le \theta_B < \theta_{B,Upper_i}, ~\theta_{C,Low_i} \le \theta_C < \theta_{C,Upper_i}]  
\end{align*} 
and all observations in each hypercube are assigned the same value, $\hat{y}_i$. The algorithm for the partitioning is a sequential branching process where branching at given points is based on the minimization of some loss function, e.g., average squared deviations between observed and fitted values. 

The end result of the procedure can be graphically displayed as an inverted tree with branches added at nodes as one proceeds down the tree. The sequential creation of the tree begins with the entire parameter space, the ``trunk'' of the tree, with subsequent branch creation at intermediate nodes that partition  the input space; e.g., if $\theta_i \le$ 10, outcomes are grouped in one branch, and if $\theta_i > 10$, outcomes go into a second branch. All observations within a given node are assigned a common value, and a loss function, e.g., mean squared deviation of the observed values and the assigned value, is calculated. If the loss function is greater than some specified value, then additional branching occurs. The partitioning ends when reductions in the loss function are considered negligible. The final tree shows terminal nodes, nodes where no additional splitting is done, also known as leaves, and all $n$ outputs will be assigned to a leaf.   

GSA can be done using the relative importance measures calculated for the input parameters. The importance of an input parameter is ``the sum of the goodness of split measures for each split for which it was the primary variable, plus goodness (adjusted agreement) for all splits in which it was a surrogate '' \citep{2023_Therneau_Atkinson}. The relative importance is this importance scaled by the sum of the importance measures for all parameters and thus a measure of sensitivity.  

\begin{figure}[h]
    \centering
    \includegraphics[width=0.8\linewidth]{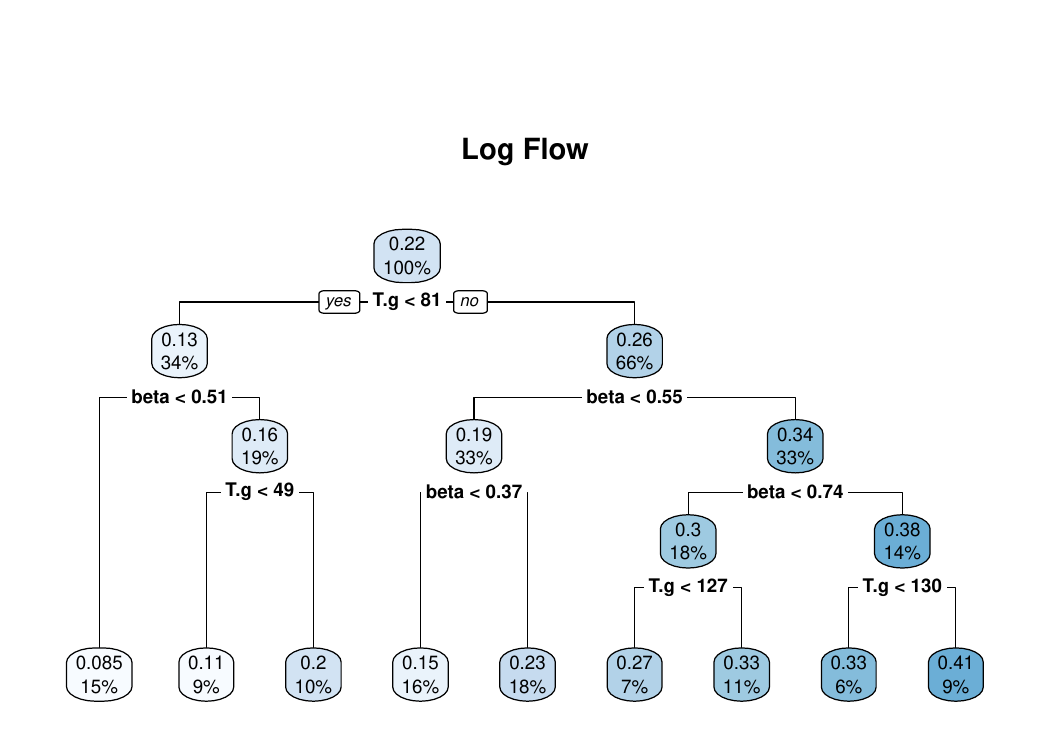}
 \caption{Example of regression tree output for log flow in SimplyP.}  
    \label{F:Method.RegTree.Tree.Example}
\end{figure}

Figure \ref{F:SimplyP.RegTree.Demo} shows the relative importance measures of the SimplyP parameters on outflow. Again, beta and Tg dominate. Figure \ref{F:Method.RegTree.Tree.Example} shows the resulting tree where the fact that Tg and beta dominate can be seen by all branching decisions, but one involving the parameter alpha, were based on those two parameters. The number at the top of each box shows the assigned log flow value assigned for that node, and the number at the bottom of each box shows the percentage of output combinations accounted for in that box.  A coarse summary is that larger outflow values result for Tg $\ge$ 85 and beta $\ge$ 55, while a more nuanced explanation is that there is an interaction effect from Tg and beta.  Heatmaps of regression tree outputs can be used to gain additional insight into correlations and interactions between input parameters, but the interpretation can be involved.  Section \ref{app.Reg.RegTree.details} provides a detailed discussion of heatmaps and their interpretation for this same example.

Regression trees are often used for prediction, i.e., for a new set of covariate values, the tree is used to predict the response. Prediction quality is often measured by a cross-validation procedure, and this then leads to a pruning of the``maximal'' tree and sometimes the complete removal of some covariates from the tree \citep{2021_James_etal}. However, as our aim was not prediction quality, but rather the determination of the relative importance of all $K$ input parameters, no pruning was done, and the maximal tree was used. 

\paragraph{Software.} Parameter combinations were generated using the function \texttt{maximinLHS} in the \texttt{R} package \texttt{lhs}. The \texttt{R} packages \texttt{rpart} and \texttt{rpart.plot} were used to fit and display the trees and to provide measures of the relative importance.  The heatmap was constructed using the package \texttt{treeheatr}.  

% \begin{figure}[h]
%     \centering
%     \includegraphics[width=0.7\linewidth]{0_Figures/SimplyP/SimplyP_Reg_Flow.pdf}
%  \caption{Example of regression tree procedure using measures of relative importance.}
%     \label{F:Method.RegTree.Example}
% \end{figure}

%% file: 2_Methods/2G_Methods_Reg_RandomForest.tex
\subsubsection{Regression-based: Random forests}  Random forests are an extension of regression trees due to \citet{2001_Breiman} (see also \citet{2021_James_etal}), where instead of producing a single regression tree, multiple regression trees are produced from $B$ bootstrap samples from the input-output space.  The predicted outputs for particular parameter combinations are averages of the individual tree predictions, and prediction quality is often improved over that of a single regression tree. The relative importance of individual input parameters can again be measured, as in regression trees, by the reduction in MSE that results when an input parameter is included in the tree. 

 Figure \ref{F:SimplyP.RF.Demo} shows the two different relative importance measures for SimplyP's log outflow. The one on the left shows how much reduction there is in the mean square error for predicted values (based on permutations ``out-of-bag'' samples, essentially test data sets) when the input parameter is included. The one on the right measures how much the leaves' ``impurity'' decreases, namely residual sums of squares, when the input parameter is included (averaged across all trees). Again,  beta and Tg dominate, but the ordering is reversed between the two measures.

\paragraph{Software.} Parameter combinations were generated using the function \texttt{maximinLHS} in the \texttt{R} package \texttt{lhs}. The function \texttt{randomForest} from the \texttt{R} package \texttt{randomForest} was used to fit the random forests, and the function \texttt{importance} was used to calculate importance measures.

% \begin{figure}[h]
%     \centering
%     \includegraphics[width=0.7\linewidth]{0_Figures/SimplyP/SimplyP_RF_Flow.pdf}
%  \caption{Example of random forest procedure using measures of relative importance.}
%     \label{F:Method.RF.Example}
% \end{figure}

%% file: 2_Methods/2H_Methods_Reg_GPR.tex
\subsubsection{\label{subsec:GPR}Regression-based: Gaussian process regression} Gaussian process regression models (GPRs) can be viewed as extensions of multiple regression models, with the extension being that, in addition to a model for the mean response as a function of covariates, the covariance between responses is modeled as well.  GPRs are based on stochastic processes (see discussion in Section \ref{subsec:Variogram}), and the indices are the covariate values. Distances between the indices are used in the modelling of the covariance, with correlations in pairs of response variables being higher for observations with more similar parameter values. Section \ref{app.Reg.GPR.details} provides more background details.

For sensitivity analysis, GPRs can be used in two different ways. One way is to model both the mean structure and the covariance as functions of input parameters.  The estimated (standardized) slope coefficients provide a measure of the relative importance of the input parameters, exactly as is done with multiple regression. The covariance model can be viewed as a means of accounting for residual variation after the mean effects have been accounted for, which depends on the mean model formulation.  The normalized inverses of the input parameter-specific range parameters are measures of the input parameters' importance in accounting for residual variations, over and above the mean response.  A second approach is to not model the mean structure as a function of the input parameters, namely, set the mean equal to a constant, and to only model the output by the covariance function of the inputs. Inputs with relatively small normalized inverses are labelled ``inert inputs'' by \citet{2016_Gu}, who found that such a method for ranking compared favorably to Sobol' sensitivities. We found that the second approach did not always provide as much information about input parameter importance; however, at least compared to other GSA measures like Sobol' sensitivities, and  here only used the first method.

% we examined both the slope coefficients (as in multiple regression) and the range parameters, or rather the normalized value of the inverse of the range parameters.  Joint examination of the magnitude of the slope coefficients and the magnitude of the range parameters for sensitivity analysis can be somewhat involved. If the absolute value of a slope coefficient (with suitable scaling as with multiple regression SRC) is large, then the parameter is having influence on the mean response. The range parameters per input parameter are reflecting the internal correlation for the input parameter in the residuals, thus after the mean effects are removed.  Smaller values of the range parameter indicate more influence in the modelling of the residuals.  If the scaled slope coefficient for an input parameter is not large but the range parameter is small, then the input parameter is having an effect on the variation in the particular model output. A single summary measure based on both the estimated slope and estimated range would be potentially helpful. 

 % Input variables with smaller values for the inverse, e.g., less than 0.1, called Inert Parameters, ``inputs that barely affect the outputs'' \citep{2018_Gu_etal}.  This corresponds directly to the observation above that inputs with larger range parameters yield similar values of $y$ and thus have less influence; e.g., $\theta_{i,1}$=10 and $\theta_{i,2}$=30 but corresponding $y_{i,1}$=2.10 and $y_{i,2}$=2.11.  

A GPR was fit to SimplyP's log outflow.  Due to computational costs, a random subsample of the LHS of size 500 was used in the fitting: computational expense increases with $N$ as an $N$ by $N$ covariance matrix is involved, and $N > 500$ sometimes would lead to a computation freeze.  The scaled absolute slope coefficients for a GPR fit to SimplyP's log outflow are shown in Figure \ref{F:SimplyP.GPR.Slope.Demo}. The parameters beta and Tg are again most important, with the alpha parameter appearing as well, and the parameter FCa (field capacity) appearing relatively important.   A linear model for the response has a relatively poor fit (as for the multiple regression) and the inverse range parameters (not shown) indicate that the Tg, FCa, and alpha are accounting for some of the residual variation.  

\paragraph{Software.} Parameter combinations were generated using the function \texttt{maximinLHS} in the \texttt{R} package \texttt{lhs}. The function \texttt{rgasp} from the \texttt{R} \texttt{RobustGaSP} package was used to fit the GPR, and the function \texttt{findInertInputs} was used for ranking the relative importance in the covariance model.

% \begin{figure}[h]
%     \centering
%     \includegraphics[width=0.7\linewidth]{0_Figures/SimplyP/SimplyP_GPR_Both_Flow.pdf}
%  \caption{Example of Gaussian Process regression procedure with measures of inverse range and absolute values of slope.}
%     \label{F:Method.GPR.Example}
% \end{figure}

%% file: 3_Simulators/3_Simulators.tex
%--------------------------------------------------
\section{\label{sec:Simulators}Three simulators}
% \tcb{Describe the 3 simulators in sufficient detail.
% Discussion of outputs that will be analysed; in particular
% whether deal with single output at a point in time (easy
% but not necessarily best nor interesting), deal with time
% series for a single output, deal with multiple outputs 
% in some combined fashion---interesting, and/or connect to
% field data---performance based SA (like using KGE).}

The three simulators used to evaluate the GSA methods are GR6J \citep{2011_Pushpalatha_etal}, SimplyP \citep{2017_Jackson_etal}, and STICS \citep{2009_Brisson}, listed in order of increasing complexity.  All are deterministic simulators.  The complexity in terms of the number of input parameters, outputs, and internal processing varies considerably between the simulators; for example, GR6J has only six input parameters while STICS has 100s of parameters, although the sensitivity analyses worked with a much reduced subset of parameters in the latter case.

General features of each simulator are described below, with more detailed descriptions in Section \ref{app:Simulators.Details}. All simulation runs, including parameter combination generation, as well as calculations of GSA measures based on simulators' outputs, were conducted in \texttt{R} running on a computer with the following hardware capabilities: an Intel(R) Core(TM) i7-10610U CPU @ 1.80GHz,  2.30 GHz.\ 32.0 GB RAM.  With the exception of the Gaussian Process Regression  calculations,   for the final results presented here, $N$ = 10,000 parameter combinations were used, a sample size that yielded consistent GSA measures.  Details of the \texttt{R} code for running the simulators from  are summarized in Section   \ref{app:sub.Rcode.Simulators}.

%\newpage 
\input{3_Simulators/3A_Sim_GR6J}

%\newpage 
\input{3_Simulators/3B_Sim_SimplyP}

%\newpage 
\input{3_Simulators/3C_Sim_STICS}

%% file: 3_Simulators/3A_Sim_GR6J.tex
\subsection{\label{subsec:GR6J.desc}GR6J - Catchment Hydrology Simulator}
 The GR6J simulator \citep{2011_Pushpalatha_etal} is a deterministic hydrological model that models water storage and water flow through a catchment.  It has been implemented worldwide for a wide variety of  hydrological applications. GR6J has been specifically proposed for low flow simulation settings \citep{2017_Caillouet_etal, 2019_Smith_etal, 2017_Crochemore_etal, 2014_Nicolle,2023_Hannaford_etal, 2017_Trudel_etal, 2011_Pushpalatha_etal}. The mathematical details of the day-by-day calculations of how water moves through the system in GR6J can be found in the appendices of \citet{2022_Sezen_Partal}. Below we sketch the general structure of GR6J with additional details given in  Section \ref{app.sec.GR6J.desc}.

GR6J is a lumped hydrological model, i.e., it has no spatial
component, with three storage compartments, Production Store, Routing Store, and Exponential Store. It operates on a daily time step with two input time series, daily precipitation ($P_d$, for day $d$) and potential evapotranspiration ($E_d$).   There are six free model parameters ($X_1$, $\ldots$, $X_6$) that need to be estimated for specific catchments. The precipitation and evapotranspiration inputs initiate a sequence of processes that are functions of model states and parameter values, which produce several intermediate stage outputs that ultimately yield daily catchment discharge on day $d$, $Q_{sim,d}$. A schematic of GR6J (Figure \ref{F:GR6J.schematic}) shows the three compartments, the daily input variables ($E$ and $P$), the six parameters ($X_1$, $\ldots$, $X_6$), and model outputs. Descriptions of the six free model parameters and default values are shown in Table \ref{T:GR6J.default}, and intermediate and final outputs are shown in  Table \ref{T:GR6J.outputs}.

For sensitivity analysis, we focused primarily on discharge, $Q_{sim,d}$, as streamflow is typically the primary calibration target in hydrological applications.  For our application of GR6J, it is the output for which the corresponding field data were available. A coarse mathematical representation of GR6J from input to discharge is the following.
\begin{align}
 Q_{sim,d}|\mathcal{\chi}_{d-1} &= F(P_{d},E_{d}, S_{d-1}, R_{1,d-1}, R_{2,d-1}, P_{r,d^-}|X_1,\ldots,X_6), ~~ d=1,\ldots, T
\end{align}
 $\mathcal{\chi}_{d-1}$ is the state of the catchment on day $d-1$ that includes current levels in the three stores and lagged values of $P_r$ (denoted by $P_{r,d^-}$). $S_{d-1}$,$R_{1,d-1}$, and $R_{2,d-1}$ are previous day volumes in the Production Store, Routing Store and Exponential Store, respectively.

 Given $Q_{sim,d}$ is a time series, multiple sensitivity analyses can be carried out on multiple days. We found that for the handful of dates that were examined, however,  the GSA results were relatively similar and we will report on results for just a single day (2018-07-15).  
 
 While our focus was primarily on $Q_{sim,d}$, in order to make clear the relative importance of the six parameters to GR6J in its entirety, we carried out limited GSA on two intermediate outputs that corresponded to the processes of production store filling and emptying ($P_r$) and passage through the hydrographs ($Q_9$).
 
 % The outputs listed approximately in order of creation within a time step and grouped into three categories: Production Store: $S$ and $P_r$, Hydrographs:$Q_9$ and $Q_1$, Outflows: $Q_{rExp}$, $Q_r$, $Q_d$, and $Q$.  Figure \ref{F:GR6J.schematic} makes obvious the influence of some parameters on some outputs and we expected SA results to identify as most important those same parameters.  For example, $X_4$ should have significant effect on the intermediate outputs $Q_9$ and $Q_1$, $X_1$, $X_2$, and $X_5$ mainly regulates stream flow amounts whereas stream flow variability is handled by a combination of $X_3$, $X_4$, and $X_6$ and, to a lesser extent, by $X_1$. 

%% file: 3_Simulators/3B_Sim_SimplyP.tex
\subsection{\label{subsec:Simulators.SimplyP.desc}SimplyP - Catchment Hydrology and Water Chemistry Simulator}

\textit{SimplyP} \citep{2017_Jackson_etal} is a dynamic, daily time step, catchment-scale model for simulating hydrology, sediment, and phosphorus (P) dynamics. Similar to GR6J, the model is conceptual, meaning the model approximates the general physical mechanisms governing hydrologic and biogeochemical processes using a set of ODEs. In contrast to GR6J, a lumped model with a single basin and no spatial structure, SimplyP is spatially semi-distributed, where the catchment is first discretized into sub-basins (based on, for example, the presence of monitoring stations or major changes in terrestrial conditions such as geology, topography, or soil type). Sub-basins are further subdivided based on land cover. Another contrast to GR6J is that SimplyP's main aim is to simulate water quality, so it simulates not only stream flow (hydrology) but also sediment and phosphorus concentrations and fluxes. The price of this additional flexibility and scope is an increase in the number of parameters. A schematic of the structure of SimplyP is shown in the supplemental material (Figure \ref{F:SimplyP.schematic}).

Similar to GR6J, exogenous forcing factors are daily time series of both precipitation and either air temperature or potential evapotranspiration (PET; if this is not provided, it is calculated from air temperature). The model also requires an estimate of P inputs to the system, including: (a) terrestrial inputs, in the form of the annual soil phosphorus balance for each land cover type (i.e., the sum of fertilizer and manure inputs minus outputs through harvested crops); (b) direct inputs to the river (point sources) from sewage and industry. Both may be assumed constant over time or provided as an input time series. 

In addition to the time series inputs and fixed parameters which describe the spatial setup (e.g. catchment areas, land cover proportions), the version of the model used here (no dynamic soil P store or dynamic vegetation cover) requires 14 parameters to constrain the hydrologic, sediment and phosphorus processes in the model; 15 when spatial variability in parameters is included (i.e. one parameter varies by land cover class). Two of these are not considered relevant in this application and were kept constant. The remaining 13 are uncertain, and model performance can be optimised by calibrating using field data. These 13 parameters were therefore used in the sensitivity analysis. Table \ref{T:SimplyP.parameters} lists the parameters and the range of values used. 

SimplyP outputs around 60 daily time series that fit into three categories: hydrology, sediment, and phosphorous. For GSA we examined four outputs:  (1) daily mean river discharge ($Q$, $m^3 s^{-1}$), (2) concentrations of suspended sediment ($SS_{conc}$), (3) total dissolved phosphorus ($TDP_{conc}$), and (4) particulate phosphorus ($PP_{conc}$), the latter three in units of $mg l^{-1}$. Table \ref{T:SimplyP.outputs} lists the selected output variables. 

An approximate mathematical representation of daily inputs to and outputs from SimplyP is the following.
\begin{align}
 \mathbf{y}_d = (Q_d, SS_{conc,d}, PP_{conc,d},  TDP_{conc}) &= F(Precip_{1:d},Temp_{1:d}|T_{S,Ar}, T_{S,SN}, f_{quick}, 
 TDP_{eff}, \ldots,\theta_{13}) 
\end{align}
where $d=1,\ldots,T$, and the subscripting for
precipitation and temperature $1:d$ denotes days 1 to $d$.
Additional details of SimplyP are given in the supplemental material (Section \ref{app.sec.SimplyP.desc}). 

We present results for the outputs for a single day of the year, chosen to correspond to a time when river discharge was low and groundwater-dominated. This was chosen as this period is the most ecologically sensitive, so understanding how to optimise model performance during these flow conditions is very relevant for model applications concerned with compliance with environmental regulations. However, the day-of-year selected for the analysis has a large effect on the ordering of the relative importance of the input parameters, as different model processes are activated under different conditions. The daily precipitation levels affect, for example, whether or not the soil water is contributing to the river flows over and above the groundwater. During the summer, for example, when daily precipitation is relatively low, effluent and groundwater TDP concentrations can be relatively important for TDP concentration, while during rainfall events, groundwater TDP effects will lessen relative to other parameters.

%% file: 3_Simulators/3C_Sim_STICS.tex
\subsection{STICS: Crop Growth Soil Biochemistry Simulator} 

STICS is a dynamic, deterministic process-based model that simulates the soil-crop-atmosphere system \citep{2009_Brisson, 2021_Wallach_etal, 2023_Beaudoin_etal}. It operates at a daily time step throughout the duration of a crop cycle, e.g., from planting to harvest, and is spatially-restricted to the ``plot scale'', meaning at an individual plant level and the pedon (the 3-dimensional body of soil surrounding the plant from the surface to a specified depth).  STICS models processes occurring in a plant, in the soil, and in cropping systems, where these processes are affected by environmental factors such as precipitation, solar radiation, temperature, and relative humidity, and by climate change impacts in general \citep{2023_Beaudoin_etal}. Plant-centered processes include plant growth, development, phenology, yield, and water and nutrient levels and balances (e.g., carbon, nitrogen) specified for multiple components of a plant, including roots, stem, and reproductive organs. Soil-centered processes include water transfers, nitrogen mineralization, nitrogen leaching, ammonia volatilization,and N$_2$O emission.  The sequencing of some of the 
processes is shown in Figure \ref{F:STICS.processes}.

STICS is by far the most complex of the three simulators, with a large staff of developers working over several decades. It has the most exogenous forcing factors, input parameters, and output time series of the three simulators considered here. The exogenous forcing factors are daily resolution, location-specific time series inputs of minimum and maximum temperatures, solar radiation, rainfall, wind speed, and relative humidity. There are around 200 general parameters in addition to location and management practices, and plant-specific input parameters and 100s of output time series (see Appendices A.1 and A.2 of \citet{2023_Beaudoin_etal}) and has been used to model many species of crops.

 We used STICS to model the growth of a barley plant and soil chemistry dynamics over a single growing season. The weather-related input time series were taken from a single location in southeast Scotland, and local fixed parameters include measures of soil characteristics.  Initially, a set of over 50 parameters was considered, which were thought relevant to eight output time series. The first iteration of GSA revealed that many of the 50 parameters had little to no effect on those eight outputs, and the focus was narrowed to 12 parameters (Table \ref{T:STICS.input.parameters}). The scope of GSA narrowed further by reducing the output set to just four outputs that are measures on the plant alone and just considering their value on the day that the barley plant is harvested (calendar day 237 in 2016). The outputs are (1) mafruit, biomass of harvested organs in tonnes/ha, (2) masec.n biomass of aboveground plant in tonnes/ha, (3) CNgrain, N concentration in fruits in \% dry weight and (4) CNplante, N concentration in the aboveground plant in \% dry weight.   
A conceptual mathematical expression for STICS is the following.
\begin{align}
\mathbf{y}_d = (\textrm{mafruit}_{d},\textrm{mascec.n}_{d},\textrm{CNgrain}_{d}, \textrm{CNplante}_{d}) & =
F(\Theta, \mathbf{w}_d, \mathbf{y}_{d-1})
\end{align}
where $\Theta$ is the vector of fixed input parameters, $\mathbf{w}_d$
are time-varying input variables, and $\mathbf{y}_d$ is a vector
of outputs on day $d$.

%% file: 4_Results/4_Results.tex
\section{\label{sec:Results}Results}
 
Robustness of the results in the GSA measures was examined by varying the number of parameter combinations from 1000 to 3000 to 10,000. For GR6J, which had the fewest parameters, the parameter rankings were quite similar for $N$=3000 and $N$=10,000, but the precision of $T_k$ and $S_{1,k}$ estimates did increase noticeably going from 3000 to 10,000. For SimplyP and STICS, which had more parameters, increasing $N$ from 3000 to 10,000 did result in changes in parameter rankings for some of the outputs.   With the exception of the GPR approach, all results presented here are based on $N$=10,000.    

Computation time falls into three categories: (1) generating $N$ parameter combinations, (2) making $N$ simulator runs, and (3) calculating GSA measures. By far, making $N$ simulator runs was the most computationally expensive, even though these simulators were particularly computationally expensive, and varied between simulators.  For GR6J, $N$=10,000 runs ranged from four to nine minutes, for SimplyP, around two minutes, and for STICS, about two hours. Generating 10,000 combinations took less than a minute, and calculations of the GSA measures took from one to two minutes, with the exception of GPR fitting. Fitting GPRs to even $N$=1000 parameter combinations could not be done with \texttt{rgasp()} as the program would hang, presumably due to the need to invert an $N$ by $N$ matrix. Instead, a random subset of $N$=500 model runs was selected, and GPR calculations took roughly 15 minutes.  

In this section, GSA results are presented for a single output on a single day ($d$): outflow for GR6J ($Q_{sim,d}$), total dissolved phosphorus ($TDP_d$) for SimplyP, and the biomass of harvested organs for a barley plant ($mafruit_d$).  The chosen day for $mafruit$ was the day of harvest. The day chosen for GR6J was relatively arbitrary, other than it being past the initial tuning period; however, the results did not change appreciably when other days were chosen. For SimplyP, as mentioned in Section \ref{subsec:Simulators.SimplyP.desc}, a single day of the year was chosen, which corresponded to summer low-flow conditions, to target parameters that control the summer TDP simulation; changing the day resulted in changed input parameter rankings.  Results for other outputs are given in Section \ref{app:Results.Details}, and for GR6J (Section \ref{app:Results.Details.GR6J}) examples of time varying GSA and GSA based on performance-based  criteria, namely KGE, are given. 

\input{4_Results/4A_Results_GR6J}

\input{4_Results/4B_Results_SimplyP}

\input{4_Results/4C_Results_STICS}

%% file: 4_Results/4A_Results_GR6J.tex
%-------------------------------------------
\subsection{\label{subsec:Results.GR6J}GR6J}
Results for $Q_{sim,d}$, outflow on $d$=2018-07-15, are presented here with results for the outputs $P_r$ and $Q_9$ in Section \ref{app:Results.Details.GR6J}. 

\begin{figure}[h]
    \centering
    \includegraphics[width=0.60\linewidth]{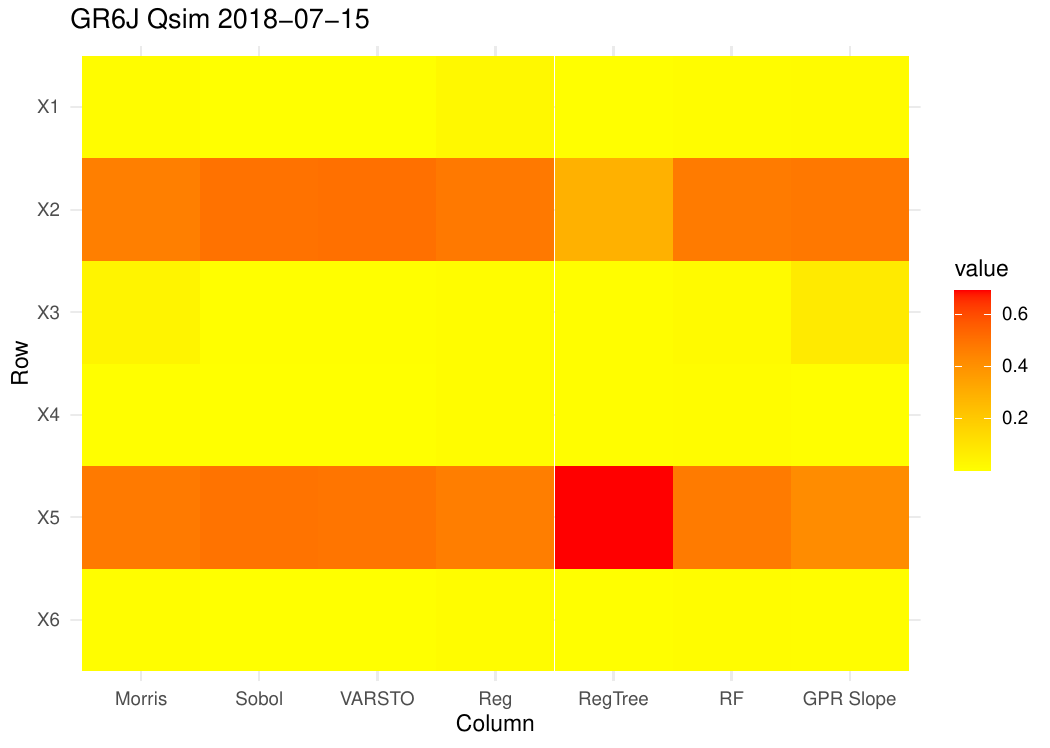}
    \caption{GR6J: Relative parameter importance for outflow on 2018-07-15 for the different GSA methods based on results in Table \ref{T:GR6J.SA.Outflow.summary}.}
    \label{F:GR6J.SA.Outflow.summary}
\end{figure}

\paragraph{Scalar measure comparisons.}
 Table \ref{T:GR6J.SA.Outflow.summary} and Figure \ref{F:GR6J.SA.Outflow.summary} show the relative measures for different GSA methods. The GSA values were scaled to sum to 1.0 per method to facilitate comparisons between methods, as well make clear the relative influence of each input parameter.  The values shown in   Table \ref{T:GR6J.SA.Outflow.summary} are based on DGSM for Morris, total sensitivity $T_k$ for Sobol', estimated $T_k$ for VARS-TO, standardized regression coefficients for multiple regression, variable importance measures for both regression trees and random forests, and the standardized regression coefficients and normalized inverses of the range parameters for Gaussian Process regression.  For nearly all the GSA measures, two parameters, $X_2$ (intercatchment exchange coefficient) and $X_5$ (intercatchment or groundwater exchange threshold; for definitions of parameters, see Table \ref{T:GR6J.default}) have the most, and with the exception of the regression trees, and quite similar influence. We note that the multiple regression model goodness of fit was quite low based on $R^2$ (0.30), and residual plots (not shown) indicate a quite poor fit with considerable nonlinearity remaining, yet the ranking of parameters was very similar to most of the other GSA measures.   Multiple regression fits were sometimes quite poor for the other outputs and the other simulators.  
\begin{table}[h]
\centering
 Outflow, $Q_{sim,d}$  \\
  \begin{tabular}{rrrrrrrrr}
   \hline
  & Morris & Sobol' & VARS-TO & Reg & RegTree & RF & \multicolumn{2}{c}{GPR} \\
  &  DGSM      & $T_i$       &  SRC  &      &  &       &  Slope &  InvRange \\ 
   \hline
$X_1$ & 0.01 & 0.00 & 0.00 & 0.04 & 0.01 & 0.02 & 0.02 & 0.01 \\ 
$X_2$ & \textbf{\tcr{0.46}} & \textbf{\tcb{0.50}} & \textbf{\tcb{0.50}} &  \textbf{\tcb{0.48}} &  \textbf{\tcr{0.27}} & \textbf{\tcr{0.45}} & \textbf{\tcb{0.48}} & \textbf{\tcr{0.38}} \\ 
$X_3$ & 0.05 & 0.00 & 0.00 & 0.01 & 0.01 & 0.02 & 0.08 & 0.16 \\ 
$X_4$ & 0.00 & 0.00 & 0.00 & 0.01 & 0.01 & 0.02 & 0.00 & 0.00 \\ 
$X_5$ &  \textbf{\tcb{0.47}} &  \textbf{\tcr{0.49}} &  \textbf{\tcr{0.49}} &  \textbf{\tcr{0.46}} & \textbf{\tcb{0.69}} & \textbf{\tcb{0.47}} &  \textbf{\tcr{0.42}} &  \textbf{\tcb{0.43}} \\ 
$X_6$ & 0.01 & 0.00 & 0.00 & 0.01 & 0.01 & 0.02 & 0.01 & 0.01 \\ 
    \hline
 \end{tabular}
\caption{GR6J: summary of results for different GSA methods applied to outflow on   day $d$=2018-07-15. Numbers in blue denote those with the largest relative value and those in red are the second largest.  Reg=multiple regression, RegTree=regression tree, RF=random forest, GPR=Gaussian Process regression. See text for explanations of the values per method.}
\label{T:GR6J.SA.Outflow.summary}
\end{table}
 \begin{figure}[h]
 \centering
     \includegraphics[width=0.8\textwidth]{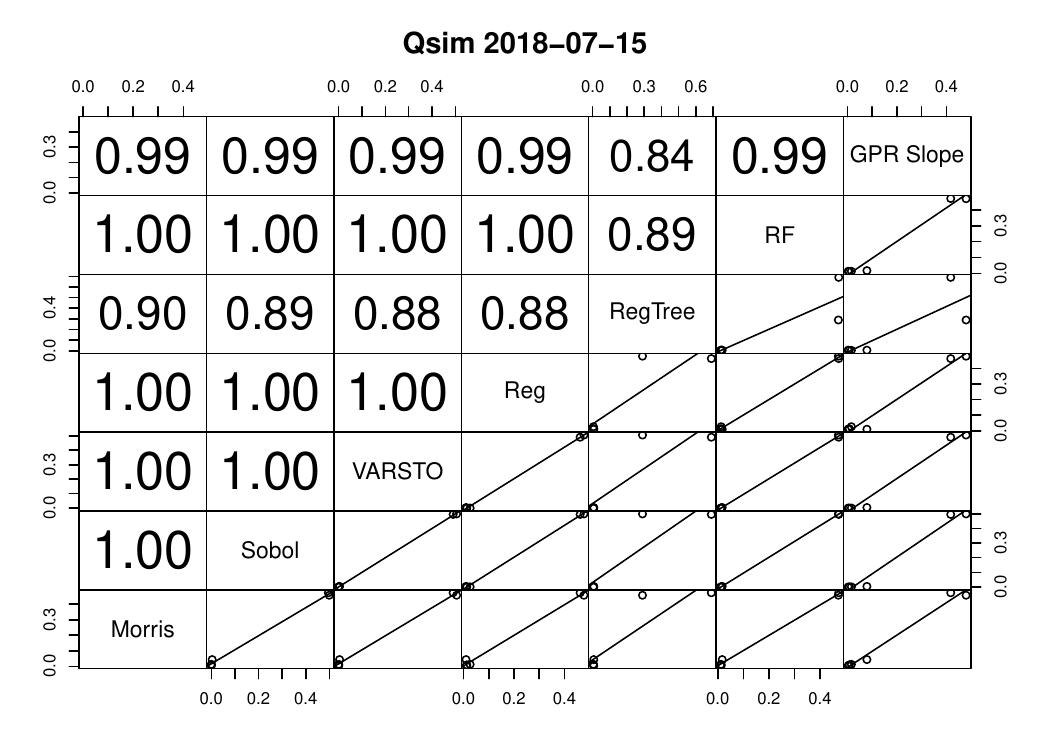}
     \caption{GR6J: pairwise scatterplots of GSA measures for the 6 parameters influence on Qsim (Outflow) for  different SA procedures along with Pearson correlation coefficients.  Reg=multiple regression, RegTree=regression tree, RF=random forest, GPR=Gaussian Process regression.}
     \label{F:GR6J.Pairwise.Qsim.SA}
 \end{figure}

These single measures alone had a high degree of consistency in the factor rankings,   as pairwise scatterplots and Pearson's correlation coefficient indicate in Figure \ref{F:GR6J.Pairwise.Qsim.SA}. The Kendall's coefficient of concordance Weight W was 0.80 with the p-value for $H_0$ of no relationship being $<0.001$. The heatmap for the scaled GSA measures (Figure \ref{F:GR6J.SA.Outflow.summary}) also indicates the high degree of similarity between methods in the evidence for $X_2$ and $X_5$ being the most influential input parameters, with the regression tree having a reversed ordering.

 \clearpage 

\begin{figure}[h]
    \centering
    \begin{subfigure}[b]{0.45\textwidth}
    \includegraphics[width=\columnwidth,height=0.28\textheight]{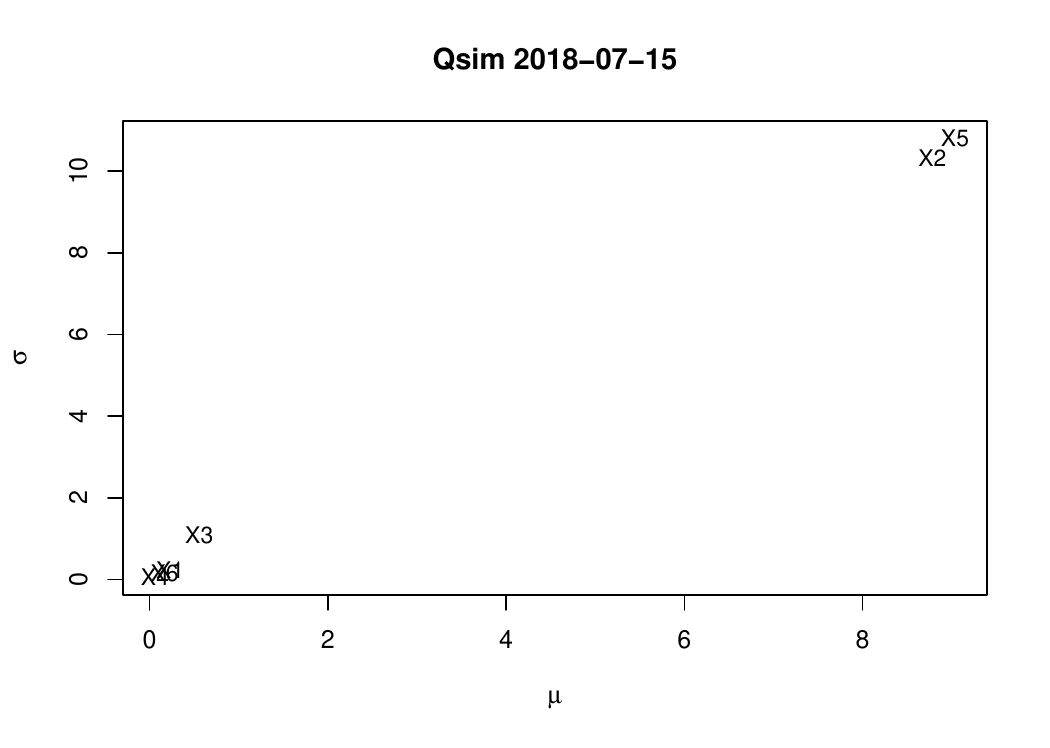}
    \caption{Morris measures, $\sigma$ versus $\mu^*$.}
    \label{F:GR6J.Morris.pair.Flow}
    \end{subfigure}
    \hfill
    \begin{subfigure}[b]{0.45\textwidth}
    \centering 
        \includegraphics[width=\columnwidth,height=0.28\textheight]{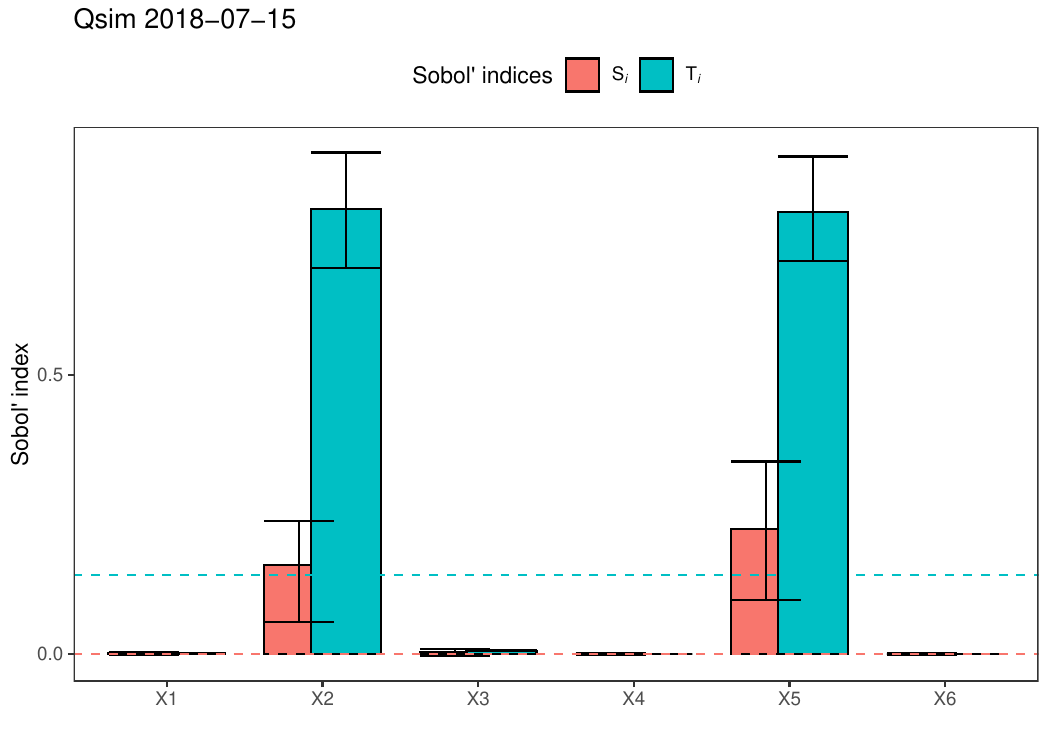}
     \caption{Sobol $S_{1,k}$ and $T_k$.}
     \label{F:GR6J.Sobol.pair.flow}
    \end{subfigure}
% --------------------------------------------------------------

   \begin{subfigure}[b]{0.45\textwidth}
    \centering
    \includegraphics[width=\columnwidth,height=0.28\textheight]{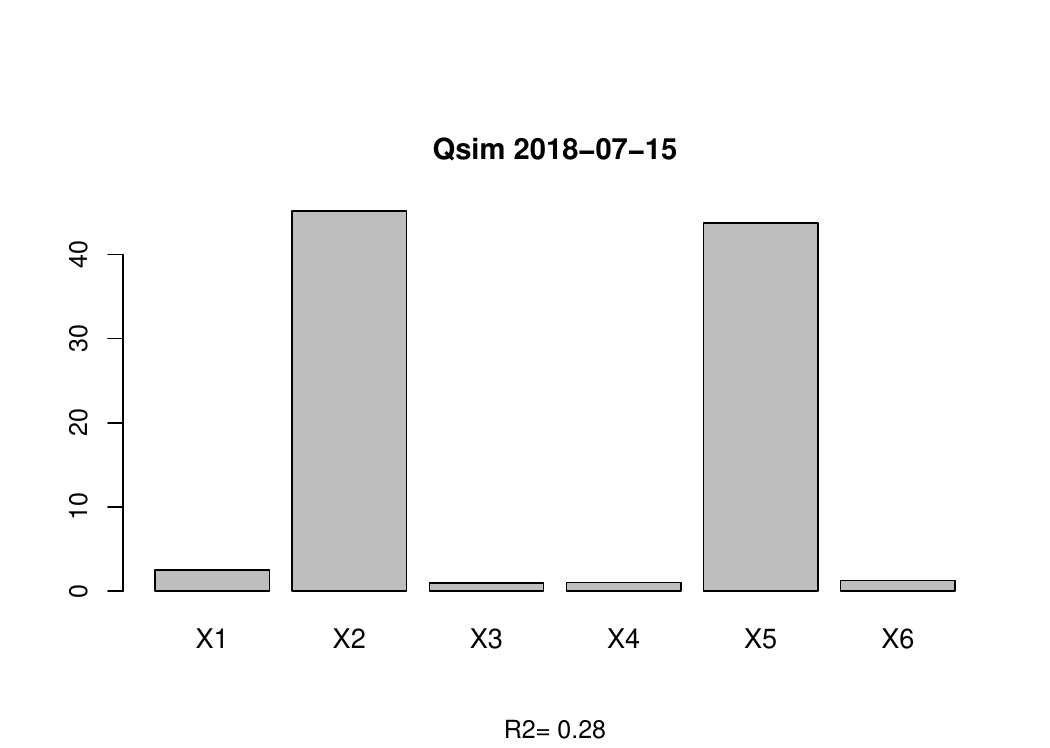}
    \caption{Multiple Regression standardized regression coefficients.} 
    \label{F:GR6J.Regression.Internal}
 \end{subfigure}   
\hfill
    \begin{subfigure}[b]{0.45\textwidth}
    \centering
    \includegraphics[width=\columnwidth,height=0.28\textheight]{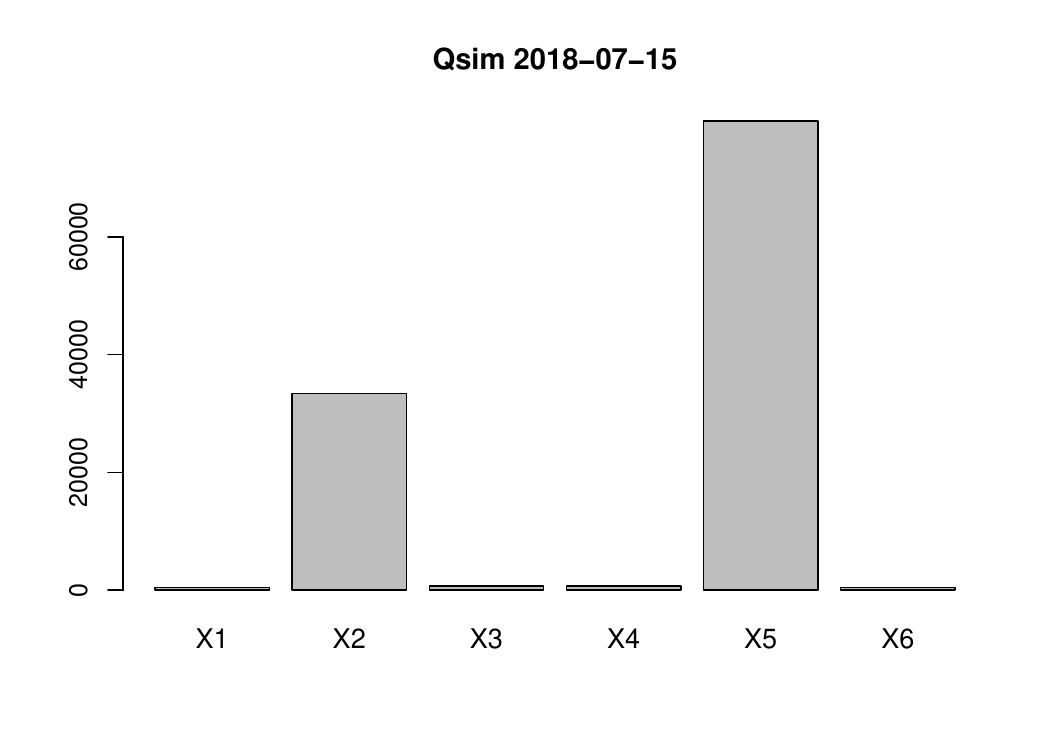} 
     \caption{Regression Tree parameter importance.}
     \label{F:GR6J.RegTree.Internal}
    \end{subfigure}   
% --------------------------------------------------------------

   \begin{subfigure}[b]{0.45\textwidth}
    \centering
    \includegraphics[width=\columnwidth,height=0.28\textheight]{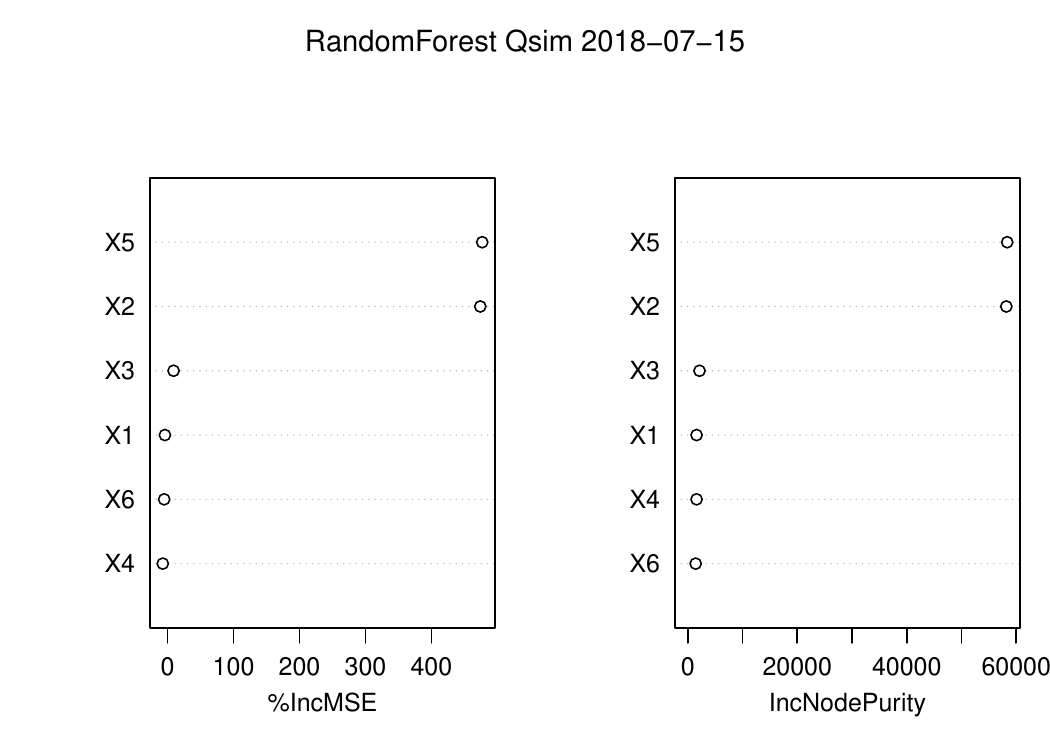}
    \caption{Random Forests parameter importance.}
    \label{F:GR6J.RF.Internal}
 \end{subfigure}   
 \hfill
   \begin{subfigure}[b]{0.45\textwidth}
    \centering
    \includegraphics[width=\columnwidth,height=0.28\textheight]{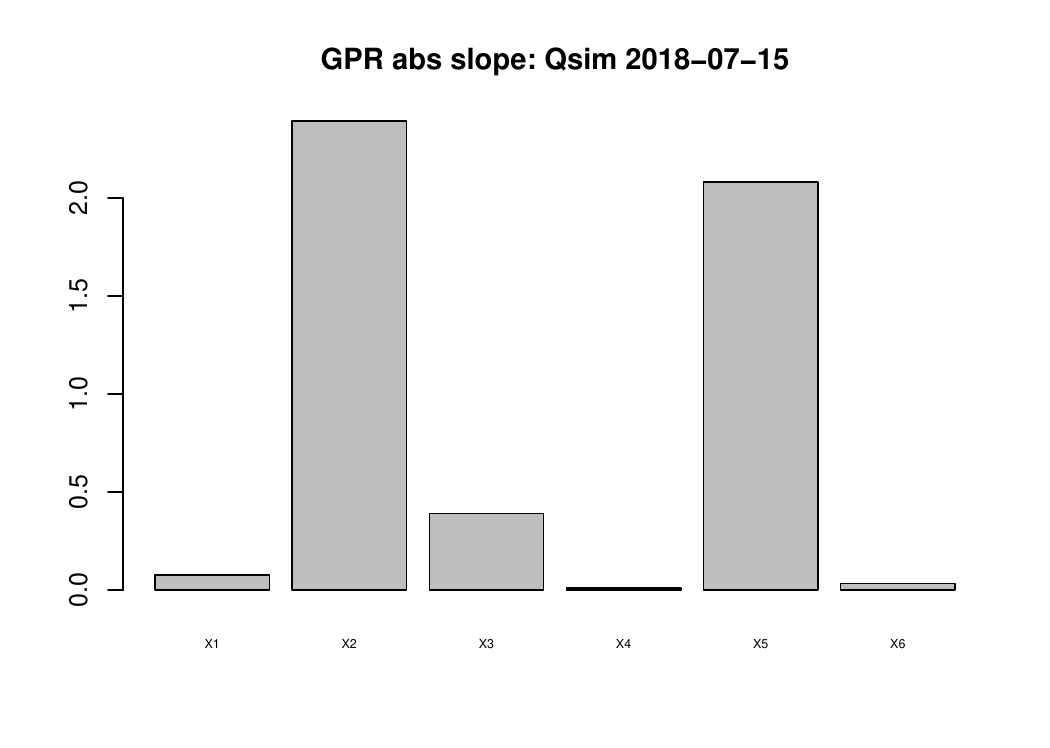}
    % \caption{GPR Inverse range and slope results.}
    \caption{GPR standardized regression coefficients.}
   \label{F:GR6J.GPR.Slope.Internal}
 \end{subfigure}   
 \caption{GR6J: Sensitivity Analyses of outflow
  (Qsim on 2018-07-15. }
 \label{F:GR6J.Qsim}
\end{figure}

\paragraph{Joint measure comparisons.} Morris and Sobol' provide two measures, and examination of both measures can provide additional insight into the influence of input parameters, in particular when interactions or higher order effects may be present. Graphical summaries, excluding VARS-TO,  which looks similar to Sobol' $T_k$, are shown in Figure \ref{F:GR6J.Qsim}. In general, as for Table \ref{T:GR6J.SA.Outflow.summary} $X_2$ and $X_5$ dominate. The Morris plot of $\sigma$ against $\mu*$ (Figure \ref{F:GR6J.Morris.pair.Flow}) indicates an interaction (or higher order effect) for these two parameters with both $\sigma$ and $\mu*$ being large. The Sobol' plots with $S_{1,k}$ and $T_k$ (Figure \ref{F:GR6J.Sobol.pair.flow}) indicate the same based on the large differences between the two. 

GPR also provides two measures, the standardized coefficients and the inverse ranges, where the latter is detecting residual influence after removing mean effects. While not shown graphically (but see Table \ref{T:GR6J.SA.Outflow.summary}), inverse range values indicate that $X_2$ and $X_5$ still have relatively high residual influence after removing the mean effects---this is probably a reflection of a linear model for the mean being inadequate, as the multiple regression results showed.  

The regression tree results show that $X_5$ has considerably more influence than $X_2$.  Closer examination of the regression tree output (Figure \ref{F:GR6J.QSim.RegTree.tree}) and a heatmap (Figure \ref{F:GR6J.QSim.RegTree.tree.heatmap}) makes clear the interaction between $X_2$ and $X_5$.  A limitation of Sobol' indices, in particular, is that when dealing with highly asymmetric distributions, variance measures may not be so informative \citep{2016_Pianosi_etal}, and the discrepancy with the regression tree results could be a reflection of the flexibility of regression trees for handling highly nonlinear relations. Further interpretation of the regression tree heatmap is given in Section   \ref{F:GR6J.QSim.RegTree.tree.heatmap}

\paragraph{TVSA and performance based GSA.}
 Section \ref{app:GR6J.Qsim.TVSA} shows the results of time-varying SA based on Sobol' sensitivities at four different dates and in all cases $X_2$ and $X_5$ dominated. A performance-based SA was carried out using Sobol' and the KGE measure for a period of one month (October 2014).  Again, $X_2$ and $X_5$ dominated.
 
\clearpage

%% file: 4_Results/4B_Results_SimplyP.tex
\newpage 
\clearpage 

\subsection{\label{subsec:Results.SimplyP}SimplyP}
Results for the sensitivity of log total dissolved phosphorus output (TDP) on day 200 of the year 2004, $TDP_{200,2004}$, to 13 input parameters are presented here. Results for three other outputs (all on log scale),  outflow ($Q$), suspended sediments ($SS$),  and particle phosphorus ($PP$), are in Section \ref{app:Results.Details.SimplyP}.  Results did vary depending on the day of year examined, as was explained previously (Section \ref{subsec:Simulators.SimplyP.desc}).  

\begin{figure}[h]
    \centering
    \includegraphics[width=0.60\linewidth]
    {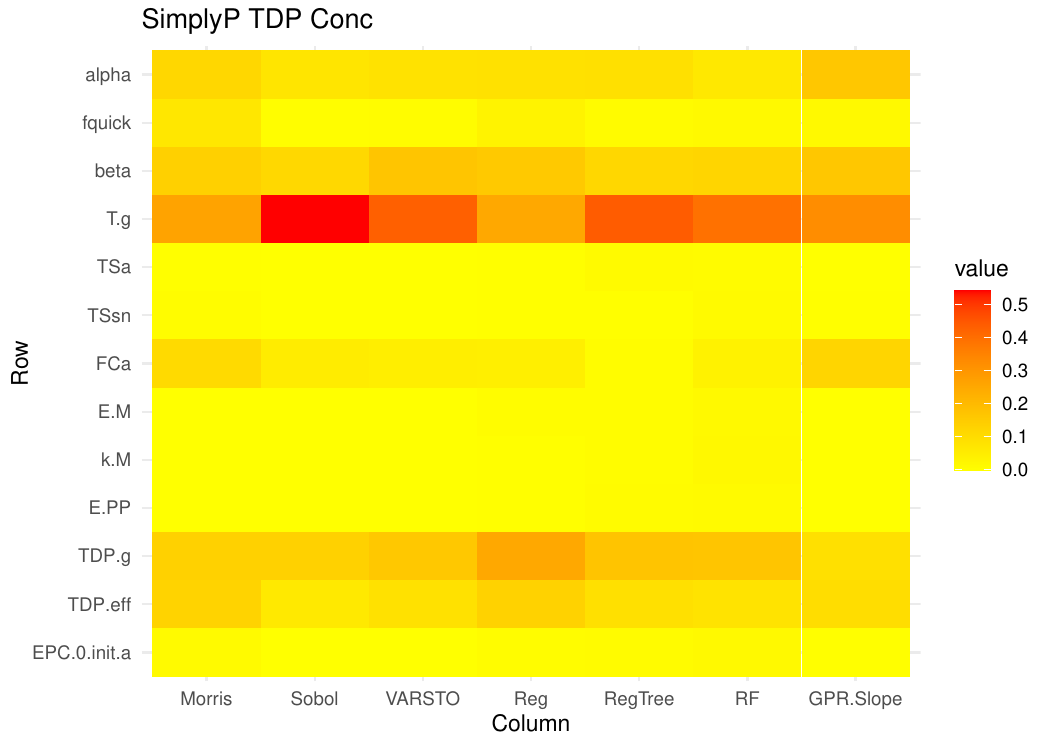}
    \caption{SimplyP: Relative parameter importance for $\log TDP_{200,2004}$ for the different SA methods.}
    \label{F:SimplyP.SA.TDP.summary}
\end{figure}

\paragraph{Scalar measure comparisons.}  
Table \ref{T:SimplyP.SA.TDP.summary} and Figure \ref{F:SimplyP.SA.TDP.summary} summarize the results of  different GSA methods, scaled as for GR6J.  For all methods, the parameter T.g (groundwater time constant) generally had the most influence, while three others, alpha (PET multiplication factor), beta (baseflow index),and TDP.g (Groundwater TDP concentration; see Table \ref{T:SimplyP.parameters}) had a sizeable influence.  

\begin{figure}[h]
 \centering
     \includegraphics[width=0.8\textwidth]{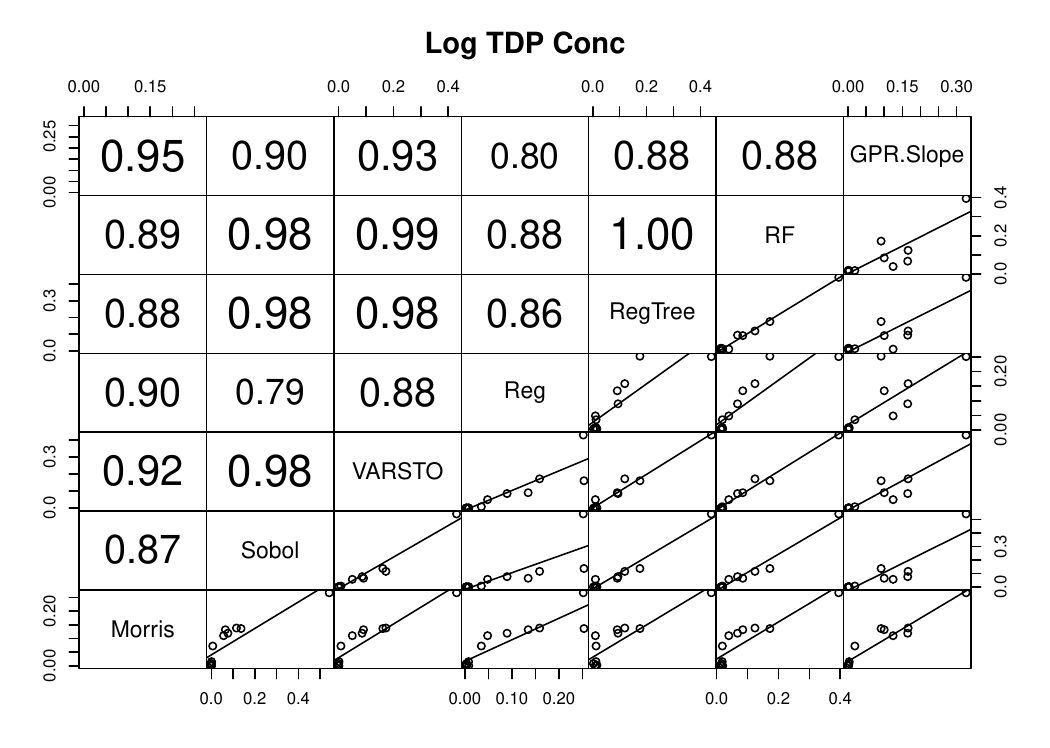}
     \caption{SimplyP: pairwise scatterplots of SA measures for the 13 parameters influence on log($TDP_{200,2004}$) for different SA procedures along with Pearson correlation coefficients. Reg=multiple regression, RegTree=regression tree, RF=random forest, GPR=Gaussian Process regression.}
     \label{F:SimplyP.Pairwise.TDP.SA}
 \end{figure}

\begin{table}[h]
\centering
 Total Dissolved Phosphorus, $TDP_{200,2004}$  \\
  \begin{tabular}{lrrrrrrrrrr}
   \hline
  &Type  & Morris & Sobol' & VARS-TO & Reg & RegTree & RF & \multicolumn{2}{c}{GPR} \\
  &        &  DGSM      &  $T_i$    &      &  &    &   &  Slope &  InvRange \\ 
   \hline
  alpha &Hyd&  0.12 & 0.08 & 0.09 & 0.09 & 0.09 & 0.07 & 0.16 & 0.15 \\ 
fquick &Hyd&  0.07 & 0.01 & 0.01 & 0.03 & 0.01 & 0.02 & 0.02 & 0.01 \\ 
 beta &Hyd&  \textbf{\tcr{0.14}} & 0.12 & \textbf{\tcr{0.17}} & 0.16 & 0.12 & 0.12 & \textbf{\tcr{0.17}}  & 0.11 \\ 
    T.g &Hyd&  \textbf{\tcb{0.27}} & \textbf{\tcb{0.54}} & \textbf{\tcb{0.43}} & \textbf{\tcb{0.25}} & \textbf{\tcb{0.44}} & \textbf{\tcb{0.35}} & \textbf{\tcb{0.33}} & \textbf{\tcb{0.52}} \\ 
TSa &Hyd&  0.00 & 0.00 & 0.00 & 0.00 & 0.02 & 0.01 & 0.00 & 0.00 \\ 
TSsn &Hyd&  0.01 & 0.00 & 0.00 & 0.00 & 0.00 & 0.02 & 0.00 & 0.00 \\ 
FCa &Hyd&  0.11 & 0.06 & 0.05 & 0.05 & 0.01 & 0.04 & 0.12 & \textbf{\tcr{0.17}} \\ 
E.M &Sed & 0.00 & 0.00 & 0.00 & 0.01 & 0.01 & 0.02 & 0.00 & 0.00 \\ 
k.M &Sed & 0.00 & 0.00 & 0.00 & 0.00 & 0.02 & 0.02 & 0.00 & 0.00 \\ 
E.PP&Pho & 0.00 & 0.00 & 0.00 & 0.00 & 0.01 & 0.02 & 0.00 & 0.00 \\ 
TDP.g &Pho& \textbf{\tcr{0.14}} & \textbf{\tcr{0.14}} & 0.16 & \textbf{\tcb{0.25}} & \textbf{\tcr{0.18}} & \textbf{\tcr{0.17}} & 0.09 & 0.00 \\ 
TDP.eff &Pho& 0.13 & 0.06 & 0.09 & 0.13 & 0.09 & 0.08 & 0.10 & 0.04 \\ 
EPC.0.init.a &Pho& 0.02 & 0.00 & 0.00 & 0.01 & 0.01 & 0.02 & 0.00 & 0.00 \\ 
     \hline
 \end{tabular}
\caption{SimplyP: summary of results for different SA methods applied to $\log TDP_{200,2004}$. Numbers in blue denote those with the largest relative value and those in red are the second largest. Reg=multiple regression, RegTree=regression tree, RF=random forest, GPR=Gaussian Process regression.}
\label{T:SimplyP.SA.TDP.summary}
\end{table}
 Pairwise comparisons between the SA methods in terms of the SA measures attached to each parameter are shown in Figure \ref{F:SimplyP.Pairwise.TDP.SA}. A positive linear relationship, of varying strength from $r$=0.79 to $r$=0.98, occurred for all pairs. The overall degree of similarity was relatively high based on the Kendall W statistic of 0.89.
 

\clearpage 

\begin{figure}[h]
    \centering
    \begin{subfigure}[b]{0.45\textwidth}
    \includegraphics[width=\columnwidth,height=0.30\textheight]{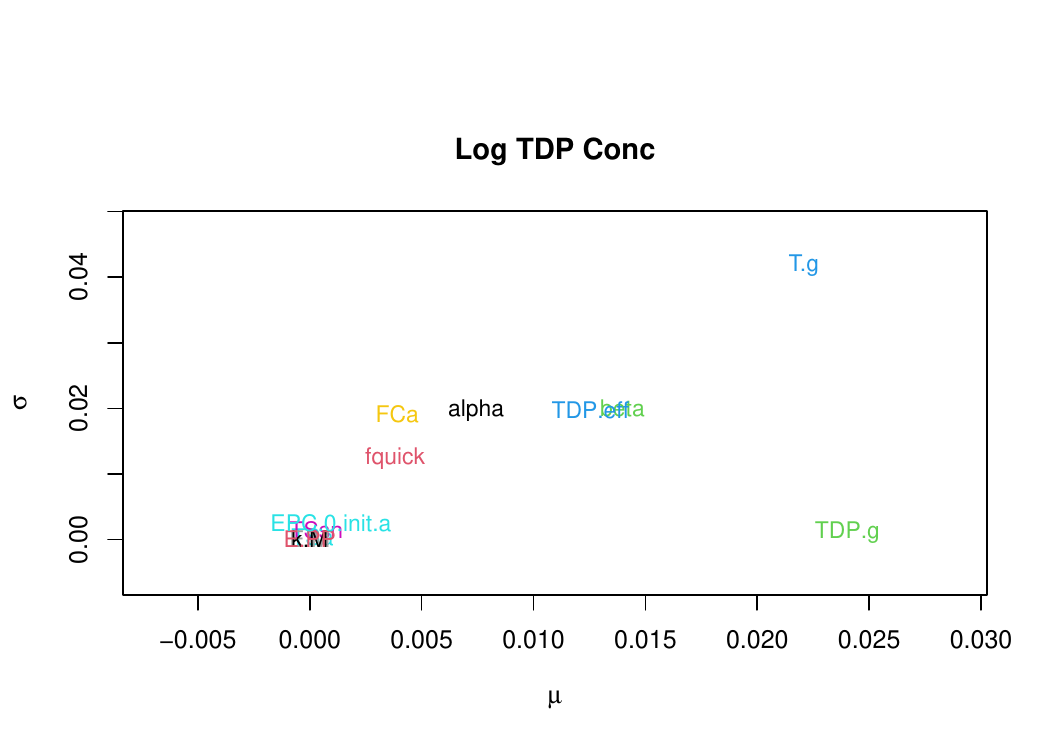}
    \caption{Morris measures, $\sigma$ versus $\mu^*$.}
    \label{F:SimplyP.Morris.TDP.Internal}
    \end{subfigure}
    \hfill
    \begin{subfigure}[b]{0.45\textwidth}
    \centering 
        \includegraphics[width=\columnwidth,height=0.30\textheight]{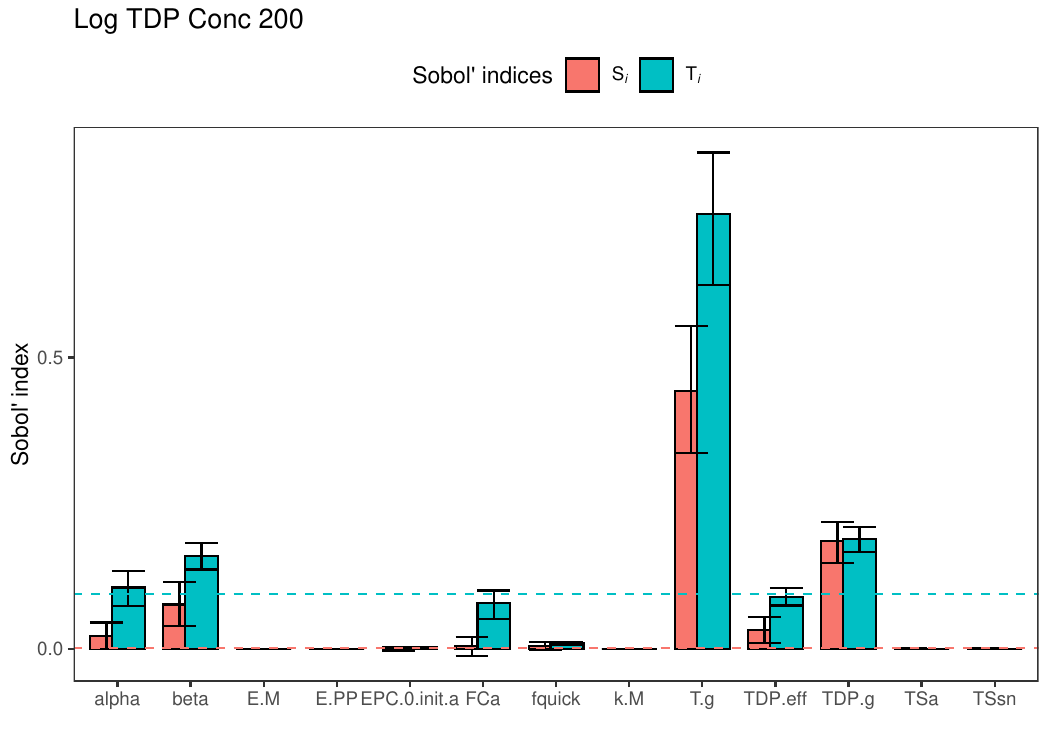}
     \caption{Sobol $S_{1,k}$ and $T_k$.}
     \label{F:SimplyP.Sobol.TDP.Internal}
    \end{subfigure}
% --------------------------------------------------------------

   \begin{subfigure}[b]{0.45\textwidth}
    \centering
    \includegraphics[width=\columnwidth,height=0.30\textheight]{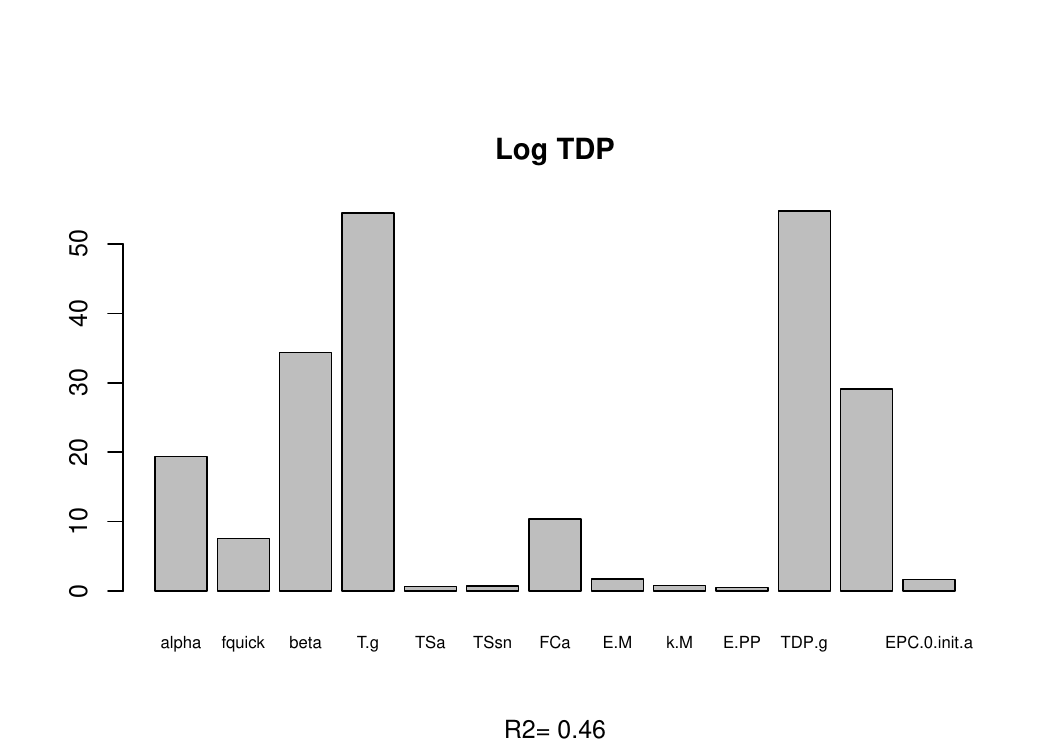}
    \caption{Multiple Regression standardized regression coefficients.}
    \label{F:SimplyP.Regression.TDP.Internal}
 \end{subfigure}   
\hfill
    \begin{subfigure}[b]{0.45\textwidth}
    \centering
    \includegraphics[width=\columnwidth,height=0.30\textheight]{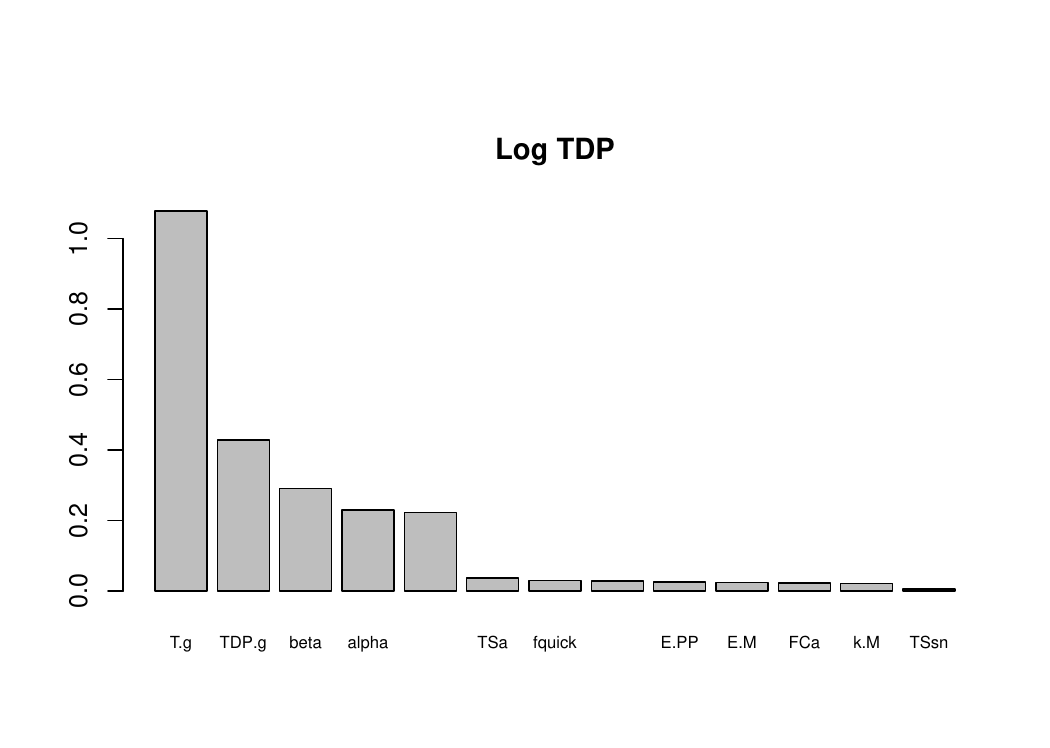} 
     \caption{Regression Tree parameter importance.}
     \label{F:SimplyP.RegTree.TDP.Internal}
    \end{subfigure}   
% --------------------------------------------------------------

   \begin{subfigure}[b]{0.45\textwidth}
    \centering
    \includegraphics[width=\columnwidth,height=0.30\textheight]{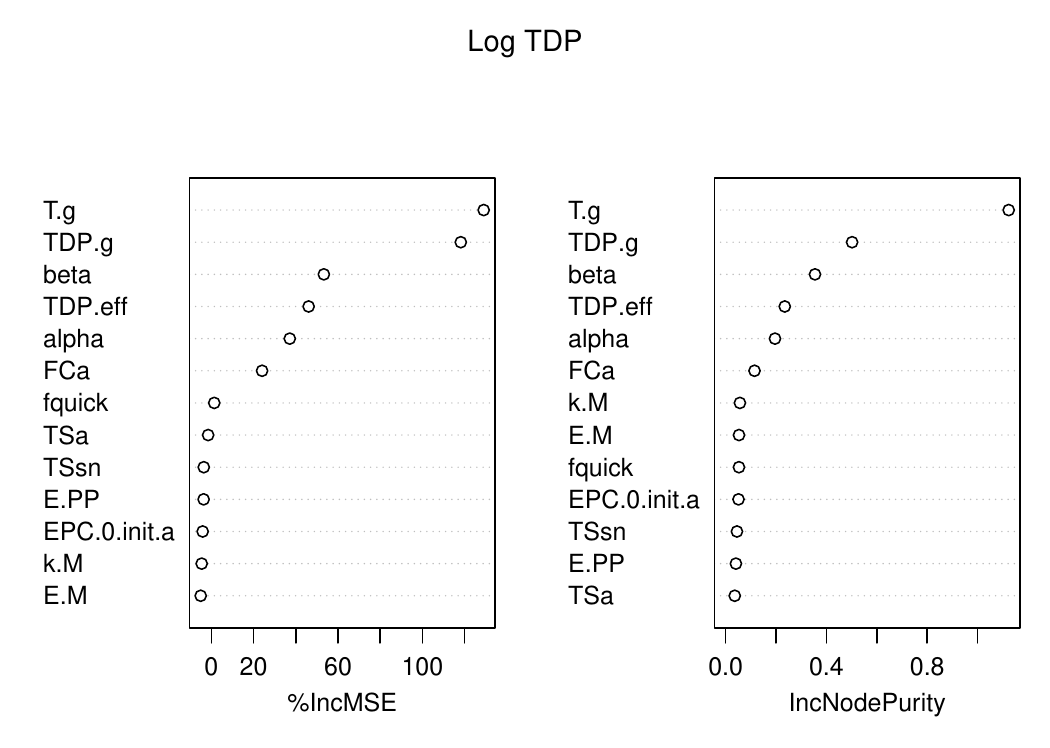}
    \caption{Random Forests parameter importance.}
    \label{F:SimplyP.RF.TDP.Internal}
 \end{subfigure}   
 \hfill
   \begin{subfigure}[b]{0.45\textwidth}
    \centering
    \includegraphics[width=\columnwidth,height=0.30\textheight]{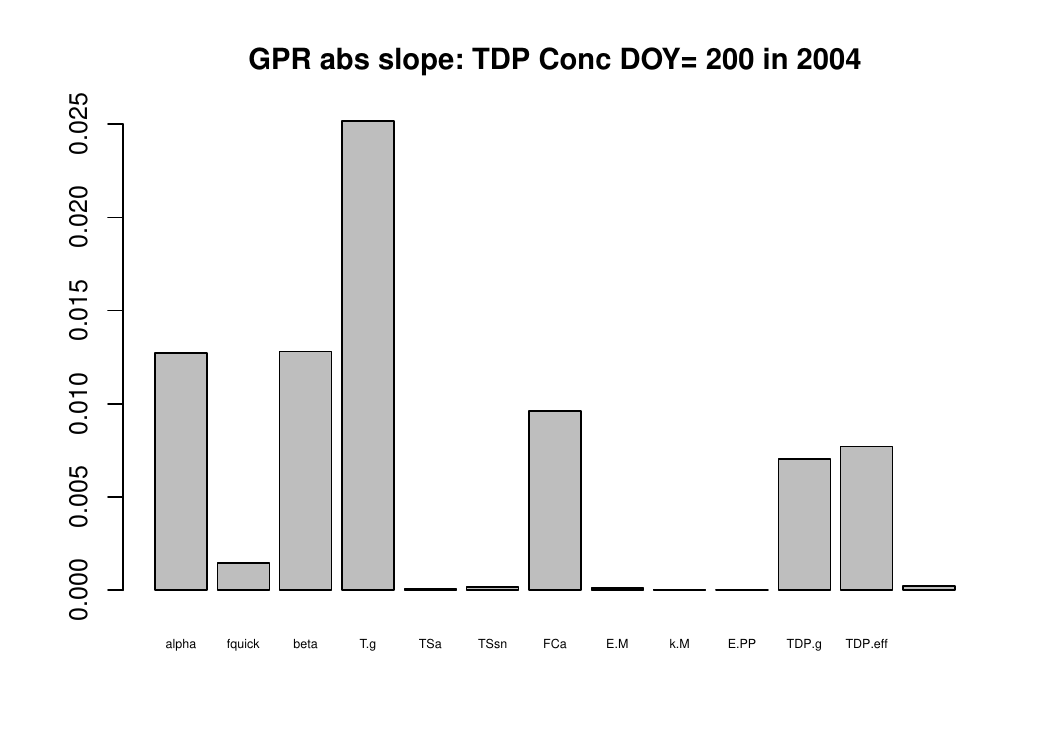}
    \caption{GPR standardized regression coefficients.}
    \label{F:SimplyP.GPR.Slope.TDP.Internal}
 \end{subfigure}   
 \caption{SimplyP: Sensitivity Analyses of
 Log TDP concentrate on Day 200 in 2004.}
 \label{F:SimplyP.TDP.Internal}
\end{figure}

%-------------------------------------------------
\paragraph{Joint measure comparisons.}
Joint GSA measures for Morris and Sobol' are shown in Figure \ref{F:SimplyP.TDP.Internal} along with results for four other methods.  Again, VARS-TO results are not shown as they are nearly identical to the Sobol $T_i$ values.   While T.g is shown to generally dominate, the Morris plot and the Sobol' plot indicate some interactions and/or higher order effects for this parameter. The regression tree heatmap (Figure \ref{F:SimplyP.TDP.RegTree.tree.heatmap}) provides insight into the nature of interactions, namely between T.g and TDP.g. 

\clearpage

%% file: 4_Results/4C_Results_STICS.tex
\subsection{\label{subsec:Results.STICS}STICS}
%\enlargethispage{11in}
%\newgeometry{paperwidth=11in,paperheight=13in}

Results of sensitivity analysis of STICS output of $mafruit$, the biomass of harvested organs in tonnes/ha, to 13 input parameters are presented in this section, with results for three other outputs, $masec.n$, $CNgrain$ and $CNplante$ in Section \ref{app:Results.Details.STICS}. 

\paragraph{Scalar measure comparisons.}

\begin{figure}[h]
    \centering
    \begin{subfigure}[b]{0.45\textwidth}
       \includegraphics[width=\columnwidth,height=0.30\textheight]{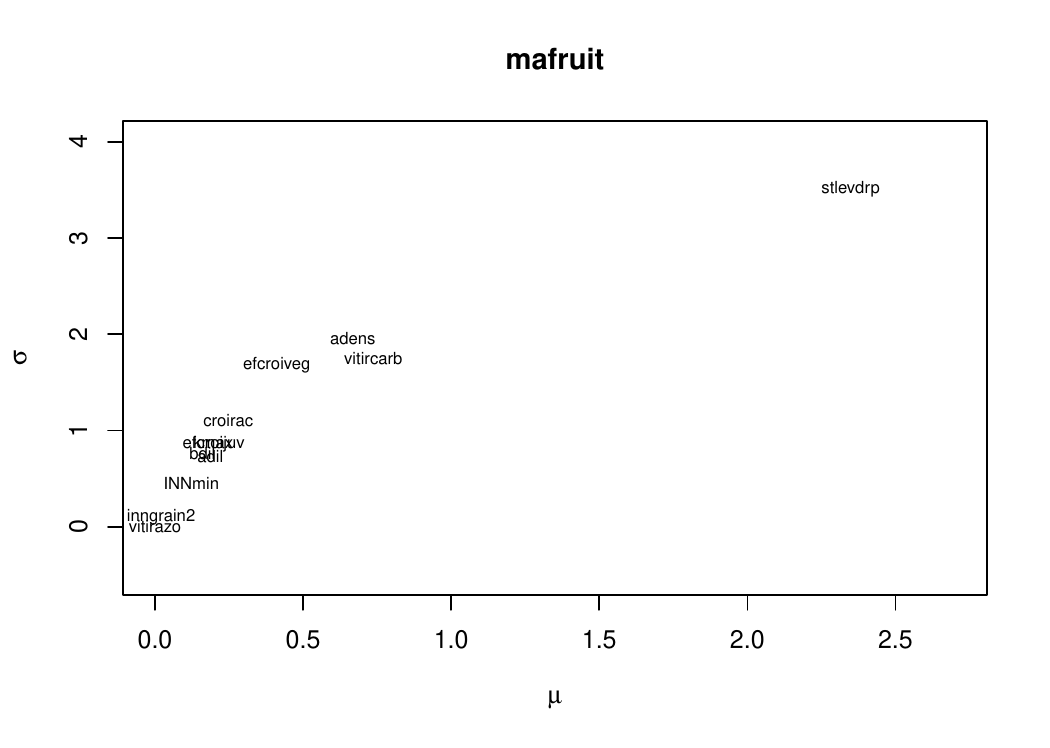}
    \caption{Morris measures, $\sigma$ versus $\mu^*$.}
    \label{F:STICS.Morris.mafruit.Internal}
    \end{subfigure}
    \hfill
    \begin{subfigure}[b]{0.45\textwidth}
    \centering 
        \includegraphics[width=\columnwidth,height=0.30\textheight]{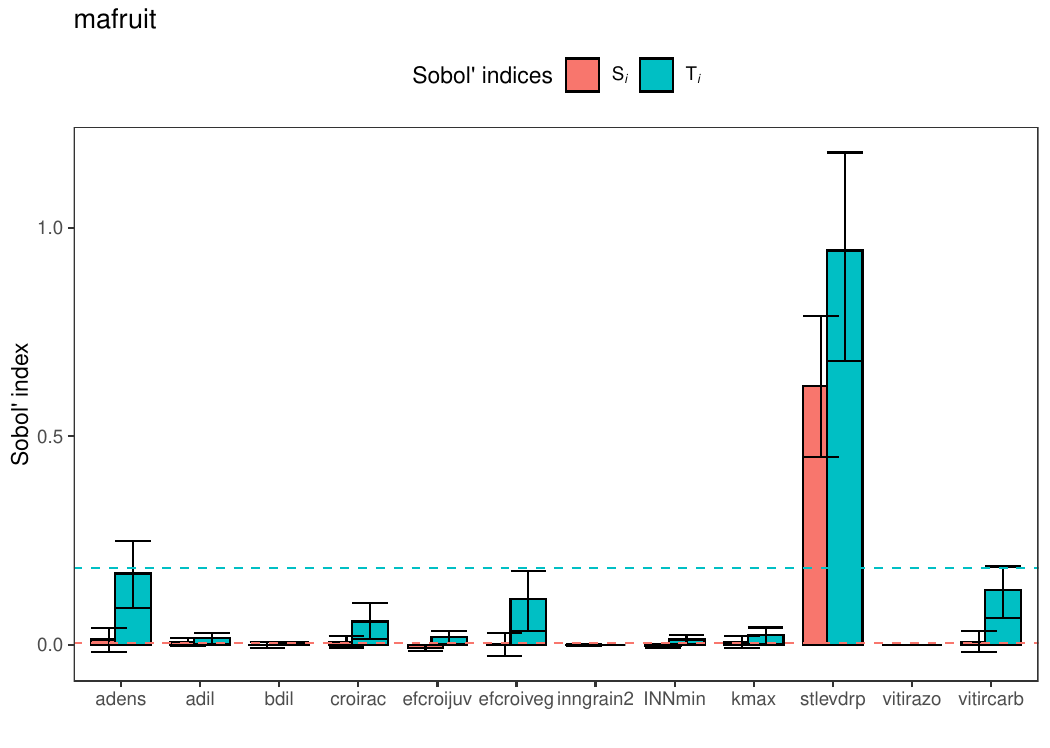}
     \caption{Sobol $S_{1,k}$ and $T_k$.}
     \label{F:STICS.Sobol.mafruit.Internal}
    \end{subfigure}
% --------------------------------------------------------------

   \begin{subfigure}[b]{0.45\textwidth}
    \centering
    \includegraphics[width=\columnwidth,height=0.30\textheight]{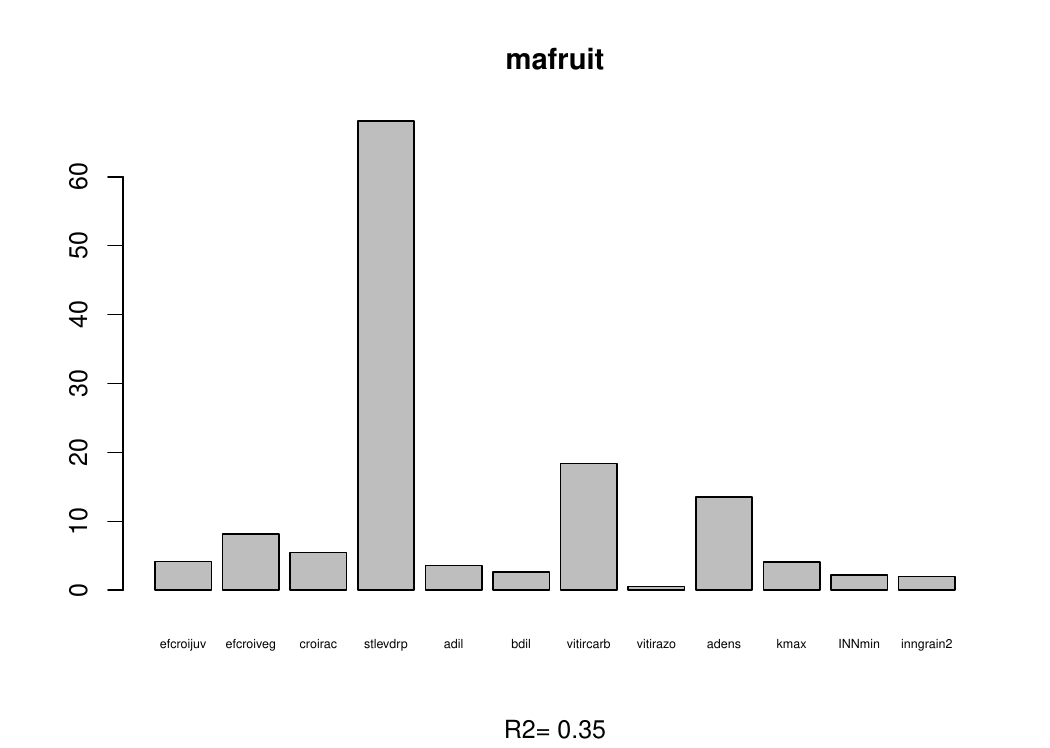}
    \caption{Multiple Regression standardized regression coefficients.}
    \label{F:STICS.Regression.mafruit.Internal}
 \end{subfigure}   
\hfill
    \begin{subfigure}[b]{0.45\textwidth}
    \centering
    \includegraphics[width=\columnwidth,height=0.30\textheight]{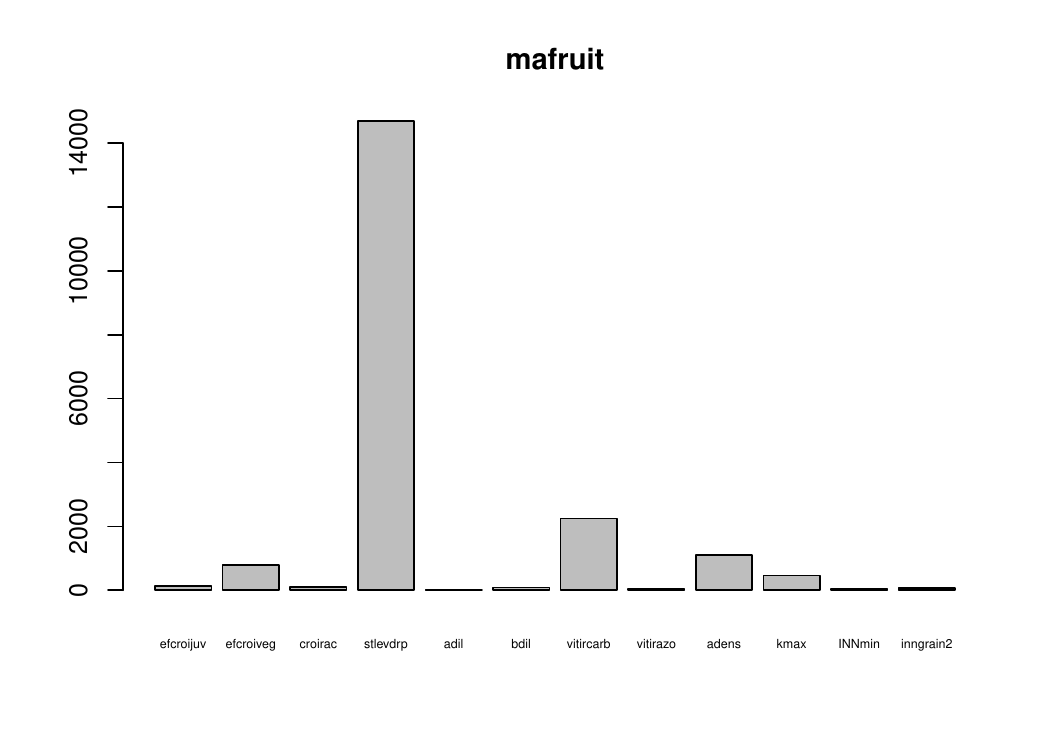} 
     \caption{Regression Tree parameter importance.}
     \label{F:STICS.RegTree.mafruit.Internal}
    \end{subfigure}   
% --------------------------------------------------------------

   \begin{subfigure}[b]{0.45\textwidth}
    \centering
    \includegraphics[width=\columnwidth,height=0.30\textheight]{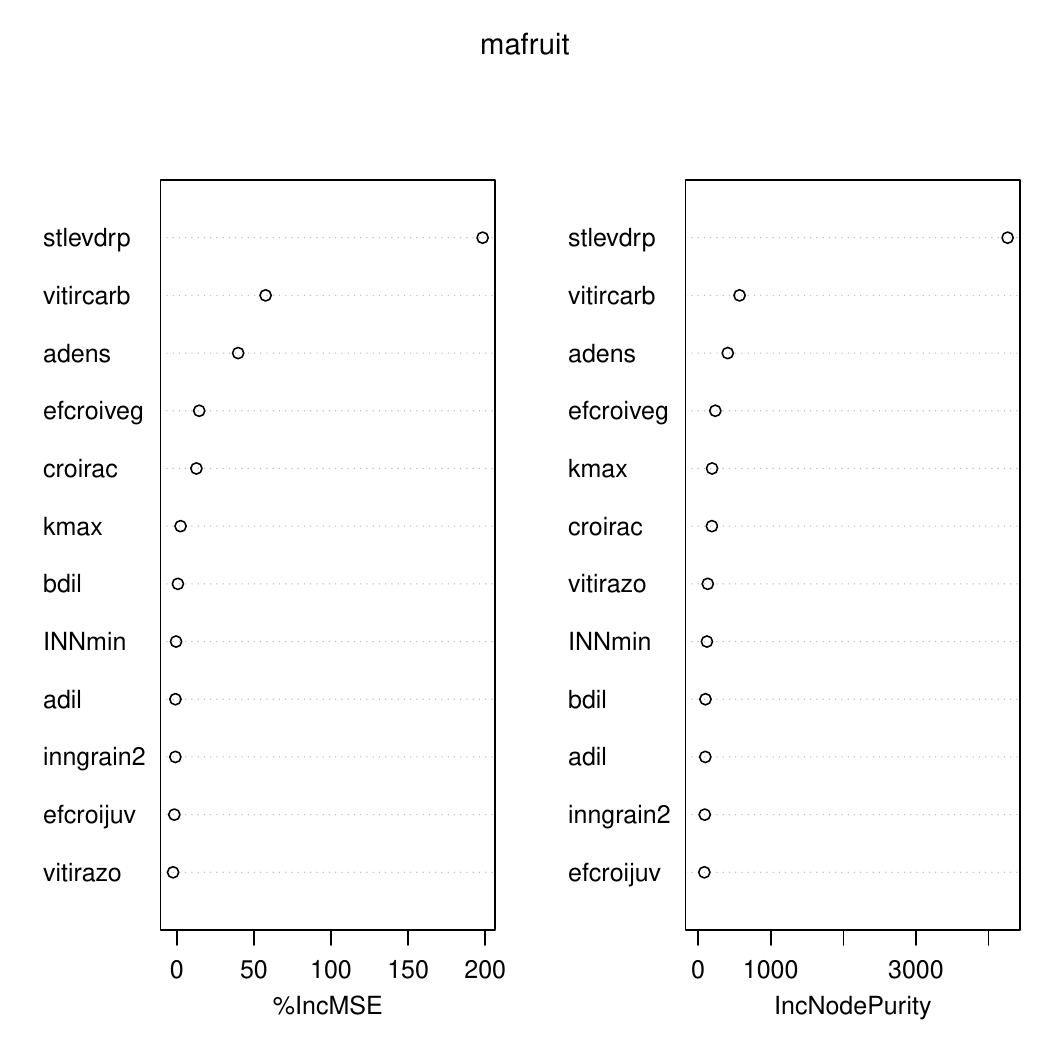}
    \caption{Random Forests parameter importance.}
    \label{F:STICS.RF.mafruit.Internal}
 \end{subfigure}   
 \hfill
   \begin{subfigure}[b]{0.45\textwidth}
    \centering
    \includegraphics[width=\columnwidth,height=0.30\textheight]{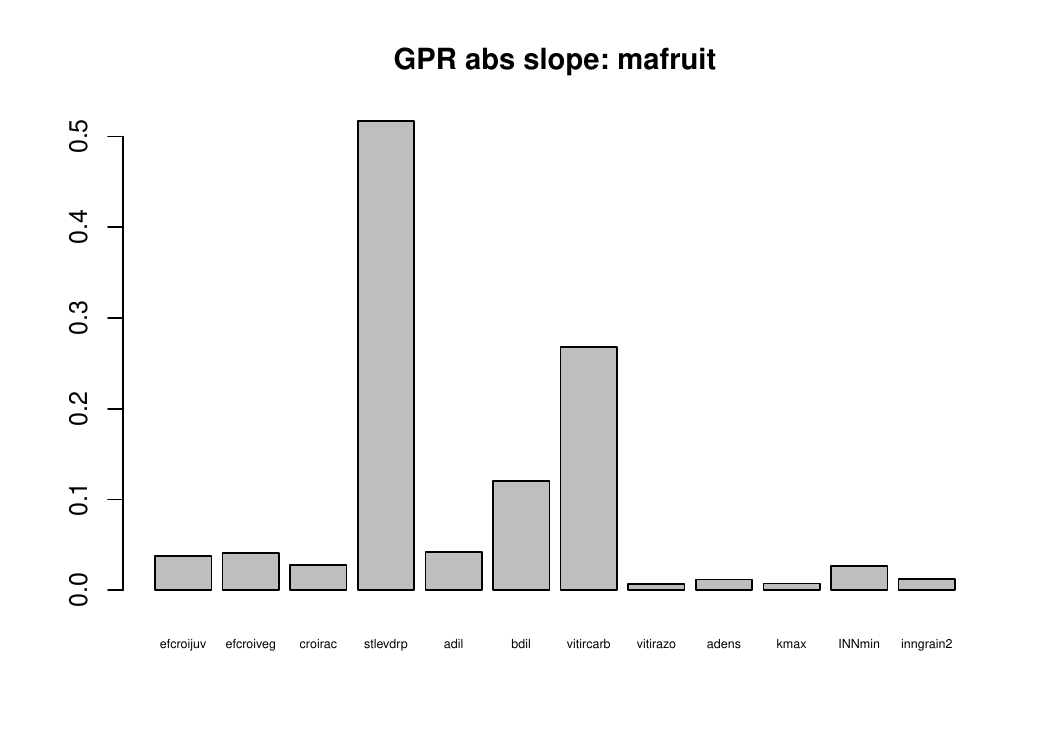}
    \caption{GPR standardized regression coefficients.}
    \label{F:STICS.GPR.Slope.mafruit.Internal}
 \end{subfigure}   
 \caption{STICS: Sensitivity Analyses of
 mafruit on day of harvest.}
 \label{F:STICS.mafruit.Internal}
  \end{figure}

Table \ref{T:STICS.SA.summary} lists the SA measures for each method across all the parameters.  For all measures, stlevdrp (cumulative thermal time between the  emergence starting date; see Table \ref{T:STICS.input.parameters} ) is consistently the most influential parameter for $mafruit$. The parameter stlevdrp equates to the amount of biomass accumulated during the period between emergence and the start of grain filling, but it will also be affected by the availability of water and nutrients.  The next most important parameters were vitircarb (rate of increase of the C harvest index vs time, g grain/ g biomass) and adens (interplant competition parameter). Visual displays of these results based on the methods are shown in Figure \ref{F:STICS.mafruit.Internal}.  

 \begin{table}[ht]
 \centering
 \begin{tabular}{lrrrrrrrrr}
   \hline
& Morris & Sobol' & VARS-TO & Reg & RegTree & RF & \multicolumn{2}{c}{GPR} \\
          & DGSM  &  $T_i$  &      &  &    &   &  Slope &  InvRange \\
   \hline
 efcroijuv & 0.06 & 0.01 & 0.01 & 0.03 & 0.01 & 0.02 & 0.03 &0.02\\ 
   efcroiveg & 0.12 & 0.07 & 0.03 & 0.06 & 0.04 & 0.04 & 0.04 & 0.04 \\ 
   croirac & 0.08 & 0.04 & 0.04 & 0.04 & 0.01 & 0.02 & 0.02 & 0.04\\ 
   stlevdrp & \textbf{\tcb{0.28}} & \textbf{\tcb{0.63}} & \textbf{\tcb{0.58}} & \textbf{\tcb{0.51}} & \textbf{\tcb{0.74}} & \textbf{\tcb{0.68}} &  \textbf{\tcb{0.46}} &  \textbf{\tcb{0.58}}\\ 
   adil &0.05 & 0.01 & 0.01 & 0.03 & 0.00 & 0.02 & 0.04 & 0.02 \\ 
   bdil & 0.05 & 0.00 & 0.00 & 0.02 & 0.00 & 0.01 & 0.11 & 0.06 \\ 
   vitircarb & 0.13  & 0.09 & \textbf{\tcr{0.19}} & \textbf{\tcr{0.14}} & \textbf{\tcr{0.11}} & \textbf{\tcr{0.09}} & \textbf{\tcr{0.24}} & \textbf{\tcr{0.10}}\\ 
   vitirazo & 0.00 & 0.00 & 0.00 & 0.00 & 0.00 & 0.01 & 0.01 & 0.00 \\ 
   adens & \textbf{\tcr{0.14}} & \textbf{\tcr{0.11}} & 0.07 & 0.10 & 0.06 & 0.07 & 0.01 & 0.09\\ 
   kmax & 0.06 & 0.02 & 0.06 & 0.03 & 0.02 & 0.02 & 0.01 &  0.02 \\ 
   INNmin & 0.03 & 0.01 & 0.01 & 0.02 & 0.00 & 0.01 & 0.02 & 0.02 \\ 
   inngrain2 & 0.01 & 0.00 & 0.00 & 0.01 & 0.00 & 0.01 & 0.01 & 0.00 \\ 
    \hline
 \end{tabular}
 \caption{STICS: summary of results for different SA methods applied to $mafruit$ on the day of harvest. Numbers in blue denote those with the largest relative value and those in red are the second largest. Reg=multiple regression, RegTree=regression tree, RF=random forest, GPR=Gaussian Process regression.}
\label{T:STICS.SA.summary}
 \end{table}

\begin{figure}[h]
    \centering
    \includegraphics[width=0.60\linewidth]{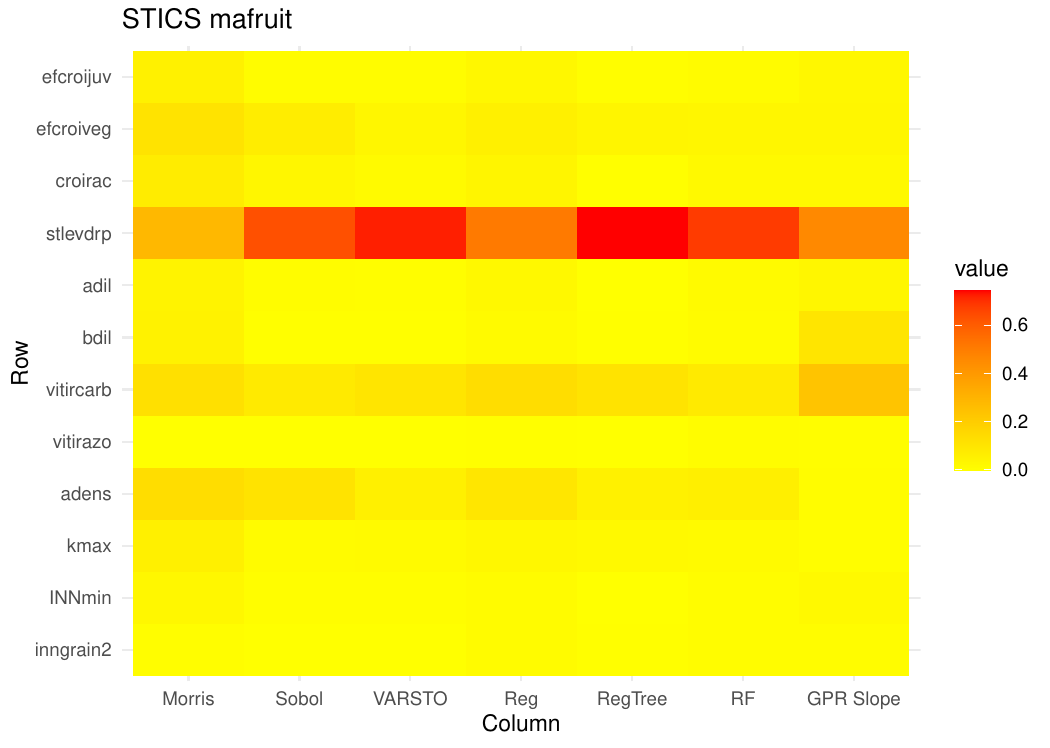}
    \caption{STICS: Relative parameter importance for mafruit for the different SA methods.}
    \label{F:STICS.SA.summary.mafruit}
\end{figure}

\begin{figure}[h]
 \centering
     \includegraphics[width=0.8\textwidth]{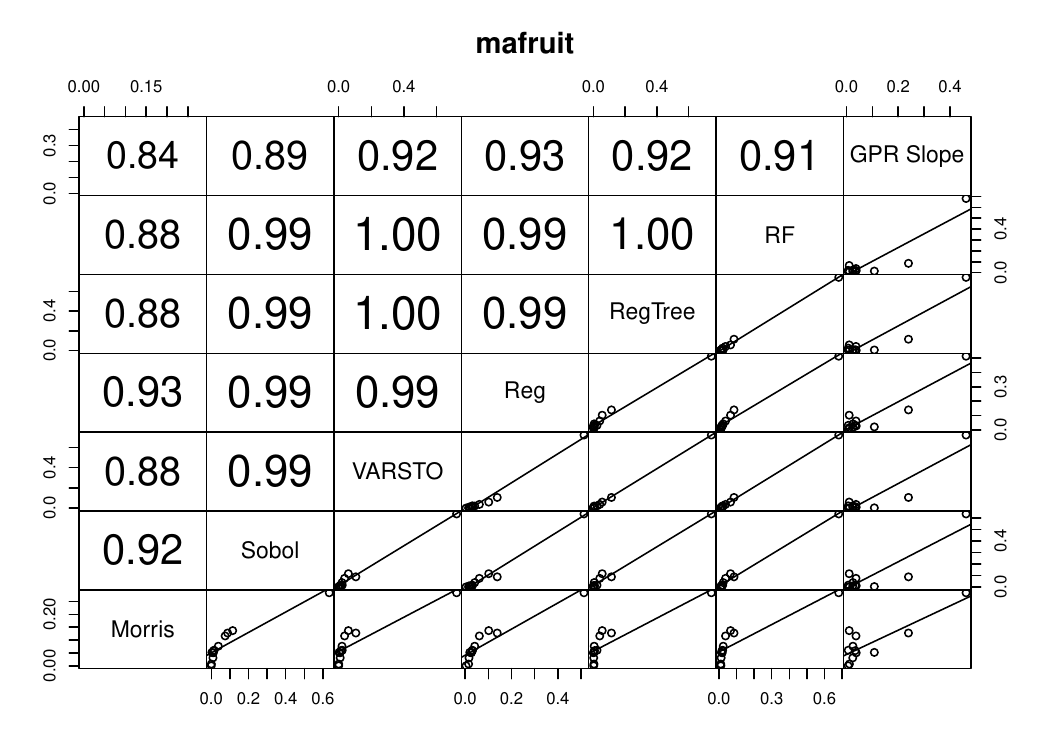}
     \caption{STICS: pairwise scatterplots of SA measures for the parameters' influence on mafruit for different SA procedures along with Pearson correlation coefficients. Reg=multiple regression, RegTree=regression tree, RF=random forest, GPR=Gaussian Process regression.}
     \label{F:STICS.Pairwise.mafruit.SA}
 \end{figure}

The pairwise similarities between the methods are shown in Figure \ref{F:STICS.Pairwise.mafruit.SA} along with Pearson correlation coefficients which ranged from 0.84 to 1.00.  The scatterplots include a linear regression line that fits the data fairly well, although the line is clearly highly influenced by the single largest values, namely those for stlevdrp. Kendall's W is 0.83, with a p-value for the null hypothesis of no concordance $<$0.001, indicating a relatively high degree of concordance across all the measures.

% \clearpage 

\paragraph{Joint measure comparisons.}
Joint GSA measures for Morris and Sobol' are shown in Figure \ref{F:STICS.mafruit.Internal} along with results for four other methods.  Again, VARS-TO results are not shown. stlevdrp dominates, followed by vitircarb and adens. The Morris plot and the Sobol' plot suggest some interactions and/or higher order effects for stlevdrp. The regression tree heatmap (Figure \ref{F:STICS.mafruit.RegTree.tree.heatmap}) provides insight into the nature of interactions, namely stlevdrp, vitircarb, and adens. 

\clearpage

%% file: 5_Discussion/5_Discussion.tex
\section{\label{sec:Discussion}Discussion}
 
 Our general motivation was to explore the application of exemplar global sensitivity analysis (GSA) methods selected from  different categories identified by \citet{2021_Razavi_etal} to simulators of differing complexity. Our target audience is scientists who are not experts in GSA but would like to apply it. We want to encourage wider routine application of GSA in the design of simulation experiments as part of best practices recommended by \citet{2021_Razavi_etal} and to carry out  sensitivity analysis properly \citep{2019_Saltelli_etal}. Our exploration included comparing the degree of similarity in Factor Ranking, ease of implementation, and ease of interpretation of results. In the course of this work, we found commonalities in the implementation of the methods, which led to a 10 step workflow for practitioners (Section \ref{sec:Methods.Implementation}).   Below, we address the questions posed in Section \ref{sec:Intro}.

\subsection{Advantages and disadvantages of different GSA methods}
 For Factor Ranking, Factor Screening, and Factor Fixing, we found Sobol' first  order sensitivity, $S_{1,k}$, and Sobol' total sensitivity, $T_k$, the easiest to comprehend, and the comparison of $T_k$ and $S_{1,k}$ the most informative measure of interactions and higher order effects. Amongst GSA practitioners, Sobol' measures have a long history and are indeed often the preferred method \citep{2024_Tarantola_etal}.  We acknowledge that Sobol' measures can perform poorly with highly skewed output and multiple modes \citep{2007_Borgonovo, 2014_borgonovo_etal}, and measures that overcome those limitations have been developed \citep{2015_Pianosi_Wagener}, but we did not include them.  
 
 For the Morris method,  the derivative global sensitivity measure, which combines the $\mu^*$ and $\sigma$ measures, was used for Factor Ranking. For assessing the presence of interactions and higher order effects, we examined plots of $\sigma$ against $\mu^*$, which is a standard practice.  We found comparing differences in Sobol' $T_k$ and $S_{1,k}$ somewhat easier than examining the Morris plot, but the computational expense can be higher than for the Morris method if Sobol' 2nd order and higher sensitivities are also evaluated. \citet{2009_Sobol_Kucherenko} and \citet{2016_Kucherenko_+_Song} make explicit links between Sobol' sensitivities and the Morris measures, with the latter discussing DGSM. 

 The variogram-based VARS-TO measure was essentially equivalent to Sobol $T_k$. We recognize that inclusion of the $IVARS_{kp}$ \citep{2016_Razavi_Gupta} would have presented a fairer picture of the power of the variogram approach, but the lack of \texttt{R} software was a disadvantage. 

 The four regression-based approaches had the advantage that general space-filling algorithms, in particular Latin Hypercube Sampling, could be used to generate the parameter combinations, in contrast to the above methods that use specialized procedures.  Adding more combinations is potentially simpler if one generates an oversample of combinations to begin with. 
 
 Of the regression-based approaches, the multiple regression approach using the standardized regression coefficients (SRCs) provides the most intuitively understandable measure. However, its suitability for quantifying the absolute importance of parameter inputs depends on the quality of the linear approximation, and in general, based on $R^2$ values, the multiple regression fits were often quite poor. The regression tree and random forests were much less restrictive, given their ability to deal with nonlinearities. In general, but not always,   regression trees and random forests  fit the output better than multiple regression, with random forests often better than regression trees (Table \ref{T.R2.summary}).  The heatmap display of a regression tree is useful for detecting interactions between input parameters, but it can be very computationally expensive to construct; e.g., with $N$=3,000 parameter combinations it could take three to four hours. The two Gaussian Process Regression-based measures, SRCs and normalized inverses of the range parameters (modelling residual effects) do, in combination, provide additional insights, but the computational expense was extremely high and the number of parameter combinations was quite restricted, e.g., $N$=300. 

\begin{table}[h]
    \centering
  \begin{tabular}{lrrr}
   \multicolumn{4}{c}{GR6J} \\
Output & Reg & RegTree & R Forest \\ \hline
Qsim & 0.28 &  0.91 & 0.99 \\
Pr   & 0.98 &  0.98 & 1.00 \\
Q9   & 0.89 &  0.92 & 0.99 \\
\\
   \multicolumn{4}{c}{SimplyP} \\
Output & Reg & RegTree & R Forest \\ \hline
TDP  & 0.46 &  0.73 & 0.86 \\
Flow & 0.89 &  0.83 & 0.98 \\
SS   & 0.82 &  0.74 & 0.96 \\ 
PP   & 0.48 &  0.73 & 0.88 \\
\\
   \multicolumn{4}{c}{STICS} \\
Output     & Reg & RegTree & R Forest \\ \hline
mafruit    & 0.35 &  0.80 & 0.90 \\
masec.n    & 0.63 &  0.72 & 0.94 \\
CNgrain    & 0.21 &  0.77 & 0.82 \\ 
CNplante   & 0.73 &  0.57 & 0.91 \\
\end{tabular}
    \caption{Adjusted $R^2$/percent of variance accounted for with different regression based analyses, multiple regression (Reg), regression tree (RegTree), and random forest (R Forest), for different outputs of the three simulators.}
    \label{T.R2.summary}
\end{table}

\subsection{Implementation difficulties and computational costs}
Some differences in implementation difficulties were just mentioned. Morris, Sobol', and variogram-based used method-specific space-filling algorithms, and  \texttt{R} code for the full range of IVARS$_{k,p}$ measures would have been helpful. 

The primary computational cost, which has little to do with the GSA procedure, is simply carrying out the simulator runs. As mentioned already, GPR  and generating heatmaps for the regression tree output were the most computationally expensive procedures. We found that in general the computational expense is more a function of the number of simulator runs, $N$, than the cost of the GSA procedure. A GSA method that works satisfactorily with low $N$ is likely to be the most computationally appealling overall, even if the SA procedure itself takes some time to run For other simulators where the computational costs are simply too high for even a relatively modest number of runs, then building emulators may be required, although the mapping of parameters in the simulator to the parameters in the emulator are often not one to one.

\subsection{Degree of similarity in Factor Rankings}
Based on a measure of inter-rater  reliability, Kendall's W, we generally found the rankings of the relative importance of the input parameters highly similar for the different SA measures. The multiple regression results were at times the least similar, as were the Gaussian process regression results. 

As Factor Ranking is a measure of relative importance,  absolute importance is more important to know for Factor Screening and Factor Fixing.  Scaling the different GSA measures to sum to 1.0 was an easy means of quantifying absolute importance. While multiple regression SRCs often had rankings similar to those of the other procedures and measures, the scaled values, and thus absolute importance, could be quite different. For example, referring to the analysis of the TDP output from SimplyP shown in Table \ref{T:SimplyP.SA.TDP.summary}, the multiple regression SRC and Sobol' $T_k$ rankings are quite similar, but Sobol' indicates that one parameter, T.g, clearly dominates, with T.g having about three times the weight, while SRC  gives equal weight to T.g and TDP.g. 

\subsection{Challenge of multiple outputs}
All the simulators produced time series outputs, and the selection of times of outputs to conduct SA on was potentially problematic. STICS was the exception in that a single time point, namely the state of the system at the time of harvest, in particular the yield, was of prime interest, although biomass over time could have been examined.   For GR6J and SimplyP, however, a handful of scalar-valued outputs at different times were examined.  An alternative is to summarize over a sequence of daily outputs, e.g., a monthly average, or in the case of external GSA, one could use a performance-based summary measure such as Kling-Gupta Efficiency (KGE) (Section \ref{sec:Methods.Implementation}). 

A complication with applying GSA multiple times to multiple points in time series output is that the relative importance of input parameters can vary, and did for some outputs from SimplyP. This can be due to the cumulative effects and current values of the daily environmental inputs, e.g., during droughts or low precipitation periods, one parameter becomes more influential, while during rainy periods, another parameter becomes more influential. Finding such differences can provide more insight into the inner workings of what may at times be somewhat of a black box simulator, and we would recommend carrying out SA on outputs at multiple points in time for this reason.

\subsection{Similarity of conclusions made by previous papers}
  \citet{2007_Cariboni_etal} in their comparison of GSA methods also stated a preference for Sobol' sensitivities when computationally feasible. They also address an issue that arose for us, of choosing the time point for SA in the case of time series outputs---noting how the importance of input parameters can vary at different time points. 
  \citet{2016_Pianosi_etal} also found important the choice of parameter bounds in the space-filling algorithms.   
  
\subsection{Summary remarks}
  For each simulator, selection of input parameters and their ranges, and the outputs to focus on was an iterative process that required guidance from subject matter specialists.  Determination of a sufficient number of parameter combinations $N$ for achieving relatively robust results was also an iterative process.   While the algorithms for generating input parameter combinations, i.e., space-filling sampling, differed between the SA methods, thanks to freely available and computationally efficient software, the computational costs were roughly equivalent for the methods, with the exception of Gaussian Process regression.  For Factor Ranking, the degree of similarity between the SA methods was relatively similar for all three simulators based on Kendall's W and pairwise scatterplots of the sensitivity measures. The simplest procedure of multiple linear regression often yielded rankings similar to more sophisticated measures. The Sobol' sensitivities, while potentially more computationally expensive, provided the most easily interpretable information about main effects of input parameters along with interactions and higher order effects, and freely available software included estimates of uncertainty in the sensitivities. Regression trees and random forests, which provided importance rankings quite similar to Sobol', are more flexible than multiple linear regression as they can include nonlinearities in the relationship between parameter inputs and simulator outputs, and close examination of the branching points, nodes, in the trees can potentially yield more insight into the simulator processes.

%% file: 6_Appendices/1_GSA_Details/6A_GSA_Details.tex
\section{\label{app.methods}Further Details on GSA Methods}

\input{6_Appendices/1_GSA_Details/6AB_Derivatives}

\input{6_Appendices/1_GSA_Details/6AC_Distribution}

\input{6_Appendices/1_GSA_Details/6AD_Variogram}

\input{6_Appendices/1_GSA_Details/6AF_Reg_RegTree}

\input{6_Appendices/1_GSA_Details/6AH_Reg_GPR}

\input{6_Appendices/1_GSA_Details/6AI_ComputerRunDetails}

%% file: 6_Appendices/1_GSA_Details/6AB_Derivatives.tex
\subsection{\label{app.Derivative.details}Derivative-based: Morris elementary effects}

The derivative-based procedure used is due to \citet{1991_Morris} with modifications from  \citet{2007_Campolongo_etal} and \citet{2009_Kucherenko_etal}. To explain the method, we begin with a numerical approximation to a partial derivative of the simulator function with respect to $\theta_k$, with the values for the remaining $K-1$ parameters fixed at ``default'' values:
\begin{align}
    \frac{d F(\Theta)}{d\theta_k}
    &\approx \frac{\Delta y}{\Delta\theta_k}
    = \frac{F(\theta_{k}+h; \Theta_{-k}^*) - 
    F(\theta_k; \Theta_{-k}^*)}{h}
\end{align}
where $\Theta^*_{-k}$ is the vector of all parameters except $\theta_k$ evaluated at the default value, and $h$ is a perturbation of $\theta_k$. This calculation is a measure of the relative change in the response around a specific value of $\theta_k$, known as an elementary effect, and is a Local SA measure.  The Morris procedure makes this global by calculating the elementary effects over a set of $p$ perturbations over the domain of $\theta_k$, $h$ $\in$ $\left [ 1/(p-1), 2/(p-1), \ldots, 1-1/(p-1)  \right ]$, thus generating a finite distribution of elementary effects.  \citet{1991_Morris} used the average value, denoted $\mu_k$, and the standard deviation, $\sigma_k$, as measures of influence where $\mu_k$ estimates the overall effect of $\theta_k$ on ,$F$ and $\sigma_k$ estimates interactions and nonlinearities. In some situations, the elementary effects may be a mix of positive and negative numbers, and $\mu_k$ may fail to accurately portray the effect, and \citet{2007_Campolongo_etal} proposed an alternative, $\mu^*_k$, which is the average of the absolute elementary effects. A commonly used way to graphically display both measures is a scatterplot of $\sigma_k$ versus $\mu^*_k$.   A summary measure combining both values  
\begin{align*}
    G_k & = \sqrt{\mu^{*,2}_k + \sigma^2_k}
\end{align*}
was proposed by \citet{2009_Kucherenko_etal} and labeled the derivative global sensitivity measure (DGSM, \citet{2022_Dela_etal}). For the case examples, we examined both the scatterplots of $\sigma_k$ vs $\mu^*_k$ as well as calculated DGSM.  We note that one of the implementation considerations is the size of the perturbation set $p$; in the applications this was set to be at least 200.

%% file: 6_Appendices/1_GSA_Details/6AC_Distribution.tex
\subsection{\label{app.Distribution.details}Distribution-based: Sobol' sensitivities}

We sketch the main ideas that underlie the general equations for the first order ($S_{1,k}$) and total ($T_k$) sensitivities (equations \ref{eq:Sobol.S1} and \ref{eq:Sobol.T}).  To begin, the simplest version of the law of total variance is the case where a random variable $Y$ is a function of a single random variable $X$: $V(Y) = E_X \left [ V(Y|X) \right ] + V_X \left [ E(Y|X) \right ]$. This general result can be applied to a function of two or more random variables. Here we consider the simplest case that does not lead to trivial components, namely, $Y$ is a function of three independent random inputs, $Y=f(X_1,X_2,X_3)$.  Without loss of generality, we focus on $X_1$, and assume it has the most influence; it dominates the variability in $Y$. We demonstrate formal calculations for the first order sensitivity, $S_1$,
and total sensitivity, $T_1$. We use the law of total variance twice, first:
\begin{align}
\label{eq:law_ttl_var_1}
V[Y] & = E_{X_1} \left [ V_{X_2,X_3} (Y|X_1) \right ]
    + V_{X_1} \left [ E_{X_2,X_3} (Y|X_1) \right ] 
\end{align}
Referring to the first term on the right-hand side of eq'n \ref{eq:law_ttl_var_1}, for a particular but arbitrary value of $X_1$, given that $X_1$ dominates variance, the term $V_{X_2,X_3}(Y|X_1)$ should be relatively small, i.e., once one knows $X_1$, there is not much variation in $Y$.  The expectation of that term over the space of $X_1$, $E_{X_1} \left [ V_{X_2,X_3} (Y|X_1) \right ]$, thus provides a ``global'' measure. Conversely, the second piece on the  right-hand side of eq'n \ref{eq:law_ttl_var_1} should be relatively large, and the relative measure, the 1st order sensitivity is:
\begin{align}
\label{eq:S1.basic}
S_{1,1} &= \frac{V_{X_1} \left [ E_{X_2,X_3} (Y|X_1) \right ]}{V(Y)}
\end{align}
For $T_1$, we use the Law of Total Probability with the outer conditioning reversed:
\begin{align}
\label{eq:law_ttl_var_2}
V[Y] & = E_{X_2,X_3} \left [ V_{X_1} (Y|X_2,X_3) \right ]
     + V_{X_2,X_3} \left [ E_{X_1} (Y|X_2,X_3) \right ]
\end{align}
Looking at the first term on the right-hand side of equation \ref{eq:law_ttl_var_2}, the inner term $V_{X_1} (Y|X_2,X_3)$ should be relatively large because knowing $X_2$ and $X_3$ does not provide much information about $Y$ given that $X_1$ dominates the variance, and again, taking the expectation over all of $X_2$ and $X_3$ provides a global measure of the variance in terms of $X_1$, and the Total Sensitivity for
$X_1$ is defined:
\begin{align}
\label{eq:T1.basic}
T_1 &= \frac{E_{X_2,X_3} \left [ V_{X_1} (Y|X_2,X_3) \right ]}{V(Y)}
\end{align}

Referring the general setting of $K$ parameters, exact or analytical calculation of $S_{1,k}$ and $T_k$, assuming independent and identical probability distributions for the $\theta_k$s, is practically impossible for most simulators.  Calculations of expectations and variances are integration problems, and thus Monte Carlo integration procedures have been developed, starting with \citet{1993_Sobol}, with many refinements developed subsequently, and a large literature on these refinements and comparisons of the algorithms. These algorithms are generally a two-stage procedure with   stage 1 being setting up two or more sets of matrices of parameter combinations, and stage 2 being Monte Carlo estimation of $S_{1,k}$ and $T_{k}$.  See Table \ref{T:SA.methods.R.code} for the \texttt{R} package and key functions for stages 1 and 2. 

%% file: 6_Appendices/1_GSA_Details/6AD_Variogram.tex
\subsection{\label{app.Variogram.details}Variogram-based: VARS-TO}

To explain the ideas underpinning the variogram approach, we begin with the simplest setting with $K$=2 inputs. Model output $Y_{\Theta}$ is a spatially indexed random variable, namely a stochastic process,  defined over a two-dimensional region, $\Theta= (\theta_1,\theta_2)$.  Spatial smoothness is assumed in the sense that the responses based on two spatially close inputs are relatively similar. Let $A$ and $B$ denote two locations with spatial coordinates $\Theta_A = (\theta_{1,A}, \theta_{2,A})$ and $\Theta_B = (\theta_{1,B}, \theta_{2,B})$. If $||\Theta_A - \Theta_B||$, a measure of distance between the coordinates, is relatively small, then $Y_A$ and $Y_B$ are expected to be relatively similar.  Conversely, if $||\Theta_A - \Theta_B||$ is relatively large, then the values for $Y_A$ and $Y_B$ are expected to be relatively far apart.

The expected value of the stochastic process at any given index is assumed to be a constant, $E[Y_\Theta]$ = $\mu$, as is the variance  at all locations, $E[(Y_\Theta-\mu)^2]$ = $\sigma^2$.   The covariance between two realizations of the process at two different locations is then a function of distances in the index set. For example,
\begin{align*}
Cov(Y_A,Y_B) = E[(Y_A-Y_B)^2] & = \sigma^2 \exp \left (\sqrt{(\theta_{1,A}-\theta_{1,B})^2+(\theta_{2,A}-\theta_{2,B})^2} \right )
\end{align*}
When the above conditions on the mean, variance, and covariance hold, the stochastic process is known as a second-order stationary process \citep{2016_Razavi_Gupta}.
These notions extend to higher dimensional index sets,  in our case to $K$ parameter simulators, and the above example covariance function can be similarly defined. We note that this stochastic process structure also underpins the Gaussian Process regression method described in Section \ref{subsec:GPR}. The assumptions about a constant mean and variance can be relaxed by fitting a model to the mean and calculating the  residuals, and then scaling the residuals such that the scaled residuals have mean 0 and constant variance. 

\citet{2016_Razavi_Gupta} work with stochastic processes with a more general stationarity assumption, known as intrinsic stationarity, which includes second-order stationarity as a special case, and is distinguished by a semi-variogram, $\gamma(\mathbf{h})$,
\begin{align*}
\gamma(\mathbf{h}) &= \frac{1}{2} Var(Y_{\Theta+\mathbf{h}} - Y_{\Theta}) = \frac{1}{2} E \left [(Y_{\Theta+\mathbf{h}} - Y_{\Theta})^2 \right ]
\end{align*}
where $\mathbf{h}$  is a vector of differences in the index vector component values. The second equality above holds under the assumption of a constant mean $\mu$. The variogram is twice the semi-variogram, namely, $2\gamma(\mathbf{h})$, but the term variogram is often loosely used for semi-variogram. An isotropic stochastic process is a special case where the variogram is a function of the distance between two vector points, $\mathbf{h}$=$||\Theta_A-\Theta_B||$, otherwise, the process is known as anisotropic.  

 The anisotropic setting is the one in which the notion of parameter sensitivities arises. In this setting, the  semi-variogram involves distance and direction, i.e., it is a measure of the covariance in one dimension (direction) of the input space, specifically a scalar-valued function, $\gamma(h_k)$, where $h_k$ is the distance (usually Euclidean) between locations $\theta_{Ak}$ and $\theta_{Bk}$.   If the stochastic process is isotropic, then the (semi-)variogram is a scalar function of the distance between the two vectors, $\gamma(||\mathbf{h}||)$, and \citet{2016_Razavi_Gupta} call this the \textit{overall variogram}, and they say that this ``characterizes some average, nondirectional covariance, and can therefore be used to normalize the directional variograms and resulting VARS-based sensitivity metrics.'' 

 A key conceptual aspect of the variogram approach, the VARS framework for sensitivity analysis, is that ``VARS links variogram analysis to the important concepts of ‘direction’ and ‘scale.’ $\ldots$ we define that a higher value of $\gamma(h_k)$ for any given $h_k$ $\ldots$ [the] higher [the] ‘rate of variability’ (or ‘sensitivity’) of the underlying response surface in the direction of the $k$th factor, at the scale represented by that $h_k$ value. Notably, this rate of variability at a particular scale in the problem domain represents the ‘scale-dependent sensitivity’ of the response surface to the corresponding factor.'' 

This means that sensitivities can vary depending on the values of $h_k$.  \citet{2016_Razavi_Gupta} define a measure that summarizes these sensitivities across a range of values of $h_k$, from 0 to some upper bound, where that upper bound is defined as some fraction of the total parameter range, $[\theta_{k,Lower},\theta_{k,Upper}]$. The summaries are the integrals of $\gamma(h_k)$ and are called Integrated Variograms. They have upper bounds that are $p$\% of the parameter range labelled $IVARS_{p}$, where for parameter $i$:
\begin{align*}
IVARS_{k,p} = \Gamma(H_{k,p}) & = \int_0^{H_{k,p}} \gamma(h_k) dh_k
\end{align*}
where $H_{k,p}$ = the $p$th percentile in the range of $\theta_k$ (which are generally scaled to be between 0 and 1). \citet{2016_Razavi_Gupta} recommend that three values of $p$ be explored: $IVARS_{10}$, $IVARS_{30}$, and $IVARS_{50}$.  

 This variogram-based approach is an important unifying perspective in that \citet{2016_Razavi_Gupta} connect this approach to both the variance of Morris' elementary effects $\sigma^2_k$ and the Sobol' total sensitivity $T_k$.  As $h_k$ goes to zero, the directional derivative goes to $\sigma^2_k$.  The directional derivative is connected to $T_k$:  
\begin{align*}
\gamma(h_k) & = T_k Var(Y) - f(Cov_{\not k},h_k)
\end{align*}
where $f(Cov_{\not k},h_k)$ is a function of the covariance of $Y$ excluding input parameter $k$ and the distance $h_k$. Thus, $T_k$ is proportional to the directional derivative:
\begin{align*}
%\label{eq:VARS-TO}
   \mbox{VARS-TO}_k =  \gamma(h_k)+f(Cov_{\not k},h_k) & \propto T_k
\end{align*}
where  VARS-TO is defined as the variance-based total order effects. We only used the VARS-TO measure in our assessment of the three simulators, primarily because of its availability in the \texttt{R} package \texttt{sensobol} (see Table \ref{T:SA.methods.R.code}), but admit that the use of $IVARS_{p}$ measures as well would have been useful based on arguments made by \citet{2016_Razavi_Gupta} for their value with some classes of simulators.

%% file: 6_Appendices/1_GSA_Details/6AF_Reg_RegTree.tex
\subsection{\label{app.Reg.RegTree.details}Regression-based: Regression Tree}

\begin{figure}
    \centering
    \includegraphics[width=0.8\linewidth]{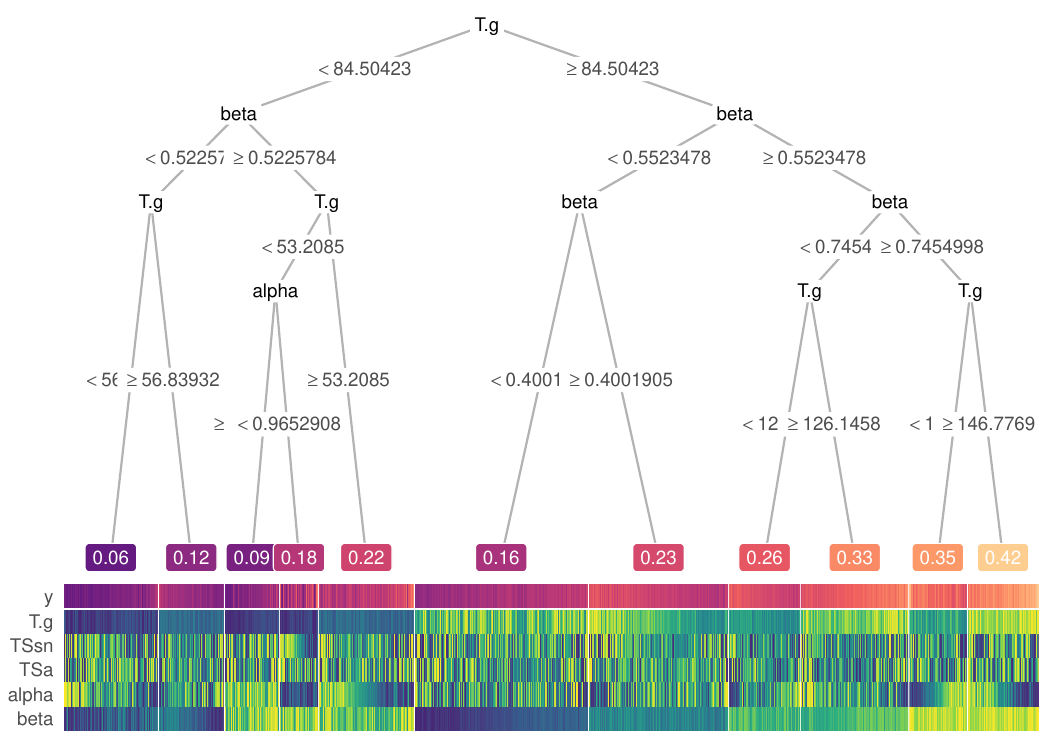}
    \caption{Regression tree and heatmap for log flow in SimplyP with the tree constructed using the only the five inputs with the largest relative importance.}
    \label{F:Method.RegTree.Heatmap.flow}
\end{figure}

An alternative display of a regression tree partitioning is a heatmap, which can provide more insight into the main effects and possible interactions of the input parameters.  Figure \ref{F:Method.RegTree.Heatmap.flow} shows the regression tree for SimplyP and the output, log flow, that was constructed using the five input parameters with the largest relative importance measures.  The tree was built using a different algorithm than that used in \texttt{rpart::rpart}, but the resulting tree is quite similar in terms of branching and terminal nodes---again, the dominance of beta and Tg in the branching is apparent, with alpha appearing once. 

Interpretation of the heat map is slightly  involved.  The top colored bar below the tree is a heat map of the assigned log outflow values below each terminal node, with darker colors, dark blue and purple, denoting lower flows, and brighter colors, orange and light red, representing higher flows.  The next five bars are color coding of the parameter values for the five most important parameters within each terminal node. Each of the $N$ simulator output values is included in the heat map construction, and when the parameter values vary within a given terminal node, the different observations can be seen by potentially many vertical columns or lines. If many vertical bands can be seen within a block under a leaf, then that indicates that the values for that parameter do not have much influence on the assigned value; for example, the block for TSsn under the far left leaf has a wide mix of colors (thus different values of TSsn).  When colors within a block for a parameter are mostly solid, then that indicates relatively similar values for that parameter in that leaf.  The more blocks with more or less solid colors, the more the relative importance of the parameter---this can be seen with Tg and beta in particular, and with alpha to a lesser degree. When the more-or-less solid color blocks of two input parameters with overall relatively high importance, such as Tg and beta, have similar patterns, that suggests a positive correlation between them \citep{2021_Le_Moore}.  For example, looking at the far left leaf with the lowest assigned outflow value, Tg is at its lowest value, and beta is relatively low.  But the combined effects of Tg and beta are complex.  Looking at the blocks for the third leaf from the left with an assigned value of 0.09, T.g is relatively low, but beta is relatively high, thus indicative of interactions between the two parameters.   \citet{2021_Le_Moore} provide more discussion of the interpretation of the heat maps. 

%% file: 6_Appendices/1_GSA_Details/6AH_Reg_GPR.tex
\subsection{\label{app.Reg.GPR.details}Regression-based: Gaussian Process Regression GSA details}

Gaussian process regression models (GPRs) can be viewed as extensions of multiple regression models, with the extension being that, in addition to a model for the mean response as a function of covariates, the covariance between responses is modeled as well. Here, we provide more details on the underlying structure of GPRs.  

For example, a multiple regression model for $n$ observations $y$ and $p$=2 predictor variables, $x_1$ and $x_2$, often assumes a multivariate normal (MVN) distribution:
\begin{align*}
\left [ \begin{array}{c} y_1 \\ y_2 \\ \vdots \\ y_n 
\end{array} \right ] & \sim \mbox{MVN} 
\left (
\left [ \begin{array}{c}
 \beta_0 + \beta_1 x_{1,1} + \beta_2 x_{2,1} \\
 \beta_0 + \beta_1 x_{1,2} + \beta_2 x_{2,2} \\
 \vdots \\
 \beta_0 + \beta_1 x_{1,n} + \beta_2 x_{2,n} 
 \end{array} \right ], ~~
\sigma^2 \left [ \begin{array}{cccc}
 1 & 0 & \ldots & 0 \\
 0 & 1 & \ldots & 0 \\
 \vdots & \vdots & \vdots & \vdots \\
 0 & 0 & \ldots & 1 \\
 \end{array}
 \right ] \right )
\end{align*}
The extension found a GPR appears in the covariance matrix
\begin{align*}
\left [ \begin{array}{c} y_1 \\ y_2 \\ \vdots \\ y_n 
\end{array} \right ] & \sim \mbox{MVN} 
\left (
\left [ \begin{array}{c}
 \beta_0 + \beta_1 x_{1,1} + \beta_2 x_{2,1} \\
 \beta_0 + \beta_1 x_{1,2} + \beta_2 x_{2,2} \\
 \vdots \\
 \beta_0 + \beta_1 x_{1,n} + \beta_2 x_{2,n} 
 \end{array} \right ], ~~
\sigma^2 \left [ \begin{array}{cccc}
 1 & \rho_{1,2} & \ldots & \rho_{1,n} \\
 \rho_{2,1} & 1 & \ldots & \rho_{2,n} \\
 \vdots & \vdots & \vdots & \vdots \\
 \rho_{n,1} & \rho_{n,2} & \ldots & 1 \\
 \end{array}
 \right ] \right )
\end{align*}
The correlations, $\rho_{i,j}$, are functions of the input parameter values, in particular, the distance between the input values. For example, with $p$=2, one such correlation function, the power exponential function, is the following.
\begin{align*}
\rho_{i,j} & = \exp \left \{ - 
\left [
\frac{\sqrt{(x_{1,i}-x_{1,j})^2}^\alpha} {\gamma_1}  +
\frac{\sqrt{(x_{2,i}-x_{2,j})^2}^\alpha} {\gamma_2}
\right ] \right \}
\end{align*}
With this correlation function, and several other commonly used correlation functions, as distances between two points in the input parameter space become smaller, the correlation between $y_i$ and $y_j$ increases.  Thus GPR   reflects the notion that if two sets of input parameters have similar values, then one expects the two corresponding outputs to have similar values. The two parameters $\gamma_1$ and $\gamma_2$, known as range parameters or length scale parameters, control the ``spatial extent'' of the correlation: as $\gamma$ increases, the correlation increases at a fixed distance $d$.

%% file: 6_Appendices/1_GSA_Details/6AI_ComputerRunDetails.tex
\subsection{\label{app.RCode.details}\texttt{R} Code details for running the simulators}

The \texttt{R} packages and functions used specifically for GSA calculations are shown in Table \ref{T:SA.methods.R.code}.
\begin{table}[h]
{\small 
\begin{tabular}{lllll}
Method & Packages & Generation & Evaluation \\ \hline 
Morris &  \texttt{sensitivity} & \texttt{x = morris(factors=N,r=p,}  & \texttt{tell(x.Morris,y)}\\  
& &\texttt{~~ design=list( } \\ 
& &\texttt{~~ type="oat",levels=20))} \\
Sobol' &  \texttt{sensobol} & \texttt{sobol\_matrices} & \texttt{sobol\_indices} \\
Variogram & \texttt{sensobol} & \texttt{vars\_matrices(star.centers,} & \texttt{vars\_to(Y,star.centers,h,params)} \\
          & & ~~ \texttt{h, params)} \\
Multiple Regression & \texttt{lhs} & \texttt{maximinLHS} & \texttt{lm(y $\sim$ ., data)} \\
Reression.Tree & \texttt{rpart} & \texttt{maximinLHS} & \texttt{RegTree.out=rpart(y $\sim$ ., data)}\\
 & & & \texttt{RegTree.out\$variable.importance} \\
 & \texttt{treeheatr}& & \texttt{heat\_tree} \\ 
Random Forest & \texttt{randomForest} & \texttt{maximinLHS} & \texttt{rf.out=randomForest(y $\sim$ ., data,importance=TRUE)}\\
& & & \texttt{importance(rf.out)} \\
Gaussian Process & \texttt{RobustGaSP} & \texttt{maximinLHS} & \texttt{rgasp(design=input,response=y,} \\
& & & \texttt{~~trend=X, nugget.est=F,
                 kernel\_type=``pow\_exp'')}\\
\end{tabular}
}
\caption{Fragments of \texttt{R} code (and packages) used for generating parameter combinations and for calculating indices.  The \texttt{lhs} package that contains the function \texttt{maximinLHS} is used for Regression, Regression Tree, Random Forest, and Gaussian Process Regression.}
\label{T:SA.methods.R.code}
\end{table}

%% file: 6_Appendices/2_Simulator_Details/6B_Simulator_Details.tex
\section{\label{app:Simulators.Details} Additional Details on Simulators}

\input{6_Appendices/2_Simulator_Details/6B1_Details_GR6J}

\input{6_Appendices/2_Simulator_Details/6B2_Details_SimplyP}

\input{6_Appendices/2_Simulator_Details/6B3_Details_STICS}

\input{6_Appendices/2_Simulator_Details/6B4_RCode_Simulators}

%% file: 6_Appendices/2_Simulator_Details/6B1_Details_GR6J.tex
\subsection{\label{app.sec.GR6J.desc}GR6J Details}

\subsubsection{Overview of GR6J with input parameters and outputs}
\begin{figure}[p]
    \centering
    \includegraphics[width=0.6\columnwidth]
         {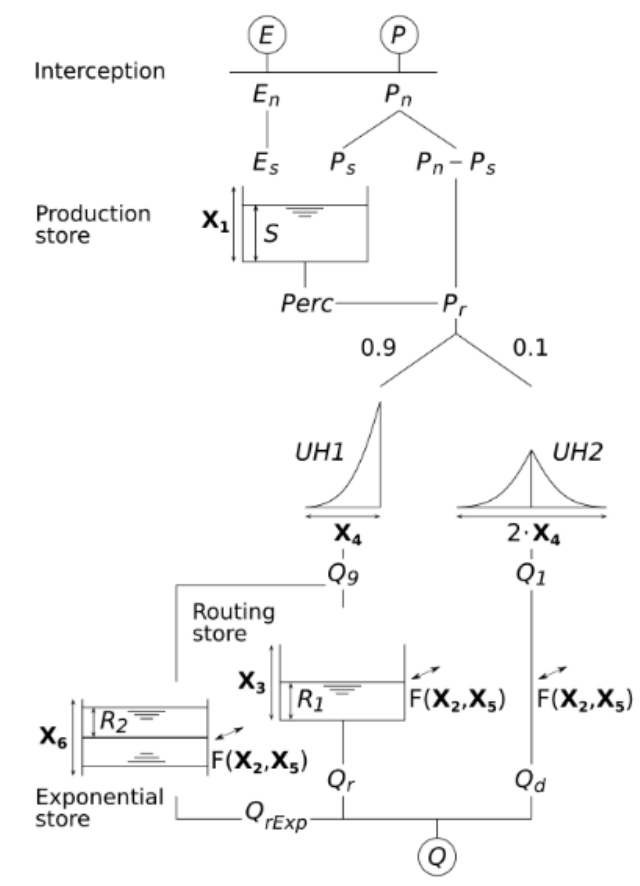}
    \caption{GR6 model structure (taken from \citet{2011_Pushpalatha_etal}). Water volumes in
    the three storage compartments of Production Storage,
Routing Storage, and Exponential Storage are indicated by
$S$, $R1$, and $R2$ and the primary output is discharge,
$Q$ $\equiv$ $Q_{sim,d}$, shown at the bottom the figure.}
    \label{F:GR6J.schematic}
\end{figure}

The schematic of GR6J shown in Figure \ref{F:GR6J.schematic} shows the three storage departments, the six input parameters, and intermediate and final outputs.  Table \ref{T:GR6J.default} provides more detail on the parameters, including default values. Table \ref{T:GR6J.outputs} describes the model outputs.

%  -----------  GR6J Parameters and Default values --------------
\begin{table}[h]
\begin{tabular}{lrrrlll}
 Parameter &Lower & Upper & ``Default'' & Description &Units & Outputs/Processes \\
         &&  & Value     &            & & Affected \\ \hline
 $X_1$  & 0.00&1460.00 & 83.9 & Production store capacity &mm & $P_s$, $E_s$, $S$, and $Perc$ \\
  $X_2$ &-1.80&   2.51 &  -6.0401e-5 & Intercatchment exchange coefficient 
  & [ND]  & $R_1$, $R_2$ \\
 $X_3$ & 0.99& 983.52& 17.2014 & Routing store capacity & mm & $R_1$, $R_2$ \\
 $X_4$ & 0.84&  19.56& 1.3970& Unit hydrograph time constant & Time & $UH1$, $UH2$, $Q_1$, $Q_9$ \\
 $X_5$ &-2.00&   2.00& 1.9980 & Intercatchment, or groundwater\\
 &&&&exchange threshold & [ND] & $R_1$, $R_2$\\
 $X_6$ & 0.31& 262.43& 34  & Exponential store depletion coefficient& mm & $Q_{rExp}$ \\
\end{tabular}
\caption{GR6J input parameters, example ``default'' values,
summary explanations, and some system states they 
directly affect. The units [ND] indicate no dimension.}
\label{T:GR6J.default}
\end{table}

%  "PotEvap"  "Precip"   
% "Prod"=S "Pn"  "Ps"  "AE"=En "Perc"      "PR"=Pr 
% "Q9"  "Q1"      
%"Rout"       "Exch"       "AExch1"     "AExch2"    
% "AExch"    "QR"         "QRExp"      "Exp"        "QD"        
% "Qsim"     

%  ----------  GR6J Outputs- Intermediate & Final -----------------
\begin{table}[h]
\centering
\begin{tabular}{lll}
Figure & Output & Description \\ 
Name   & Name   & \\ \hline 
$S$  & Prod  & Production store level  [mm]    \\
$P_n$& Pn  & net rainfall = $\max(P-E,0)$, mm/d  \\
$P_s$& Ps  & If $P_n>0$, a function of $P_n$, else 0, mm/d] \\
$E_n$   & AE & actual evapotranspiration = $\max(E-P,0)$, mm/d \\
$Perc$ &Perc  &percolation = f($S$), mm/d   \\
$P_r$  & PR   &$P_r$=($P_n$-$P_s$)+$Perc$, mm/d   \\
$Q_9$ & Q9   & UH1 outflow, $f(0.9*P_r)$, mm/d   \\
$Q_1$ & Q1   & UH2 outflow, $f(0.1*P_r)$, mm/d   \\
$R_1$ & Rout & Routing store level, $f(0.6*Q_9)$, mm   \\
$R_2$ & Exp  & Exponential store level, $f(0.4*Q_9)$, mm \\ 
$Q_{rExp}$&QRExp & Exponential store outflow, mm/d \\
$Q_r$ & QR   & routing store outflow, $f(R_1)$, mm/d  \\
$Q_d$ & QD	 & direct flow from UH2 after exchange, $f(Q_1)$, mm/d   \\
$Q_{sim}$   &Qsim  & discharge, $Q_sim=Q_r+Q_{rExp}+Q_d$, mm/d  \\
\end{tabular}
\caption{GR6J model outputs: The Figure Name column contains output 
labels shown on Figure \ref{F:GR6J.schematic}, the Output Name column has
program output names, and  the Description column indicates a functional operation.}
\label{T:GR6J.outputs}
\end{table}

\subsubsection{Process details within GR6J}
A detailed sequencing of the processes,
within a single day, that link the inputs to the ultimate
discharge output $Q_{sim}$ is provided here. Notation largely matches that
shown in Figure \ref{F:GR6J.schematic}.  Functional
operations are denoted generically by $f(\cdot)$.

\paragraph{Step 1. Calculation of Production Store Level and Input
to Hydrographs.}
\begin{itemize}
    \item Net precipitation ($P_n$) and net evapotranspiration
  ($E_n$) are calculated from
$P$ and $E$:  $P_n$=$\max(P-E,0)$ and $E_n$=$\max(E-P,0)$. 
 \item Production Store content is then either 
 incremented by $P_s$, $f(P_n|S_{t-1},X_1)$, if $P_n>0$, or
 decremented by $E_s$, $f(E_n|S_{t-1},X_1)$, if $E_n>0$, which
 can be written:
 \begin{align*}
     S^* &= S_{t-1} + P_s - E_s
 \end{align*}
 Note: if $P_n>0$ then $E_s$=0 and if $E_n>0$, then $P_s$=0.
  \item Next water ``leaks'' from the production store,
  percolates out, the amount is denoted $Perc$, a
  $f(S^*,X_1)$.  Store content $S$ is thus decremented by
  this loss: 
\begin{align*}
    S &= S^* - Perc
\end{align*} 
   \item The percolation and surplus from $P_n$, $\max(P_n-P_s,0)$ are summed to yield a volume, $P_r$, reaching the 
   routing part: 
\begin{align*}
    P_r &= Perc + \max(P_n-P_s,0)
\end{align*} 
\end{itemize}
\paragraph{Step 2. Passage through Hydrographs.}
The above output $P_r$ is then divided and input to two 
different hydrographs.
The reaction time of the catchment (flow delays)
is expressed using  a one-sided hydrograph $UH1$ (for
routed flows) and a two-sided hydrograph $UH2$ (for direct
flows). Parameter $X_4$ controls the base time of both 
hydrographs, accounting for flow delays.
  
In particular, two weighted combinations of
current and lagged values of $Pr$ are calculated that
quantify passage through the two different hydrographs. 90\% 
of one combination is routed through $UH1$  and the amount 
passing through is denoted $Q_9$.   10\% of another combination
is routed through $UH2$  and the amount passing through 
is $Q_1$.  
\paragraph{Step 3. Branching of $Q_9$ to Routing and Exponential Stores.}
The above output $Q_9$ serves as input to these stores,
with 60\% going to the Routing Store and 40\% to the Exponential 
Store\footnote{The Exponential Store reproduces long recessions and low flows.}.  
\begin{itemize}
  \item An intermediate calculation is $F$:
\begin{align*}
F &= X_2 \left ( \frac{R_{1,t-1}}{X_3} - X_5 \right )
\end{align*}
The parameters, $X_2$ 
(multiplicative effect) and $X_5$ (additive effect)
are dimensionless and contribute
to adjust the catchment's water balance by controlling the 
quantity of 
 water that is considered lost, or gained, from groundwater
aquifers or neighboring catchments.
\item The amount ending up in the Routing Store depends
on the value of $F$ relative to $R_{t-1}$ and $Q_9$: 
\begin{align*}
R_1^* &= \left \{ 
\begin{array}{ll}
R_{t-1}+0.6 Q_9 + F, & \mbox{if $R_{1,t-1}+0.6 Q_9 + F>0$} \\
0, & \mbox{else}
\end{array}
\right .
\end{align*}  
$R_1^*$ is then divided into two parts, one the routing store
output $Q_r$ and the rest staying in the routing store:
\begin{align*}
    Q_r &= R_1^* \left \{ 1 - 
    \left [ 1 + \left ( \frac{R_1^*}{X_3} \right )^{4} 
    \right ]^{\frac{-1}{4}} \right \} \\
    R_1 &= R_1^* - Q_r
\end{align*}
\item The 40\% of $Q_9$ going to the Exponential 
Store is
calculated by first incrementing the current level of the
Exponential Store: 
\begin{align*}
R_2^* &= R_{2,t-1} + 0.4 Q_9 + F
\end{align*}
Then one portion of $R^2$ is output at $Q_{rExp}$ and the
remainder stays in the store:
\begin{align*}
Q_{rExp} &=  X_6 \log \left ( 1 + \exp \left ( 
 \frac{R_2^*}{X_6} \right ) \right ) \\
R_2 &= R_2^* - Q_{rExp}
\end{align*}  
\end{itemize} 
\paragraph{Step 4. Direct outflow from $UH2$'s $Q_1$.}
Using the intermediate value $F$ calculated above
and $Q_1$ from $UH2$, the direct outflow is a simple 
calculation:
\begin{align*}
Q_d &= \left \{
\begin{array}{ll}
 Q_1 - F, & \mbox{if $Q_1+F > 0$} \\
2 Q_1, & \mbox{else}
\end{array} 
\right .
\end{align*}

\paragraph{Step 5. Ultimate outflow calculation.}
Outflow is simply the sum of three previous flow outputs. 
\begin{align}
Q (\equiv Q_{sim}) & = Q_r + Q_{rExp} + Q_d
\end{align}

\subsubsection{Running, calibrating, and evaluating GR6J}
The daily precipitation and temperature data needed to run the model came from  the gridded (1 km × 1 km) HadUK datasets \citep{2019_Hollis_etal} and potential evapotranspiration (E) was estimated  using the  Priestly-Taylor method \citep{1972_Priestley_Taylor}. 

To calibrate, validate, and evaluate the model output,
discharge ($Q \equiv Q_{sim}$, Figure \ref{F:GR6J.schematic}), daily observed
streamflow data from gauging locations in Tarland (Could and 
Aboyne) for the period 2013-2018 were used (The James Hutton 
Institute). The model is implemented in the open-source 
\texttt{R} package \texttt{airGR} \citep{2017_Coron_etal}
and calibrated using an in-built optimization algorithm 
from \citet{1987_Michel}. 

%% file: 6_Appendices/2_Simulator_Details/6B2_Details_SimplyP.tex
\subsection{\label{app.sec.SimplyP.desc}SimplyP Details}

We used SimplyP v0.4.2 implemented in Mobius v1 (for later versions of SimplyP, implemented in Mobius2, see https://nivanorge.github.io/Mobius2/).

\subsubsection{Process details}
\begin{figure}[h]
    \centering
    \includegraphics[width=0.99\columnwidth]{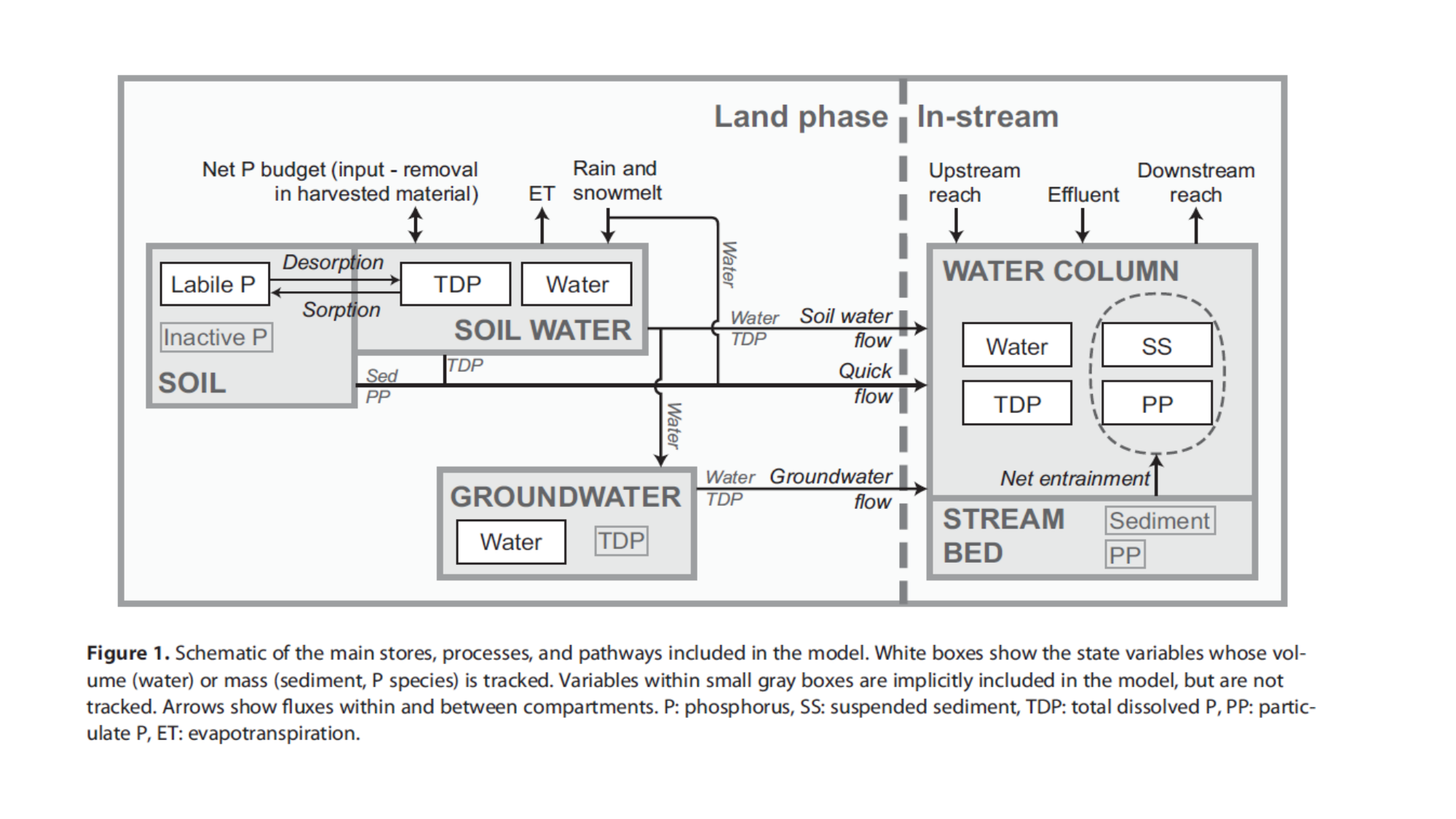}
    \caption{Schematic of SimplyP (copied with permission
    from \citet{2017_Jackson_etal}).}
    \label{F:SimplyP.schematic}
\end{figure}

To simulate the transport of dissolved and particulate P under varying flow conditions, a slightly more complex representation of terrestrial hydrology is needed than is used in GR6J (see Figure \ref{F:SimplyP.schematic}). In SimplyQ (the SimplyP flow module), precipitation is partitioned at the soil surface into quick flow, which flows directly to the stream without a time lag, and into water which enters the soil water store. This partitioning is determined by the $f_{quick}$  parameter, the proportion of precipitation that contributes to  quick flow. The change in soil water volume over time depends on the balance of inputs (precipitation minus quick flow) and outputs (actual evapotranspiration (AET) and soil water flow). AET is calculated from the PET input time series, limiting ET when soil water volume approaches field capacity ($FC$). Water flows out of the soil box when the soil water volume is above $FC$. This flow is proportional to the soil water volume above $FC$, with $1/T_s$ (the soil water time constant) as the constant of proportionality. A fraction, $beta$ (the baseflow index) of the soil water flow percolates to the groundwater store; The remainder enters the stream. Groundwater flow to the stream is proportional to the groundwater volume with a constant of proportionality $1/T_{gw}$. To allow concentrations and fluxes from the reach to be calculated, SimplyQ estimates reach volume and discharge using a simple mass balance approach, incorporating an estimate of water velocity derived using Manning's equation.
 
Phosphorus processes in SimplyP are usually calculated separately for two land classes, agricultural and semi-natural, and separating P into two fractions, particulate P (PP) and total dissolved P (TDP). For PP, agricultural land is often further split into high erodibility (e.g., arable) and low erodibility (e.g., improved grassland) classes. Within the soil compartment, P is present in three forms: TDP in soil water, labile soil P, and inactive soil P. The initial soil water TDP content is set equal to the user-supplied EPC0 parameter (the equilibrium TDP concentration at which there is no net exchange of P between the soil water and labile soil). In the simplest model application, EPC0 may be set to be constant throughout the model run, meaning a constant soil water TDP concentration. Otherwise, a dynamic EPC0 is used, whereby EPC0 changes according to a simple linear sorption relationship relating soil total P concentration and EPC0 (see Supplementary Information of \citep{2017_Jackson_etal} for details). Inactive soil P is constant over time. TDP is transferred out of the soil water store along with soil water, entering the reach via quick flow and soil water flow. Groundwater flow is assumed to have a constant (user-defined) TDP concentration, $TDP_{g, conc}$. Sewage effluent TDP inputs to the reach are supplied via a user-supplied annual input, $TDP_{eff}$, in the simplest case. Particulate P (PP) in SimplyP is assumed to be sediment-bound, so PP inputs to the reach are assumed to be proportional to sediment inputs. These sediment inputs are calculated in a simplistic manner, where inputs from all sources are lumped together and estimated as a function of in-stream discharge, as $SS_{conc} = E_{sus} Q_r^{k_M}$, where $Q_r$ is in-stream discharge, $E_{sus}$ is the sediment-discharge rating coefficient and $k_M$ is the sediment-discharge non-linear coefficient. $E_{sus}$ varies spatially according to reach and sub-catchment slope, a land cover factor (all of which can be based on data or literature), and a sediment input scaling factor, $E_M$, which is calibrated. An optional sediment reduction factor may be included as a simple means of carrying out scenario analyses. The mass of PP input to the stream is simply the mass of sediment transported to the stream from each land use class, multiplied by the P content of the soil in that class. Multiplying by an optional enrichment factor, $E_{PP}$, accounts for the selective transport to the stream of finer sediment particles, which tend to be P-rich compared to source soils.

In summary, SimplyP solves around 19 ODEs simultaneously for each reach in the catchment. Initial conditions are required for each ODE, which are defined using three user-supplied parameters, of which the initial total soil P content and soil water TDP concentration in agricultural land (given by the $EPC_{0,init}$ parameter) are particularly important. All other initial conditions are derived from these parameters or using simple assumptions. The model requires a number of GIS-derived parameters, which are likely to be well-constrained (e.g., sub-catchment areas and slopes). The remaining model parameters, around 23 (24-27 when spatial variability between land-use classes is taken into account), are less well constrained. At least 8 of these are optional (before taking spatial variability into account), and plausible ranges for the majority of parameters can be extracted from measured data or literature. Only four parameters must be determined purely through calibration, including the sediment input scaling factor, $E_M$, soil water time constants (likely split by land class), and the proportion of precipitation routed to quick flow, $f_{quick}$.

In this setup, we applied SimplyP in a medium-sized mixed land-use catchment in northeast Scotland. PET was provided as an input time series, soil water TDP was assumed to be constant over time, as was soil erodibility. We assume P inputs (to land and effluent inputs to the reach) are constant over time and treat them as model parameters.

% \tcb{To add: Finalise what we include in the SA and justify the choice here. My suggestion: use the whole catchment draining to Coull as the spatial setup, i.e. drop the sub-catchments (for simplicity, given this paper has lots in it). Set EPC0 to be constant. Same for erodibility over the year. Then all of the following parameters should be important, and all are uncertain to some extent: $T_{s,A}$ and $T_{s,S}$, $f_{quick}$ (probably just one value for all land classes), baseflow index, $T_g$, $E_m$, $k_m$, $EPC0_{init}$, $TDP_{eff}$, $TDP_g$, $E_PP$. Could drop $TDP_{eff}$ as I expect high covariance between $TDP_{eff}$ and $TDP_g$, depending on what you would want to show with the analysis. If that's way too many, we can cut down. Then parameters to fix: snow stuff, PET reduction factor (alpha), minimum groundwater flow, instream flow params, vegetation cover factor, initial total soil P content, net annual P input to the soil, soil mass per m2 (the last 2 wouldn't be used anyway if EPC0 is set to be constant over the run)}

\subsubsection{Input parameters and output variables for SA}
Parameters used in sensitivity analysis are shown in Table \ref{T:SimplyP.parameters}, those that were kept fixed are in Table \ref{T:SimplyP.fixed.values}, and the four outputs evaluated are in Table \ref{T:SimplyP.outputs}. 

\begin{table}[h]
    \centering
\begin{tabular}{lllrr}
Type &Parameter Description & Label & Lower & Upper \\ \hline 
%
% 1
Hyd & PET multiplication factor      &  alpha &0.750    &1.25 \\
% 2
Hyd & Proportion of precipitation that contributes \\
& \hspace{0.2 in} to quick flow &    $f_{quick}$ & 0.000   &     0.10\\
% 3
Hyd & Baseflow index        &   beta  & 0.200   &     0.90\\
% 4
Hyd & Groundwater time constant   &  $T_g$ & 20.000   &   200.00\\
% 5 and 6
Hyd & Soil water time constant    &  $T_S$$^*$  (Ar)  &0.100   &  3.00\\
Hyd & Soil water time constant    & $T_S$$^*$  (IG)  & 0.100   &  3.00\\
% 7 
Hyd & Soil water time constant   &  $T_S$  (SN)  & 3.000  &  10.00\\
% 8, 9 and 10 
Hyd & Soil field capacity         &  $FC$$^{**}$  (Ar)  &50.000  &  500.00 \\
Hyd & Soil field capacity        &  $FC$$^{**}$  (IG)  &50.000  &  500.00\\
Hyd & Soil field capacity        &  $FC$$^{**}$  (SN)  &50.000  &  500.00\\
% 11
Sed & Reach sediment input scaling factor   & $E_M$ & 2.000 &  6.00\\
% 12
Sed & Sediment input non-linear coefficient     &   $k_M$ & 1.500  &  3.00\\
% 13 
Pho & Particulate P enrichment factor     &  $E_{PP}$ & 1.000 &   5.00\\
% 14 
Phy & Groundwater TDP concentration     &  $TDP_{g}$ & 0.005   &  0.03\\
% 15
Pho & Reach effluent TDP inputs         & $TDP_{eff}$& 0.050   & 0.20 \\
% 16 and 17 
Pho & Initial soil water TDP concentration and EPC0 &  $EPC_{0 init}$$^{***}$ (Ar) &0.030      & 0.20 \\
Pho & Initial soil water TDP concentration and EPC0 &  $EPC_{0 init}$$^{***}$ (IG)& 0.030   &0.20 \\  
\end{tabular}
\caption{SimplyP parameters used in sensitivity analysis. Parameters linked together are noted by $^*$, $^{**}$, or $^{***}$. AR=Arable, IG=Improved Grassland; SN=Semi-Natural. Hyd=Hydrology, Sed=Sediment, Pho=Phosphorus. }
\label{T:SimplyP.parameters}
\end{table}

\begin{table}
\begin{tabular}{llrr}

Parameter name&   Value \\ \\
% \multicolumn{2}{c}{Thornthwaite PET Module}\\ \hline
% Latitude & 60 \\
% PET multiplication factor & 1 \\ \\
%
\multicolumn{2}{c}{SimplySnow Module} \\ \hline 
Initial snow depth as water equivalent & 0 \\
Degree-day factor for snowmelt & 2.74 \\ \\
\multicolumn{2}{c}{SimplyQ Module} \\ \hline 
Manning's coefficient & 0.04 \\
Catchment area & 5.17 \\
Reach length & 10000 \\
Reach slope & 0.02
Initial in-stream flow & 1 \\
Land use proportions (Ar,IG,SN) & 0.25, 0.25, 0.50 \\ \\
\multicolumn{2}{c}{SimplySed Module} \\ \hline 
Mean slope of land in the subcatchment (Ar,IG,SN) & 4, 4, 10 \\
Vegetation land cover (Ar,IG,SN)& 0.2, 0.09, 0.021 \\
Reduction of load in sediment (Ar,IG,SN) & 0, 0, 0 \\
Dynamic erodibility (Ar,IG,SN) & false, false, false \\
% Day of year when soil erodibility $\ldots$ spring-grown crops (Ar,IG,SN)& 60, 60, 60 \\ 
% Day of year when soil erodibility $\ldots$ autumn-grown crops (Ar,IG,SN)& 304, 304, 304 \\ 
% Proportion of spring grown crops (Ar,IG,SN)& 0.65, 0.65, 0.65 \\ \\
%
\multicolumn{2}{c}{SimplyP Module} \\ \hline 
Dynamic soil water EPC0, TDP and soil labile P &  false \\
Run in calibration mode & true \\
% Soil mass per m2 & 95 \\
Phosphorus sorption coefficient & 0.00585 \\
% SRP fraction & 0.7 \\
% Net annual P input to soil (Ar,IG,SN)& 10 10, 0 \\
Initial total soil P content (Ar,IG,SN) & 1458, 1458, 873 \\
Inactive soil P content  (Ar,IG,SN) & 873, 873, 873 \\   
\end{tabular}
\caption{SimplyP parameters that were not varied as they control processes which were not included in the simplified model setup used in this study, such as long-term variation in soil water TDP concentration and within-year variation in soil erodibility), and their value. AR=Arable, IG=Improved Grassland; SN=Semi-Natural. }
\label{T:SimplyP.fixed.values}
\end{table}

\begin{table}[h]
    \centering
    \begin{tabular}{llll}
    Category & Name & Abbreviation & Units \\ \hline 
    Hydrology &  Reach flow & $Q$ &  daily mean, $m^3/sec$      \\
    Sediment     & Reach suspended sediment concentration & $SS_{conc}$ & mg/l \\
    Phosphorus & Reach TDP concentration & $TDP_{conc}$ & mg/l\\
    Phosphorus & Reach PP concentration & $PP_{conc}$ & mg/l\\
    \end{tabular}
    \caption{SimplyP outputs used in sensitivity analysis.}
    \label{T:SimplyP.outputs}
\end{table}
 
\newpage

%% file: 6_Appendices/2_Simulator_Details/6B3_Details_STICS.tex
\clearpage 

\subsection{\label{app.sec.STICS.desc}STICS Details}

\begin{figure}[h]
    \centering
    \includegraphics[width=0.8\columnwidth]{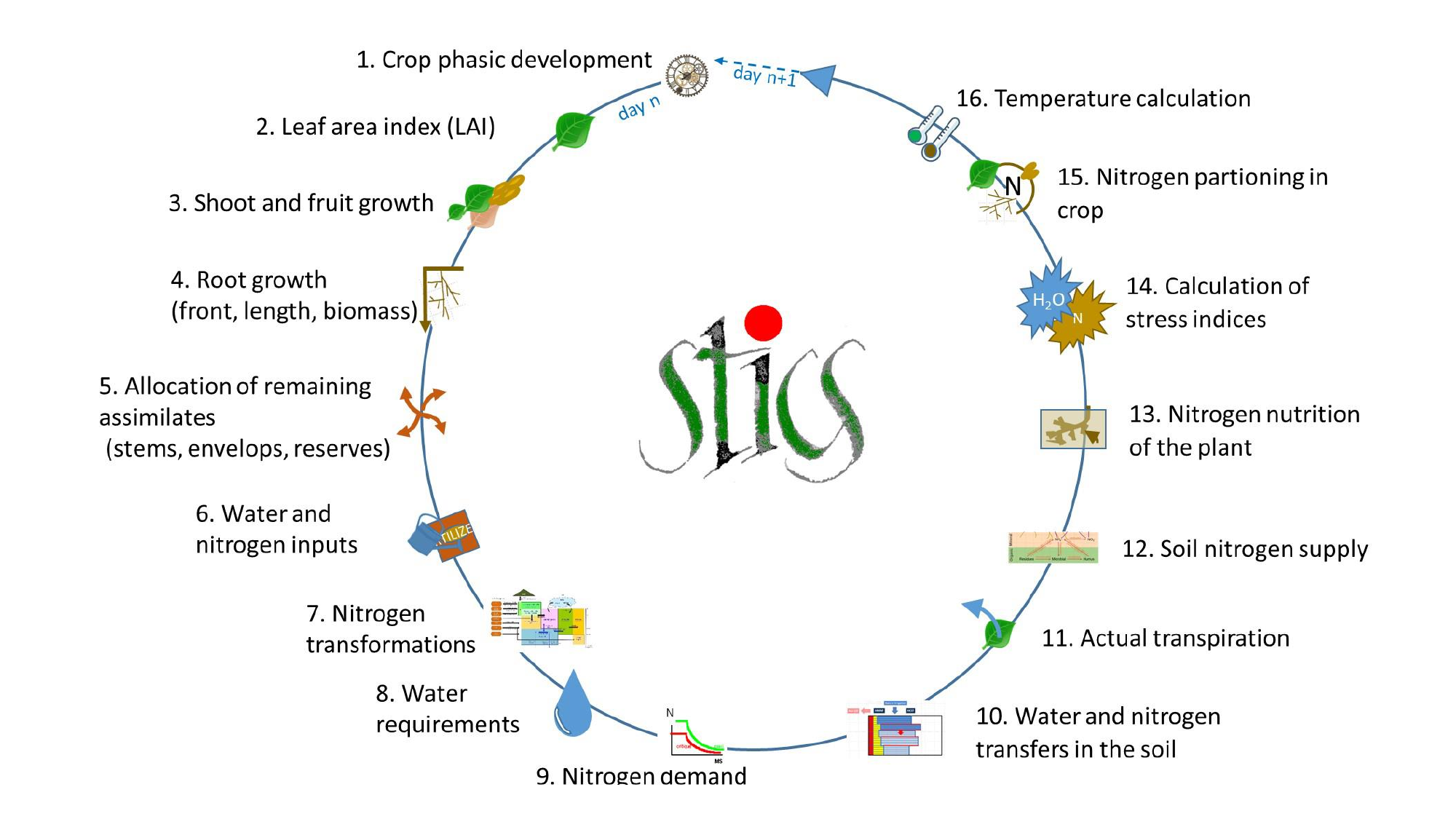}
    \caption{Process sequence in STICS (from \citet{2023_Beaudoin_etal}).}
    \label{F:STICS.processes}
\end{figure}

The STICS model \cite{1998_Brisson_etal} is a generic, process-based soil–crop model that simulates the effects of climate, soil conditions, and management practices on crop yield and quality. It is capable of representing a wide range of arable crops within cropping sequences (Yin etal 2020a, Yin etal 2020b).
%\cite{2020a_Yin_etal}, \cite{2020b_Yin_etal}. 
Operating on a daily time step, STICS describes crop phenological development and the flows of water, carbon, and nitrogen throughout the cropping cycle. There are more than 200 parameters in the model, which define the crop growth and soil carbon and nitrogen flows \citep{2012_varella_etal}. For each site-year combination, the climate, the initial soil conditions, and the management practices are defined \citep{2023_Beaudoin_etal}. The crop and soil processes in the model are simulated on a daily basis (Figure \ref{F:STICS.processes}). The phenological development of the crop is dependent on accumulated crop temperature, the day length, and the vernalisation requirements of the specific variety. The crop growth is calculated using the concept of radiation use efficiency, which combines the processes of photosynthesis and respiration. Radiation use efficiency is computed from the quantity of radiation intercepted by the green leaf area, atmospheric carbon dioxide levels, and is modified by the degree of abiotic stress experienced by the crop. These stresses relate to temperature and the availability of water and nitrogen. The crop uptake of nitrogen is dependent on the supply of nitrogen from the soil and the requirements of the crop for nitrogen. The model simulates the nitrogen cycle, the losses that occur through leaching, and the emissions of nitrous oxide. Residue return from the crop production cycle, and the addition of manure, have an impact on soil carbon stocks within the model. The soil water available to the crop is a function of the quantity of daily rainfall and irrigation, and the losses through evapotranspiration. The water flow in the profile is based on the concept of the tipping bucket \citep{2023_Beaudoin_etal}.

 The sequencing of 16 of the 
processes of STICS is shown in Figure \ref{F:STICS.processes}.  The 12 input parameters used for sensitivity analysis are shown in Table \ref{T:STICS.input.parameters}.  
The four output variables are measures made on the plant and are listed in Table \ref{T:STICS.outputs}.

\begin{table}[h] 
\centering 
\begin{tabular}{llll}
Parameter & Description & Lower & Upper \\ \hline 
efcroijuv & maximum radiation use efficiency during the vegetative stage &1 &7\\ 
efcroiveg & maximum radiation use efficiency during the juvenile phase &1 &10 \\ 
croirac &  elongation rate of the root apex &$\epsilon$&0.5 \\ 
stlevdrp &  cumulative thermal time between the  emergence starting date\\
& of filling of harvested organs & $\epsilon$& 6000\\ 
adil & parameter of the critical dilution curve & 1.0& 7.0 \\ 
bdil &   parameter of the critical dilution curve &0.01& 0.8 \\ 
vitircarb &  rate of increase of the Carbon harvest index vs time& 0.0010& 0.02\\ 
vitirazo & rate of increase of the Nitrogen harvest index vs time& 0.0010& 0.04 \\ 
adens & interplant competition parameter& -2& -$\epsilon$  \\ 
kmax &   maximum crop coefficient for water requirements&0.5& 4 \\ 
INNmin &  minimum value of INN possible for the crop & $\epsilon$& 1\\ 
inngrain2 & INN minimal for null net absorption of N during grain filling & 0.03& 2.0 \\
\end{tabular}
\caption{STICS: input parameter.  Definitions from Java STICS user guide, 2017, see https://stics.inra.fr. $\epsilon$=0.01.}
\label{T:STICS.input.parameters}
\end{table}

\begin{table}
\centering
\begin{tabular}{ll}
Output & Definition \\ \hline 
mafruit  & biomass of harvested organs, tonnes/ha \\
masec.n  & biomass of aboveground plant, tonnes/ha \\
CNgrain  & N concentration in fruits,\% dry weight \\
CNplante & N concentration in the aboveground plant, \% dry weight \\ \hline 
\end{tabular}
\caption{STICS: outputs. Descriptions of these variables and the parameters are taken from the Java STICS user guide, 2017, see https://stics.inra.fr).}
\label{T:STICS.outputs}    
\end{table}

% \subsubsection*{Details of running STICS} \tcb{This is not
% for the paper---just a reminder regarding the necessary files
% and what they contain (input parameters
% and input variables, etc.} 
% \begin{itemize}
% \item Unit of Simulation (USM) file.  This contains the local parameters and
% describes a cropping situation combining a plant,soil type and 
% weather conditions. The files associated to a USM are found in the
% user’s working directories (e.g., BarleyWorkspace).
% \begin{verbatim}
%     <usm nom = "BarleyN2016">
%     <datedebut>66</datedebut>
%     <datefin>365</datefin>
%     <finit>mais_ini.xml</finit>
%     <nomsol>Boghall</nomsol>
%     <fstation>Boghall_sta.xml</fstation>
%     <fclim1>Gogar.2016</fclim1>
%     <fclim2>Gogar.2016</fclim2>
%     <culturean>1</culturean>
%     <nbplantes>1</nbplantes>
%     <codesimul>0</codesimul>
%     <plante dominance = "1">
%       <fplt>proto_barley_plt.xml</fplt>
%       <ftec>BarleyN2016_tec.xml</ftec>
%       <flai>mais.lai</
% \end{verbatim}
% %
% \item The files of general and plants parameters are not attached to the local context of the USM.  e.g., proto\_barley\_plt.xml 
% located in STICS-plant directory.
% %
% \item The local model inputs (i.e. time and space related input variables) include (a) soil characteristics (e.g., Boghall), (b) 
% daily precipitation and temperature (e.g., Gogar2016).
% Also Boghall\_sta.xml has weather station name with latitude,
% climate. 

% \end{itemize}

%% file: 6_Appendices/2_Simulator_Details/6B4_RCode_Simulators.tex
\subsection{\label{app:sub.Rcode.Simulators}\texttt{R} code for running simulators for given input parameter combination}

\begin{enumerate}
    \item GR6J. 
An \texttt{R} package called \texttt{airGR} specifically created to run GR6J was used with a function called \texttt{RunModel}, which in turn relied on output from two other package functions, \texttt{CreateInputsModel} and \texttt{CreateRunOptions}.  An example snippet of code is shown below.
\begin{verbatim}
z <- airGR::RunModel(
   InputsModel = InputsModel,    #Created above by CreateInputsModel
   RunOptions  = RunOptions,     #Created above by CreateRunOptions
   Param       = Param.opt,      
   FUN_MOD     = RunModel_GR6J) 
\end{verbatim}

\item SimplyP. The code for SimplyP is written in C++ and an \texttt{R} wrapper function has been  created for compiling the code (Personal communication, Magnus Norling).
\begin{verbatim}
sourceCpp(paste0(Mobius.dir,'RWrapper/mobius_r.cpp')) 
## This brings in the following functions
##  "mobius_setup_from_parameter_and_input_file"; 
##  "mobius_setup_from_parameter_file_and_input_series"
##  "mobius_run_model"; "mobius_get_result_series"; "mobius_set_parameter_double"   
##  "mobius_set_parameter_uint"; "mobius_set_parameter_bool"; "mobius_set_parameter_time"
##  "mobius_run_with"; "mobius_print_result_structure"   

# Load RData file which is a dataframe (includes temperature, precipitation, etc)
load(file=paste0(Data.dir,"Tarland_data_matrix.RData"))

# set-up to Run SimplyP  
mobius_setup_from_parameter_and_input_file(
    ParameterFileName=Tarland_Par_File_Name,
    InputFileName=Tarland_Inputs_File_Name)

# Run SimplyP
mobius_run_model()
\end{verbatim}

\item STICS.  The STICS code, written in Java, is contained in a separate folder (labelled \texttt{STICS} below) that includes two key directories, one with plant specific information and another where calculations will be made (the workspace).
\begin{verbatim}
#STICS program directories
STICS.root           <- "D:/STICS/"
STICS.Workspace.path <- paste0(STICS.root,"Barley_Workspace/")
STICS.Plants.path    <- paste0(STICS.root,"plant/") # will get proto_barley_sens1_plt.xml, etc   

# Creates a model input template, called a USM, to which different parameter combinations 
# will be passed
source(file=paste0(Rcode.path,
  "Fixed_Values_Loading/4_USM_File_Creation.R"))

# Creating the multiple plant files (for different parameter combinations) in the plant directory
alter.stics.plt(usmspecs.basic = usmspecs.basic,
                alterpars.plt  = alterpars.plt.df,
                file.usms      = "usms.xml",
                Workspace.path = STICS.Workspace.path,
                Plants.path    = STICS.Plants.path,
                verbose=FALSE)

# Running STICS 
system(paste0("JavaSticsCmd.exe --run ",STICS.Workspace.path))
\end{verbatim}

\end{enumerate}

%% file: 6_Appendices/3_Results_Details/6C_Results_Details.tex
\section{\label{app:Results.Details}Additional Results}
Additional details of the results of the SA applied to the three models are included in this Section.

\input{6_Appendices/3_Results_Details/6C1_Results_GR6J_Details}

\input{6_Appendices/3_Results_Details/6C2_Results_SimplyP_Details}

\input{6_Appendices/3_Results_Details/6C3_Results_STICS_Details}

%% file: 6_Appendices/3_Results_Details/6C1_Results_GR6J_Details.tex
\subsection{\label{app:Results.Details.GR6J}GR6J}

\subsubsection{\label{app.GR6J.Qsimdetails}Outflow, $Q_{sim,d}$, regression tree details}
\begin{figure}[h]
    \centering
    \includegraphics[width=0.5\linewidth]{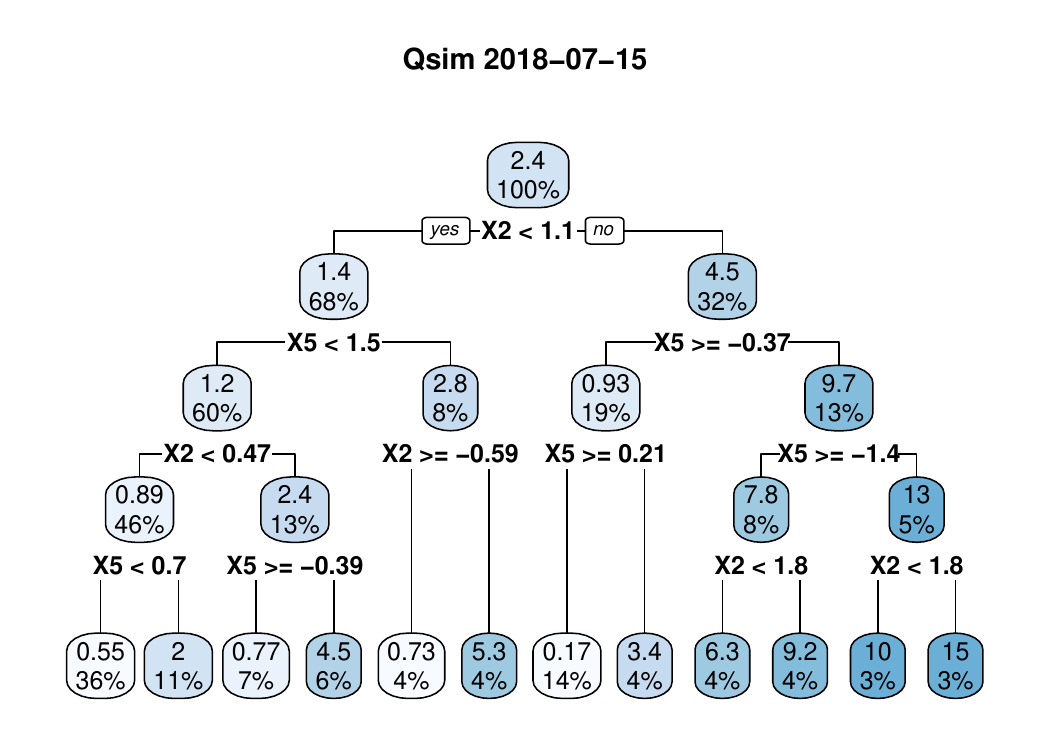}
    \caption{GR6J: Regression tree for QSim}
    \label{F:GR6J.QSim.RegTree.tree}
\end{figure}

\begin{figure}[h]
    \centering
    \includegraphics[width=0.6\linewidth]{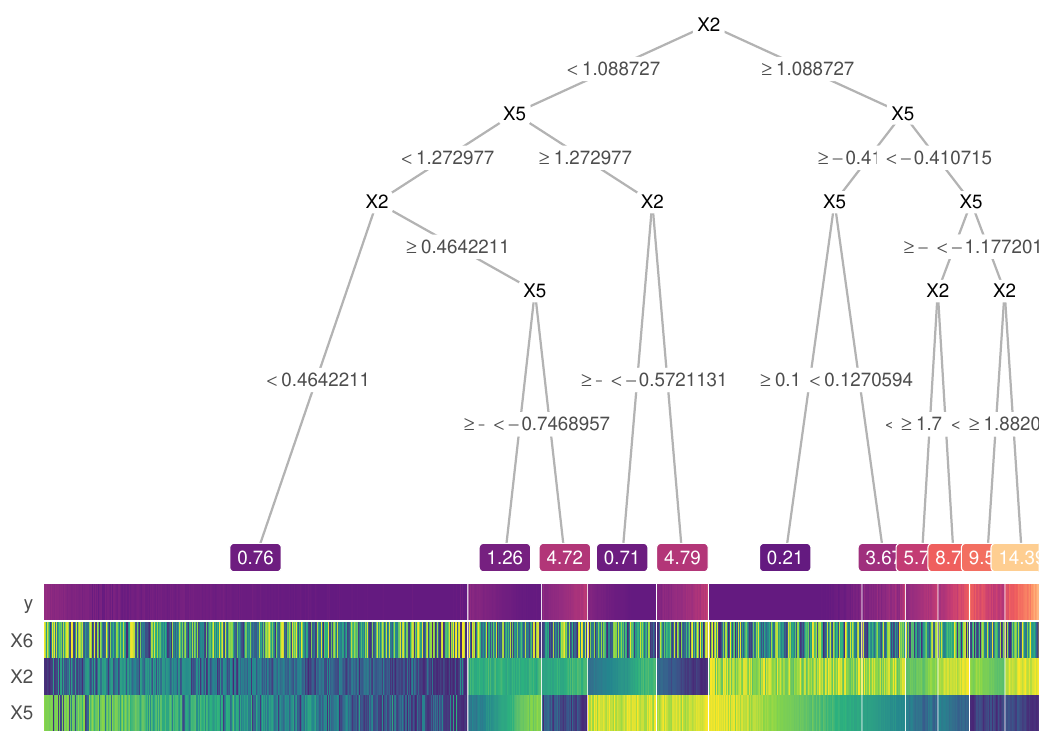}
    \caption{GR6J: Regression tree for QSim based on top 3 inputs with heatmap.}
    \label{F:GR6J.QSim.RegTree.tree.heatmap}
\end{figure}

The final regression tree for $Q_{sim,d}$ is shown in Figure \ref{F:GR6J.QSim.RegTree.tree}. X2 and X5 dominate as all branching decisions are based solely on the values for those two parameters.   Figure \ref{F:GR6J.QSim.RegTree.tree.heatmap} shows the tree based only on X2, X5, and X6. The lack of influence by X6 is clear in the heatmap, given the mixing of values within each of the $y$ outcome blocks (many vertical and different colored lines).  The importance of X2 and X5 is apparent by the solid parameter within some of the $y$ outcome blocks (e.g., assigned values for $Q_{sim}$ of 4.72 and 4.79), while the reversing of input parameter values for X2 and X5 for those same blocks with very similar assigned $Q_{sim}$ values indicates interactions between the two parameters.   

\subsubsection{\label{app:GR6J.Qsim.TVSA}$Q_{sim,d}$ time varying SA and external SA}
\begin{figure}[h]
    \centering
    \includegraphics[width=0.5\linewidth]{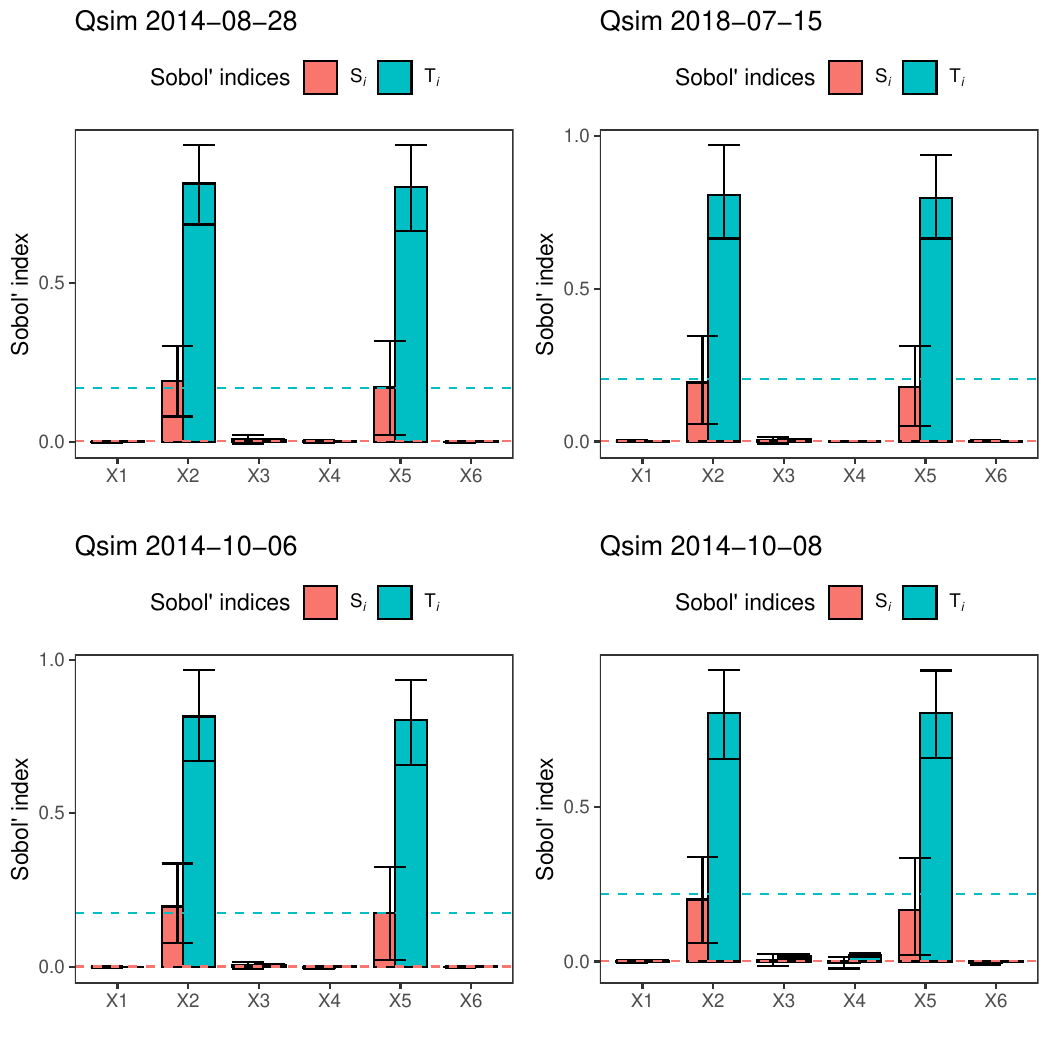}
    \caption{GR6J: Consistency of Sobol' results for $Q_{sim,d}$ over multiple days.}
    \label{F:GR6J.QSim.Sobol.Multipledays}
\end{figure}

\begin{figure}[h]
    \centering
    \includegraphics[width=0.5\linewidth]{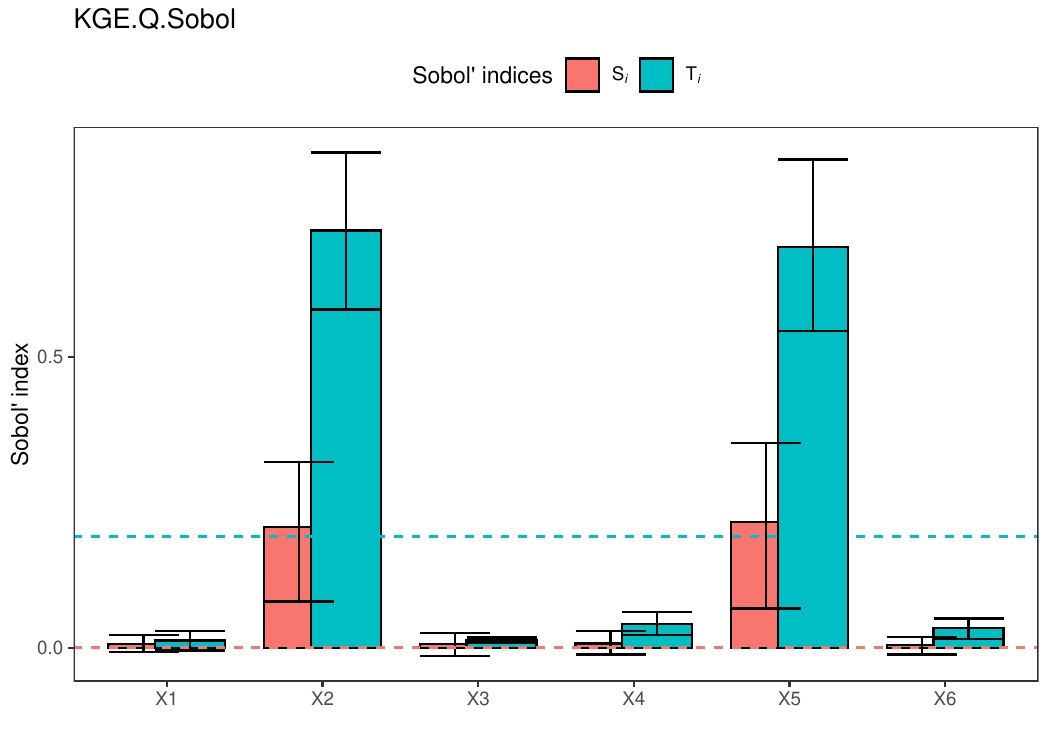}
       \caption{GR6J: External SA using Sobol' for KGE measure of $Q_{sim,d}$ for all of October 2014.}
    \label{F:GR6J.QSim.SobolKGE}
\end{figure}

Time varying SA of the Sobol' measures is shown in Figure \ref{F:GR6J.QSim.Sobol.Multipledays}, where the relative importance of the input parameters is shown to be quite similar for four different days.  A demonstration of External SA using the KGE measure (calculated for $Q_{sim}$ across all 31 days in October 2014 is shown in Figure \ref{F:GR6J.QSim.SobolKGE}, and again the parameters X2 and X5 dominate.
\clearpage

\subsubsection{Details on $P_r$ and $Q_9$}
Regarding the intermediate outputs of $P_r$ and $Q_9$, the relative importance of the input parameters matched what would be expected based upon the connections shown in Figure \ref{F:GR6J.schematic}. For $P_r$, only $X_1$ had any effect, and for $Q_9$, $X_1$ and $X_4$ were the most influential. Details are provided below. 

\subsubsection{$P_r$}
\begin{table}[h]
\centering
 Production store filling and emptying, $P_r$  \\
  \begin{tabular}{rrrrrrrrr}
   \hline
 & Morris & Sobol' & VARS-TO & Reg & RegTree & RF & \multicolumn{2}{c}{GPR} \\
  &  DGSM &  $T_i$  &         &      &  &       &  Slope &  InvRange \\
   \hline
$X_1$ & \textbf{\tcb{1.00}} & \textbf{\tcb{1.00}} & \textbf{\tcb{1.00}} & \textbf{\tcb{1.00}} & \textbf{\tcb{0.99}} & \textbf{\tcb{0.97}} & \textbf{\tcb{1.00}} & \textbf{\tcb{1.00}} \\ 
$X_2$ & 0.01 & 0.00 & 0.00 & 0.02 & 0.01 & 0.01 & 0.00 & 0.00 \\ 
$X_3$ & 0.00 & 0.00 & 0.00 & 0.00 & 0.00 & 0.01 & 0.00 & 0.00 \\ 
$X_4$ & 0.00 & 0.00 & 0.00 & 0.00 & 0.01 & 0.01 & 0.00 & 0.00 \\ 
$X_5$ & 0.00 & 0.00 & 0.00 & 0.00 & 0.01 & 0.01 & 0.00 & 0.00 \\ 
$X_6$ & 0.00 & 0.00 & 0.00 & 0.00 & 0.00 & 0.01 & 0.00 & 0.00 \\ 
    \hline
 \end{tabular}
\caption{GR6J: summary of results for different SA methods applied to $P_r$ (production store filling and emptying) on  2018-07-15. See text for explanations of the values shown. The most influential parameter per SA method is indicated by blue color.}
\label{T:GR6J.SA.PR.summary}
\end{table}

 \begin{figure}[h]
    \centering
    \includegraphics[width=0.60\linewidth]{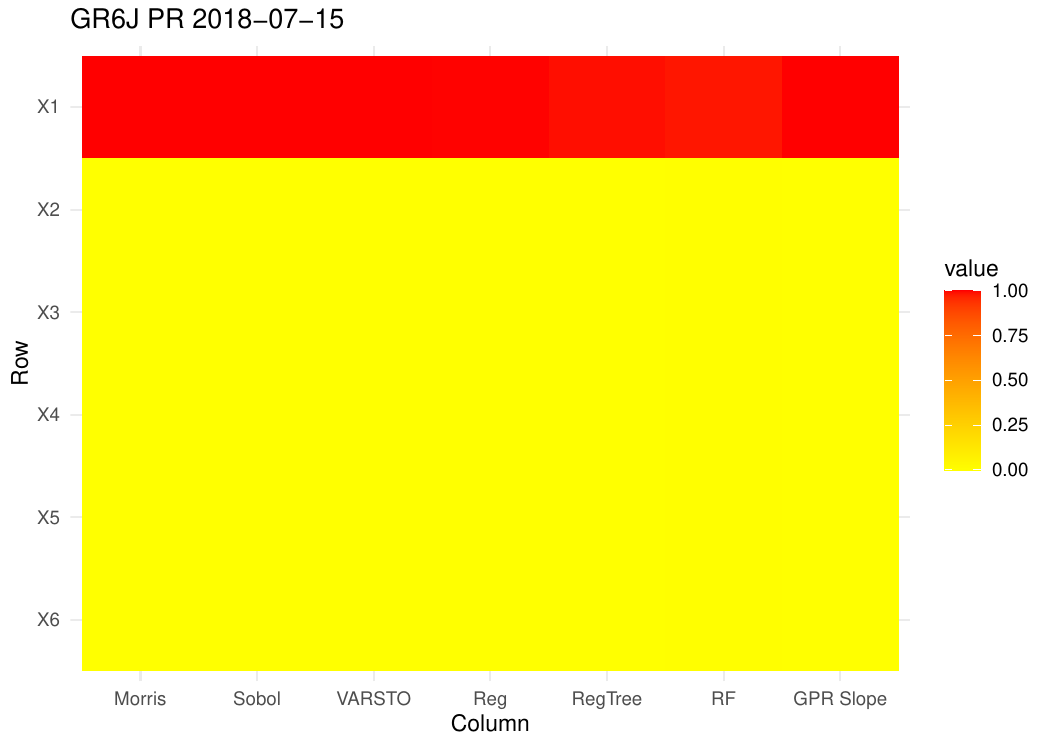}
    \caption{GR6J: Relative parameter importance for $P_r$ on 2018-07-15 for the different SA methods.}
    \label{F:GR6J.SA.PR.summary}
\end{figure}
 
\begin{figure}[h]
    \centering
    \begin{subfigure}[b]{0.45\textwidth}
    \includegraphics[width=\columnwidth,height=0.30\textheight]{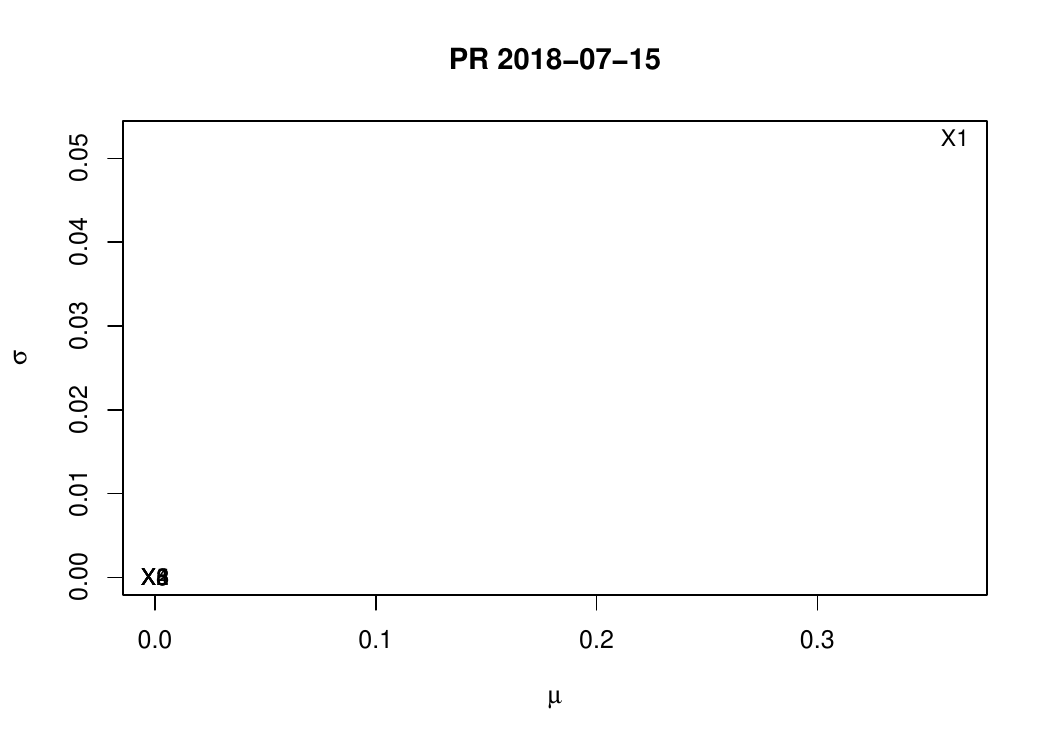}
    \caption{Morris $\sigma$ versus $\mu^*$.}
    \label{F:GR6J.Morris.Pr.Internal}
    \end{subfigure}
    \hfill
    \begin{subfigure}[b]{0.45\textwidth}
    \centering 
        \includegraphics[width=\columnwidth,height=0.30\textheight]{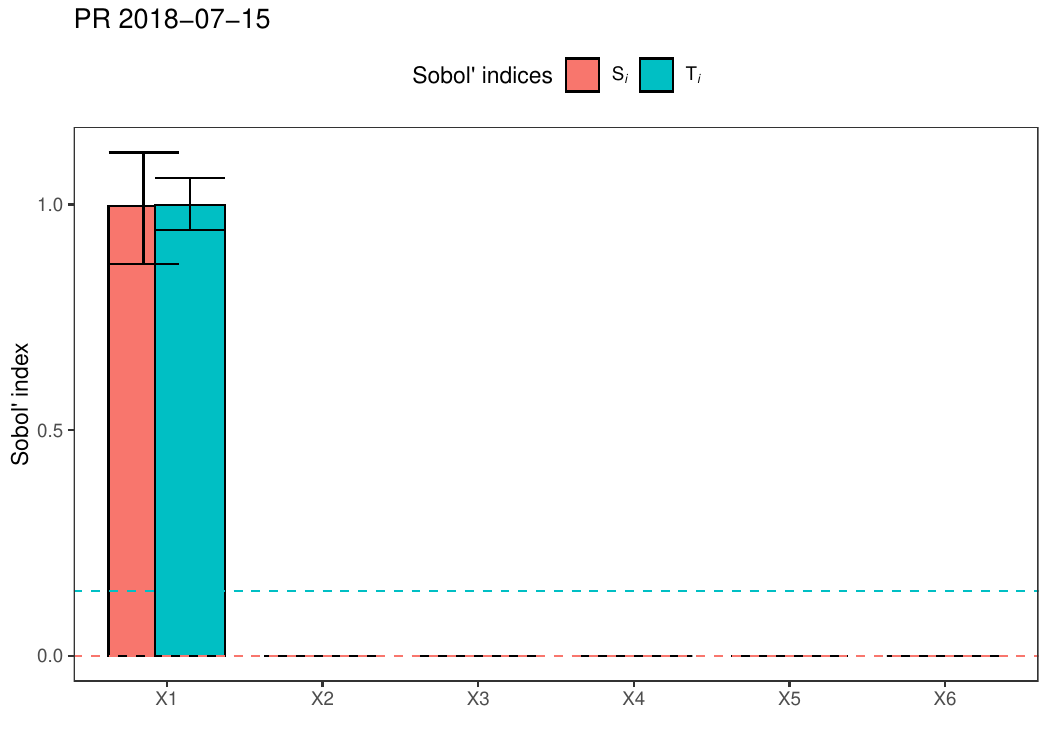}
     \caption{Sobol $S_{1,k}$ and $T_k$.}
     \label{F:GR6J.Sobol.Pr.Internal}
    \end{subfigure}
% --------------------------------------------------------------

   \begin{subfigure}[b]{0.45\textwidth}
    \centering
    \includegraphics[width=\columnwidth,height=0.30\textheight]{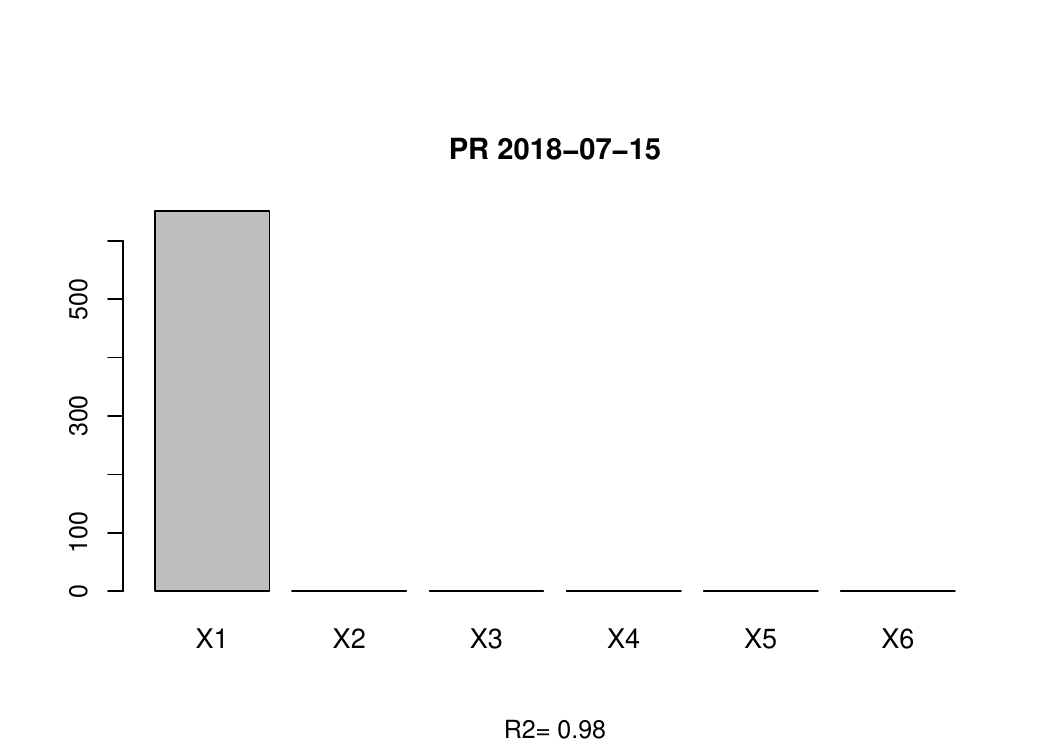}
    \caption{Multiple Regression standardized regression coefficients.}
    \label{F:GR6J.Regression.Pr.Internal}
 \end{subfigure}   
\hfill
    \begin{subfigure}[b]{0.45\textwidth}
    \centering
    \includegraphics[width=\columnwidth,height=0.30\textheight]{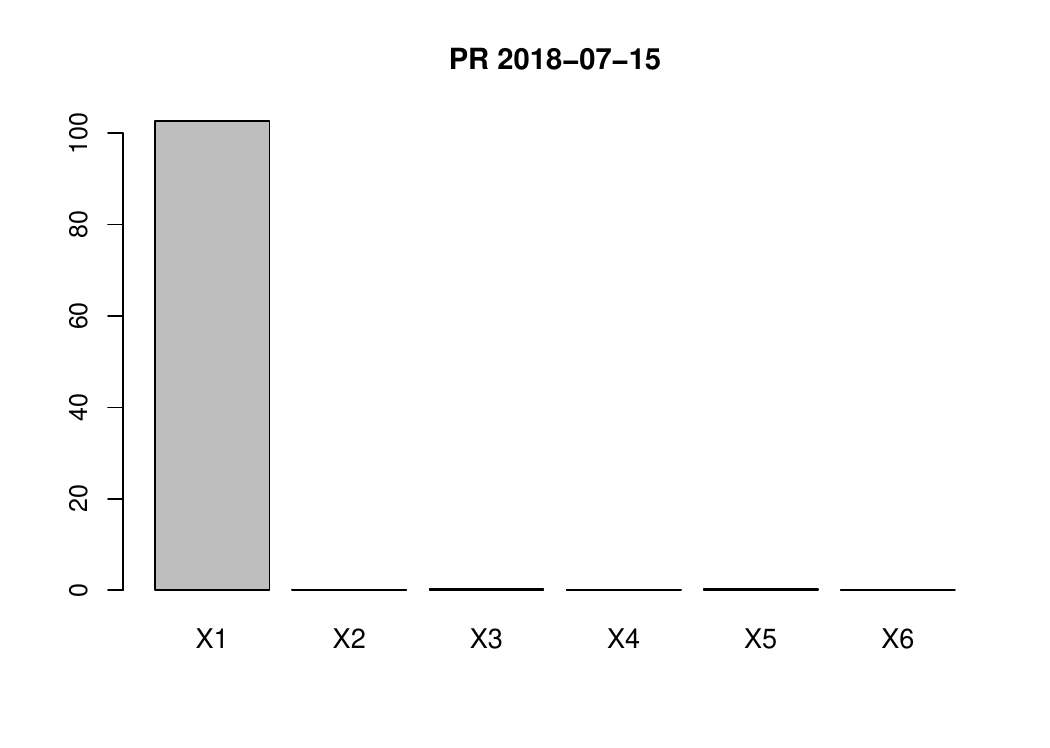} 
     \caption{Regression Tree parameter importance.}
     \label{F:GR6J.RegTree.Pr.Internal}
    \end{subfigure}   
% --------------------------------------------------------------

   \begin{subfigure}[b]{0.45\textwidth}
    \centering
    \includegraphics[width=\columnwidth,height=0.30\textheight]{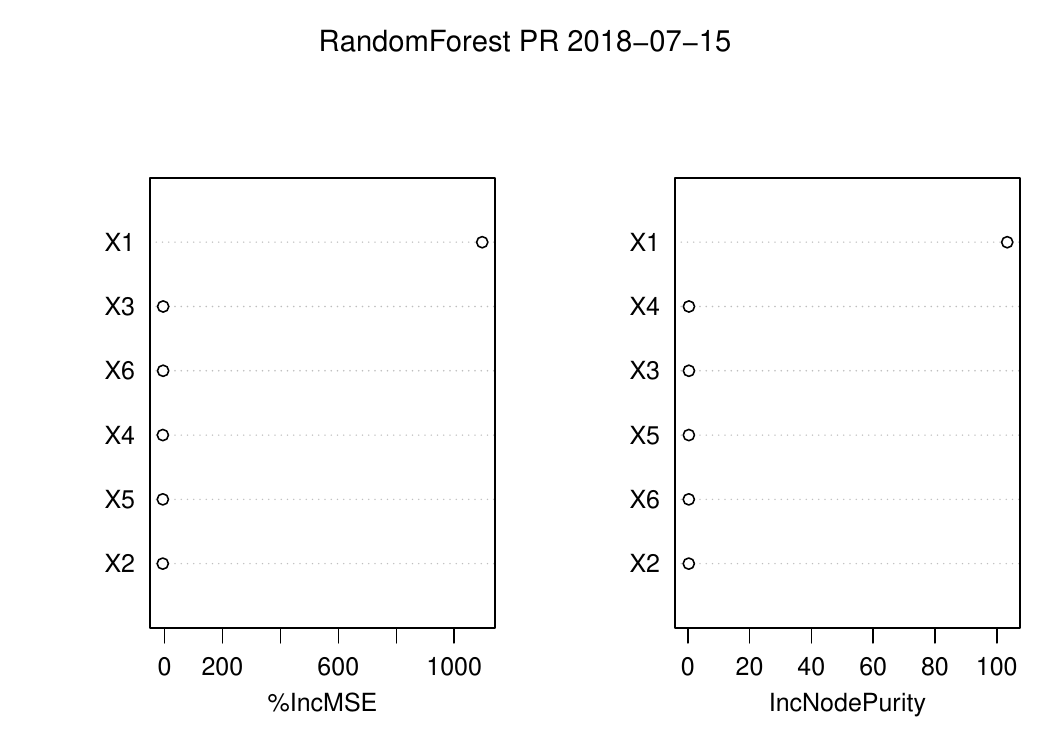}
    \caption{Random Forests parameter importance.}
    \label{F:GR6J.RF.Pr.Internal}
 \end{subfigure}   
 \hfill
   \begin{subfigure}[b]{0.45\textwidth}
    \centering
    \includegraphics[width=\columnwidth,height=0.30\textheight]{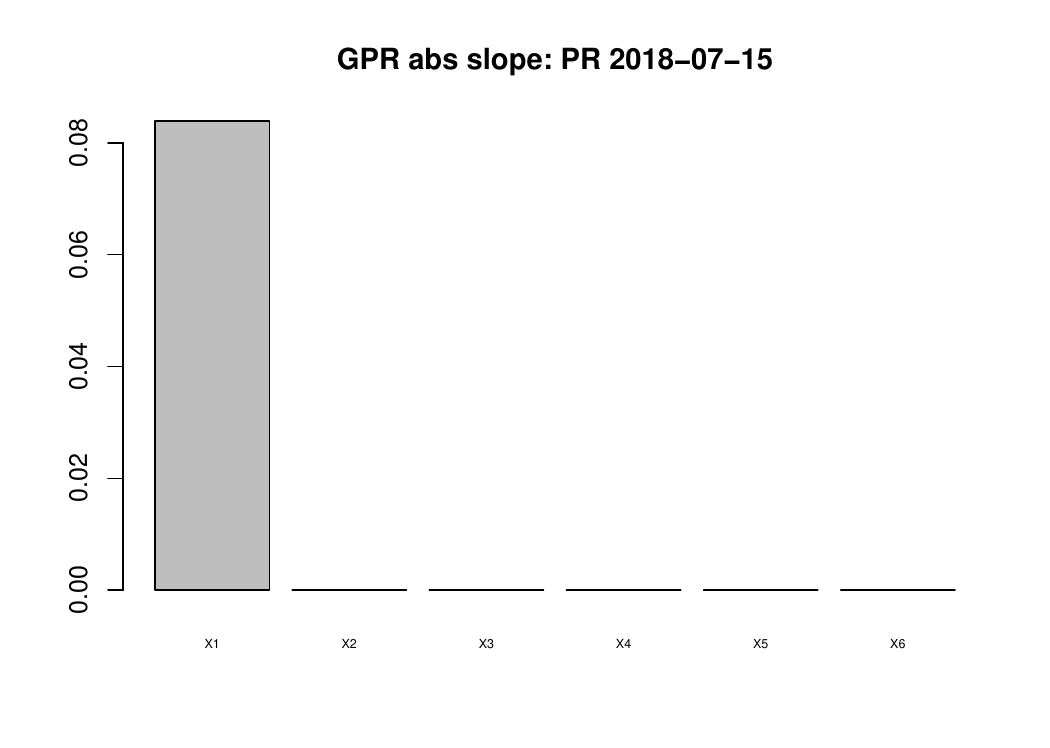}
    \caption{GPR standardized regression coefficient.}
    \label{F:GR6J.GPR.Slope.Pr.Internal}
 \end{subfigure}   
 \caption{GR6J: Sensitivity Analyses of
 $P_r$ (production store filling and emptying)  on 2015-07-15.}
 \label{F:GR6J.PR}
\end{figure}

\begin{figure}[h]
 \centering
     \includegraphics[width=0.8\textwidth]{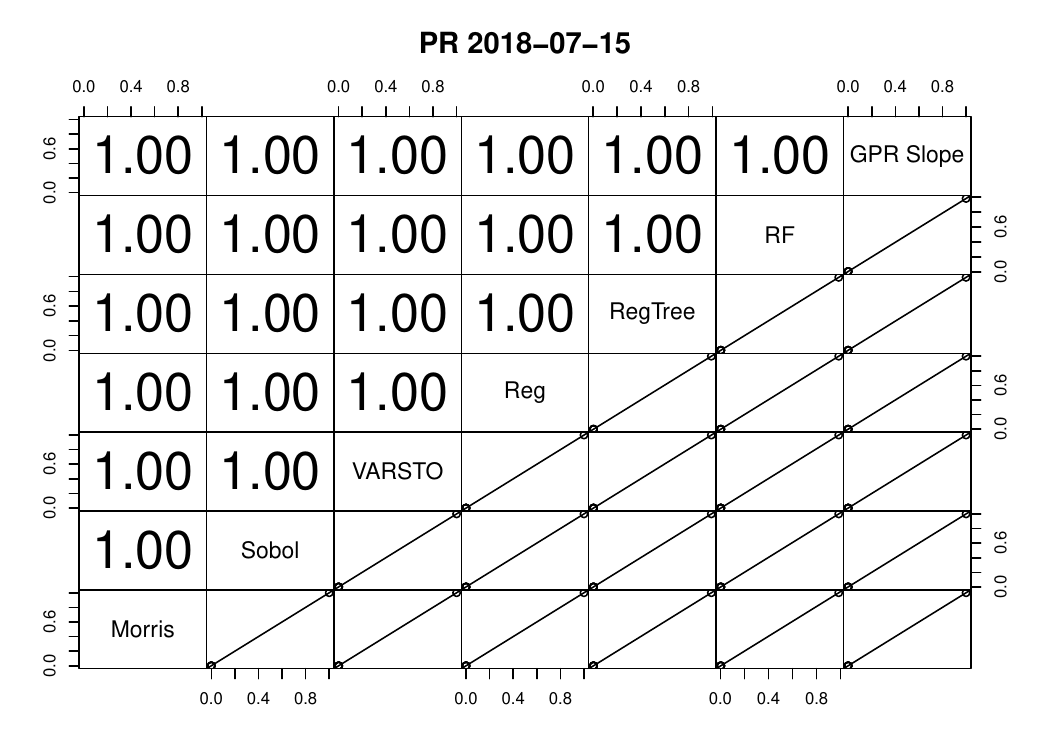}
     \caption{GR6J: pairwise scatterplots of SA measures for the 6 parameters influence on $P_r$ (production store filling and emptying) for  different SA procedures along with Pearson correlation coefficients.}
     \label{F:GR6J.Pairwise.PR.SA}
 \end{figure}

Figure \ref{F:GR6J.Pairwise.PR.SA} shows all pairwise scatterplots of the SA measures for $P_r$ (production store filling and emptying) and Pearson correlation coefficients. The Kendall's coefficient of concordance Weight W was 0.59 with the p-value for $H_0$ of no relationship being $<0.001$. 

 %--------------------------------------------------------------------------------
 \clearpage 
\subsubsection{$Q_9$}
\begin{figure}[h]
    \centering
    \begin{subfigure}[b]{0.45\textwidth}
    \includegraphics[width=\columnwidth,height=0.30\textheight]{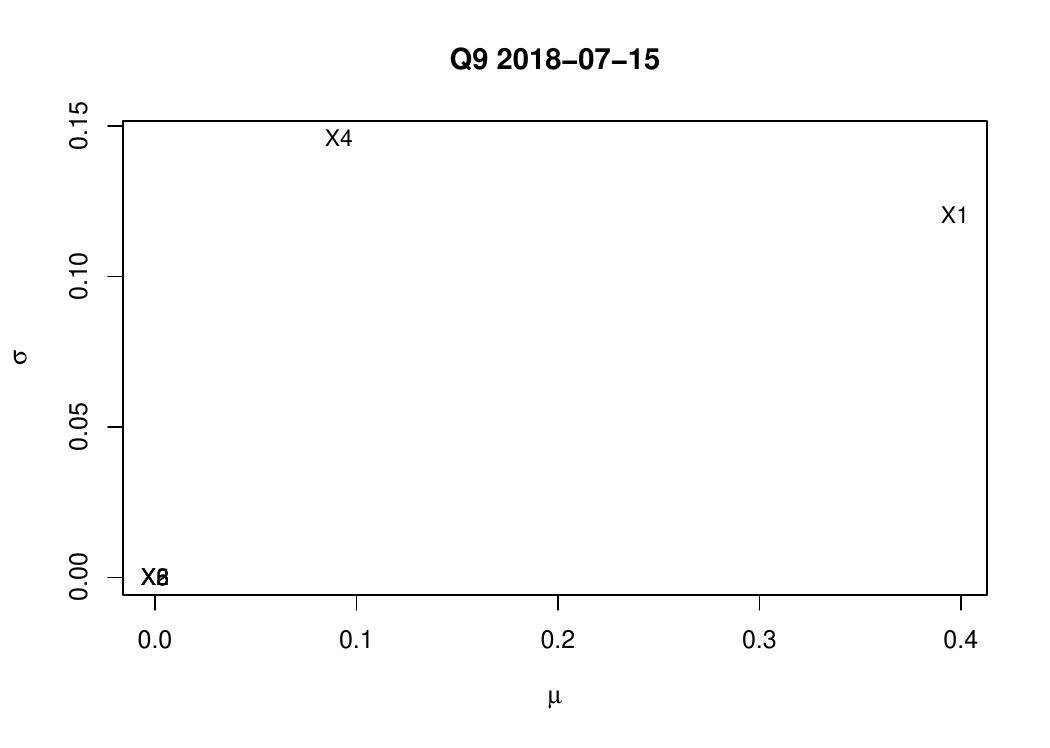}
    \caption{Morris $\sigma$ versus $\mu^*$.}
    \label{F:GR6J.Morris.Q9.Internal}
    \end{subfigure}
    \hfill
    \begin{subfigure}[b]{0.45\textwidth}
    \centering 
        \includegraphics[width=\columnwidth,height=0.30\textheight]{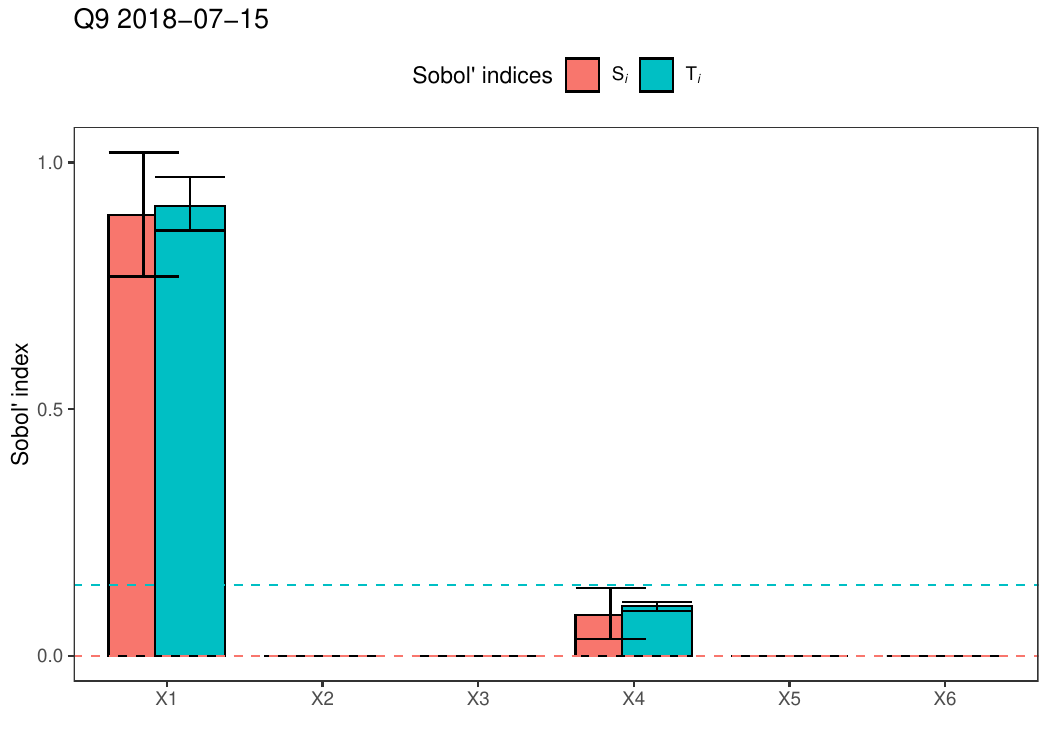}
     \caption{Sobol $S_{1,k}$ and $T_k$.}
     \label{F:GR6J.Sobol.Q9.Internal}
    \end{subfigure}
% ------------------------------------------------

   \begin{subfigure}[b]{0.45\textwidth}
    \centering
    \includegraphics[width=\columnwidth,height=0.30\textheight]{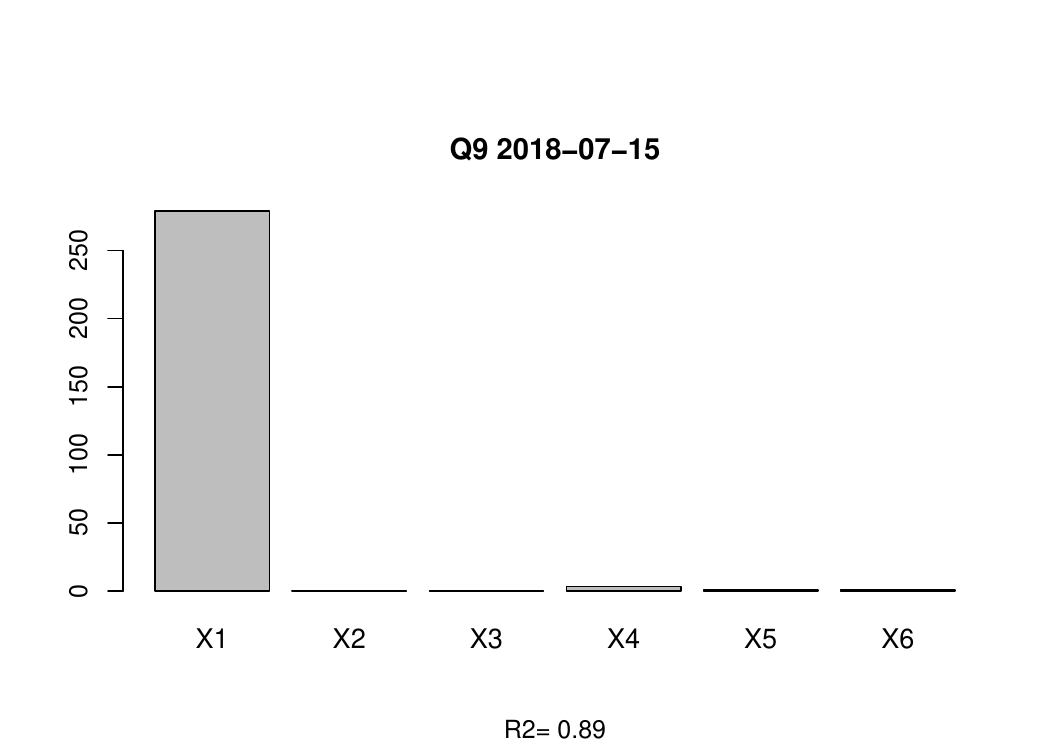}
    \caption{Multiple Regression standardized regression coefficients.}
    \label{F:GR6J.Regression.Q9.Internal}
 \end{subfigure}   
\hfill
    \begin{subfigure}[b]{0.45\textwidth}
    \centering
    \includegraphics[width=\columnwidth,height=0.30\textheight]{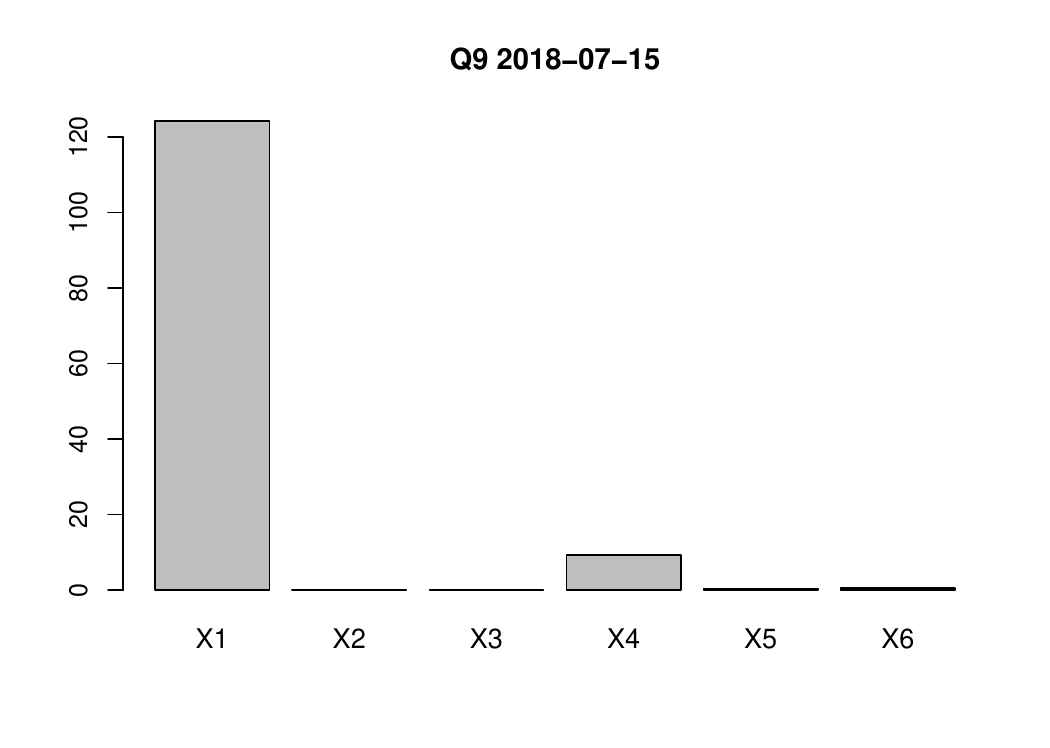} 
     \caption{Regression Tree parameter importance.}
     \label{F:GR6J.RegTree.Q9.Internal}
    \end{subfigure}   
% --------------------------------------------------------------

   \begin{subfigure}[b]{0.45\textwidth}
    \centering
    \includegraphics[width=\columnwidth,height=0.30\textheight]{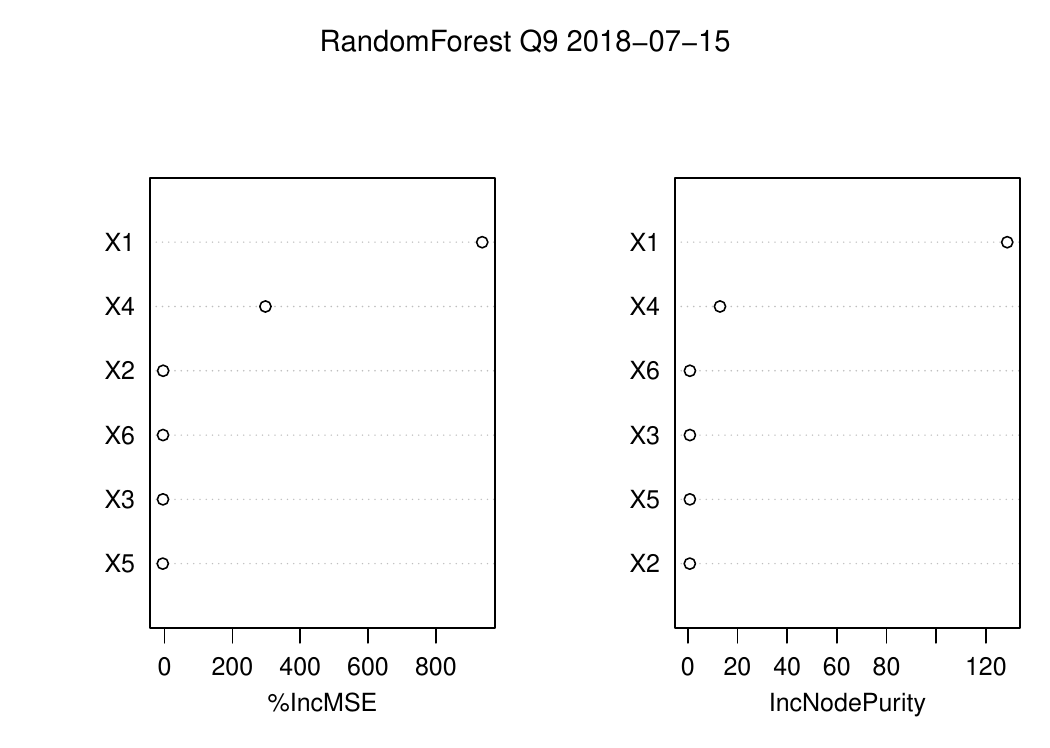}
    \caption{Random Forests parameter importance.}
    \label{F:GR6J.RF.Q9.Internal}
 \end{subfigure}   
 \hfill
   \begin{subfigure}[b]{0.45\textwidth}
    \centering
    \includegraphics[width=\columnwidth,height=0.30\textheight]{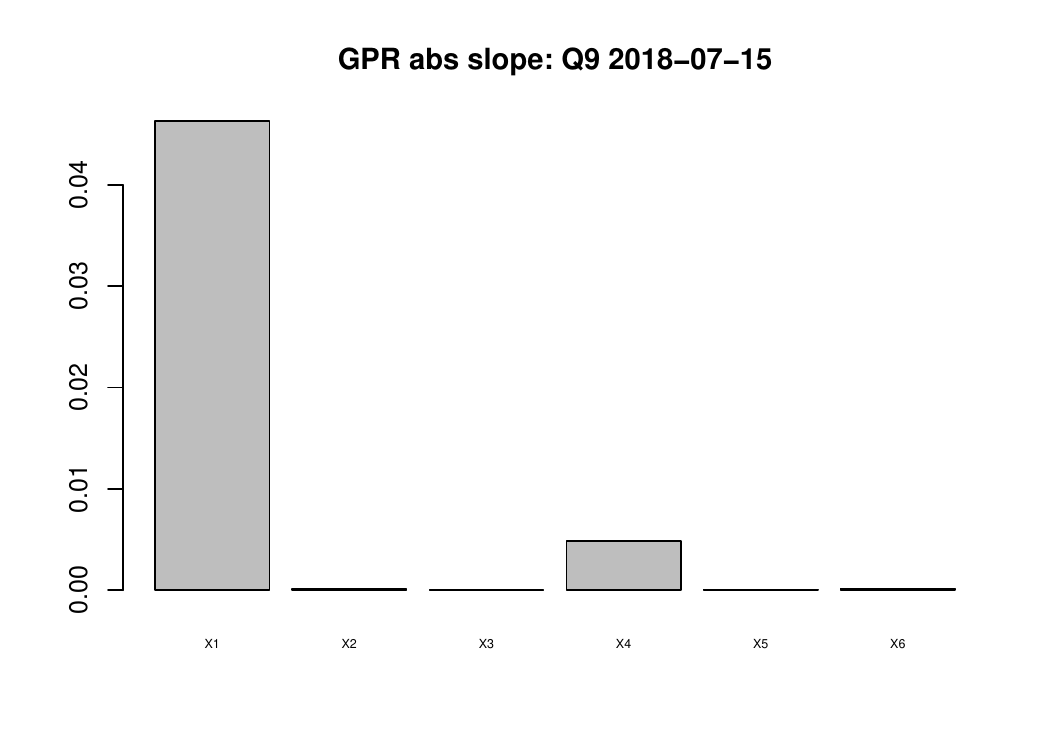}
    \caption{GPR standardized regression coefficients.}
    \label{F:GR6J.GPR.Slope.Q9.Internal}
 \end{subfigure}   
 \caption{GR6J: Sensitivity Analyses of
 $Q_9$ (production store filling and emptying)  on 2015-07-15.}
 \label{F:GR6J.Q9}
\end{figure}

With the exception of the GPR Inverse Range results (Table \ref{T:GR6J.SA.Q9.summary} and Figure \ref{F:GR6J.Q9}) $X_1$ has the most influence on $Q_9$, passage through the hydrographs, followed by $X_4$. The ordering is reversed for GPR Inverse Range, which accounts for input parameter effects on the residuals after accounting for the effects on the mean. 

\begin{table}[h]
\centering
 Passage through the hydrographs, $Q_9$  \\
  \begin{tabular}{rrrrrrrrr}
   \hline
  & Morris & Sobol' & VARS-TO & Reg & RegTree & RF & \multicolumn{2}{c}{GPR} \\
  & DGSM   &  $T_i$  &         &      &  &       &  Slope &  InvRange \\
   \hline
$X_1$ & \textbf{\tcb{0.70}} & \textbf{\tcb{0.90}} & \textbf{\tcb{0.92}} & \textbf{\tcb{0.97}} & \textbf{\tcb{0.92}} & \textbf{\tcb{0.87}} & \textbf{\tcb{0.90}} & \textbf{\tcr{0.28}} \\ 
$X_2$ & 0.00 & 0.00 & 0.00 & 0.00 & 0.01 & 0.01 & 0.00 & 0.00 \\ 
$X_3$ & 0.00 & 0.00 & 0.00 & 0.01 & 0.01 & 0.01 & 0.00 & 0.00 \\ 
$X_4$ & \textbf{\tcr{0.30}} & \textbf{\tcr{0.10}} & \textbf{\tcr{0.08}} & 0.01 & \textbf{\tcr{0.05}} & \textbf{\tcr{0.09}} & \textbf{\tcr{0.09}} & \textbf{\tcb{0.72}} \\ 
$X_5$ & 0.00 & 0.00 & 0.00 & 0.00 & 0.01 & 0.01 & 0.00 & 0.00 \\ 
$X_6$ & 0.00 & 0.00 & 0.00 & 0.00 & 0.01 & 0.01 & 0.00 & 0.00 \\ 
    \hline
 \end{tabular}
\caption{GR6J: summary of results for different SA methods applied to $Q_9$ (passage through the hydrographs) on  2018-07-15. See text for explanations of the values shown. The most influential parameter per SA method is indicated by blue color and next most influential in red color.}
\label{T:GR6J.SA.Q9.summary}
\end{table}

 \begin{figure}[h]
    \centering
    \includegraphics[width=0.60\linewidth]{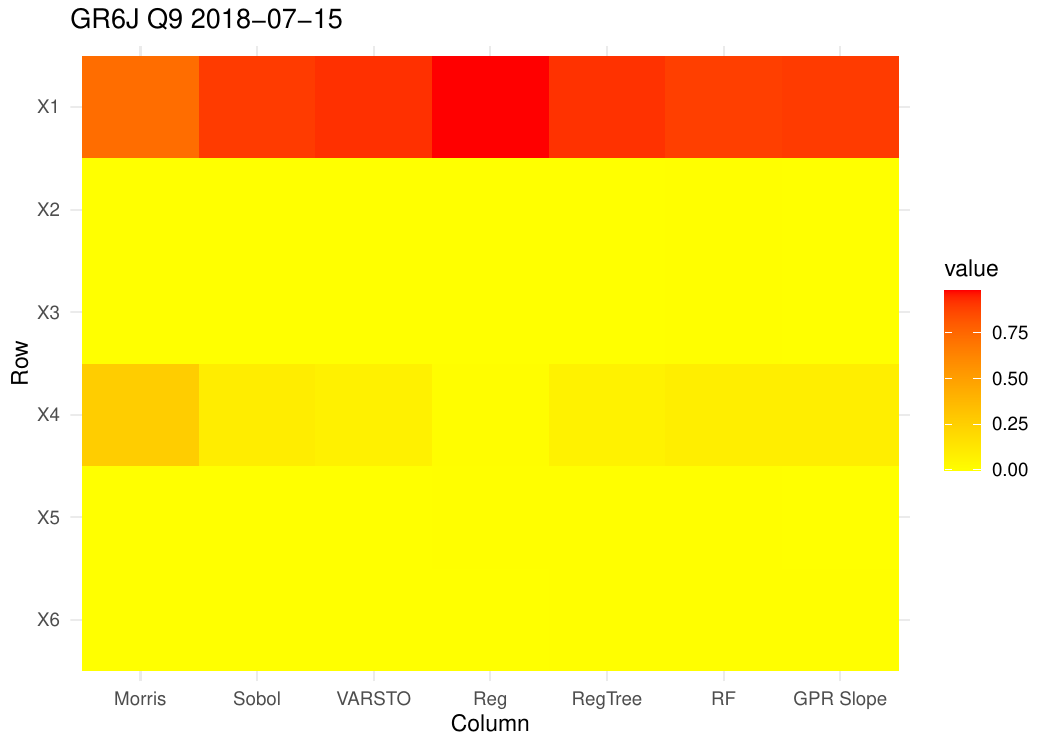}
    \caption{GR6J: Relative parameter importance for $Q_9$ on 2018-07-15 for the different SA methods.}
    \label{F:GR6J.SA.Q9.summary}
\end{figure} 

\begin{figure}[h]
 \centering
     \includegraphics[width=0.8\textwidth]{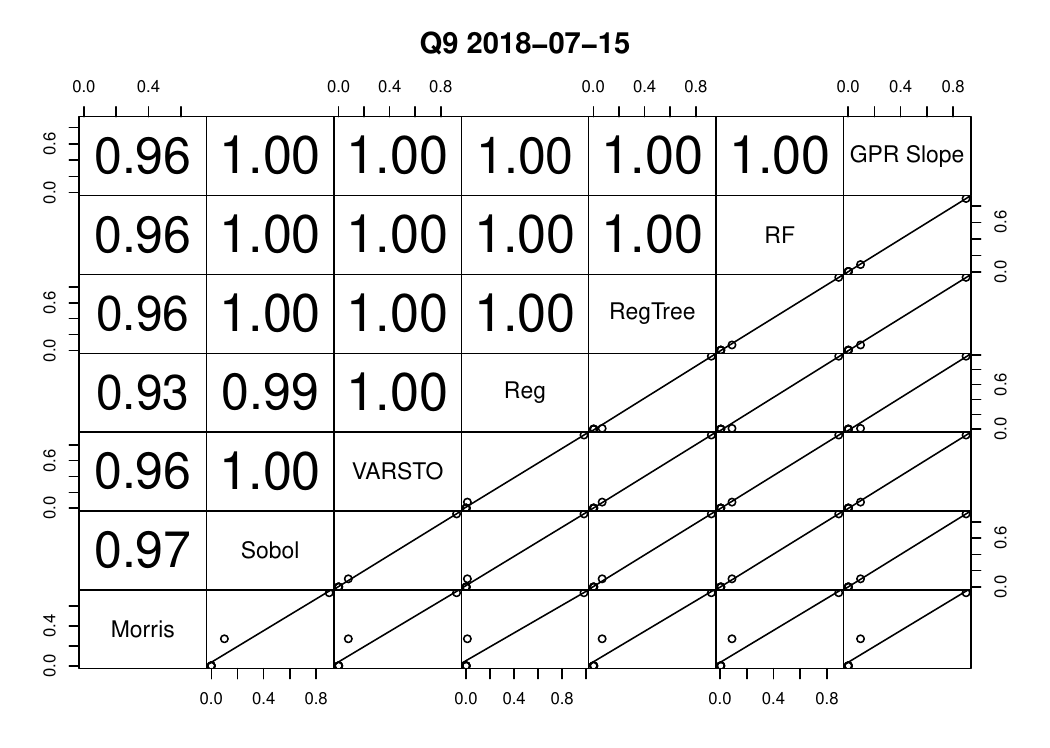}
     \caption{GR6J: pairwise scatterplots of SA measures for the 6 parameters influence on $Q_9$ (passage through the hydrographs) for  different SA procedures along with Pearson correlation coefficients.}
     \label{F:GR6J.Pairwise.Q9.SA}
 \end{figure}

Figure \ref{F:GR6J.Pairwise.Q9.SA} shows all pairwise scatterplots of the SA measures for $Q_9$ (passage through the hydrographs) and Pearson correlation coefficients. The Kendall's coefficient of concordance Weight W was 0.84 with the p-value for $H_0$ of no relationship being $<0.001$. 
 
\clearpage 

\subsubsection{GR6J: Summary of parameter importance across multiple outputs}
\begin{figure}[h]
    \centering
    \includegraphics[width=0.7\linewidth]{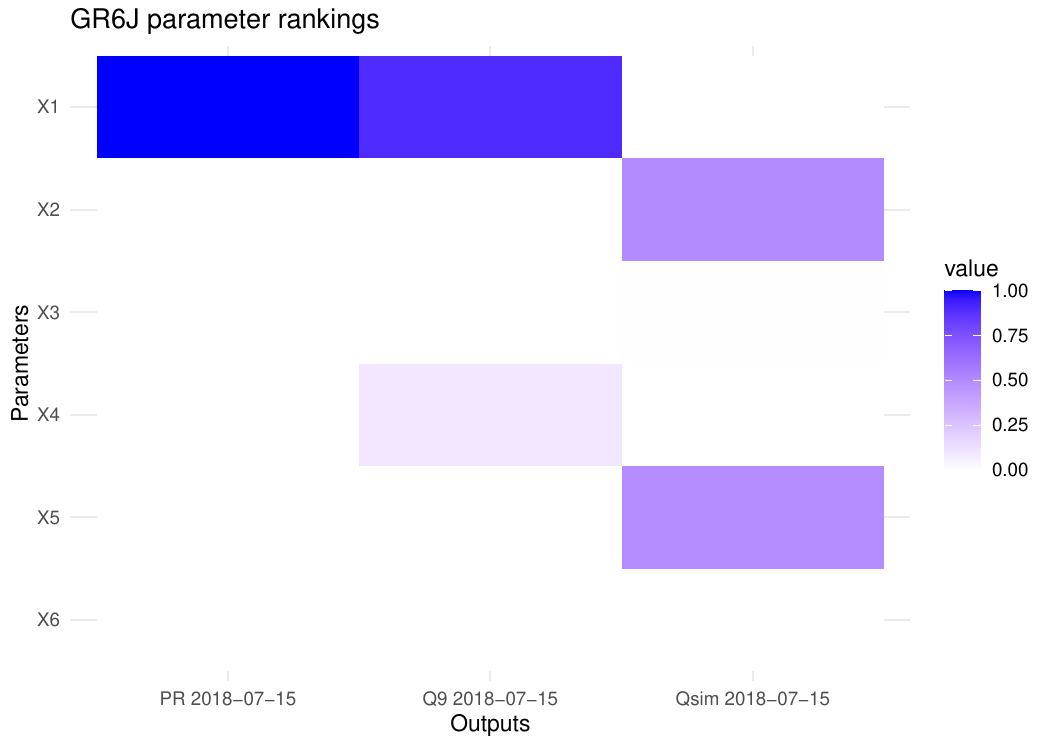}
    \caption{GR6J: Relative parameter importance for three outputs based on Sobol' $T_i$.}
    \label{F:GR6J.Importance.Summary}
\end{figure}

The relative importance of the GR6J parameters for the three outputs considered here is summarized in both Table \ref{T:GR6J.Importance.Summary} and Figure \ref{F:GR6J.Importance.Summary}.
\begin{table}[h]
\centering
\begin{tabular}{lrrr}
    Parameter & Pr & Q9 & QSim \\ \hline
    X1 & \tcb{1.00} & \tcb{0.90} & 0.00 \\
    X2 & 0.00 & 0.00 & \tcb{0.50} \\
    X3 & 0.00 & 0.00 & 0.00 \\
    X4 & 0.00 & \tcb{0.10} & 0.00 \\
    X5 & 0.00 & 0.00 & \tcb{0.49} \\
    X6 & 0.00 & 0.00 & 0.00 \\
\end{tabular}
\caption{GR6J: Relative parameter importance for three outputs based on Sobol' $T_i$.}
\label{T:GR6J.Importance.Summary}
\end{table}

\clearpage 

%% file: 6_Appendices/3_Results_Details/6C2_Results_SimplyP_Details.tex
\subsection{\label{app:Results.Details.SimplyP}SimplyP}
Results for sensitivity of SimplyP's output on day 200 of the year 2004 to the input parameters for three outputs, outflow ($Q$), suspended sediments ($SS$), and particle phosphorus ($PP$), all on log scale, are shown here.

\subsubsection{\label{app.SimplyP.TDPdetails}$TDP$ regression tree details}

\begin{figure}[h]
    \centering
    \includegraphics[width=0.6\linewidth]{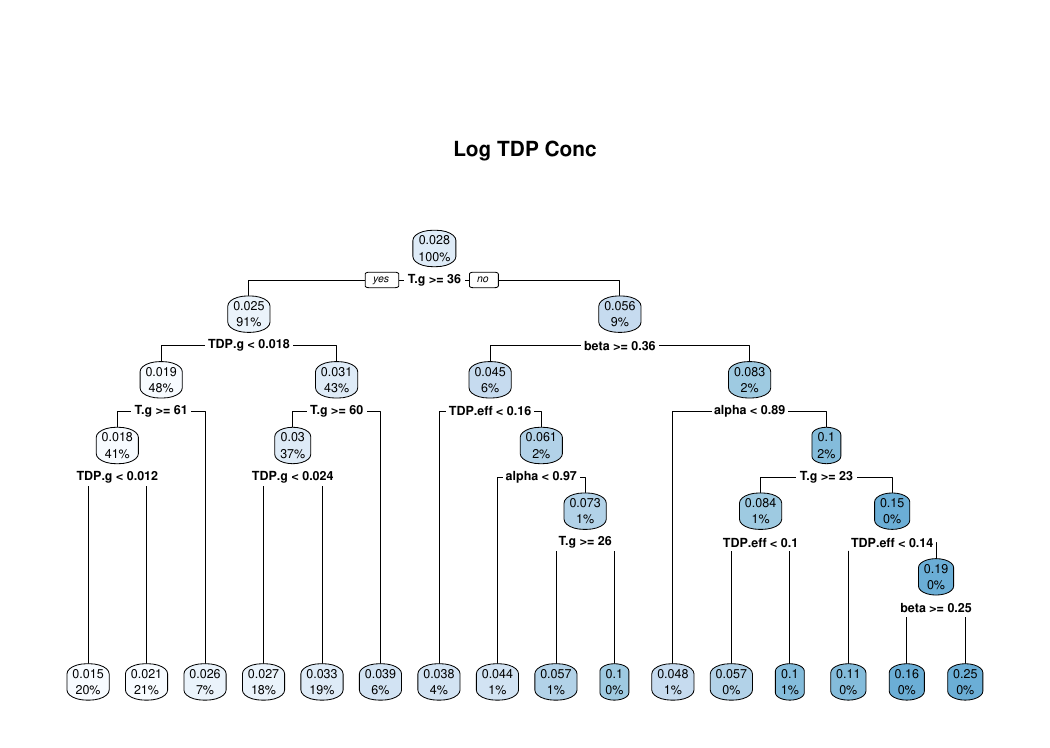}
    \caption{SimplyP: Regression tree for TDP}
    \label{F:SimplyP.TDP.RegTree.tree}
\end{figure}
\begin{figure}[h]
    \centering
    \includegraphics[width=0.7\linewidth]{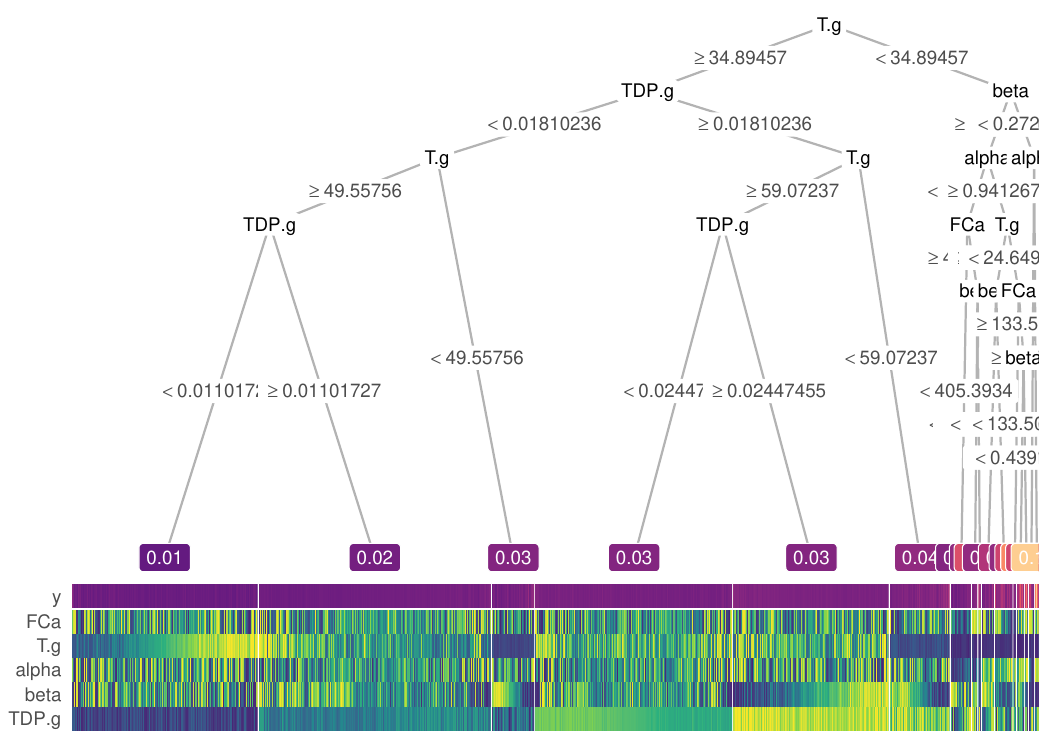}
    \caption{SimplyP: Regression tree for TDP based on top 5 inputs with heatmap.}
    \label{F:SimplyP.TDP.RegTree.tree.heatmap}
\end{figure}

% ******************************************
\subsubsection{Log of $Q_{200,2004}$}
\begin{figure}[h]
    \centering
    \includegraphics[width=0.60\linewidth]{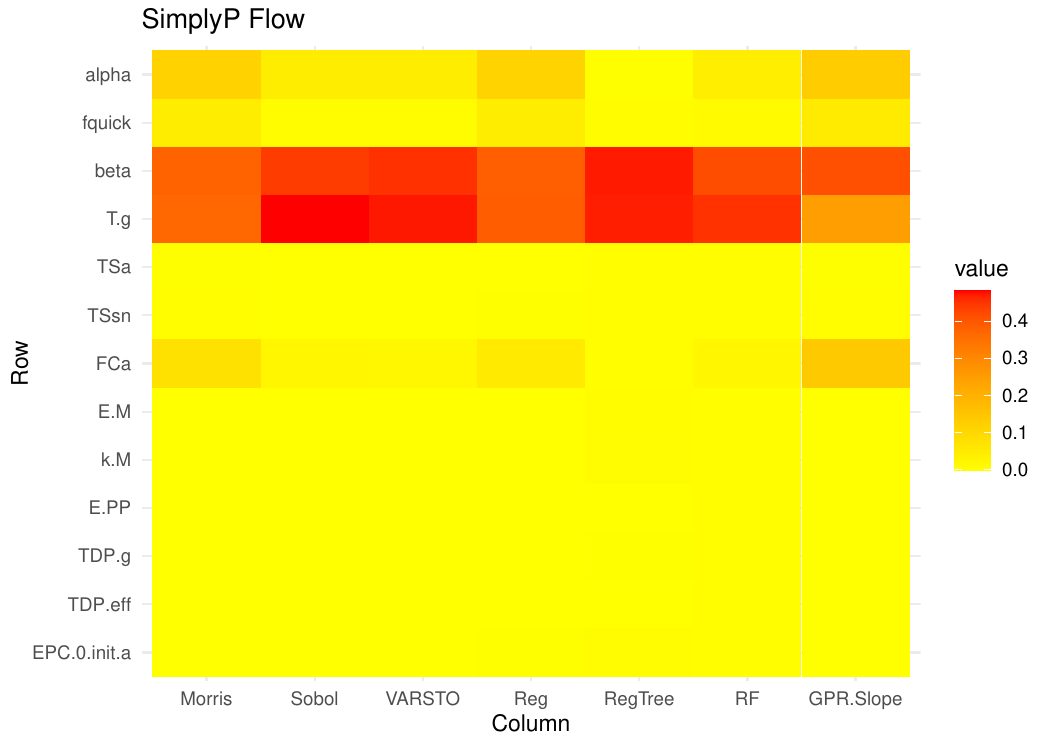}
    \caption{SimplyP: Relative parameter importance for $\log Flow_{200,2004}$ for the different SA methods.}
    \label{F:SimplyP.SA.Flow.summary}
\end{figure}

\begin{figure}[h]
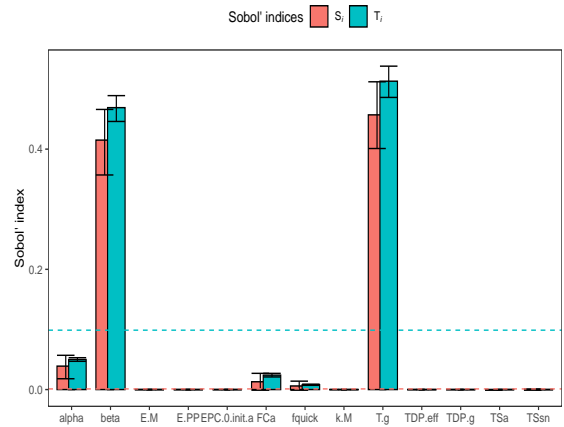
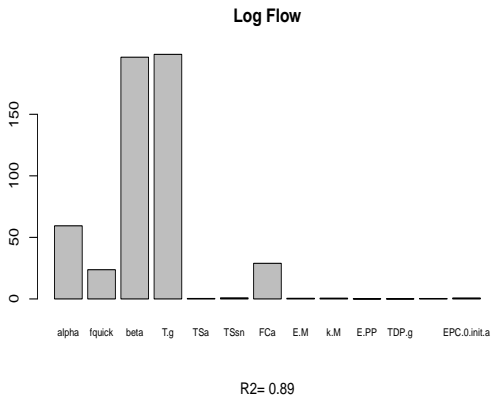
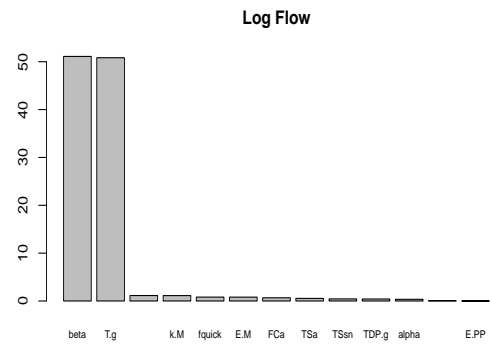
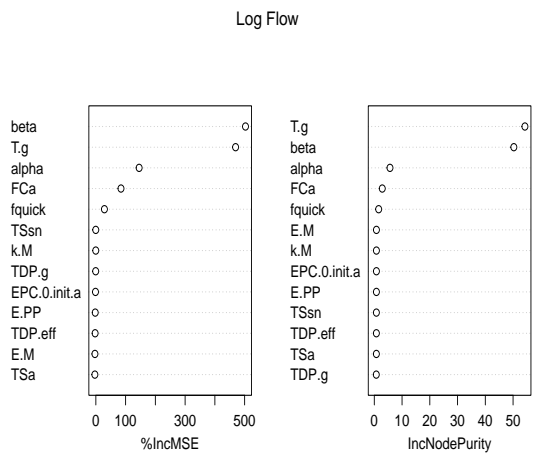
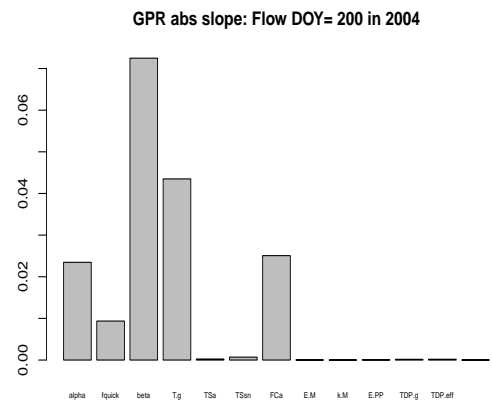

    \centering
    \begin{subfigure}[b]{0.45\textwidth}
    \includegraphics[width=\columnwidth,height=0.28\textheight]{0_Figures/SimplyP/SimplyP_Morris_mustar_sigma_Log_Flow_200.pdf}
    \caption{Morris $\sigma$ versus $\mu^*$.}
    \label{F:SimplyP.Morris.Flow}
    \end{subfigure}
    \hfill
    \begin{subfigure}[b]{0.45\textwidth}
    \centering 
        \includegraphics[width=\columnwidth,height=0.28\textheight]{0_Figures/SimplyP/SimplyP_Sobol_Log_Flow_200.pdf}
     \caption{Sobol $S_{1,k}$ and $T_k$.}
     \label{F:SimplyP.Sobol.Flow}
    \end{subfigure}
% ------------------------------------------

   \begin{subfigure}[b]{0.45\textwidth}
    \centering
    \includegraphics[width=\columnwidth,height=0.28\textheight]{0_Figures/SimplyP/SimplyP_Reg_Flow_200.pdf}
    \caption{Multiple Regression standardized regression coefficients.}
    \label{F:SimplyP.Regression.Flow}
 \end{subfigure}   
\hfill
    \begin{subfigure}[b]{0.45\textwidth}
    \centering
    \includegraphics[width=\columnwidth,height=0.28\textheight]{0_Figures/SimplyP/SimplyP_RegTree_Flow_200.pdf} 
     \caption{Regression Tree parameter importance.}
     \label{F:SimplyP.RegTree.Flow}
    \end{subfigure}   
% ---------------------------------------------

   \begin{subfigure}[b]{0.45\textwidth}
    \centering
    \includegraphics[width=\columnwidth,height=0.28\textheight]{0_Figures/SimplyP/SimplyP_RF_Flow_200.pdf}
    \caption{Random Forests results.}
    \label{F:SimplyP.RF.Flow}
 \end{subfigure}   
 \hfill
   \begin{subfigure}[b]{0.45\textwidth}
    \centering
\includegraphics[width=\columnwidth,height=0.28\textheight]{0_Figures/SimplyP/SimplyP_GPR_Slope_Flow_200.pdf}
    \caption{GPR standardized regression coefficients.}
    \label{F:SimplyP.GPR.Slope.Flow}
 \end{subfigure}   
 \caption{SimplyP: Six methods of Sensitivity Analyses applied to Log Flow on Day 200 in 2004.}
 \label{F:SimplyP.Q.Internal}
\end{figure}

Table \ref{T:SimplyP.SA.Flow.summary} and Figure \ref{F:SimplyP.SA.Flow.summary} summarize the results of different SA methods, scaled to sum to 1.0 per method, for assessing the relative effects of the 13 SimplyP parameters on $Q_{200,2004}$. See Section \ref{subsec:Results.GR6J} for descriptions of the values shown per method. The plots of the SA results for $Q$ based on six of the methods that were shown previously in Figure \ref{F:SimplyP.Q.Demonstration} are included here again for convenience (Figure \ref{F:SimplyP.Q.Internal}). As both the table and figure indicate, for most methods, the parameters beta and T.g have the most influence. GPR inverse range is detecting the importance of the inputs on the residuals of the regression model, and indicates that T.g is accounting for variation over and above its effect on the mean, and two other parameters, alpha and FCa are having effects, which is also indicated in the Sobol' plot.  

\begin{table}[h]
\centering
 log Outflow, $Q_{200,2004}$  \\
  \begin{tabular}{lrrrrrrrrrr}
   \hline &Type  & Morris & Sobol' & VARS-TO & Reg & RegTree & RF & \multicolumn{2}{c}{GPR} \\
 &        & DGSM &  $T_i$ &      &  &    &   &  Slope &  InvRange \\ 
   \hline
 alpha & Hyd &  0.12 & 0.05 & 0.05 & 0.12 & 0.00 & 0.05 & 0.13 &  \textbf{\tcr{0.25}}  \\ 
fquick & Hyd &0.05 & 0.01 & 0.01 & 0.05 & 0.01 & 0.01 & 0.05 & 0.07 \\ 
  beta & Hyd &\textbf{\tcb{0.38}} & \textbf{\tcr{0.45}} & \textbf{\tcb{0.47}} & \textbf{\tcb{0.38}} & \textbf{\tcb{0.45}} & \textbf{\tcr{0.41}} & \textbf{\tcb{0.41}} & 0.12   \\ 
T.g    & Hyd &\textbf{\tcr{0.37}} & \textbf{\tcb{0.48}} & \textbf{\tcr{0.47}} & \textbf{\tcb{0.39}} & \textbf{\tcr{0.47}} & \textbf{\tcb{0.45}} & \textbf{\tcr{0.25}}  & \textbf{\tcb{0.29}} \\  
   TSa & Hyd & 0.00 & 0.00 & 0.00 & 0.00 & 0.01 & 0.01 & 0.00& 0.00  \\ 
TSsn   & Hyd & 0.00 & 0.00 & 0.00 & 0.00 & 0.00 & 0.01 & 0.00& 0.01 \\ 
FCa    & Hyd & 0.08 & 0.02 & 0.02 & 0.06 & 0.01 & 0.02 & 0.14& \textbf{\tcr{0.25}}  \\ 
E.M    & Sed & 0.00 & 0.00 & 0.00 & 0.00 & 0.01 & 0.01 & 0.00 &0.00 \\ 
k.M    & Sed & 0.00 & 0.00 & 0.00 & 0.00 & 0.01 & 0.01 & 0.00&0.00 \\ 
E.PP   & Pho & 0.00 & 0.00 & 0.00 & 0.00 & 0.00 & 0.01 & 0.00&0.00 \\ 
TDP.g  & Pho & 0.00 & 0.00 & 0.00 & 0.00 & 0.00 & 0.01 & 0.00&0.00 \\ 
TDP.eff& Pho & 0.00 & 0.00 & 0.00 & 0.00 & 0.00 & 0.01 & 0.00&0.00 \\ 
EPC.0.init.a &Pho & 0.00 & 0.00 & 0.00 & 0.00 & 0.01 & 0.01 & 0.00&0.00 \\ 
    \hline
 \end{tabular}
\caption{SimplyP: summary of results for different SA methods applied to log $Q_{200,2004}$. Numbers in blue denote those with the largest relative value and those in red are the second largest.   See text in Section \ref{subsec:Results.GR6J} for explanations of the values shown. Reg=regression, RegTree=regression tree, RF=random forest, GPR=Gaussian Process regression.}
\label{T:SimplyP.SA.Flow.summary}
\end{table}

\clearpage

\clearpage 

\begin{figure}[h]
 \centering
     \includegraphics[width=0.8\textwidth]{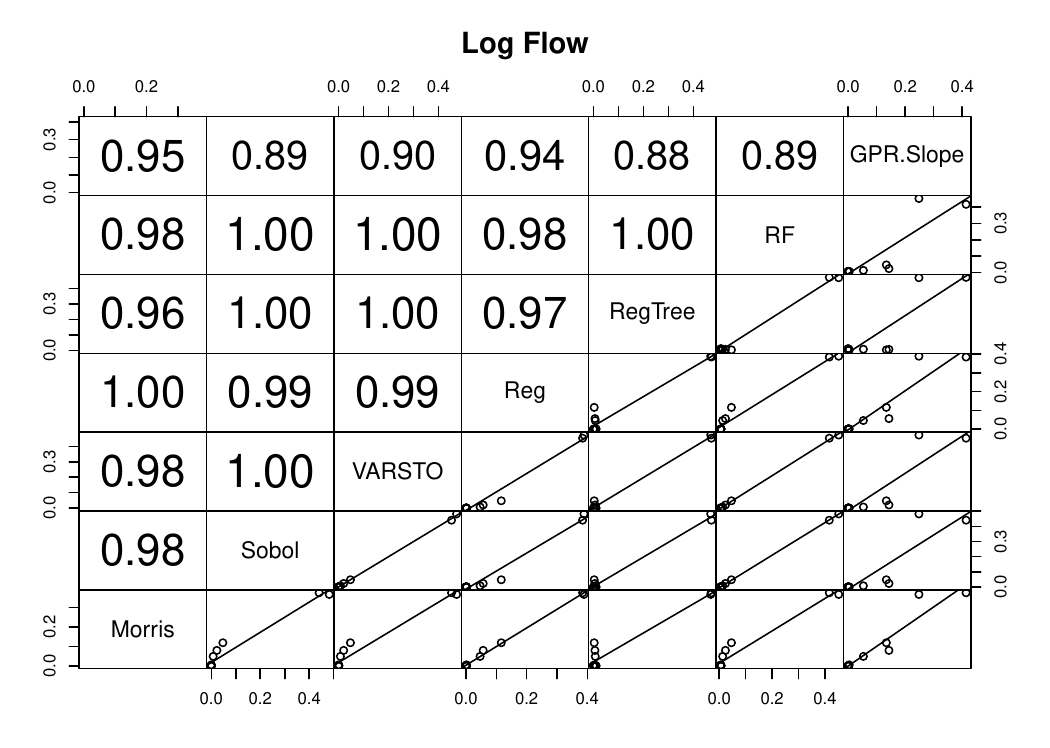}
     \caption{SimplyP: pairwise scatterplots of SA measures for the 13 parameters influence on log($Q_{200,2004}$) for different SA procedures along with Pearson correlation coefficients. Reg=regression, RegTree=regression tree, RF=random forest, GPR=Gaussian Process regression.}
     \label{F:SimplyP.Pairwise.Q.SA}
 \end{figure}

 Pairwise comparisons between the SA methods in terms of the SA measures attached to each parameter are shown in Figure \ref{F:SimplyP.Pairwise.Q.SA}. A positive linear relationship  occurred for all pairs. The overall degree of similarity was relatively high based on the Kendall W statistic of 0.84.
 
\clearpage 

\begin{figure}[h]
    \centering
    \includegraphics[width=0.6\linewidth]{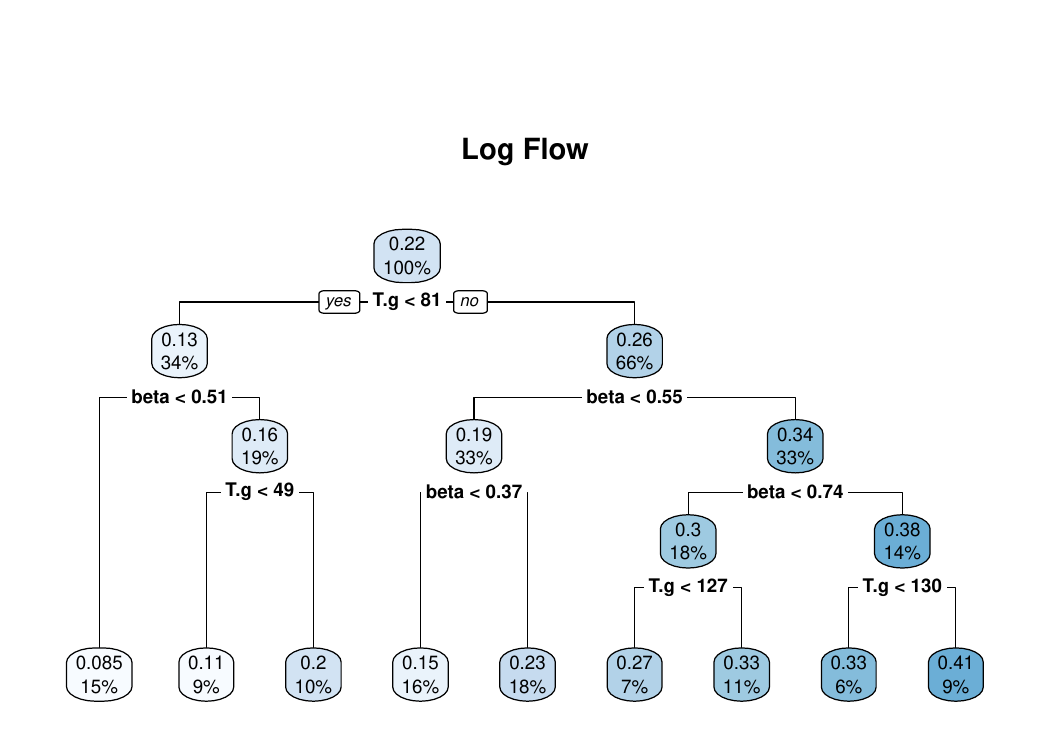}
    \caption{SimplyP: Regression tree for log flow.}
    \label{F:SimplyP.flow.RegTree.tree}
\end{figure}

\begin{figure}[h]
    \centering
    \includegraphics[width=0.6\linewidth]{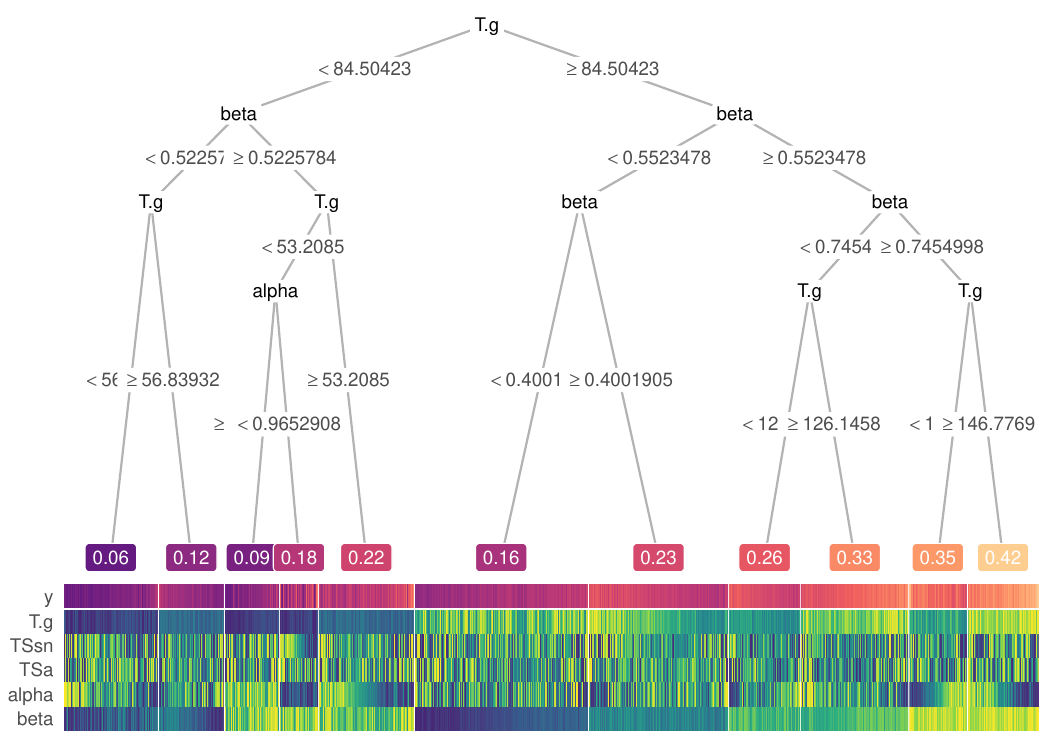}
    \caption{SimplyP: Regression tree for log flow based on top 4 inputs with heatmap.}
\label{F:SimplyP.flow.RegTree.tree.heatmap}
\end{figure}

 Further details on the regression tree for log flow are shown in Figure \ref{F:SimplyP.flow.RegTree.tree},  which shows beta and T.g  influencing most branching decisions.   Figure \ref{F:SimplyP.flow.RegTree.tree.heatmap} shows the tree based only on the top four parameters which again shows beta and T.g dominating, but with possible interactions amongst the four. 

\clearpage 
% ************************************
\subsubsection{Log of $SS_{200,2004}$}

\begin{figure}[h]
    \centering
    \begin{subfigure}[b]{0.45\textwidth}
    \includegraphics[width=\columnwidth,height=0.30\textheight]{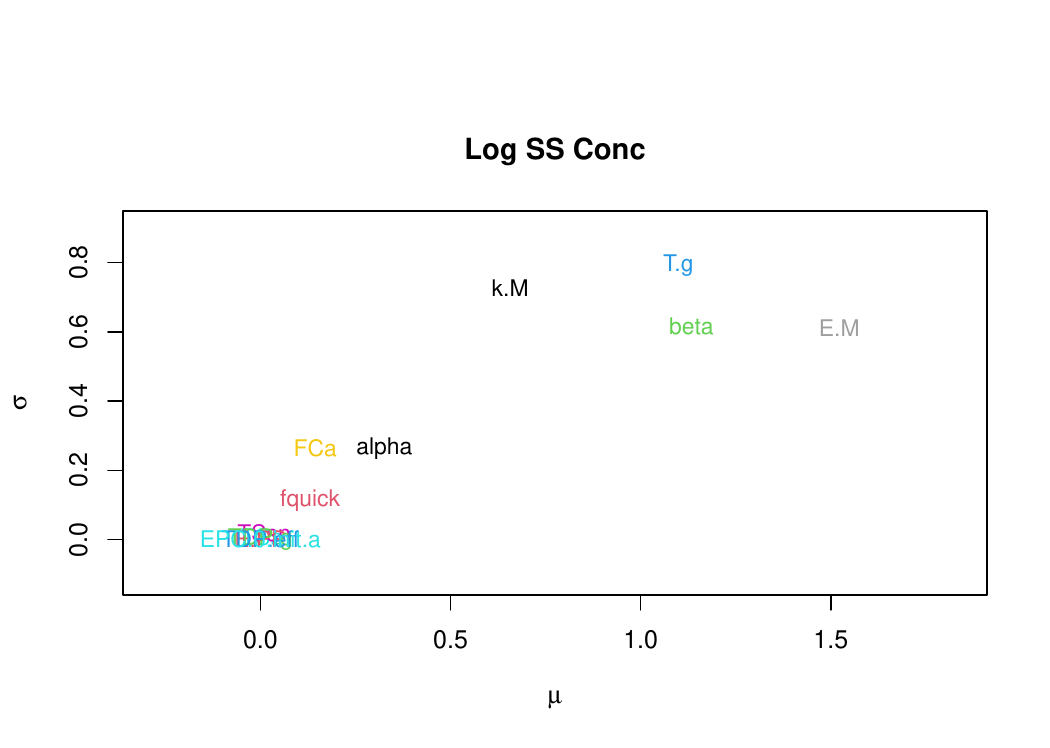}
    \caption{Morris $\sigma$ versus $\mu^*$.}
    \label{F:SimplyP.Morris.SS.Internal}
    \end{subfigure}
    \hfill
    \begin{subfigure}[b]{0.45\textwidth}
    \centering 
        \includegraphics[width=\columnwidth,height=0.30\textheight]{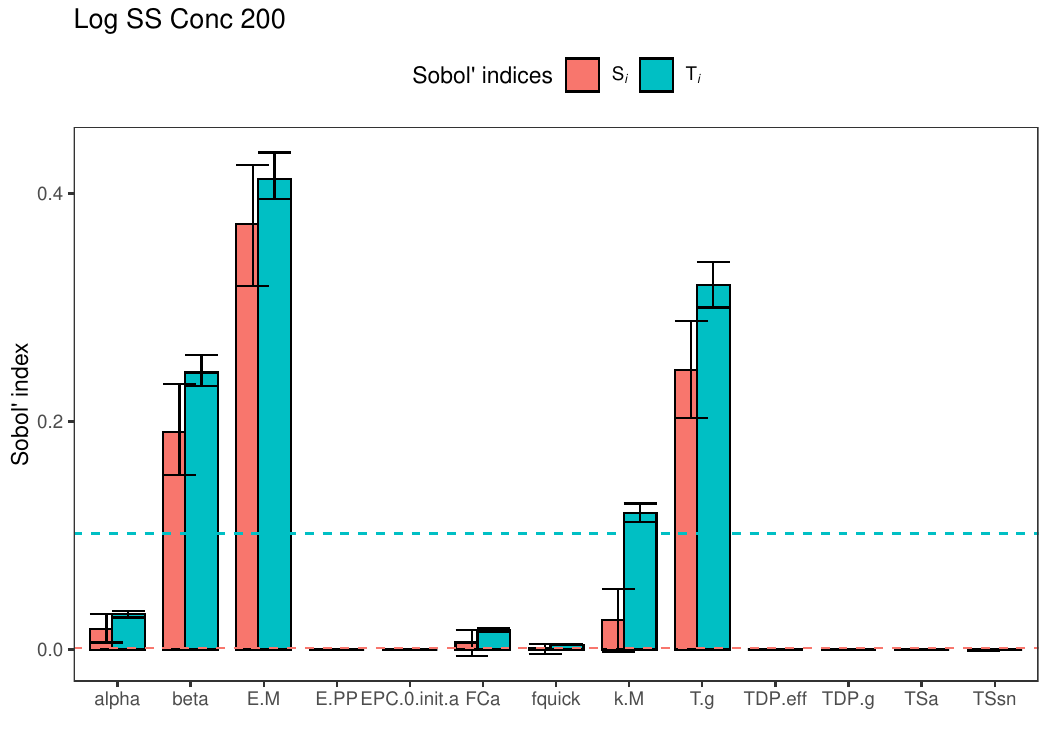}
     \caption{Sobol' $S_{1,k}$ and $T_k$.}
     \label{F:SimplyP.Sobol.SS.Internal}
    \end{subfigure}
% ------------------------------------

   \begin{subfigure}[b]{0.45\textwidth}
    \centering
    \includegraphics[width=\columnwidth,height=0.30\textheight]{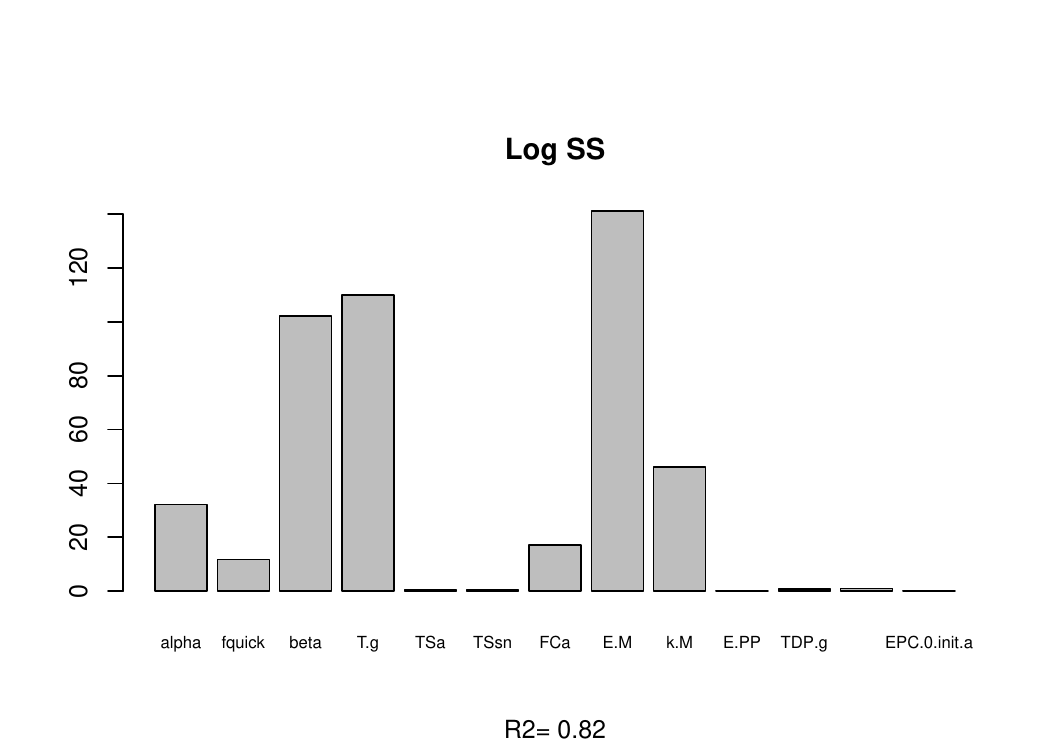}
    \caption{Multiple Regression standardized regression coefficients.}
    \label{F:SimplyP.Regression.SS.Internal}
 \end{subfigure}   
\hfill
    \begin{subfigure}[b]{0.45\textwidth}
    \centering
    \includegraphics[width=\columnwidth,height=0.30\textheight]{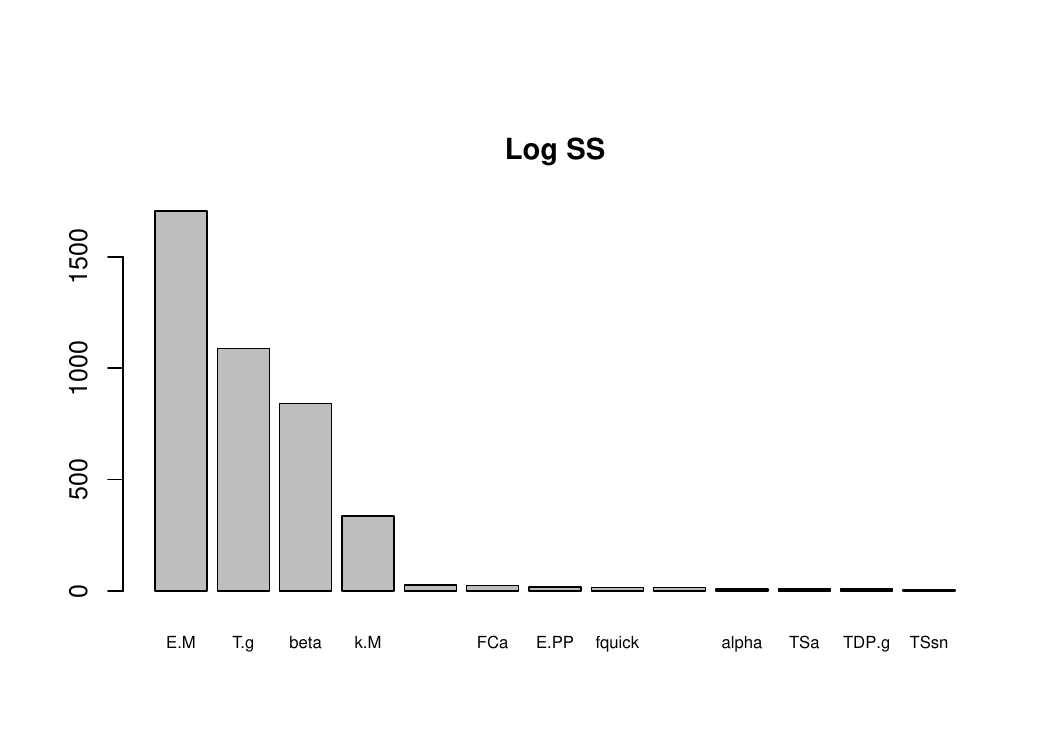} 
     \caption{Regression Tree parameter importance.}
     \label{F:SimplyP.RegTree.SS.Internal}
    \end{subfigure}   
% ---------------------------------------

   \begin{subfigure}[b]{0.45\textwidth}
    \centering
    \includegraphics[width=\columnwidth,height=0.30\textheight]{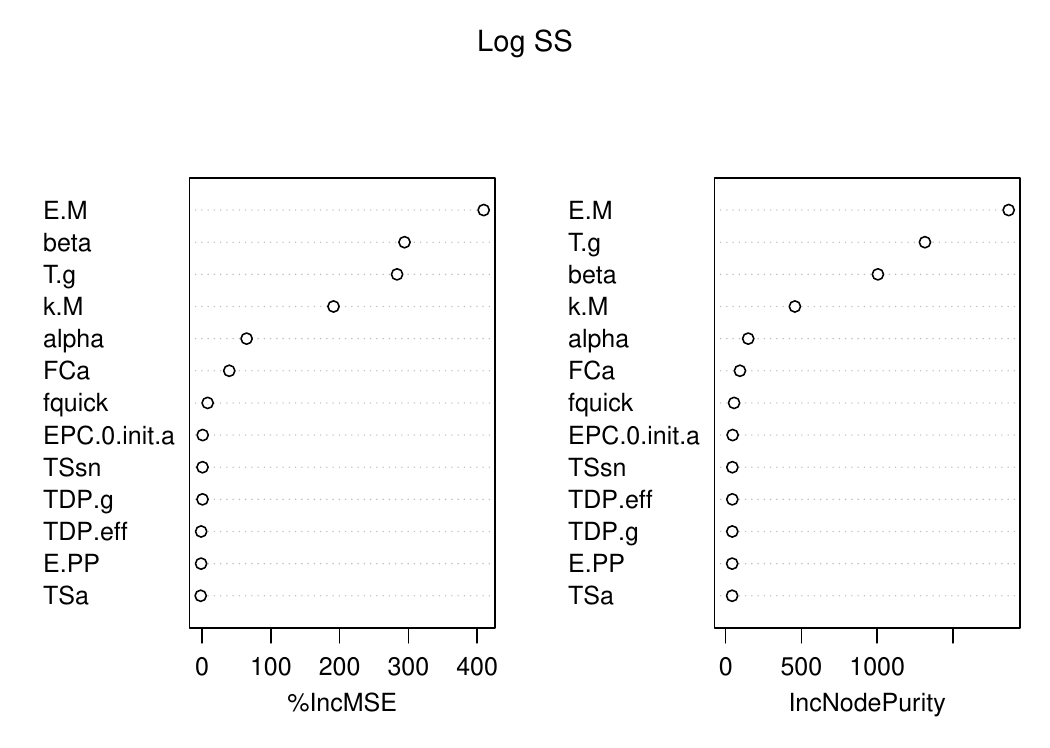}
    \caption{Random Forests parameter importance.}
    \label{F:SimplyP.RF.SS.Internal}
 \end{subfigure}   
 \hfill
   \begin{subfigure}[b]{0.45\textwidth}
    \centering
    \includegraphics[width=\columnwidth,height=0.30\textheight]{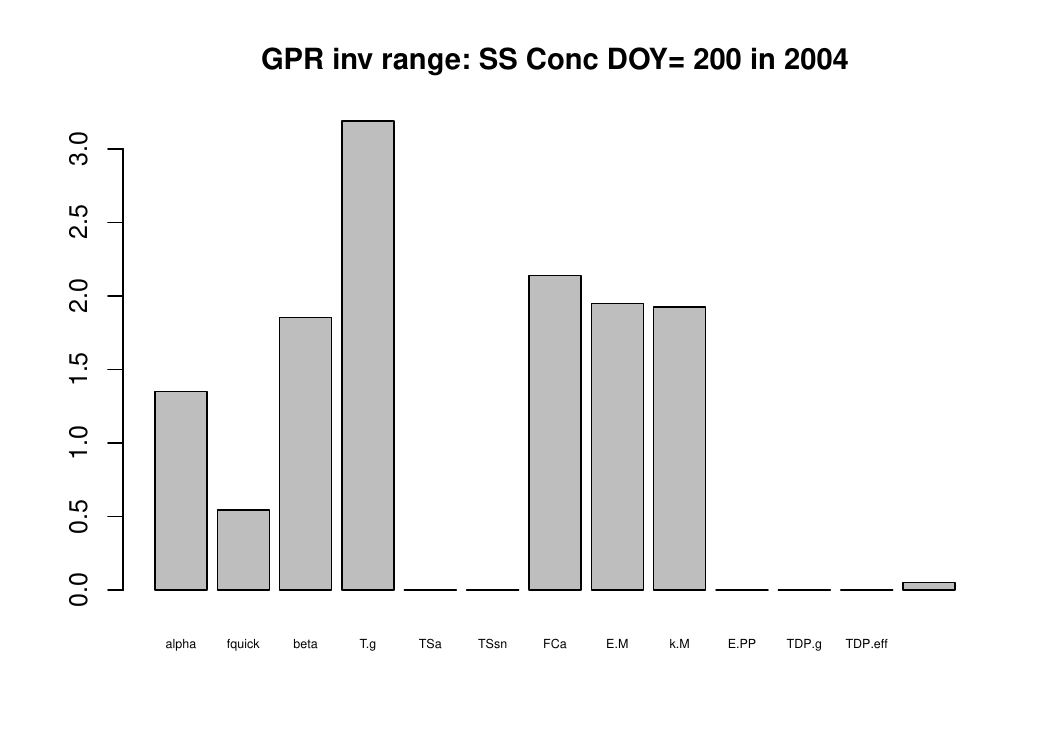}
    \caption{GPR standardized regression coefficients.}
    \label{F:SimplyP.GPR.Slope.SS.Internal}
 \end{subfigure}   
 \caption{SimplyP: Sensitivity Analyses of
 Log SS on Day 200 in 2004.}
 \label{F:SimplyP.SS.Internal}
\end{figure}

Table \ref{T:SimplyP.SA.SS.summary} summarizes the results of the different SA methods, each scaled to sum to 1.0 per method, for assessing the relative effects of the 13 SimplyP parameters on log $SS_{200,2004}$. Plots of the SA results for $SS$ based on six of the methods are shown in Figure \ref{F:SimplyP.SS.Internal}, in particular, Morris, Sobol, multiple regression, regression tree, random forests, and GP regression slopes. The VARS-TO results are again nearly identical to the Sobol $T_i$ values. As shown in the table and the figure,  the parameters $E_M$ and $T.g$ had the most influence, while two other parameters, beta and k.M, also appear influential. The combined outputs from GPR indicates the same overall importance of these four parameters.  
\begin{table}[h]
\centering
 Suspended Sediment, $SS_{200,2004}$  \\
  \begin{tabular}{lrrrrrrrrr}
   \hline
   \hline &Type  & Morris & Sobol' & VARS-TO & Reg & RegTree & RF & \multicolumn{2}{c}{GPR} \\
 &        &  DGSM      &   $T_i$      &      &  &    &   &  Slope &  InvRange \\ 
   \hline 
    alpha &Hyd& 0.07 & 0.03 & 0.02 & 0.07 & 0.00 & 0.03 & 0.12& 0.10 \\ 
    fquick &Hyd& 0.03 & 0.00 & 0.00 & 0.03 & 0.00 & 0.01 & 0.04 &0.04\\ 
    beta &Hyd& .21 & 0.21 & 0.22 & 0.22 & 0.21 & 0.19 & \textbf{\tcr{0.22}} & 0.14 \\ 
    T.g &Hyd& \textbf{\tcr{0.22}} & \textbf{\tcr{0.28}} & \textbf{\tcr{0.26}} & \textbf{\tcr{0.24}} & \textbf{\tcr{0.26}} & \textbf{\tcr{0.25}} &  0.11  &\textbf{\tcb{0.25}}\\ 
    TSa &Hyd& 0.00 & 0.00 & 0.00 & 0.00 & 0.00 & 0.01 & 0.00 & 0.00 \\ 
    TSsn &Hyd& 0.00 & 0.00 & 0.00 & 0.00 & 0.00 & 0.01 & 0.00 & 0.00 \\ 
    FCa &Hyd& 0.05 & 0.01 & 0.01 & 0.04 & 0.01 & 0.02 & 0.12  & 0.16 \\ 
    E.M &Sed& \textbf{\tcb{0.27}} & \textbf{\tcb{0.36}} & \textbf{\tcb{0.37}} & \textbf{\tcb{0.30}} & \textbf{\tcb{0.42}} & \textbf{\tcb{0.36}} & \textbf{\tcb{0.29}} & \textbf{\tcr{0.15}} \\ 
    k.M &Sed& 0.14 & 0.10 & 0.11 & 0.10 & 0.08 & 0.09 & 0.09 & 0.15 \\ 
    E.PP &Pho& 0.00 & 0.00 & 0.00 & 0.00 & 0.00 & 0.01 & 0.00 & 0.00 \\ 
    TDP.g &Pho& 0.00 & 0.00 & 0.00 & 0.00 & 0.00 & 0.01 & 0.00 &0.00 \\ 
    TDP.eff &Pho&0.00 & 0.00 & 0.00 & 0.00 & 0.00 & 0.01 & 0.00 & 0.00 \\ 
    EPC.0.init.a &Pho& 0.00 & 0.00 & 0.00 & 0.00 & 0.01 & 0.01 & 0.00 & 0.00 \\ 
    \hline
 \end{tabular}
\caption{SimplyP: summary of results for different SA methods applied to $\log SS_{200,2004}$. See text in Section \ref{subsec:Results.GR6J} for explanations of the values shown. Reg=regression, RegTree=regression tree, RF=random forest, GPR=Gaussian Process regression.}
\label{T:SimplyP.SA.SS.summary}
\end{table}
 
 \begin{figure}[h]
    \centering
    \includegraphics[width=0.60\linewidth]{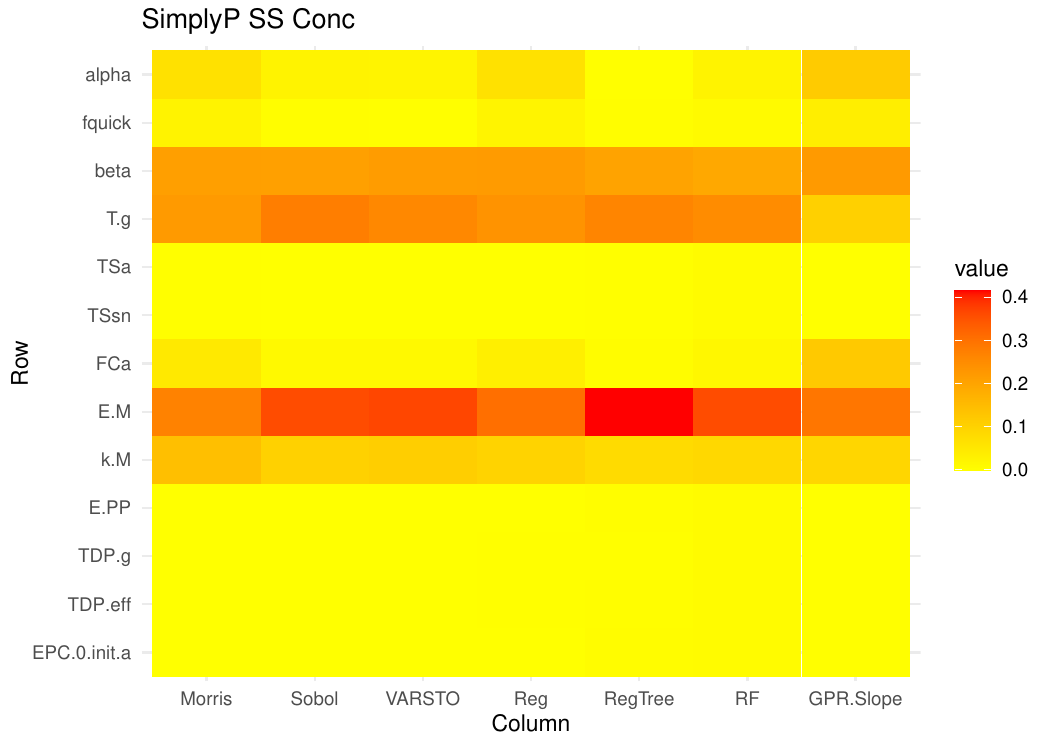}
    \caption{SimplyP: Relative parameter importance for $\log SS_{200,2004}$ for the different SA methods.}
    \label{F:SimplyP.SA.SS.summary}
\end{figure}
\clearpage

\clearpage 

 Pairwise comparisons between the SA methods in terms of the SA measures attached to each parameter are shown in Figure \ref{F:SimplyP.Pairwise.Q.SA}. A positive linear relationship, of varying strength from $r$=0.84 to $r$=1.00, occurred for all pairs. The overall degree of similarity was relatively high based on the Kendall W statistic of 0.95.
 \begin{figure}[h]
 \centering
     \includegraphics[width=0.8\textwidth]{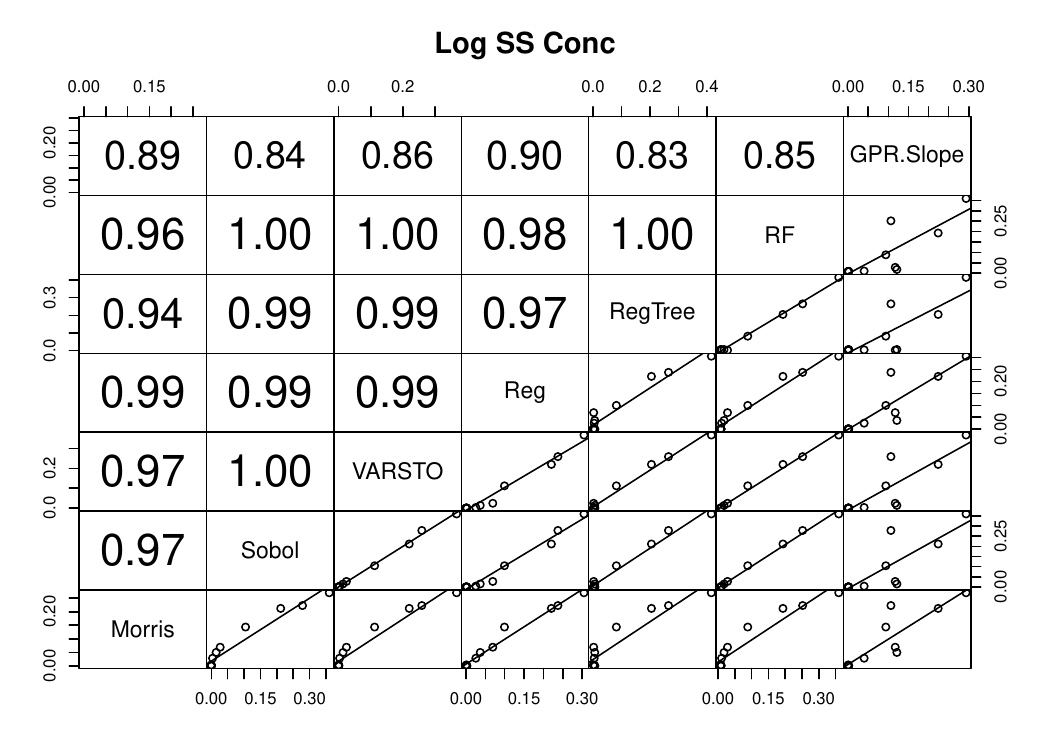}
     \caption{SimplyP: pairwise scatterplots of SA measures for the 13 parameters influence on log($SS_{200,2004}$) for different SA procedures along with Pearson correlation coefficients. Reg=regression, RegTree=regression tree, RF=random forest, GPR=Gaussian Process regression.}
     \label{F:SimplyP.Pairwise.SS.SA}
 \end{figure}

\clearpage 

\begin{figure}[h]
    \centering
    \includegraphics[width=0.6\linewidth]{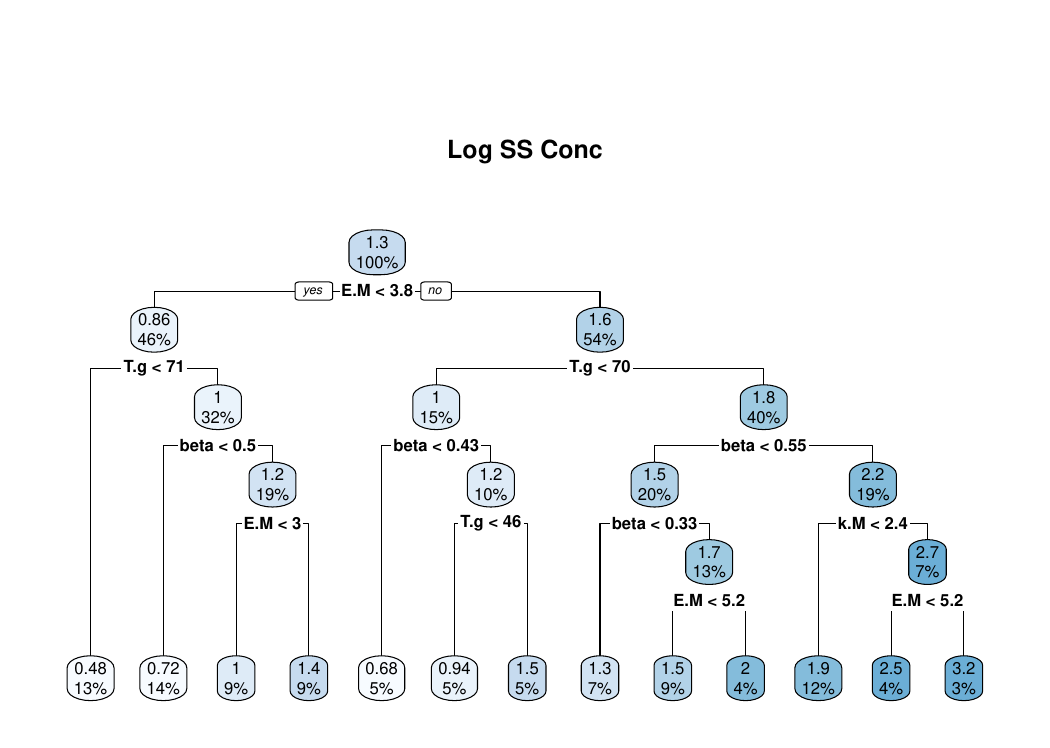}
    \caption{SimplyP: Regression tree for log SS.}
    \label{F:SimplyP.SS.RegTree.tree}
\end{figure}

\begin{figure}[h]
    \centering
    \includegraphics[width=0.6\linewidth]{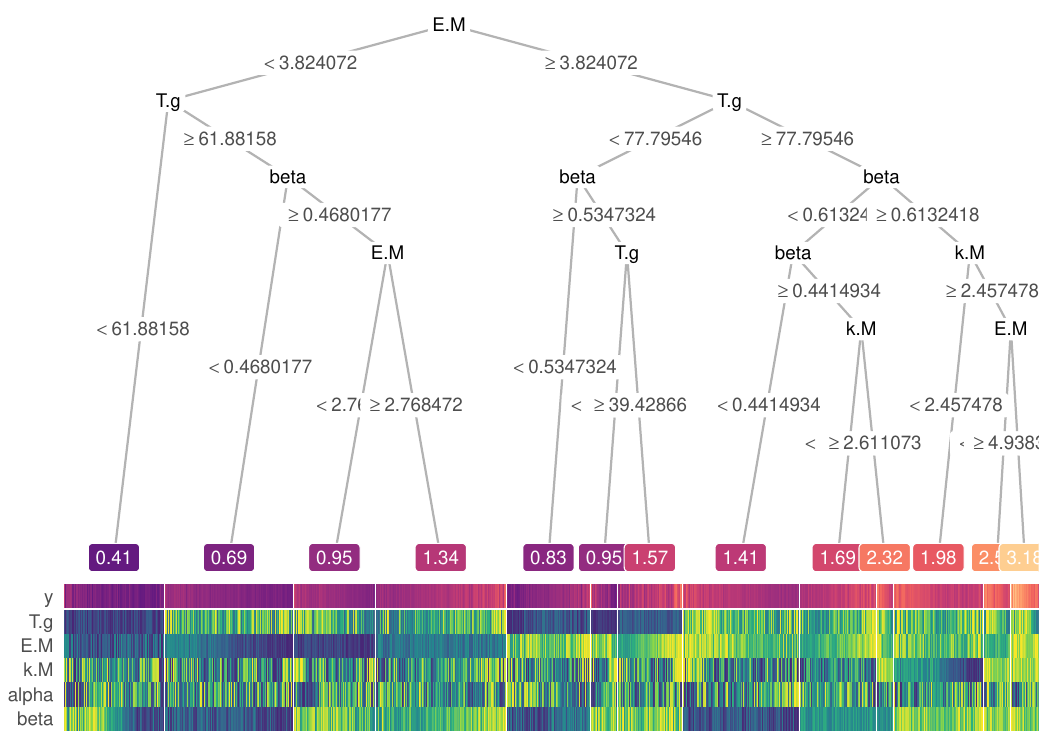}
    \caption{SimplyP: Regression tree for log SS based on top 5 inputs with heatmap.}
\label{F:SimplyP.SS.RegTree.tree.heatmap}
\end{figure}

 Further details on the regression tree for log SS are shown in Figure \ref{F:SimplyP.SS.RegTree.tree},  which shows four to five parameters  influencing most branching decisions. Figure \ref{F:SimplyP.SS.RegTree.tree.heatmap} shows the tree based only on the top five parameters, that shows several possible interactions amongst the five. 

\clearpage 
 
%-----------------SimplyP log  PP ********-
\subsubsection{Log of $PP_{200,2004}$}

\begin{figure}[h]
    \centering
    \begin{subfigure}[b]{0.45\textwidth}
    \includegraphics[width=\columnwidth,height=0.30\textheight]{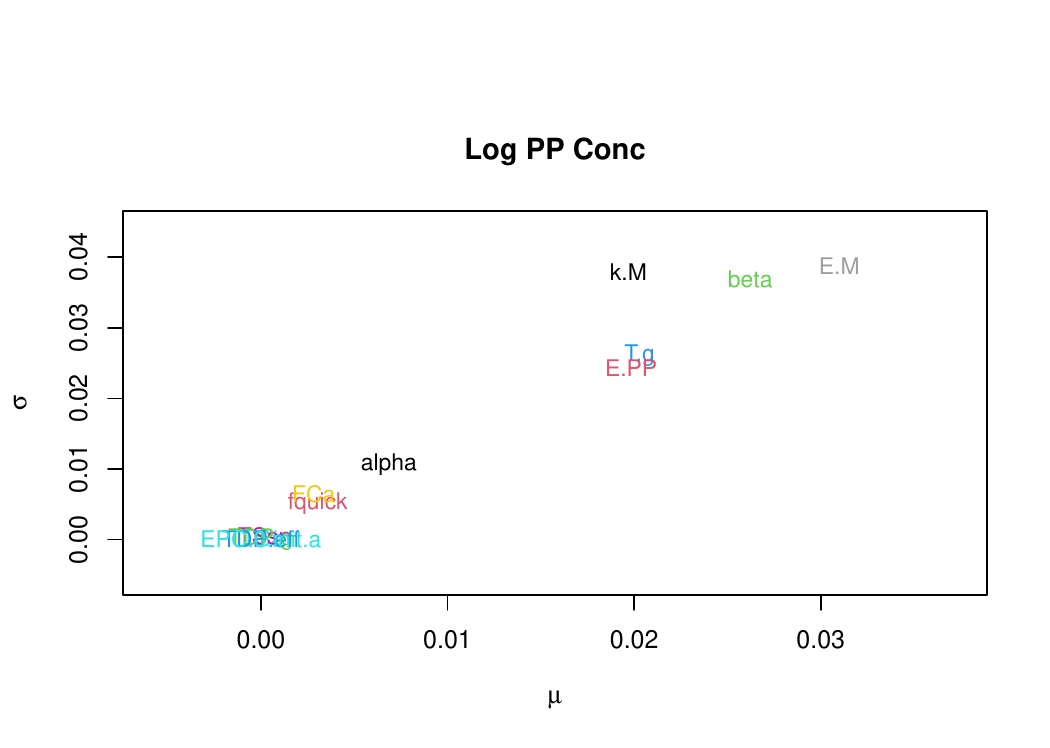}
   \caption{Morris $\sigma$ versus $\mu^*$.}
    \label{F:SimplyP.Morris.PP.Internal}
    \end{subfigure}
    \hfill
    \begin{subfigure}[b]{0.45\textwidth}
    \centering 
        \includegraphics[width=\columnwidth,height=0.30\textheight]{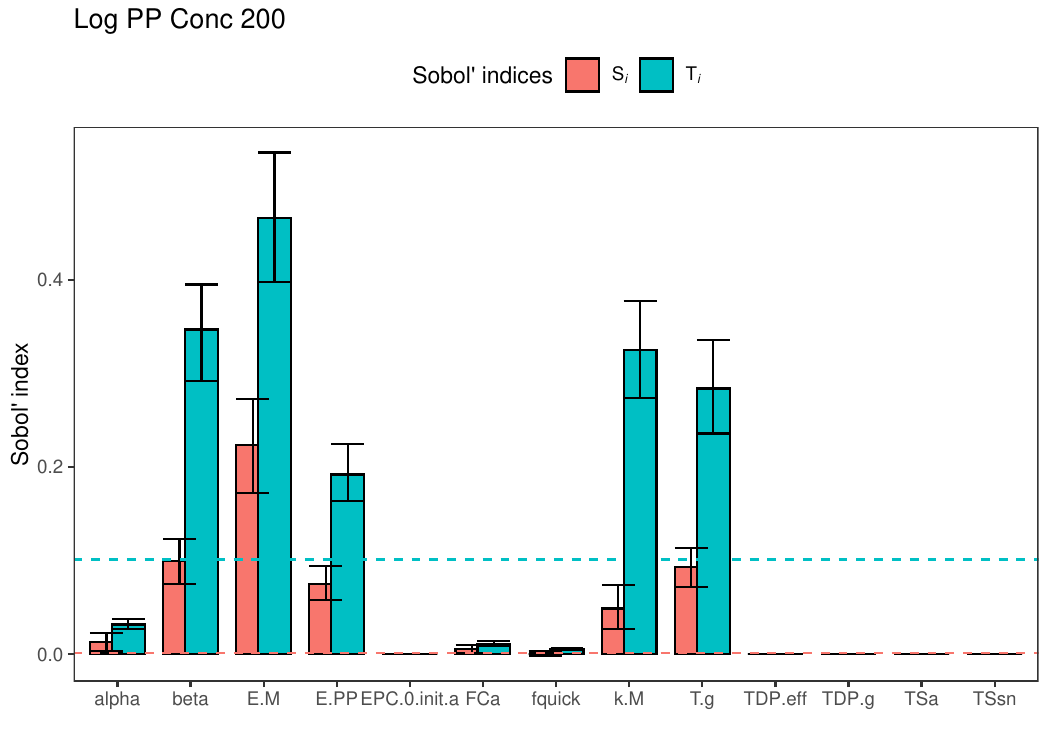}
     \caption{Sobol $S_{1,k}$ and $T_k$.}
     \label{F:SimplyP.Sobol.PP.Internal}
    \end{subfigure}
% ----------------------------------

   \begin{subfigure}[b]{0.45\textwidth}
    \centering
\includegraphics[width=\columnwidth,height=0.30\textheight]{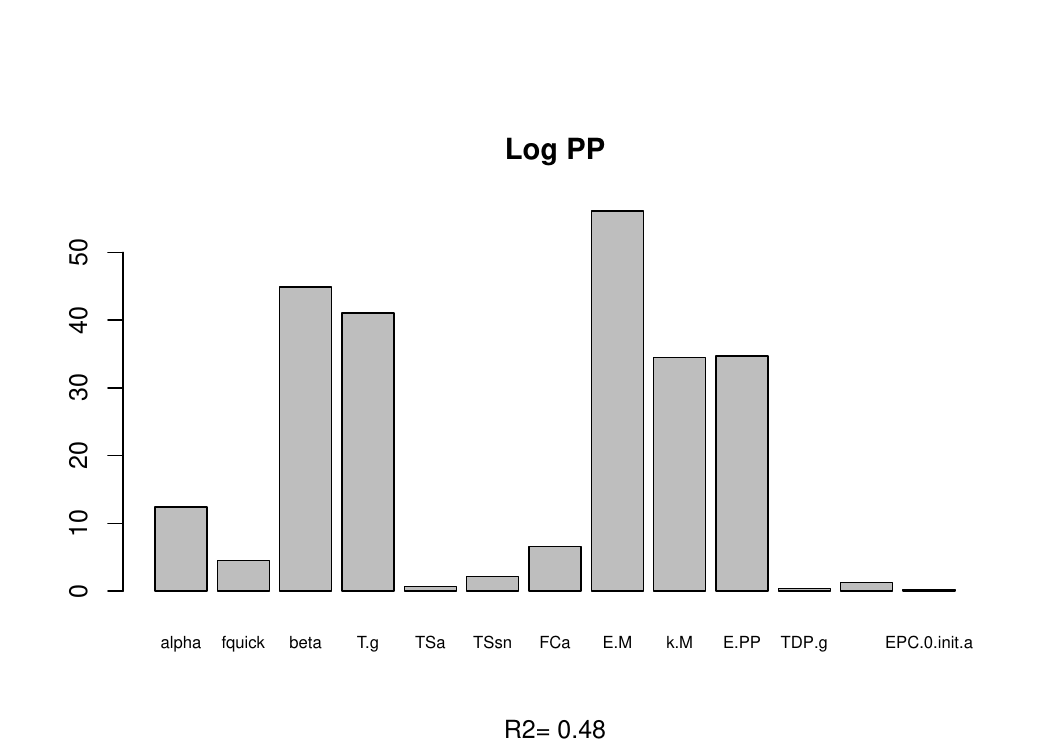}
    \caption{Multiple Regression standardized regression coefficients.}
    \label{F:SimplyP.Regression.PP.Internal}
 \end{subfigure}   
\hfill
    \begin{subfigure}[b]{0.45\textwidth}
    \centering
    \includegraphics[width=\columnwidth,height=0.30\textheight]{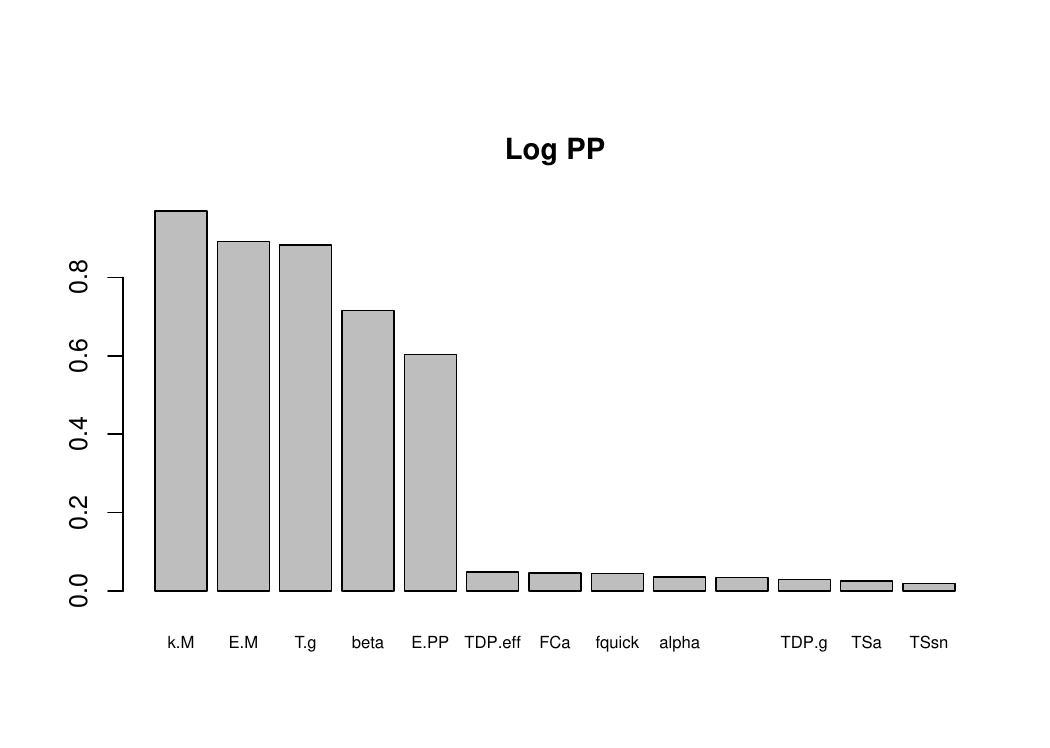} 
     \caption{Regression Tree parameter importance.}
     \label{F:SimplyP.RegTree.PP.Internal}
    \end{subfigure}   
% --------------------------------------------

   \begin{subfigure}[b]{0.45\textwidth}
    \centering
    \includegraphics[width=\columnwidth,height=0.30\textheight]{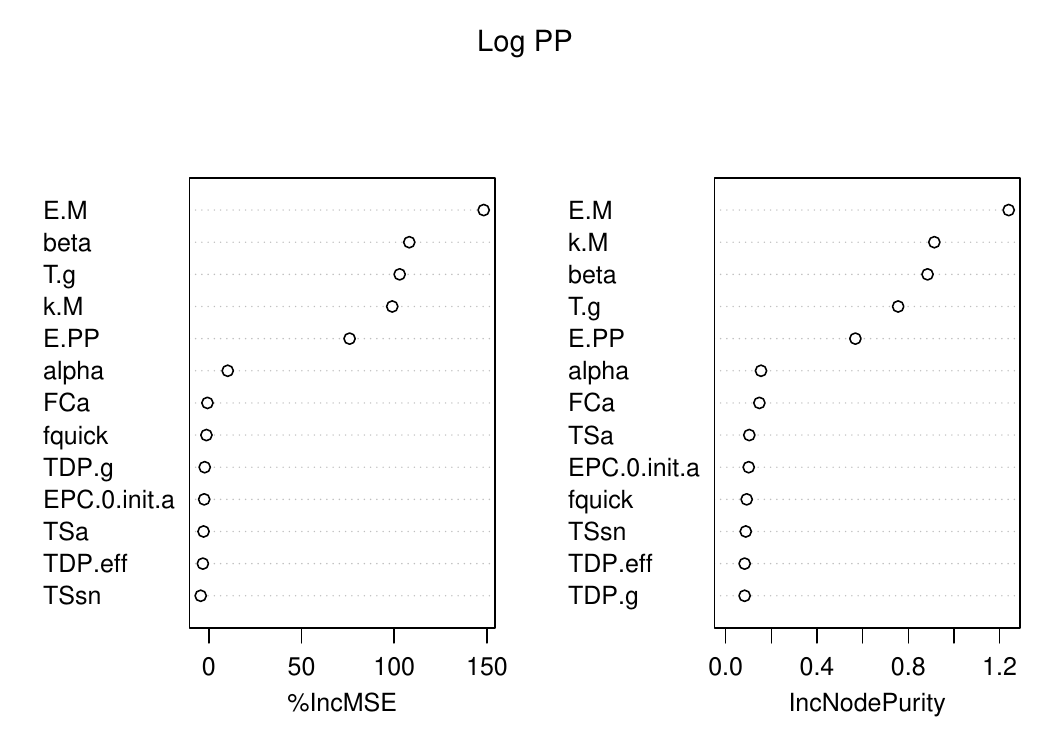}
    \caption{Random Forests results.}
    \label{F:SimplyP.RF.PP.Internal}
 \end{subfigure}   
 \hfill
   \begin{subfigure}[b]{0.45\textwidth}
    \centering
    \includegraphics[width=\columnwidth,height=0.30\textheight]{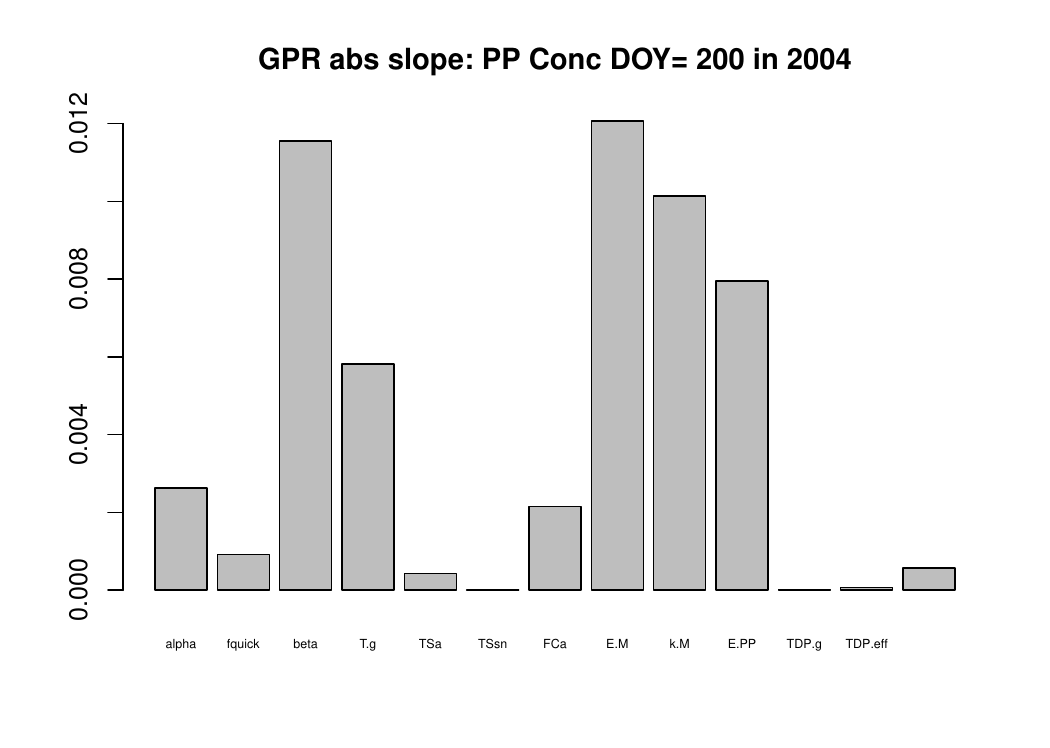}
    \caption{GPR standardized regression coefficients.}
    \label{F:SimplyP.GPR.Slope.PP.Internal}
 \end{subfigure}   
 \caption{SimplyP: Sensitivity Analyses of
 Log $PP$ on Day 200 in 2004.}
 \label{F:SimplyP.PP.Internal}
\end{figure}

Table \ref{T:SimplyP.SA.PP.summary} summarizes the results of the different SA methods, scaled to sum to 1.0 per method, for assessing the relative effects of the 13 SimplyP parameters on log $PP_{200,2004}$.  Plots of the SA results for $PP$ based on six of the methods are shown in Figure \ref{F:SimplyP.PP.Internal}.  As shown in the table and the figure,  five of the parameters had a sizeable influence. For all methods, the parameter $E_M$ had the most, or second most, influence, while beta, $k.M$,  $T.g$ and $E_{PP}$ had a sizeable influence.   
 
\begin{table}[h]
\centering
 Log $PP_{200,2004}$  \\
  \begin{tabular}{lrrrrrrrrr}    \hline
    &Type  & Morris & Sobol' & VARS-TO & Reg & RegTree & RF & \multicolumn{2}{c}{GPR} \\
 &        & DGSM &  $T_i$ &      &  &    &   &  Slope &  InvRange \\ 
   \hline 
 alpha &Hyd& 0.06 & 0.02 & 0.02 & 0.05 & 0.01 & 0.03 & 0.05  & 0.07 \\ 
 fquick &Hyd& 0.03 & 0.00 & 0.00 & 0.02 & 0.01 & 0.02 & 0.02 & 0.07 \\ 
 beta &Hyd& \textbf{\tcb{0.20}} & \textbf{\tcr{0.21}} & 0.18 & \textbf{\tcr{0.19}} & 0.16 & \textbf{\tcr{0.17}} & \textbf{\tcr{0.21}} &  0.16  \\ 
 T.g &Hyd& 0.14 & 0.17 & \textbf{\tcr{0.16}} & 0.17 & 0.20 & 0.14 & 0.11 & 0.15 \\
 TSa &Hyd& 0.00 & 0.00 & 0.00 & 0.00 & 0.01 & 0.02 & 0.01 & 0.04 \\ 
 TSsn &Hyd& 0.00 & 0.00 & 0.00 & 0.01 & 0.00 & 0.02 & 0.00 & 0.00 \\ 
 FCa &Hyd& 0.03 & 0.01 & 0.01 & 0.03 & 0.01 & 0.03 & 0.04 & 0.03 \\ 
 E.M &Sed& \textbf{\tcr{0.22}} & \textbf{\tcb{0.28}} & \textbf{\tcb{0.33}} & \textbf{\tcb{0.23}} & \textbf{\tcr{0.21}} & \textbf{\tcb{0.24}} & \textbf{\tcb{0.22}} & \textbf{\tcr{0.18}} \\ 
 k.M &Sed& 0.19 & \textbf{\tcr{0.20}} & 0.21 & 0.14 & \textbf{\tcb{0.22}} & 0.18 &  0.19  & \textbf{\tcb{0.21}} \\ 
 E.PP &Pho& 0.0.14 & 0.12 & 0.09 & 0.14 & 0.14 & 0.11 & 0.15 & 0.12 \\ 
 TDP.g &Pho& 0.00 & 0.00 & 0.00 & 0.00 & 0.01 & 0.02 & 0.00 & 0.00 \\ 
 TDP.eff & Pho& 0.00 & 0.00 & 0.00 & 0.01 & 0.01 & 0.02 & 0.00 & 0.00 \\ 
 EPC.0.init.a &Pho& 0.00 & 0.00 & 0.00 & 0.00 & 0.01 & 0.02 & 0.01 & 0.00 \\ 
   \hline
 \hline   
 \end{tabular}
\caption{SimplyP: summary of results for different SA methods applied to $\log PP_{200,2004}$. See text in Section \ref{subsec:Results.GR6J} for explanations of the values shown. Reg=regression, RegTree=regression tree, RF=random forest, GPR=Gaussian Process regression.}
\label{T:SimplyP.SA.PP.summary}
\end{table}

 \begin{figure}[h]
    \centering
    \includegraphics[width=0.60\linewidth]{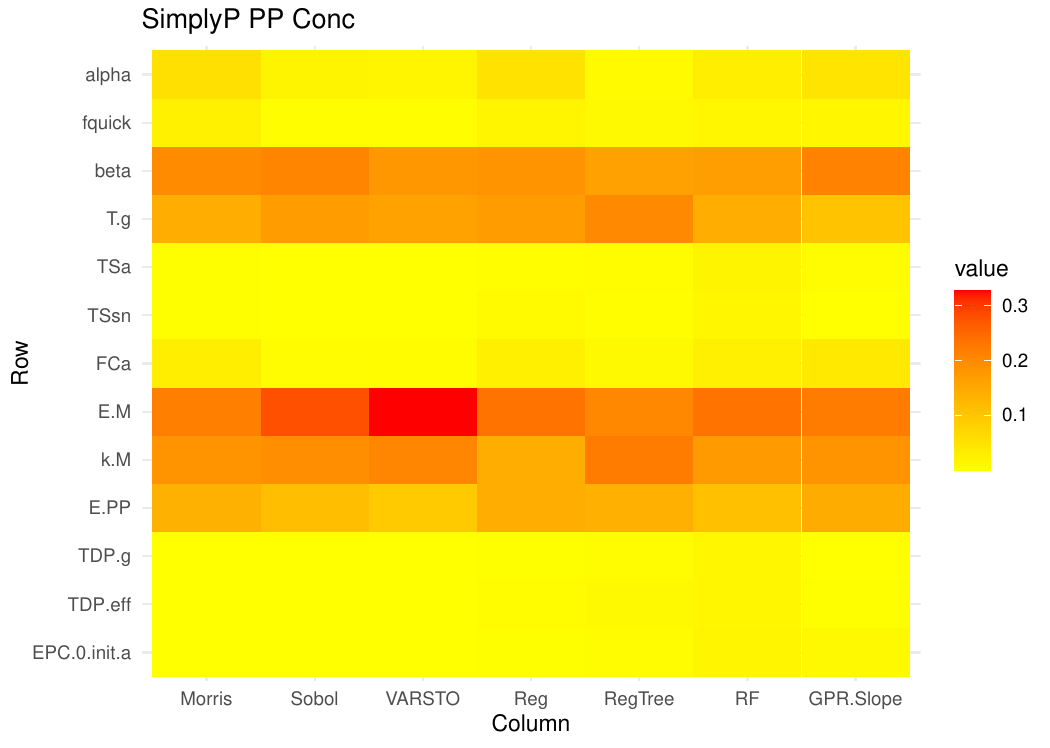}
    \caption{SimplyP: Relative parameter importance for $\log PP_{200,2004}$ for the different SA methods.}
    \label{F:SimplyP.SA.PP.summary}
\end{figure}
\clearpage

\clearpage 

\begin{figure}[h]
 \centering
     \includegraphics[width=0.8\textwidth]{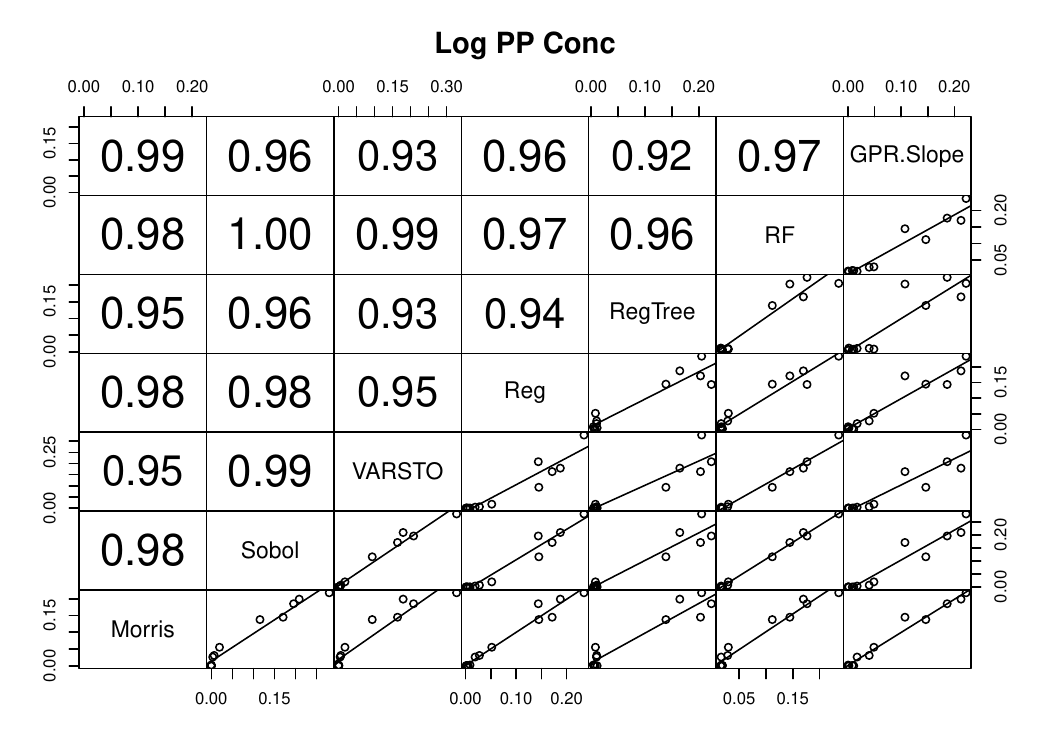}
     \caption{SimplyP: pairwise scatterplots of SA measures for the 13 parameters influence on log($PP_{200,2004}$) for different SA procedures along with Pearson correlation coefficients. Reg=regression, RegTree=regression tree, RF=random forest, GPR=Gaussian Process regression.}
     \label{F:SimplyP.Pairwise.PP.SA}
 \end{figure}

Pairwise comparisons between the SA methods in terms of the SA measures attached to each parameter are shown in Figure \ref{F:SimplyP.Pairwise.PP.SA}. A positive linear relationship, of varying strength from $r$=0.77 to $r$=0.98, occurred for all pairs. The overall degree of similarity was relatively high based on the Kendall W statistic of 0.89.

\clearpage     
 
\begin{figure}[h]
    \centering
    \includegraphics[width=0.6\linewidth]{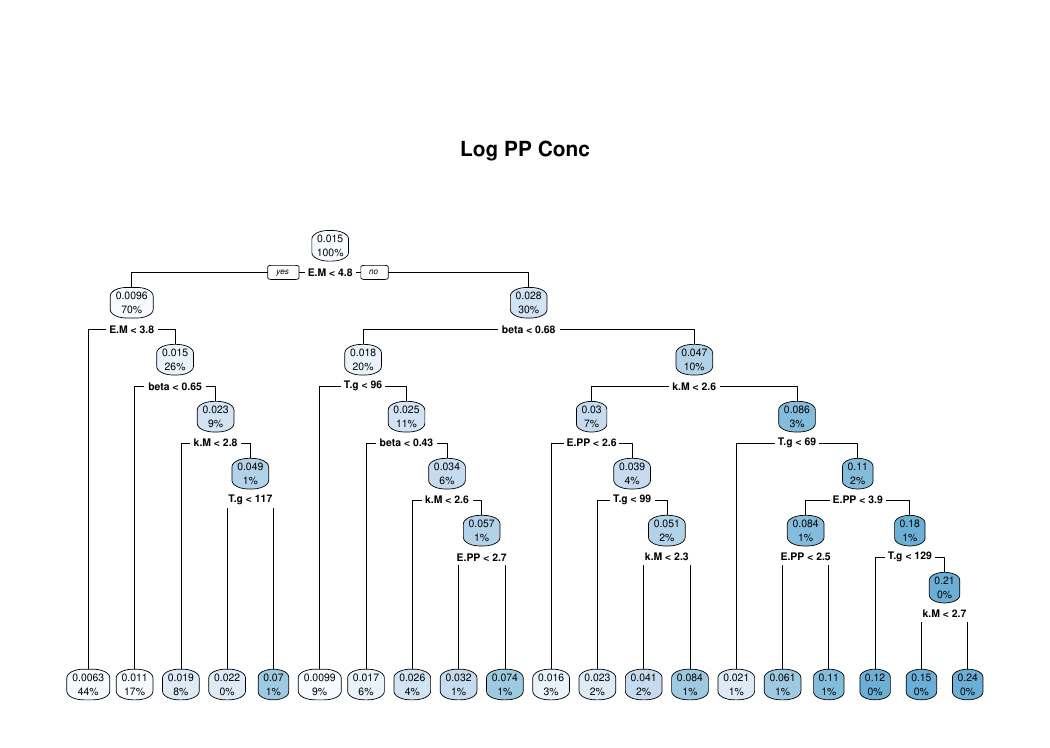}
    \caption{SimplyP: Regression tree for log PP.}
    \label{F:SimplyP.PP.RegTree.tree}
\end{figure}

\begin{figure}[h]
    \centering
    \includegraphics[width=0.6\linewidth]{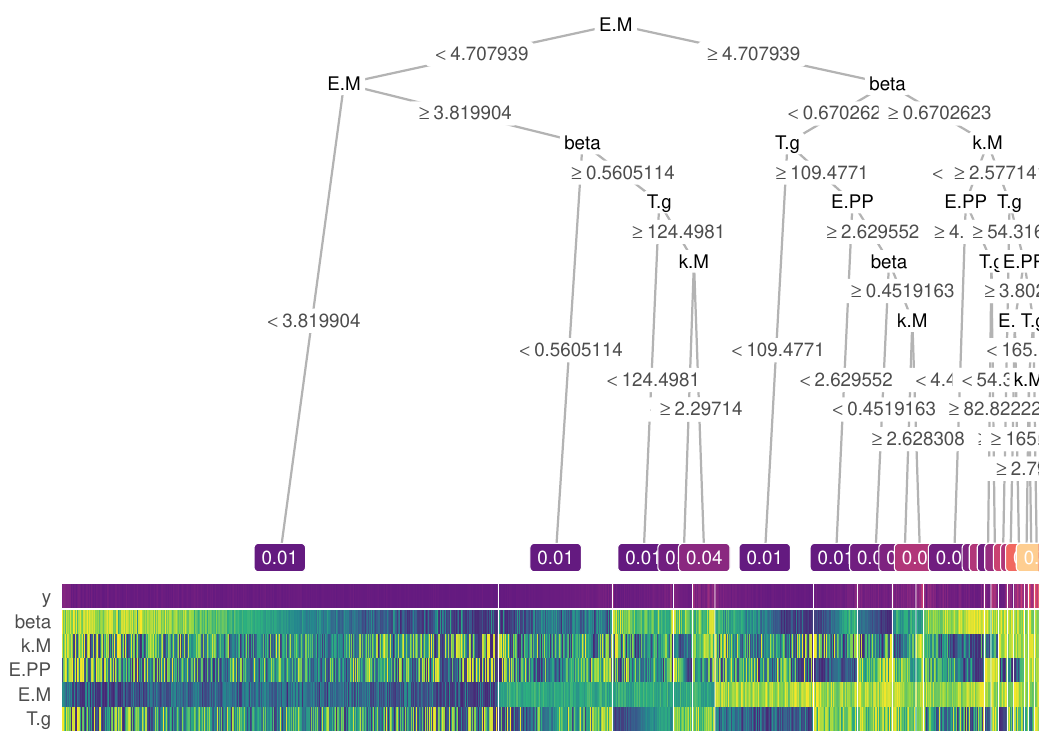}
    \caption{SimplyP: Regression tree for log PP based on top 5 inputs with heatmap.}
\label{F:SimplyP.PP.RegTree.tree.heatmap}
\end{figure}

Further details on the regression tree for log PP are shown in Figure \ref{F:SimplyP.PP.RegTree.tree},  which shows four to five parameters  influencing most branching decisions. Figure \ref{F:SimplyP.PP.RegTree.tree.heatmap} shows the tree based only on the top five parameters, that shows several possible interactions amongst the five. 

\clearpage 
%---------------------------------------------
\subsubsection{SimplyP: Summary of parameter importance across multiple outputs}
\begin{figure}[h]
    \centering
    \includegraphics[width=0.5\linewidth]{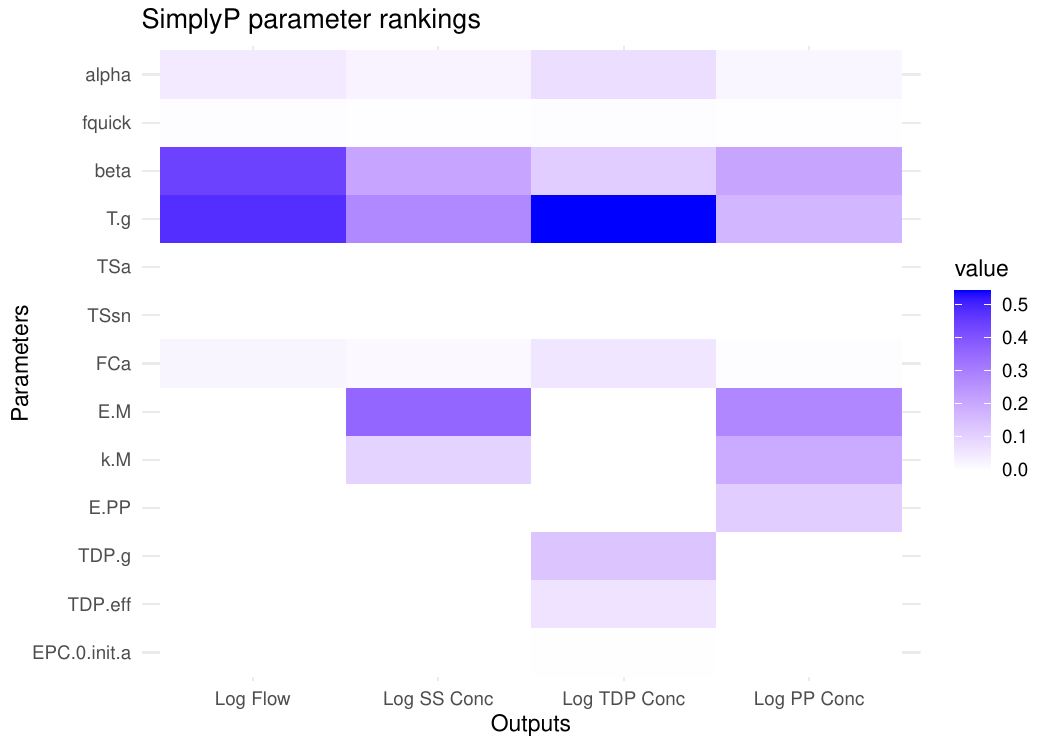}
    \caption{SimplyP: Relative parameter importance for four outputs based on Sobol' $T_i$.}
    \label{F:SimplyP.Importance.Summary}
\end{figure}

The relative importance of the SimplyP parameters for the four outputs, based on Sobol' $T_i$, is summarized in both Table \ref{T:SimplyP.Importance.Summary} and Figure \ref{F:SimplyP.Importance.Summary}.
\begin{table}[h]
\centering
\begin{tabular}{lrrrrr}
Parameter & Type &   Log Flow & Log SS Conc & Log TDP Conc & Log PP Conc \\ \hline
alpha  &Hyd& 0.05 & 0.03 & 0.08 & 0.02   \\ 
fquick &Hyd& 0.01 & 0.00 & 0.01 & 0.00  \\ 
beta   &Hyd& \textbf{\tcr{0.44}} &  0.21 & 0.12 & \textbf{\tcr{0.21}}   \\ 
T.g    &Hyd& \textbf{\tcb{0.48}} & \textbf{\tcr{0.28}} &  \textbf{\tcb{0.54}} & 0.17 \\ 
TSa    &Hyd& 0.00 & 0.00 & 0.00 & 0.00  \\ 
TSsn   &Hyd& 0.00 & 0.00 & 0.00 & 0.00 \\ 
FCa    &Hyd& 0.02 & 0.01 & 0.06 & 0.01  \\ 
E.M    &Sed& 0.00 & \textbf{\tcb{0.36}} & 0.00 &  \textbf{\tcb{0.28}}  \\ 
k.M    &Sed& 0.00 & 0.10 & 0.00 &  \textbf{\tcr{0.20}}   \\ 
E.PP   &Pho& 0.00 & 0.00 & 0.00 & 0.12 \\ 
TDP.g  &Pho&  0.00 & 0.00 & \textbf{\tcr{0.14}} & 0.00  \\ 
TDP.eff &Pho& 0.00 & 0.00 & 0.06 & 0.00 \\ 
EPC.0.init.a &Pho& 0.00 & 0.00 & 0.00 & 0.00 \\ 
   \hline
\end{tabular}
\caption{SimplyP: Relative parameter importance for four outputs based on Sobol' $T_i$.}
\label{T:SimplyP.Importance.Summary}
\end{table}

\clearpage

%% file: 6_Appendices/3_Results_Details/6C3_Results_STICS_Details.tex
\subsection{\label{app:Results.Details.STICS}STICS}
Section \ref{subsec:Results.STICS} summarized the GSA results for $mafruit$, and here the same summaries are presented for the other three outputs of interest: $masec.n$, $CNgrain$, and $CNplante$. For additional details on terms used, see Section \ref{subsec:Results.STICS}.

\subsubsection{\label{app.STICS.mafruit.details}Regression tree details for mafruit}
\begin{figure}[h]
    \centering
       \includegraphics[width=0.6\linewidth]{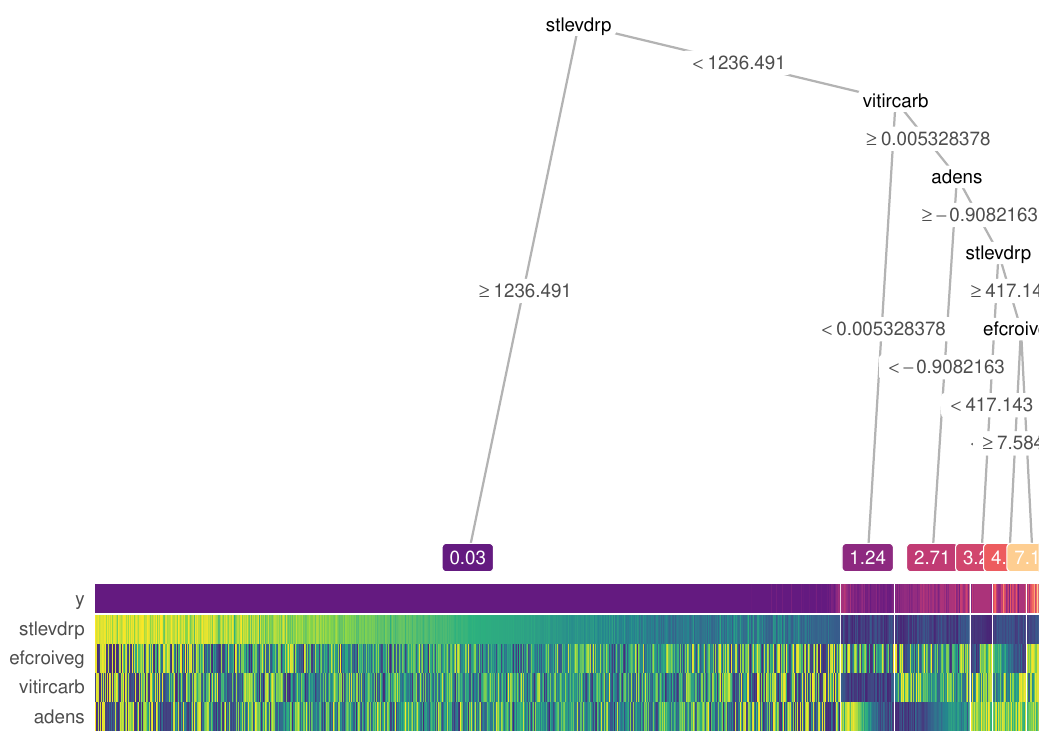}
    \caption{STICS: Regression tree for mafruit based on top 4 inputs with heatmap.}
    \label{F:STICS.mafruit.RegTree.tree.heatmap}
\end{figure}

The final regression tree for mafruit is shown in Figure F:STICS.mafruit.RegTree.tree,  which shows the dominance of stlevdrp. Figure \ref{F:STICS.mafruit.RegTree.tree.heatmap} shows the tree based only on the top four input parameters.  This again shows the dominance of stlevdrp but also the importance of vitircarb and adens, which may have some degree of interaction between these two and with stlevdrp. 

\begin{figure}[h]
    \centering
    \includegraphics[width=0.6\linewidth]{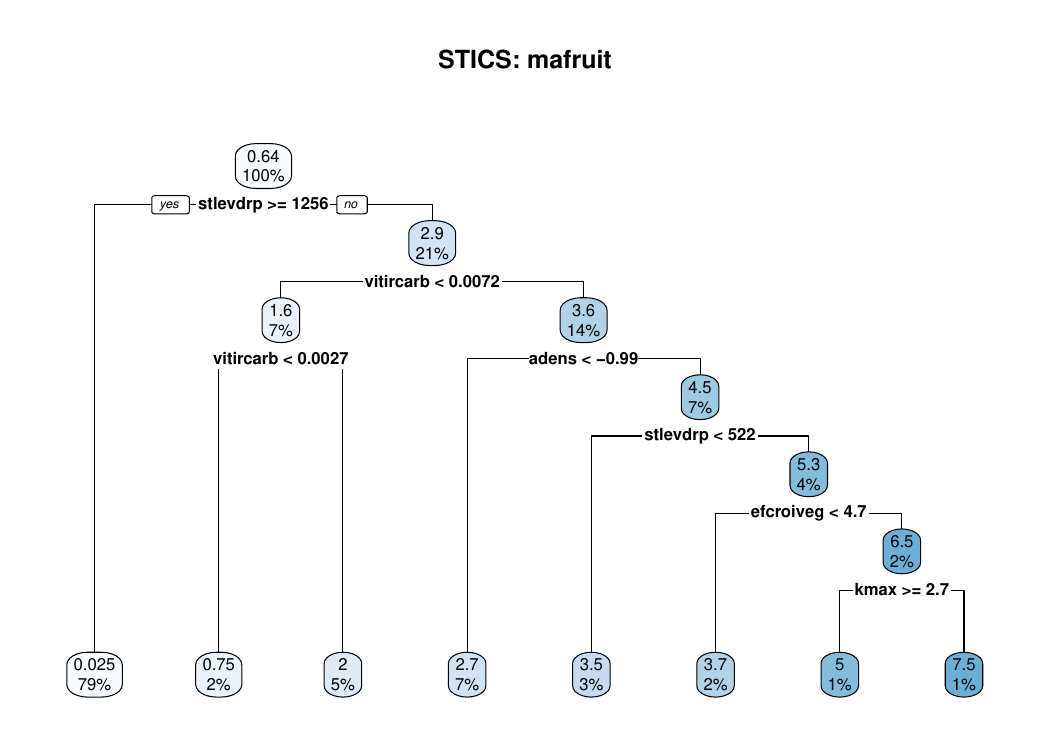}
    \caption{STICS: Regression tree for mafruit}
    \label{F:STICS.mafruit.RegTree.tree}
\end{figure}
\clearpage

% ******************************* %
\subsubsection{$masec.n$}
\begin{figure}[h]
    \centering
    \includegraphics[width=0.60\linewidth]{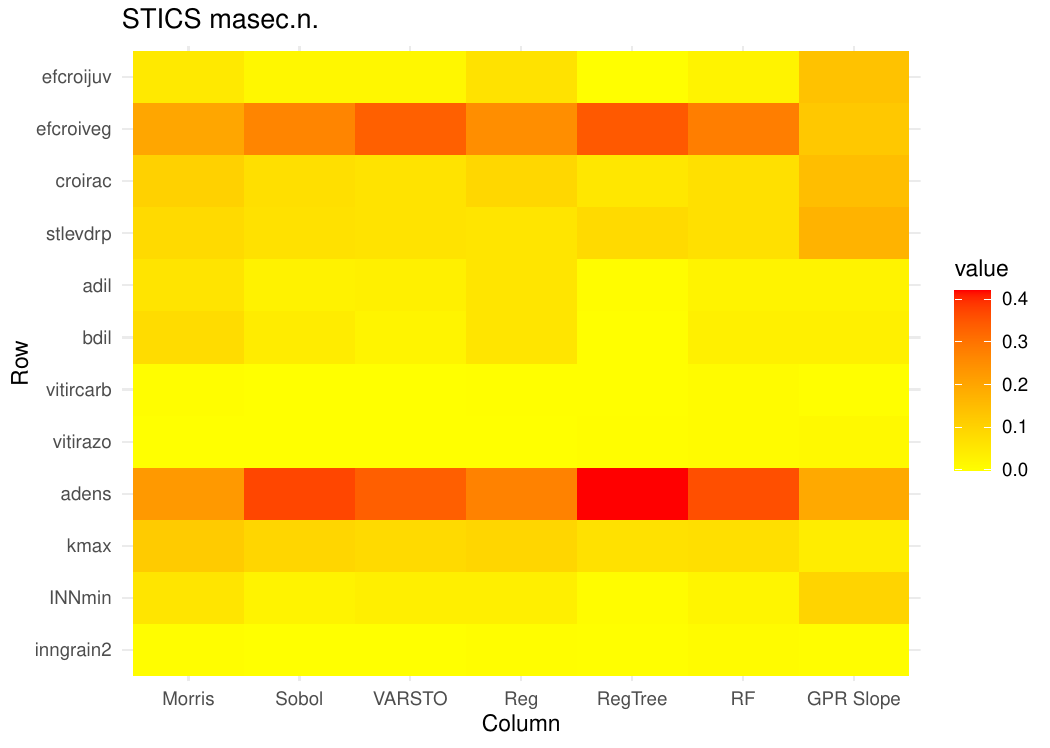}
    \caption{STICS: Relative parameter importance for masec.n for the different SA methods.}
    \label{F:STICS.masecn.summary}
\end{figure}

\begin{figure}[h]
    \centering
    \begin{subfigure}[b]{0.45\textwidth}
    \includegraphics[width=\columnwidth,height=0.25\textheight]{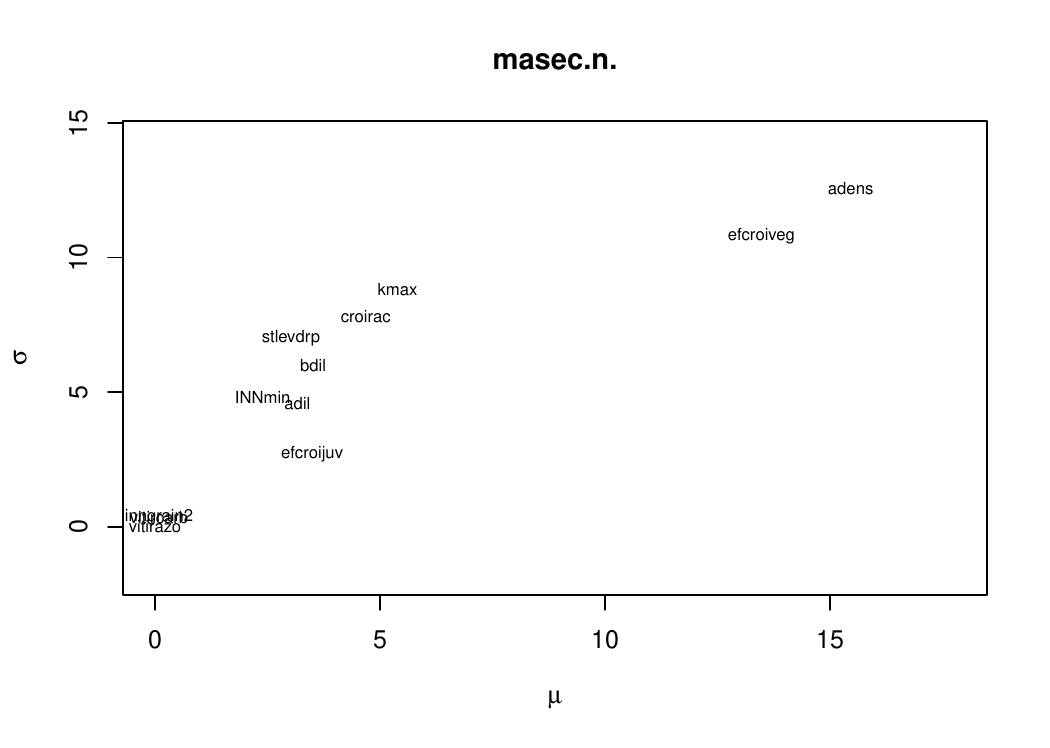}
    \caption{Morris  $\sigma$ versus $\mu^*$.}
    \label{F:STICS.Morris.masec.Internal2}
    \end{subfigure}
    \hfill
    \begin{subfigure}[b]{0.45\textwidth}
    \centering 
        \includegraphics[width=\columnwidth,height=0.25\textheight]{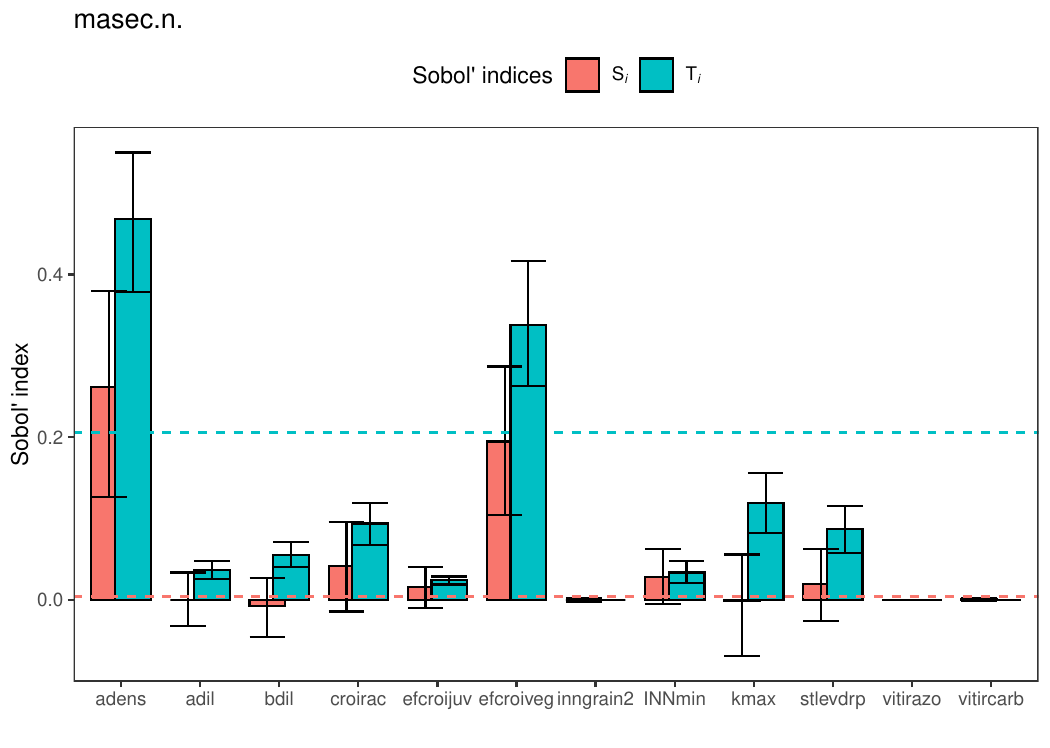}
     \caption{Sobol $S_{1,k}$ and $T_k$.}
     \label{F:STICS.Sobol.masec.Internal2}
    \end{subfigure}
% ------------------------------------------------
   \begin{subfigure}[b]{0.45\textwidth}
    \centering
    \includegraphics[width=\columnwidth,height=0.25\textheight]{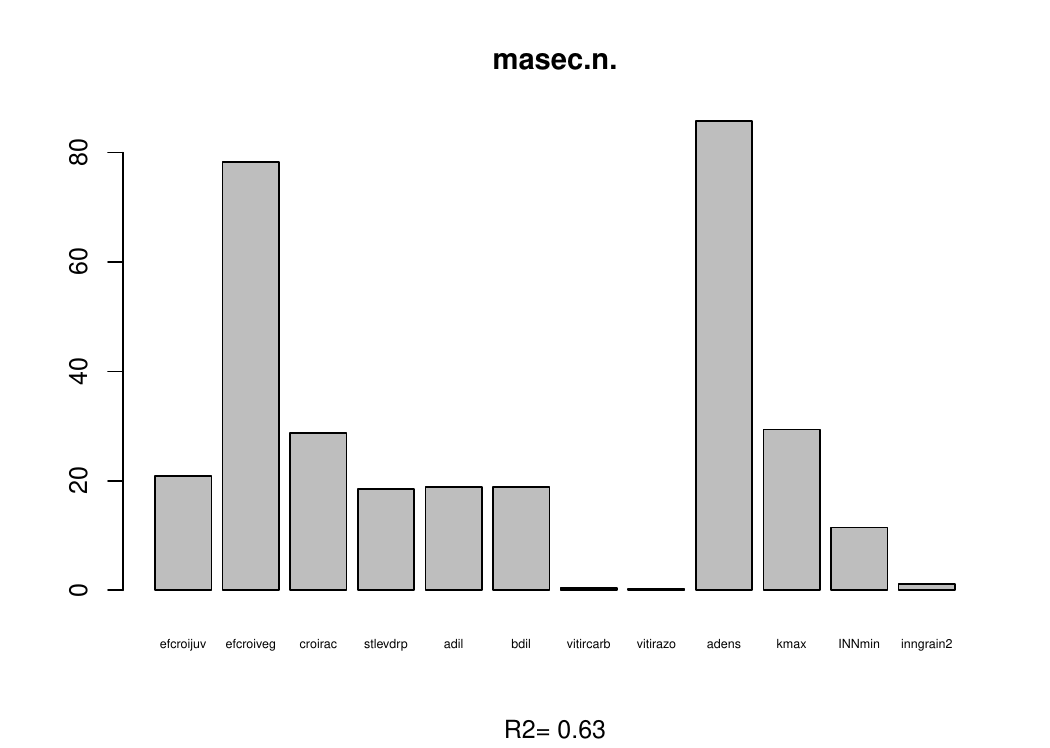}
    \caption{Multiple Regression standardized regression coefficients.}
    \label{F:STICS.Regression.masec.Internal2}
 \end{subfigure}   
\hfill
    \begin{subfigure}[b]{0.45\textwidth}
    \centering
    \includegraphics[width=\columnwidth,height=0.25\textheight]{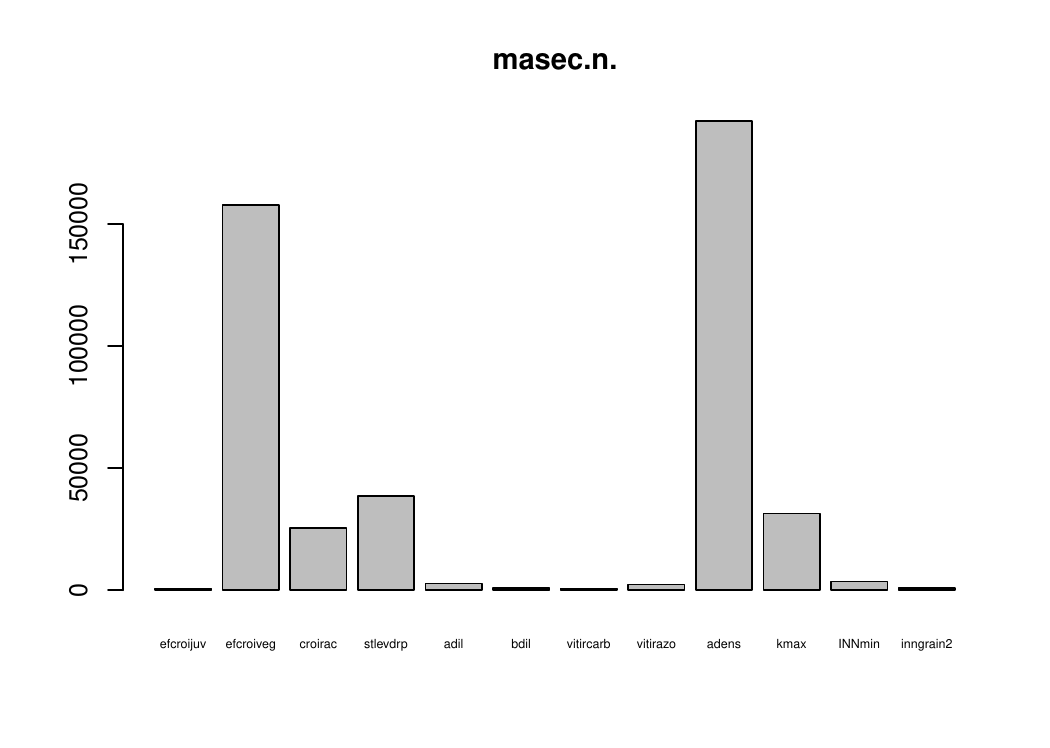} 
     \caption{Regression Tree parameter importance.}
     \label{F:STICS.RegTree.masec.Internal2}
    \end{subfigure}   
% --------------------------------------------------------------

   \begin{subfigure}[b]{0.45\textwidth}
    \centering
    \includegraphics[width=\columnwidth,height=0.25\textheight]{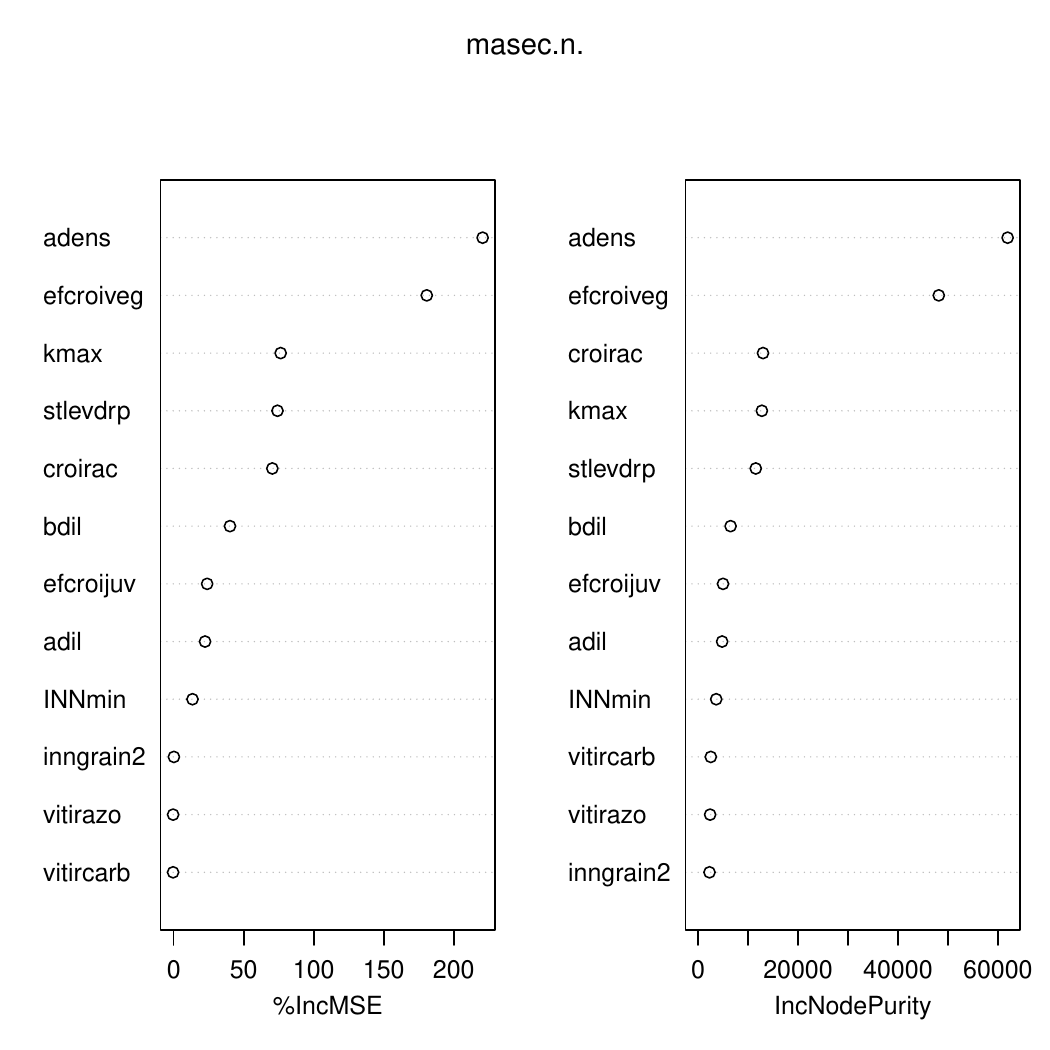}
    \caption{Random Forests results.}
    \label{F:STICS.RF.masec.Internal2}
 \end{subfigure}   
 \hfill
   \begin{subfigure}[b]{0.45\textwidth}
    \centering
    \includegraphics[width=\columnwidth,height=0.25\textheight]{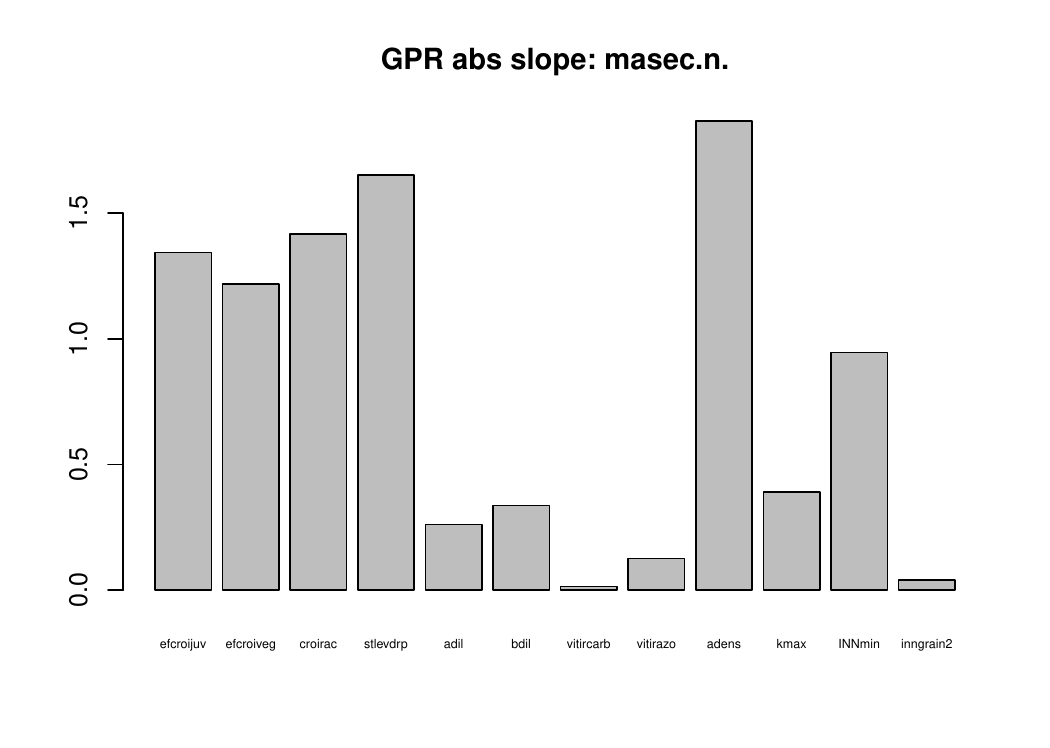}
    \caption{GPR standardized regression coefficients.}
    \label{F:STICS.GPR.Both.Internal2}
 \end{subfigure}   
 \caption{STICS: Sensitivity Analyses of
 $masec.n$ on day of harvest.}
 \label{F:STICS.masec.Internal}
  \end{figure}

The parameter most influencing $masec.n$ based on all the SA methods was adens followed by efcroiveg, with one exception being stlevdrp being second with GPR slope (Table \ref{T:STICS.SA.summary.masecn}, Figures \ref{F:STICS.masecn.summary} and \ref{F:STICS.masec.Internal}).  
 
 \begin{table}[h]
\centering
 $masec.n$ \\
 \begin{tabular}{lrrrrrrrr}   \hline
    & Morris & Sobol' & VARS-TO & Reg & RegTree & RF & \multicolumn{2}{c}{GPR} \\
     & DGSM &  $T_i$ &      &  &    &   &  Slope &  InvRange \\ 
   \hline
 efcroijuv & 0.05 & 0.02 & 0.02 & 0.06 & 0.01 & 0.03 & 0.14 & 0.03 \\ 
   efcroiveg & \textbf{\tcr{0.19}} & \textbf{\tcr{0.29}} & \textbf{\tcr{0.31}} & \textbf{\tcr{0.25}} & \textbf{\tcr{0.32}} & \textbf{\tcr{0.28}} & 0.13 &  \textbf{\tcr{0.16}} \\ 
   croirac & 0.12 & 0.09 & 0.08 & 0.09 & 0.07 & 0.07 & 0.15 & 0.11 \\ 
   stlevdrp & 0.09 & 0.06 & 0.06 & 0.06 & 0.07 & 0.07 &   \textbf{\tcr{0.17}} & 0.15 \\ 
   adil & 0.07 & 0.03 & 0.03 & 0.05 & 0.01 & 0.03 & 0.03 & 0.00 \\ 
   bdil & 0.07 & 0.04 & 0.01 & 0.07 & 0.02 & 0.04 &  0.04 & 0.09 \\ 
   vitircarb & 0.01 & 0.00 & 0.00 & 0.00 & 0.02 & 0.01 & 0.00&  0.00 \\ 
   vitirazo & 0.00 & 0.00 & 0.00 & 0.00 & 0.01 & 0.01 &  0.01 & 0.00 \\ 
   adens & \textbf{\tcb{0.25}} & \textbf{\tcb{0.36}} & \textbf{\tcb{0.38}} & \textbf{\tcb{0.28}} & \textbf{\tcb{0.39}} & \textbf{\tcb{0.35}} & \textbf{\tcb{0.19}} & \textbf{\tcb{0.24}}  \\ 
   kmax & 0.10 & 0.09 & 0.10 & 0.09 & 0.07 & 0.07 & 0.04 & 0.08   \\ 
   INNmin & 0.05 & 0.03 & 0.01 & 0.03 & 0.01 & 0.02 & 0.10 & 0.06 \\ 
   inngrain2 & 0.01 & 0.00 & 0.00 & 0.00 & 0.00 & 0.01 & 0.00&  0.00 \\ 
    \hline
 \end{tabular}
 \caption{STICS: summary of results for different SA methods applied to $masec.n$ on the day of harvest. See text in Section \ref{subsec:Results.GR6J} for explanations of the values shown. The largest measure is indicated by bold blue typeface, with the second largest in red.  Reg=regression, RegTree=regression tree, RF=random forest, GPR=Gaussian Process regression.}
\label{T:STICS.SA.summary.masecn}
\end{table}

\begin{figure}[h]
 \centering
     \includegraphics[width=0.9\textwidth]{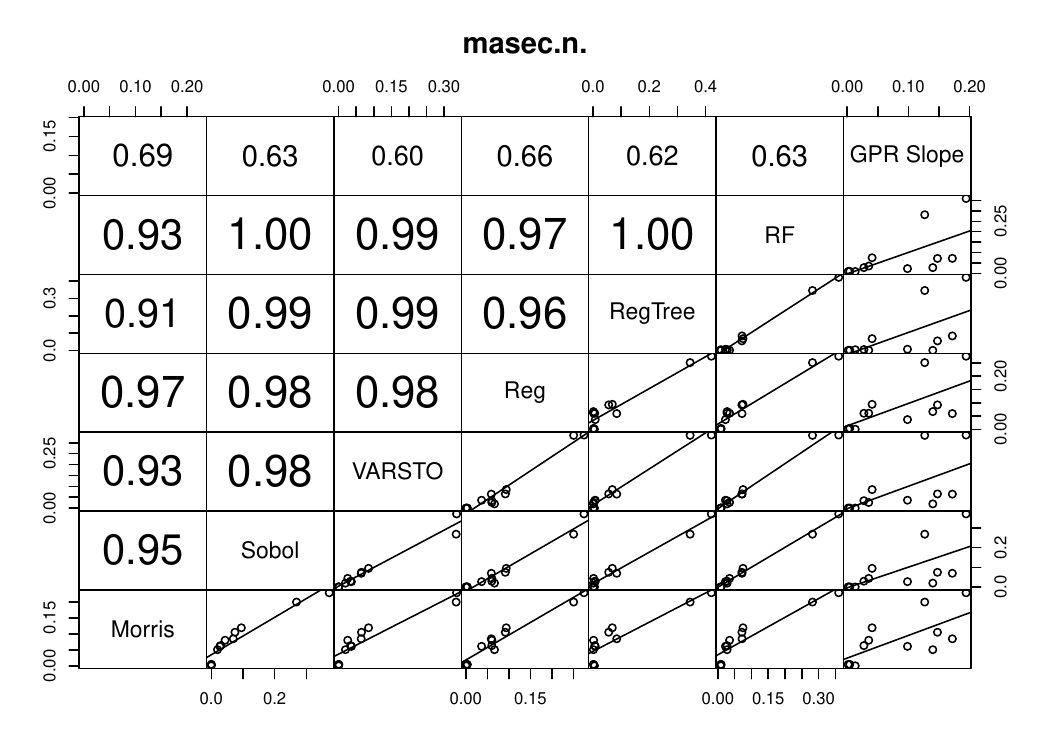}
     \caption{STICS: pairwise scatterplots of SA measures for the   parameters influence on $masec.n$ for different SA procedures along with Spearman correlation coefficients.  Reg=regression, RegTree=regression tree, RF=random forest, GPR=Gaussian Process regression.}
     \label{F:STICS.Pairwise.masec.SA}
 \end{figure}

The pairwise similarities between the methods are shown in Figure \ref{F:STICS.Pairwise.masec.SA} along with Pearson correlation coefficients, which ranged from 0.75 to 1.00.  The scatterplots include a linear regression line that fits the data fairly well. Kendall's W is 0.87, with a p-value for the null hypothesis of no concordance $<$0.001, indicating a relatively high degree of concordance across all the measures.
 \clearpage 

\begin{figure}[h]
    \centering
    \includegraphics[width=0.6\linewidth]{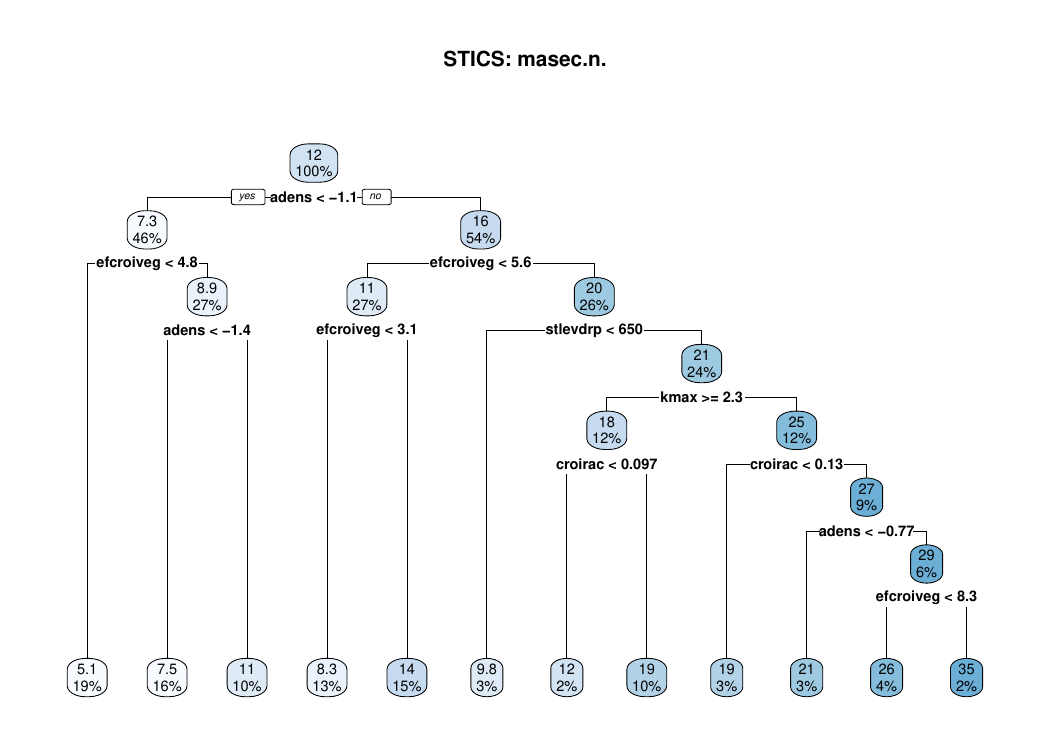}
    \caption{STICS: Regression tree for masec.n}
    \label{F:STICS.masec.n.RegTree.tree}
\end{figure}

\begin{figure}[h]
    \centering
    \includegraphics[width=0.6\linewidth]{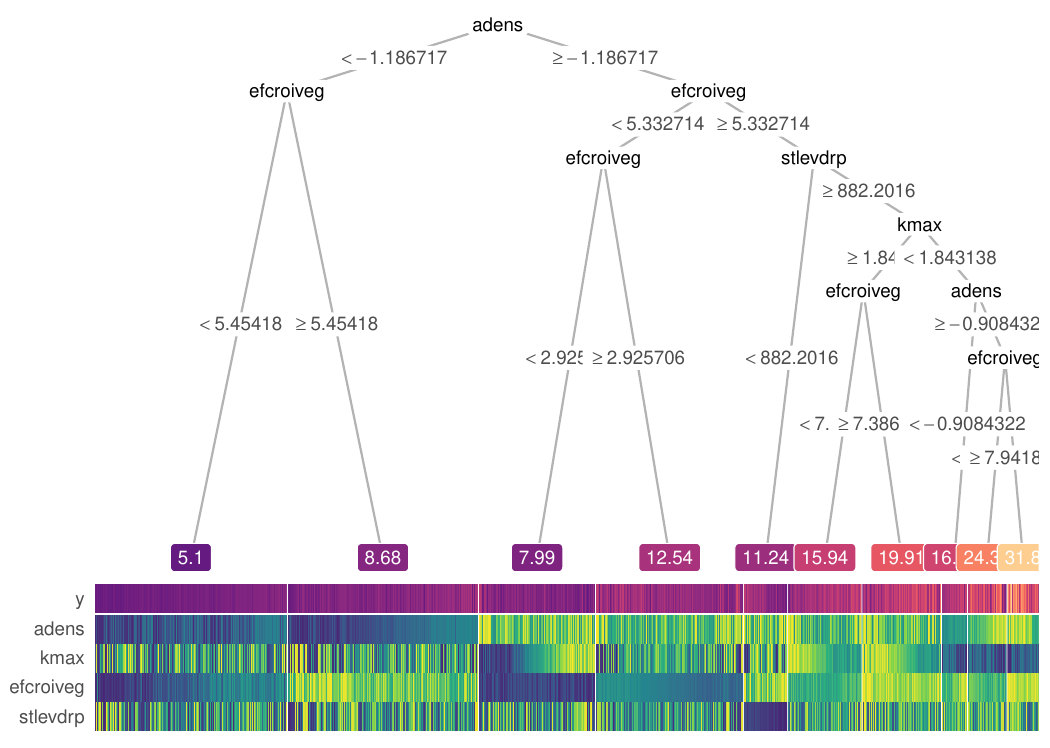}
    \caption{STICS: Regression tree for masec.n based on top 4 inputs with heatmap.}
    \label{F:STICS.masec.n.RegTree.tree.heatmap}
\end{figure}

Further details on the regression tree for masec.n are shown in Figure \ref{F:STICS.masec.n.RegTree.tree}, which shows that adens and efcroiveg dominate as all branching decisions are based largely on the values for those two input parameters.   Figure \ref{F:STICS.masec.n.RegTree.tree.heatmap} shows the tree based only on the top four inputs with possible interaction between efcroiveg and adens indicated. 

 %------------------------------------------------
 \newpage
 \clearpage 
% *****************CNgrain********************** %
 \subsubsection{$CNgrain$}
\begin{figure}[h]
    \centering
    \includegraphics[width=0.60\linewidth]{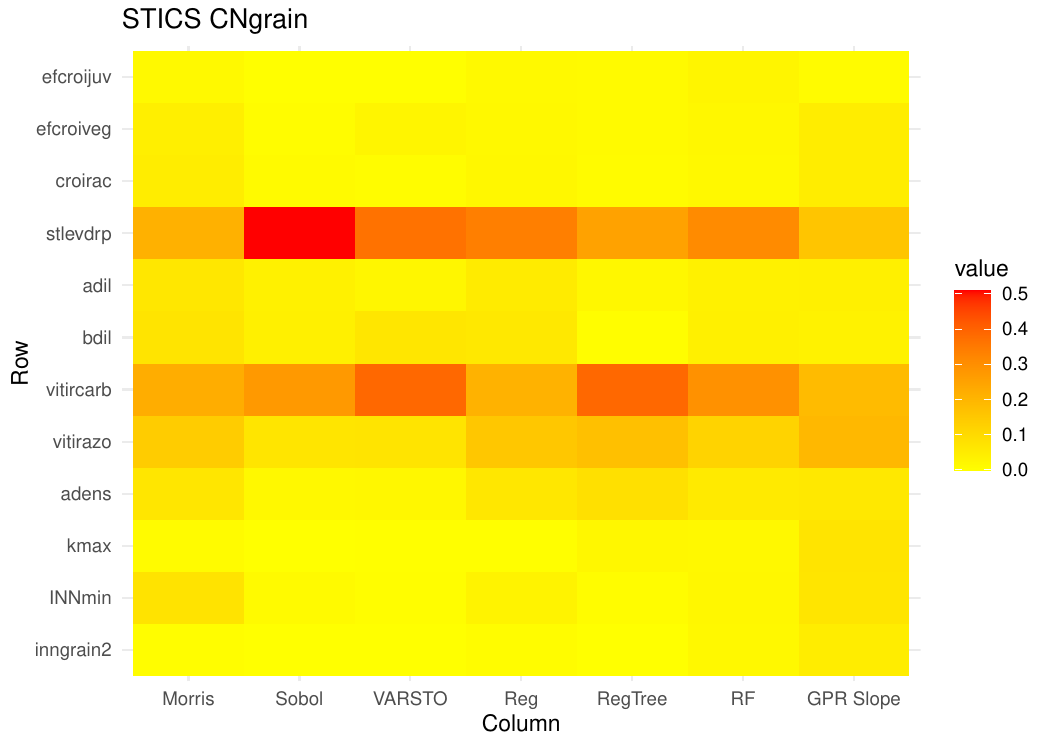}
    \caption{STICS: Relative parameter importance for CNgrain for the different SA methods.}
    \label{F:STICS.SA.summary.CNgrain}
\end{figure}

\begin{figure}[h]
    \centering
    \begin{subfigure}[b]{0.45\textwidth}
    \includegraphics[width=\columnwidth,height=0.30\textheight]{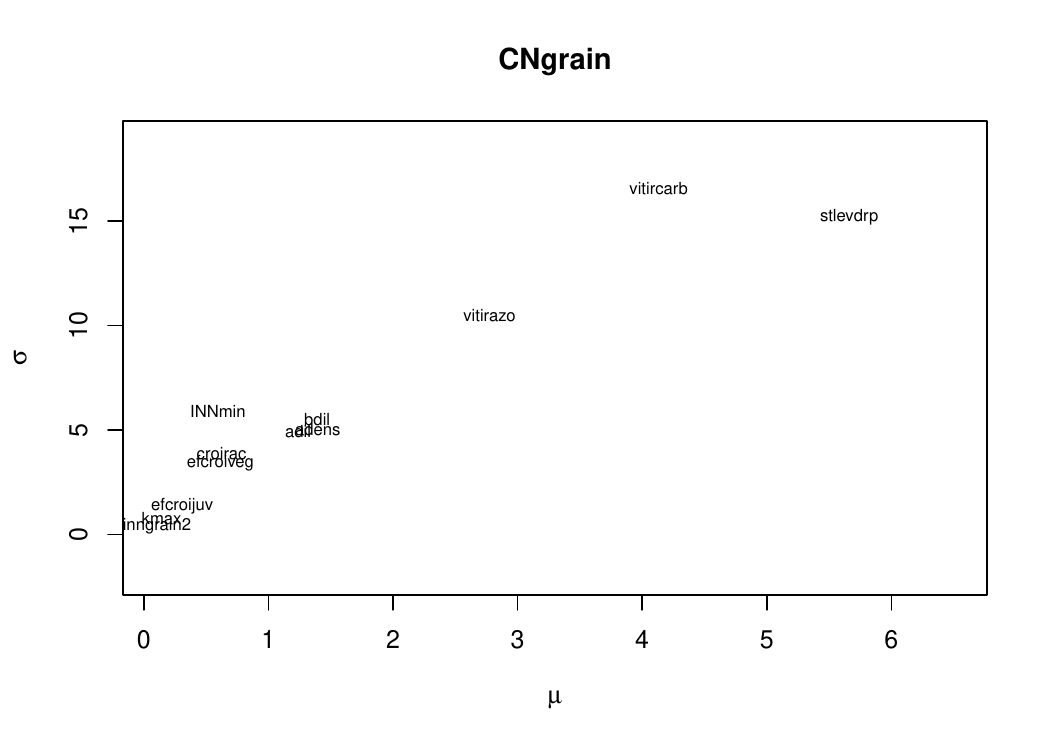}
    \caption{Morris $\sigma$ versus $\mu^*$.}
    \label{F:STICS.Morris.CNgrain.Internal3}
    \end{subfigure}
    \hfill
    \begin{subfigure}[b]{0.45\textwidth}
    \centering 
        \includegraphics[width=\columnwidth,height=0.30\textheight]{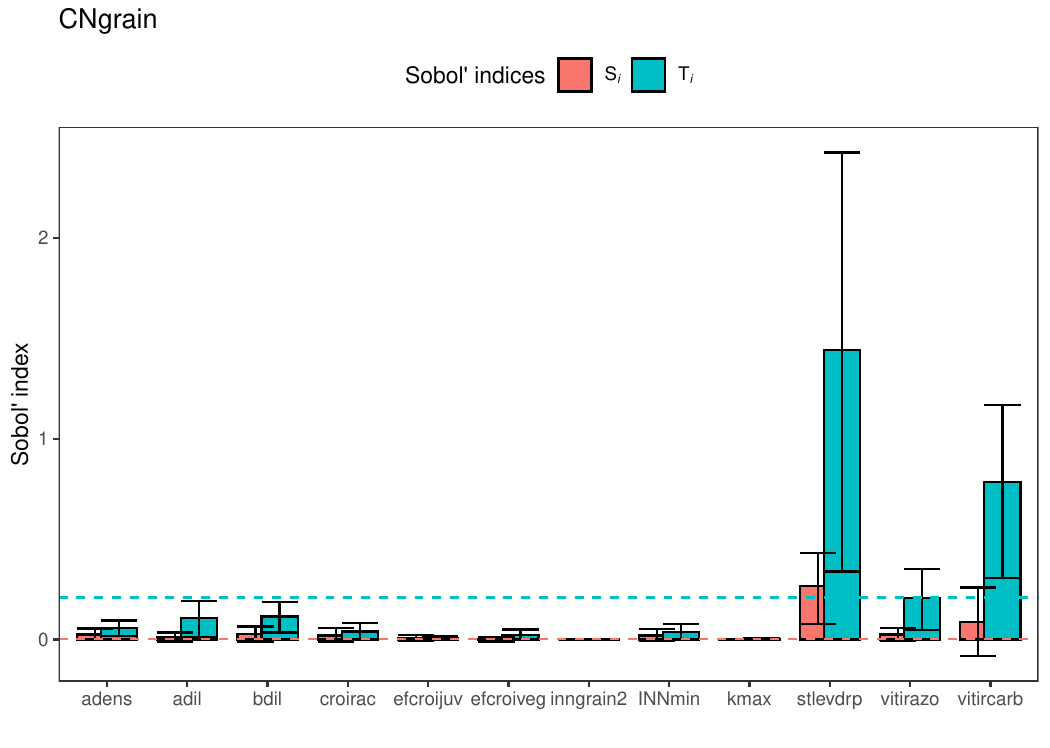}
     \caption{Sobol $S_{1,k}$ and $T_k$.}
     \label{F:STICS.Sobol.CNgrain.Internal3}
    \end{subfigure}
% -----------------------------------------------

   \begin{subfigure}[b]{0.45\textwidth}
    \centering
    \includegraphics[width=\columnwidth,height=0.30\textheight]{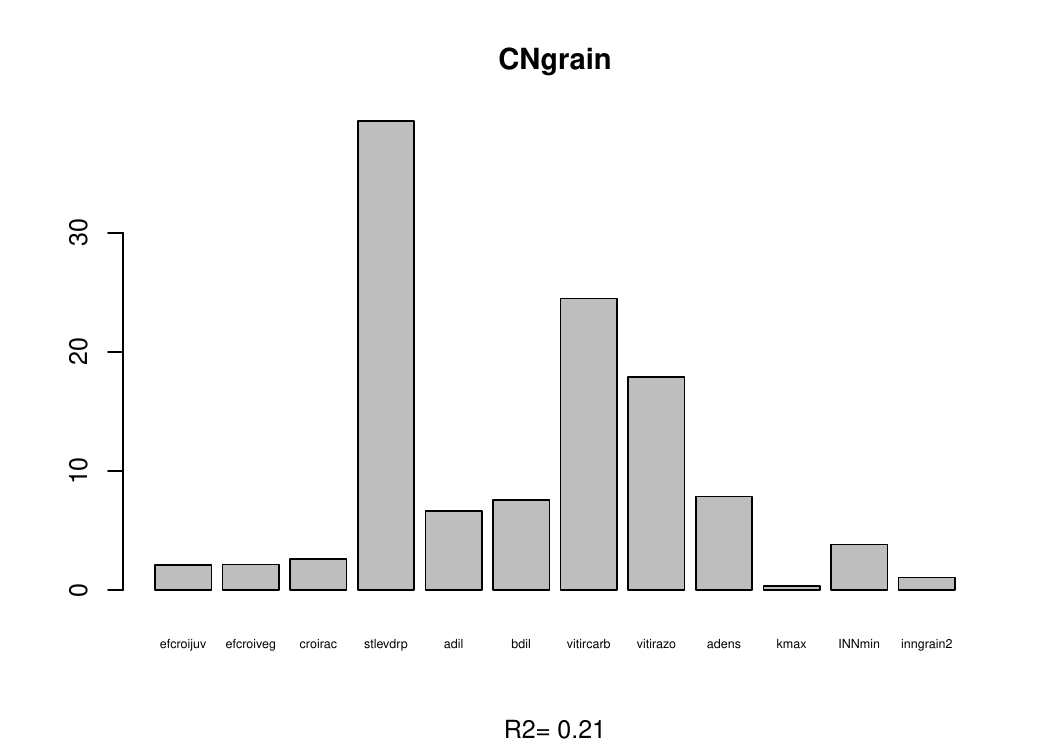}
    \caption{Multiple Regression standardized regression coefficients.}
    \label{F:STICS.Regression.CNgrain.Internal3}
 \end{subfigure}   
\hfill
    \begin{subfigure}[b]{0.45\textwidth}
    \centering
  \includegraphics[width=\columnwidth,height=0.30\textheight]{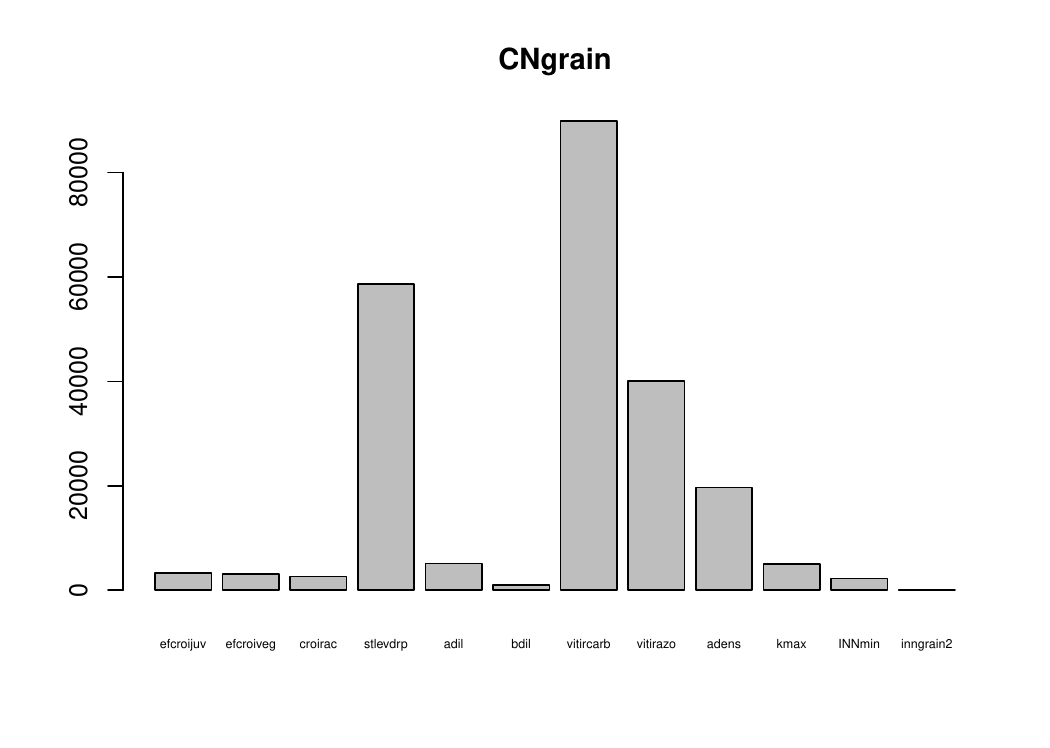} 
     \caption{Regression Tree parameter importance.}
     \label{F:STICS.RegTree.CNgrain.Internal3}
    \end{subfigure}   
% --------------------------------------------------------------

   \begin{subfigure}[b]{0.45\textwidth}
    \centering
    \includegraphics[width=\columnwidth,height=0.30\textheight]{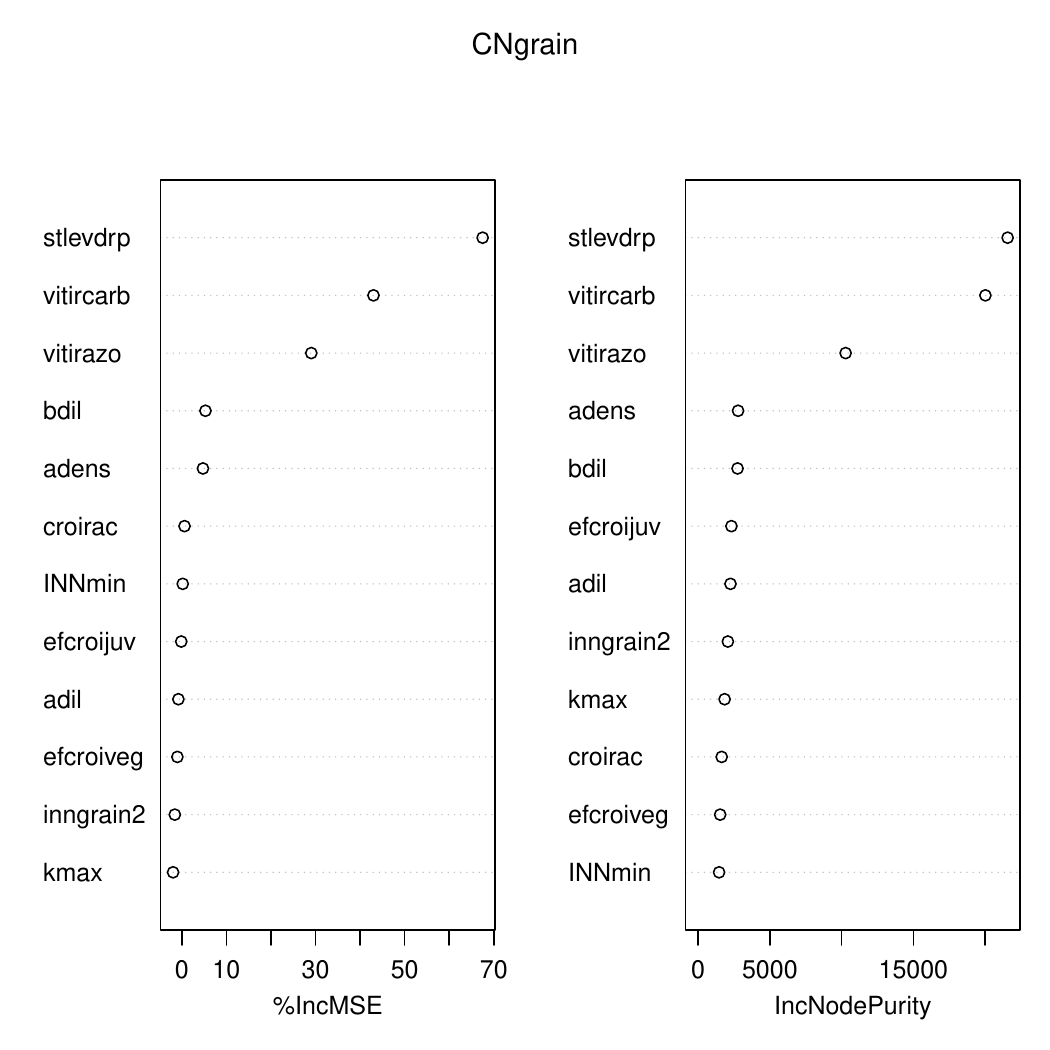}
    \caption{Random Forests results.}
    \label{F:STICS.RF.CNgrain.Internal3}
 \end{subfigure}   
 \hfill
   \begin{subfigure}[b]{0.45\textwidth}
    \centering
    \includegraphics[width=\columnwidth,height=0.30\textheight]{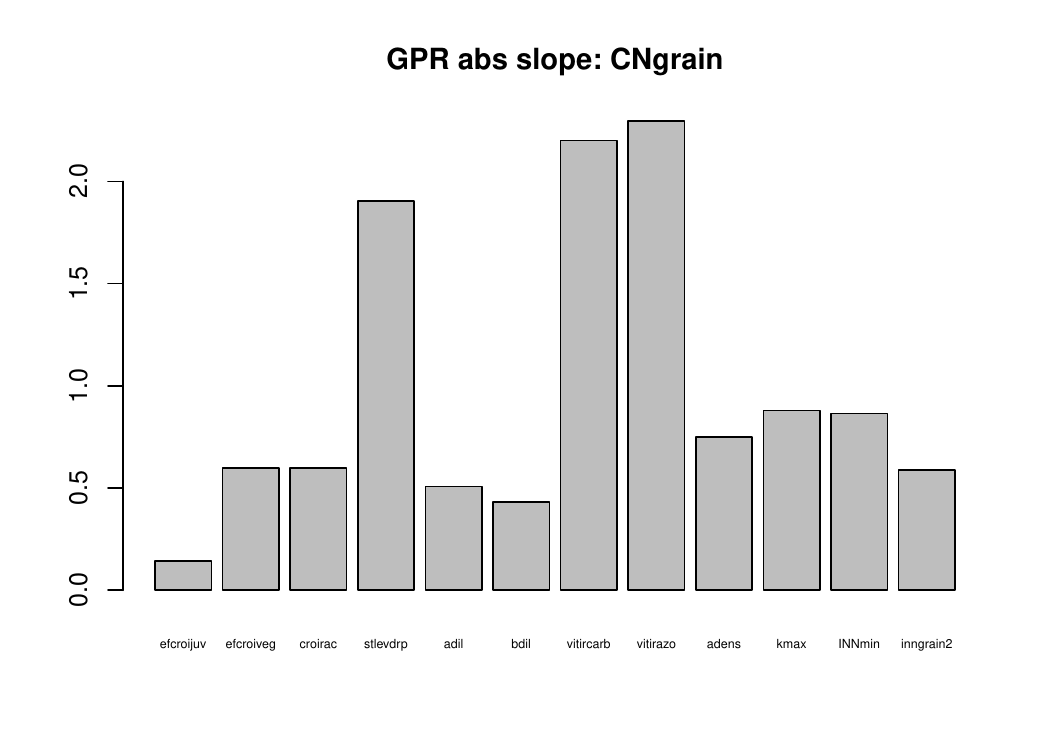}
    \caption{GPR standardized regression coefficients.}
    \label{F:STICS.GPR.Both.CNgrain.Internal3}
 \end{subfigure}   
 \caption{STICS: Sensitivity Analyses of
 $CNgrain$ on day of harvest.}
 \label{F:STICS.CNgrain.Internal}
  \end{figure}

The parameter most influencing $CNgrain$ based on all SA methods but the regression tree and GPR slope was stlevdrp (Table \ref{T:STICS.SA.summary.CNgrain}, Figures \ref{F:STICS.SA.summary.CNgrain} and \ref{F:STICS.CNgrain.Internal}). The second and third most influential parameters alternated between vitircarb and virtirazo.  The \texttt{R} regression tree  package \texttt{rpart} yields a ``missing value'' (NA) for the parameter INNmin, but whether that affected the other results is not known.

 \begin{table}[h]
\centering
 $CNgrain$ \\
 \begin{tabular}{lrrrrrrrr}   \hline
  & Morris & Sobol' & VARS-TO & Reg & RegTree & RF & \multicolumn{2}{c}{GPR} \\
     & DGSM &  $T_i$ &      &  &    &   &  Slope &  InvRange \\ 
 \hline
 efcroijuv & 0.02 & 0.00 & 0.01 & 0.01 & 0.04 & 0.03 & 0.01 &0.02\\ 
   efcroiveg & 0.03 & 0.00 & 0.00 & 0.03 & 0.03 & 0.03 & 0.05 & 0.01 \\ 
   croirac & 0.02 & 0.01 & 0.00 & 0.04 & 0.02 & 0.03 & 0.05 & 0.04 \\ 
   stlevdrp & \textbf{\tcb{0.23}} & \textbf{\tcb{0.45}} & \textbf{\tcb{0.55}} & \textbf{\tcb{0.36}} & \textbf{\tcr{0.26}} & \textbf{\tcb{0.31}} &  0.16  & \textbf{\tcb{0.32}} \\ 
   adil & 0.06 & 0.01 & 0.02 & 0.04 & 0.00 & 0.03 & 0.04 & 0.02 \\ 
   bdil & 0.09 & 0.06 & 0.05 & 0.06 & 0.01 & 0.04 & 0.04 & 0.01 \\ 
   vitircarb & 0.22 & \textbf{\tcr{0.37}} & \textbf{\tcr{0.30}} & \textbf{\tcr{0.22}} & \textbf{\tcb{0.34}} & \textbf{\tcr{0.27}} & \textbf{\tcr{0.19}}  & \textbf{\tcb{0.32}}\\ 
   vitirazo & \textbf{\tcb{0.23}} & 0.05 & 0.06 & 0.17 & 0.25 & 0.15 & \textbf{\tcb{0.20}} & 0.06   \\ 
   adens & 0.05 & 0.03 & 0.01 & 0.07 & 0.02 & 0.04 & 0.06 & 0.04   \\ 
   kmax & 0.03 & 0.00 & 0.00 & 0.00 & 0.01 & 0.02 & 0.07 & 0.06   \\ 
   INNmin & 0.01 & 0.01 & 0.00 & 0.01 & 0.00 & 0.02 & 0.07 & 0.04  \\ 
   inngrain2 & 0.01 & 0.00 & 0.00 & 0.00 & 0.02 & 0.03 & 0.05 & 0.04 \\ 
    \hline
 \end{tabular}
 \caption{STICS: summary of results for different SA methods applied to $CNgrain$ on the day of harvest. See text in Section \ref{subsec:Results.GR6J} for explanations of the values shown. The largest measure is indicated by bold blue typeface, with the second largest in red.  Reg=regression, RegTree=regression tree, RF=random forest, GPR=Gaussian Process regression.}
\label{T:STICS.SA.summary.CNgrain}
 \end{table}

\begin{figure}[h]
 \centering
     \includegraphics[width=0.9\textwidth]{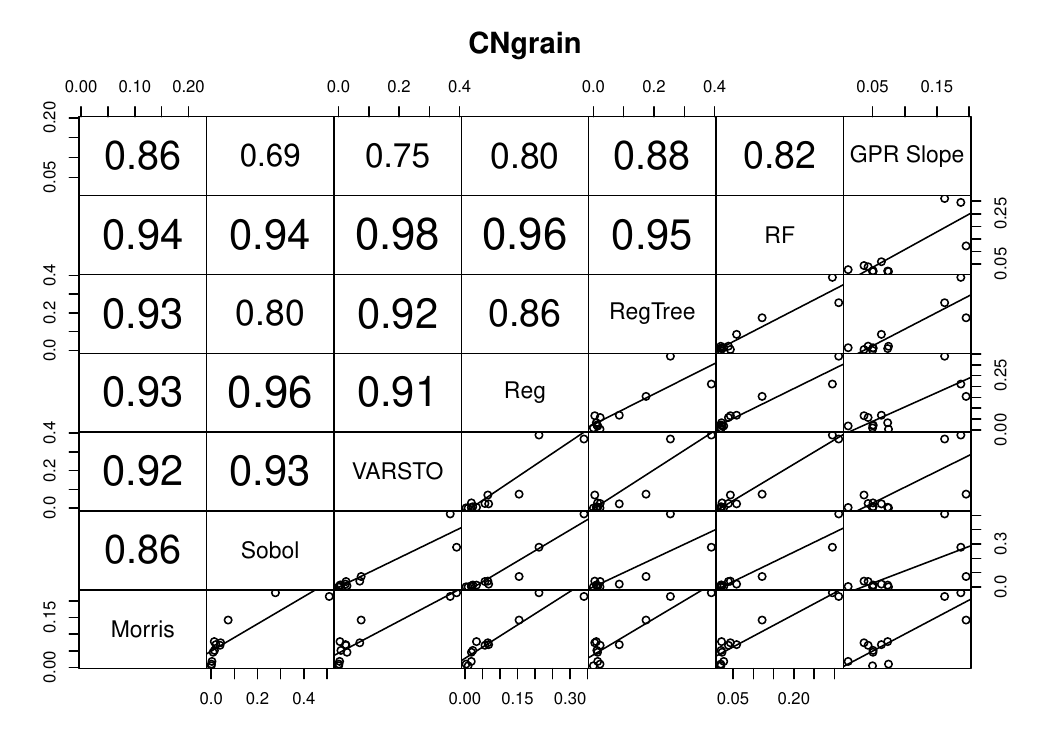}
     \caption{STICS: pairwise scatterplots of SA measures for the  parameters influence on $masec.n$ for different SA procedures along with Spearman correlation coefficients. Reg=regression, RegTree=regression tree, RF=random forest, GPR=Gaussian Process regression.}
     \label{F:STICS.Pairwise.CNgrain.SA}
 \end{figure}

On a pairwise basis, the methods were generally highly correlated    (Figure \ref{F:STICS.Pairwise.CNgrain.SA}). The overall concordance based on Kendall's W was 0.77  with p-value $<$0.001. 

\clearpage 

\begin{figure}[h]
    \centering
    \includegraphics[width=0.6\linewidth]{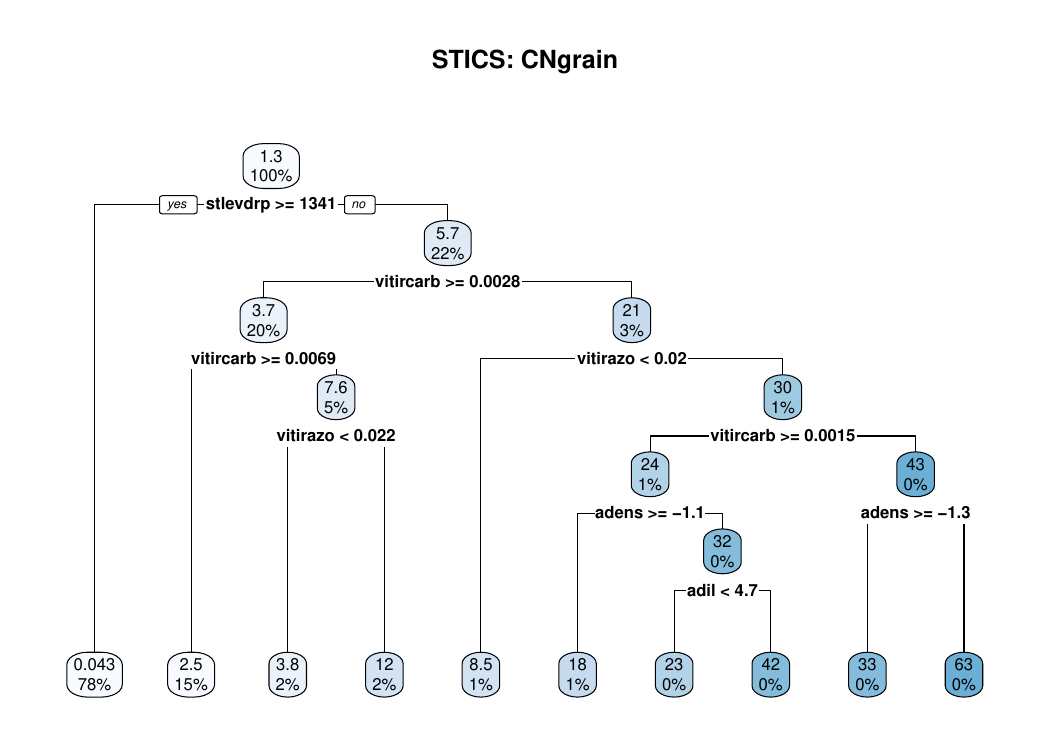}
    \caption{STICS: Regression tree for CNgrain}
    \label{F:STICS.CNgrain.RegTree.tree}
\end{figure}

\begin{figure}[h]
    \centering
    \includegraphics[width=0.6\linewidth]{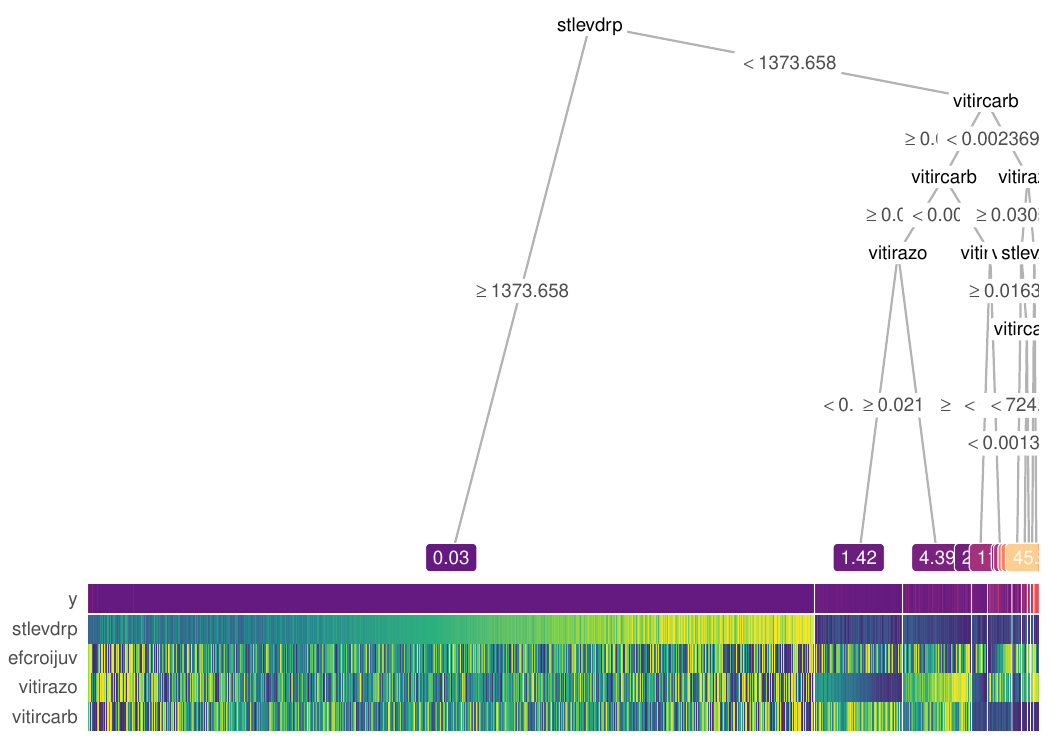}
    \caption{STICS: Regression tree for CNgrain based on top 4 inputs with heatmap.}
    \label{F:STICS.CNgrain.RegTree.tree.heatmap}
\end{figure}

Further details on the regression tree for CNgrain are shown in Figure \ref{F:STICS.CNgrain.RegTree.tree}, which shows that stlevdrp dominates with 78\% of the combinations in the larger values of stlevdrp ($\ge$ 1341).    Figure \ref{F:STICS.CNgrain.RegTree.tree.heatmap} shows the tree based only on the top four input parameters, again showing dominance of stlevdrp and potential interactions with vitirazo and vitircarb. 

 %------------CNplante--------------------
 \newpage
 \clearpage 
 \subsubsection{$CNplante$} 
\begin{figure}[h]
    \centering
    \includegraphics[width=0.60\linewidth]{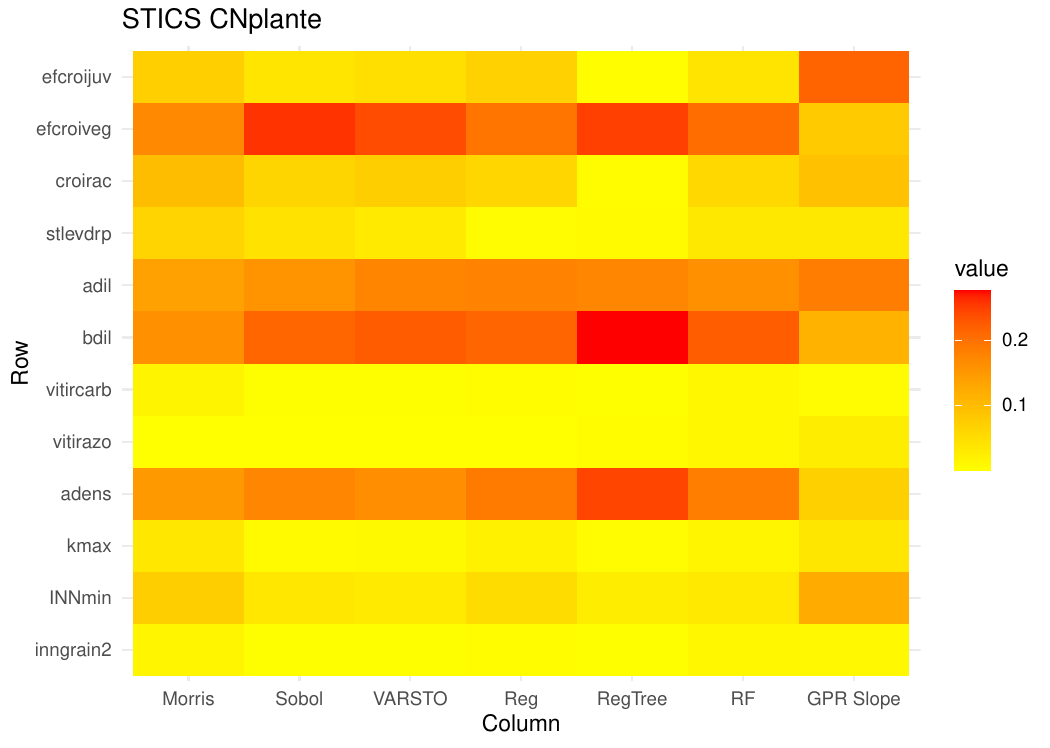}
    \caption{STICS: Relative parameter importance for CNplante for the different SA methods.}
    \label{F:STICS.SA.summary.CNplante}
\end{figure}

\begin{figure}[h]
    \centering
    \begin{subfigure}[b]{0.45\textwidth}
    \includegraphics[width=\columnwidth,height=0.30\textheight]{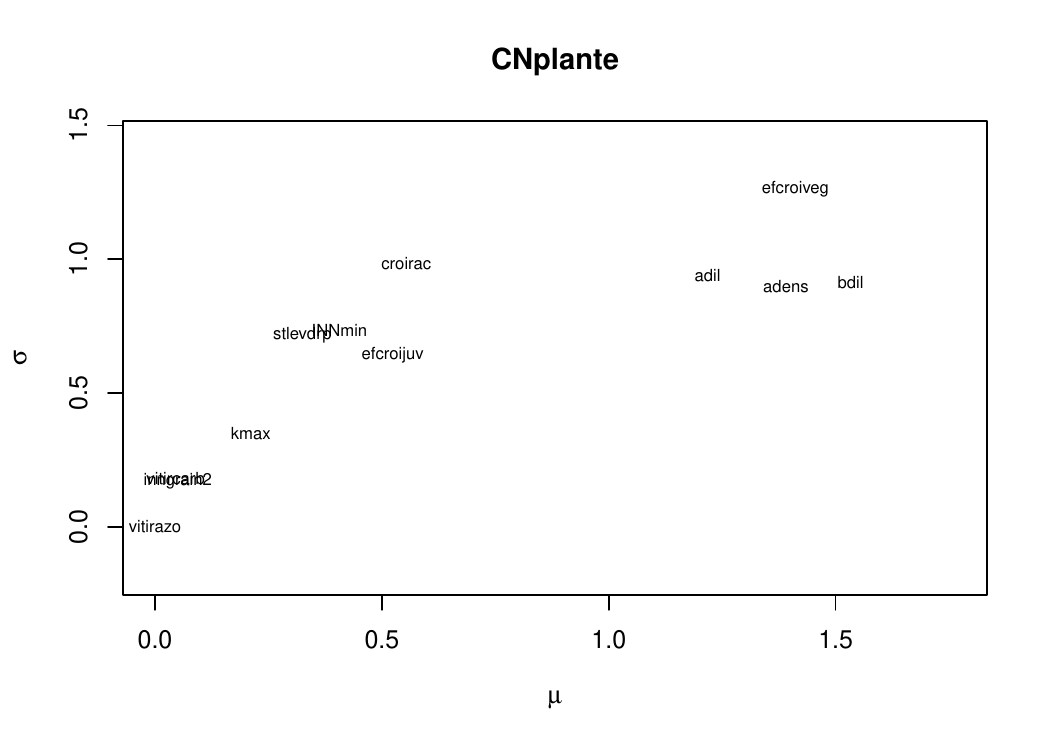}
    \caption{Morris measures, $\sigma$ versus $\mu^*$.}
    \label{F:STICS.Morris.Internal4}
    \end{subfigure}
    \hfill
    \begin{subfigure}[b]{0.45\textwidth}
    \centering 
        \includegraphics[width=\columnwidth,height=0.30\textheight]{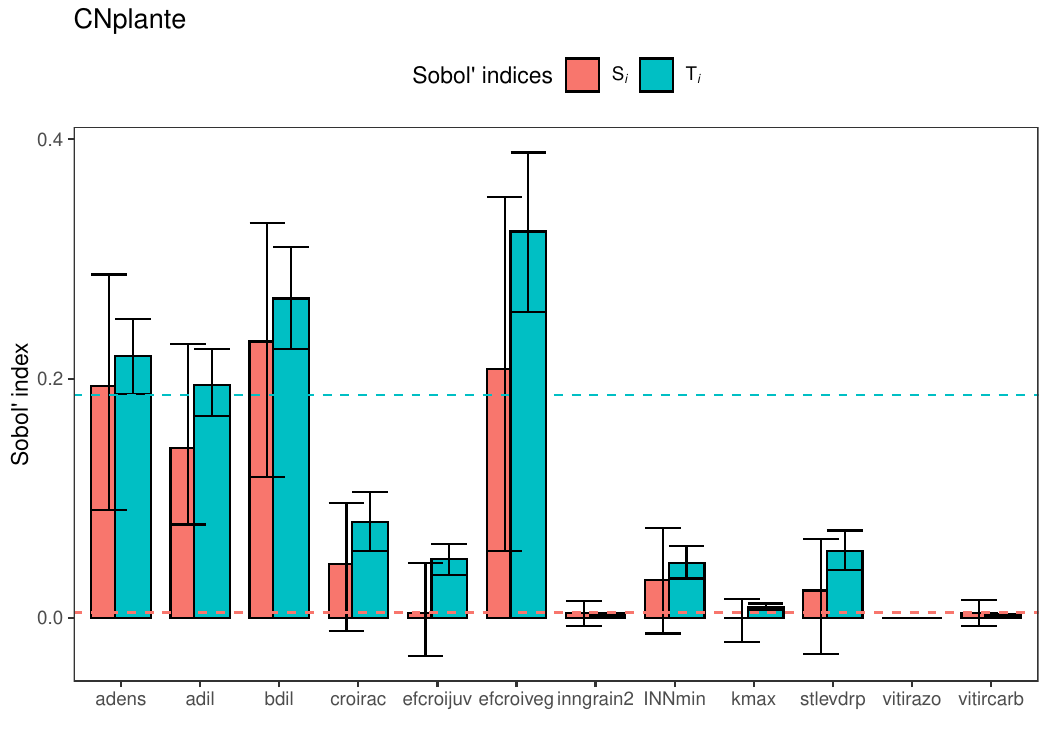}
     \caption{Sobol $S_{1,k}$ and $T_k$.}
     \label{F:STICS.Sobol.CNplante.Internal4}
    \end{subfigure}
% ---------------------------------------------

   \begin{subfigure}[b]{0.45\textwidth}
    \centering
    \includegraphics[width=\columnwidth,height=0.30\textheight]{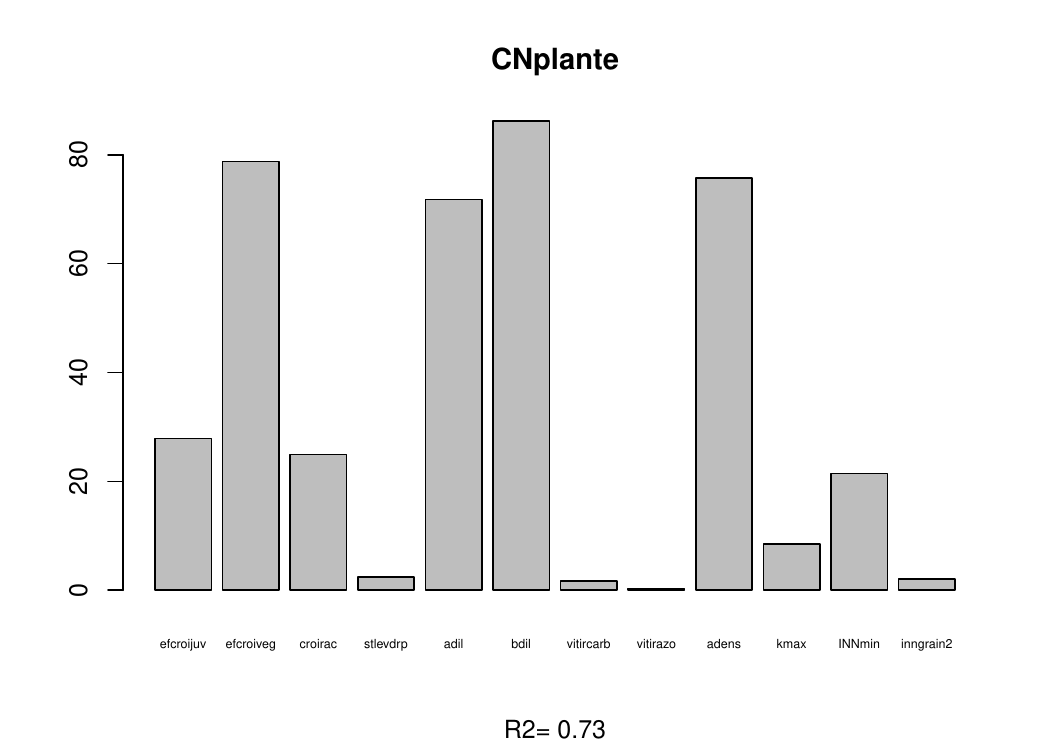}
    \caption{Multiple Regression standardized regression coefficients.}
    \label{F:STICS.Regression.CNplante.Internal4}
 \end{subfigure}   
\hfill
    \begin{subfigure}[b]{0.45\textwidth}
    \centering
    \includegraphics[width=\columnwidth,height=0.30\textheight]{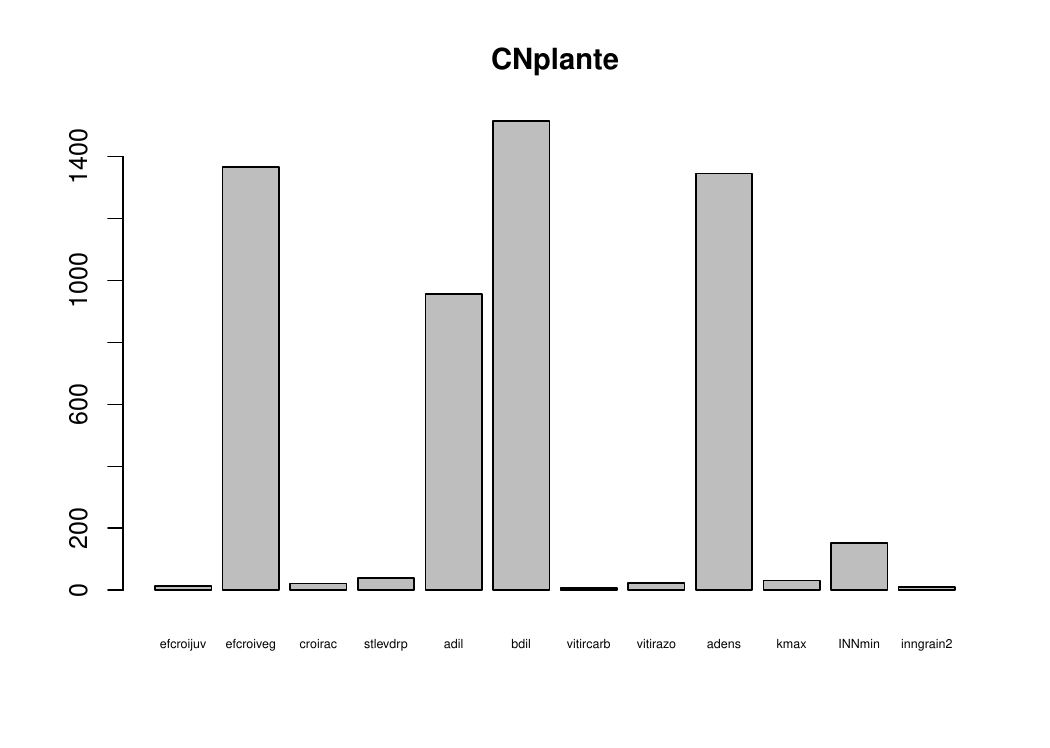} 
     \caption{Regression Tree parameter importance.}
     \label{F:STICS.RegTree.CNplante.Internal4}
    \end{subfigure}   
% ------------------------------------------------

   \begin{subfigure}[b]{0.45\textwidth}
    \centering
    \includegraphics[width=\columnwidth,height=0.30\textheight]{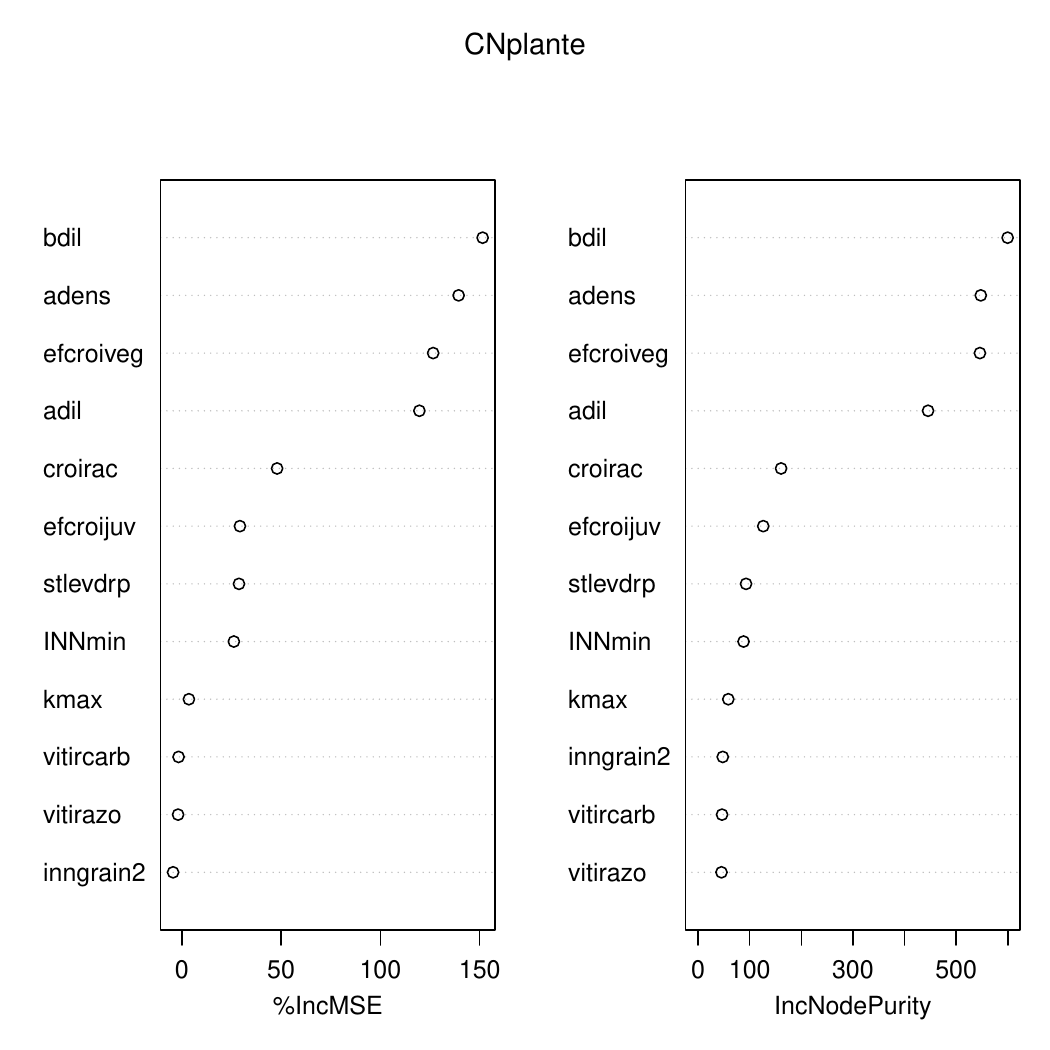}
    \caption{Random Forests results.}
    \label{F:STICS.RF.CNplante.Internal4}
 \end{subfigure}   
 \hfill
   \begin{subfigure}[b]{0.45\textwidth}
    \centering
    \includegraphics[width=\columnwidth,height=0.30\textheight]{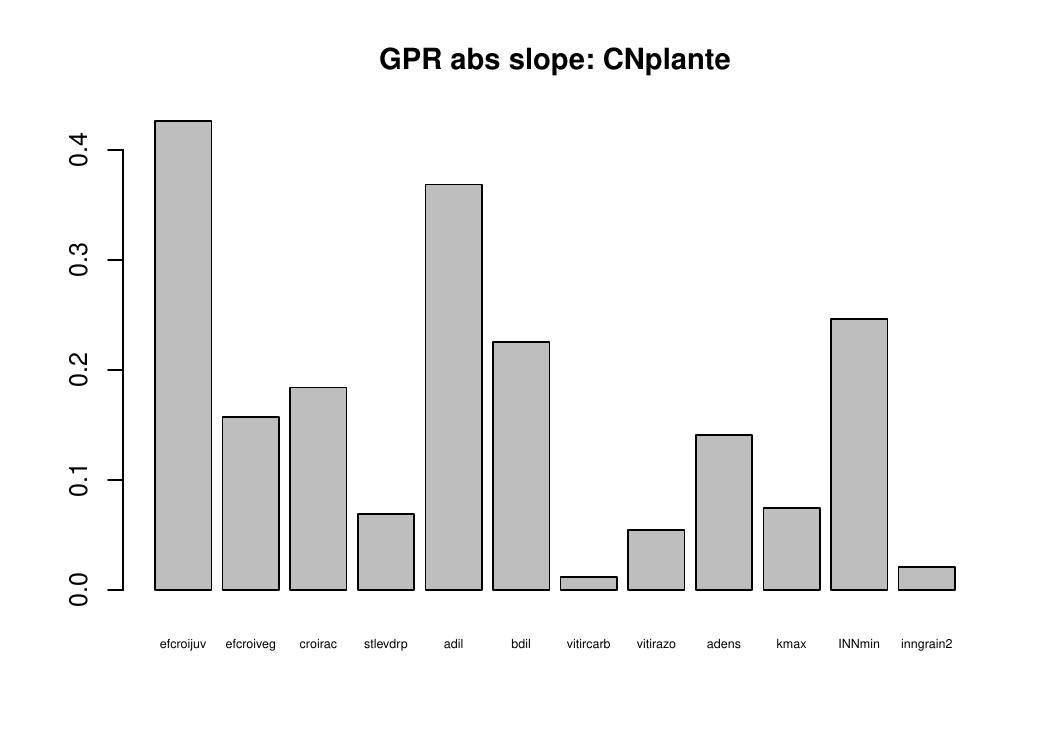}
    \caption{GPR standardized regression coefficients.}
    \label{F:STICS.GPR.Both.CNplante.Internal4}
 \end{subfigure}   
 \caption{STICS: Sensitivity Analyses of
 $CNplante$ on day of harvest.}
 \label{F:STICS.CNplante.Internal}
  \end{figure}

 There were four parameters that most influenced $CNgrain$, bdil, efcroiveg, adens, and adil with the ordering varying somewhat between measures (Table \ref{T:STICS.SA.summary.CNplante}, Figures \ref{F:STICS.SA.summary.CNplante} and \ref{F:STICS.CNplante.Internal}).

 \begin{table}[h]
\centering
 $CNplante$ \\
 \begin{tabular}{lrrrrrrrr}   \hline
  & Morris & Sobol' & VARS-TO & Reg & RegTree & RF & \multicolumn{2}{c}{GPR} \\
     & DGSM &  $T_i$ &      &  &    &   &  Slope &  InvRange \\  
 \hline
 efcroijuv & 0.08 & 0.03 & 0.04 & 0.07 & 0.02 & 0.05 & \textbf{\tcb{0.22}} & 0.12 \\ 
efcroiveg & \textbf{\tcr{0.17}} & \textbf{\tcr{0.26}} & \textbf{\tcb{0.25}} & 0.19 & 0.23 & 0.19 & 0.08 &\textbf{\tcr{0.18}} \\ 
   croirac & 0.10 & 0.04 & 0.05 & 0.06 & 0.01 & 0.06 &  0.09 & \textbf{\tcb{0.14}} \\ 
   stlevdrp & 0.07 & 0.03 & 0.02 & 0.01 & 0.01 & 0.03 & 0.03 & 0.11 \\ 
   adil & 0.15 & 0.17 & 0.18 & 0.18 & 0.17 & 0.16 & \textbf{\tcr{0.19}} & 0.10 \\ 
   bdil & \textbf{\tcb{0.16}} & \textbf{\tcb{0.23}} & \textbf{\tcr{0.24}} & \textbf{\tcb{0.21}} & \textbf{\tcb{0.25}} & \textbf{\tcb{0.21}} & 0.11 &  0.08 \\ 
   vitircarb & 0.02 & 0.00 & 0.00 & 0.00 & 0.01 & 0.02 & 0.01 &0.01\\ 
   vitirazo & 0.00 & 0.00 & 0.00 & 0.01 & 0.01 & 0.02 & 0.03 & 0.03 \\ 
   adens & 0.15 & 0.19 & 0.17 & \textbf{\tcr{0.19}} & \textbf{\tcr{0.25}} & \textbf{\tcr{0.20}} & 0.07 & 0.11 \\ 
   kmax & 0.03 & 0.01 & 0.02 & 0.02 & 0.01 & 0.02 & 0.04 & 0.04   \\ 
   INNmin & 0.06 & 0.03 & 0.02 & 0.05 & 0.03 & 0.03 &  0.12 & 0.07 \\ 
   inngrain2 & 0.01 & 0.00 & 0.00 & 0.00 & 0.00 & 0.02 & 0.01& 0.01 \\ 
    \hline
 \end{tabular}
\caption{STICS: summary of results for different SA methods applied to $CNplante$ on the day of harvest. See text in Section \ref{subsec:Results.GR6J} for explanations of the values shown. The largest measure is indicated by bold blue typeface, with the second largest in red. Reg=regression, RegTree=regression tree, RF=random forest, GPR=Gaussian Process regression.}
\label{T:STICS.SA.summary.CNplante}
 \end{table}

\begin{figure}[h]
 \centering
     \includegraphics[width=0.9\textwidth]{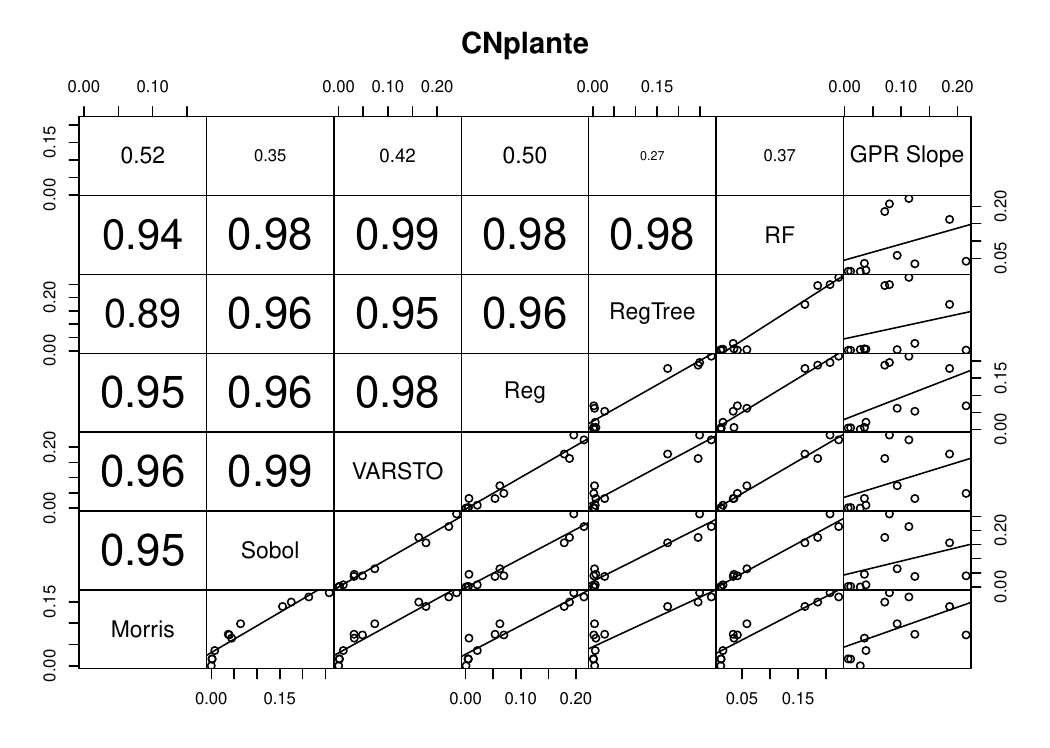}
     \caption{STICS: pairwise scatterplots of SA measures for the parameters influence on $CNplante$ for different SA procedures along with Spearman correlation coefficients. Reg=regression, RegTree=regression tree, RF=random forest, GPR=Gaussian Process regression.}
     \label{F:STICS.Pairwise.CNplante.SA}
 \end{figure}

 On a pairwise basis, the methods were highly correlated with the exception of the GPR slope (Figure \ref{F:STICS.Pairwise.CNplante.SA}). The overall concordance based on Kendall's W was 0.84 with p-value $<$0.001.   
 
 \clearpage 

\begin{figure}[h]
    \centering
    \includegraphics[width=0.6\linewidth]{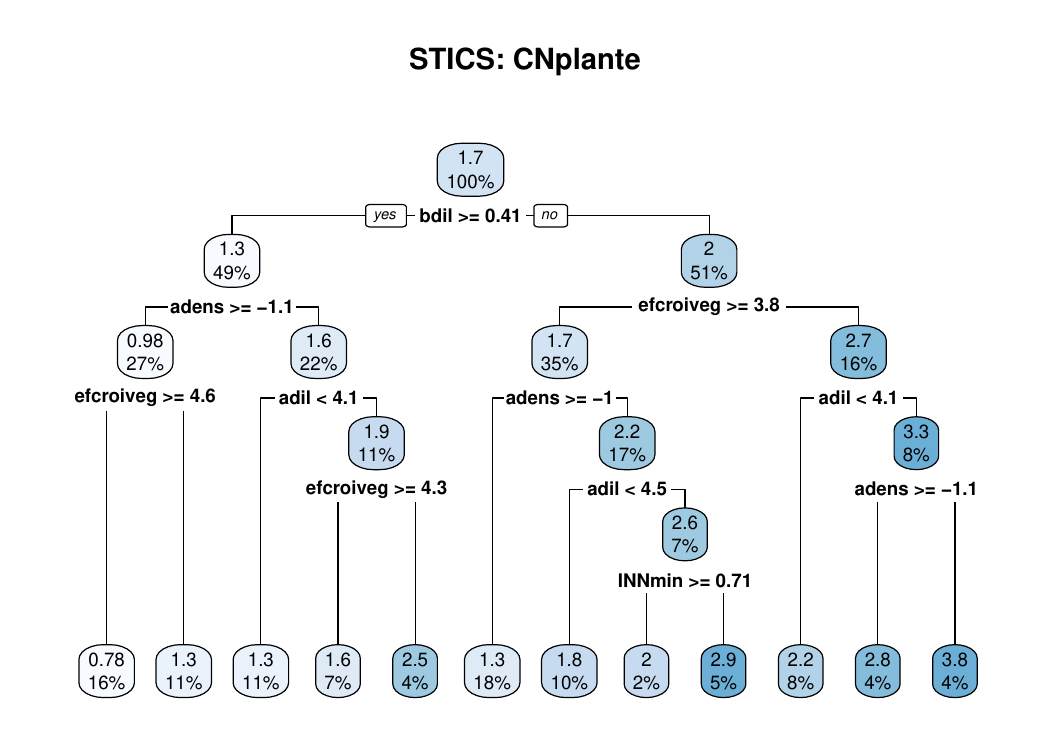}
    \caption{STICS: Regression tree for CNplante}
    \label{F:STICS.CNplante.RegTree.tree}
\end{figure}

\begin{figure}[h]
    \centering
    \includegraphics[width=0.6\linewidth]{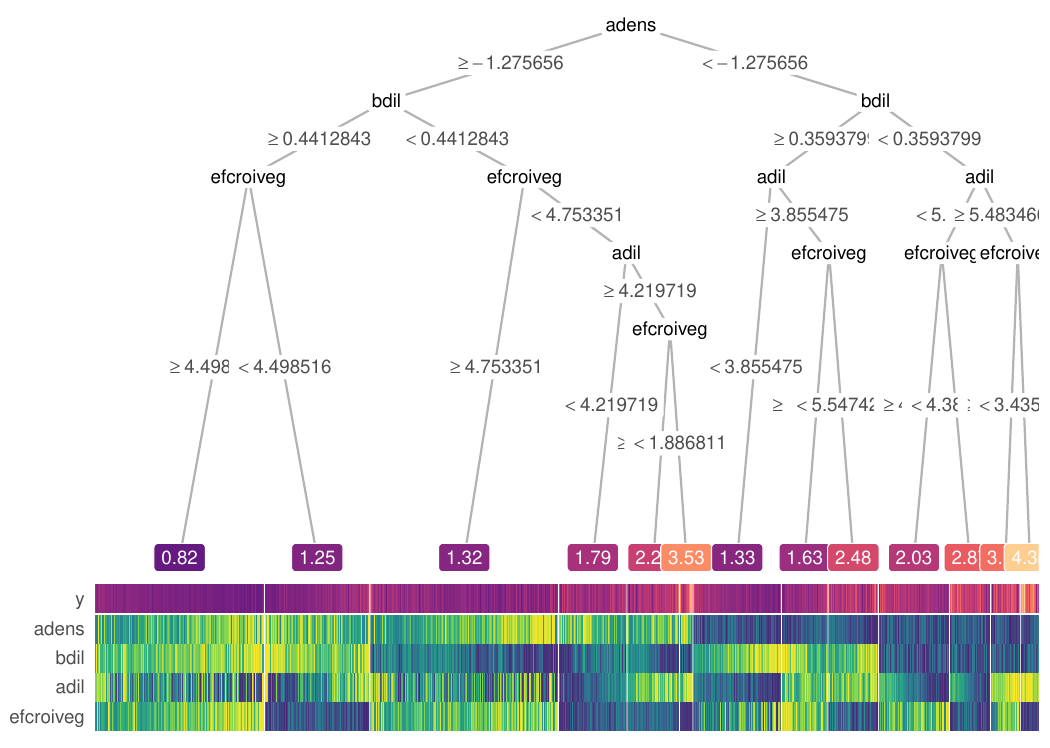}
    \caption{STICS: Regression tree for CNplante based on top 4 inputs with heatmap.}
    \label{F:STICS.CNplante.RegTree.tree.heatmap}
\end{figure}

Further details on the regression tree for CNplante are shown in Figure \ref{F:STICS.CNplante.RegTree.tree}, which shows that several parameters are influencing the branching decisions.   Figure \ref{F:STICS.CNplante.RegTree.tree.heatmap} shows the tree based only on the top four parameters, and that interactions amongst the four are apparent. 

\clearpage 
\subsubsection{STICS: Summary of parameter importance across multiple outputs}
\begin{figure}[h]
    \centering
    \includegraphics[width=0.60\linewidth]{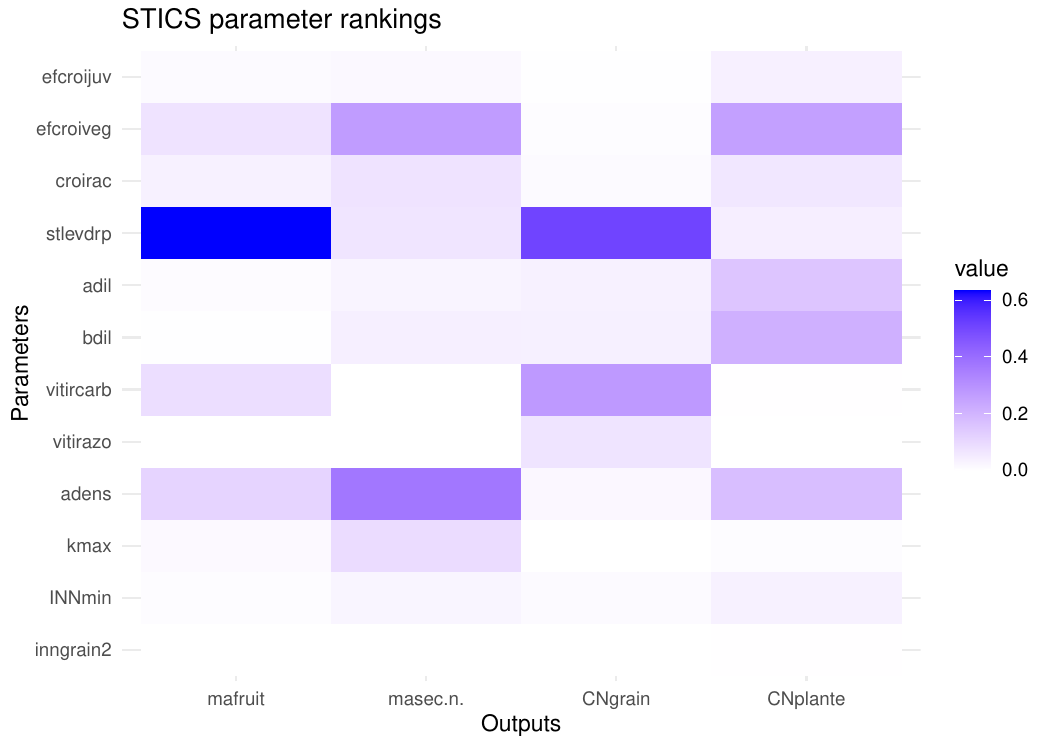}
    \caption{STICS: Relative parameter importance for four outputs based on Sobol' $T_i$.}
    \label{F:STICS.Importance.Summary}
\end{figure}

The relative importance of the STICS parameters for the four outputs, based on Sobol' $T_i$, is summarized in both Table \ref{T:STICS.Importance.Summary} and Figure \ref{F:STICS.Importance.Summary}.
\begin{table}[h]
\centering
\begin{tabular}{lrrrrr}
Parameter &  mafruit & mascec.n & CNgrain & CNplante \\ \hline
efcroijuv & 0.01 & 0.02 & 0.00 & 0.04 \\ 
efcroiveg & 0.07 & \textbf{\tcr{0.27}} & 0.01 & \textbf{\tcr{0.26}}   \\ 
croirac   & 0.04 & 0.07 & 0.01 & 0.06 \\ 
stlevdrp  &  \textbf{\tcb{0.63}} & 0.07 & \textbf{\tcb{0.51}} & 0.04  \\ 
adil      & 0.01 & 0.03 & 0.04 & 0.16 \\ 
bdil      & 0.00 & 0.04 & 0.04& \textbf{\tcb{0.21}} \\ 
vitircarb & \textbf{\tcr{0.09}} & 0.00 & \textbf{\tcr{0.28}} & 0.00 \\
vitirazo  &  0.00 & 0.00 & 0.07 & 0.00   \\ 
adens     & 0.11 &  \textbf{\tcb{0.37}}  & 0.02 & 0.18   \\ 
kmax      & 0.02 & 0.09 & 0.00 & 0.01   \\ 
INNmin    & 0.01 & 0.03 & 0.01 & 0.04  \\ 
inngrain2 & 0.00 & 0.00 & 0.00 & 0.00  \\ 
\hline
\end{tabular}
\caption{Relative parameter importance for four STICS outputs based on Sobol' $T_i$.}
\label{T:STICS.Importance.Summary}
\end{table}  
